\documentclass[a4paper,11pt]{article}
\pdfoutput=1 
\date{}
\usepackage{jheppub}
\usepackage{amsmath,amssymb,amsfonts,amsxtra, mathrsfs,graphics,graphicx,amsthm,epsfig, youngtab,bm,longtable,float,tikz,empheq,nccmath}
\usepackage[margin=0.5cm]{caption}

\newcommand*\widefbox[1]{\fbox{\hspace{2em}#1\hspace{2em}}}
\newcommand*\vwidefbox[1]{\fbox{\hspace{4em}#1\hspace{4em}}}
\allowdisplaybreaks[1]
\usetikzlibrary{positioning}
\newcommand{\be}{\begin{equation}}
\newcommand{\ee}{\end{equation}}
\newcommand{\ba}{\begin{array}}
\newcommand{\ea}{\end{array}} 
\newcommand{\bi}{\begin{itemize}}
\newcommand{\ei}{\end{itemize}}
\def\vec#1{\bm{#1}}
\def\bea#1\eea{\allowdisplaybreaMs \begin{align}#1\end{align}}
 \newcommand{\ben}{\begin{enumerate}}
\newcommand{\een}{\end{enumerate}}
\newcommand{\bean}{\begin{eqnarray*}}
\newcommand{\eean}{\end{eqnarray*}}
\newcommand{\eref}[1]{(\ref{#1})}

\newcommand{\nn}{\nonumber}

\newcommand{\tr}{\mathrm{Tr}}

\newcommand{\tq}{\widetilde{q}}
\newcommand{\tE}{\widetilde{E}}
\newcommand{\tx}{\widetilde{x}}
\newcommand{\wh}{\widehat}

\newcommand{\BC}{\mathbb{C}}
\newcommand{\BR}{\mathbb{R}}

\newcommand{\BZ}{\mathbb{Z}}

\newcommand{\sI}{\mathscr{I}}

\newcommand{\comment}[1]{}
\newcommand{\CF}{{\cal F}}
\newcommand{\CL}{{\cal L}}
\newcommand{\CS}{{\cal S}}
\newcommand{\CG}{{\cal G}}
\newcommand{\CT}{{\cal T}}
\newcommand{\CD}{{\cal D}}
\newcommand{\CM}{{\cal M}}
\newcommand{\CO}{{\cal O}}

\newcommand{\CN}{{\cal N}}

\newcommand{\CA}{{\cal A}}
\newcommand{\CP}{{\cal P}}

\newcommand{\CH}{{\cal H}}
\newcommand{\CZ}{{\cal Z}}
\newcommand{\CR}{{\cal R}}
\newcommand{\CI}{{\cal I}}
\newcommand{\CU}{{\cal U}}
\newcommand{\CQ}{{\cal Q}}

\newcommand{\de}{\mathrm{d}}

\newcommand{\tti}{\widetilde{t}}

\newcommand{\ta}{\widetilde{a}}
\newcommand{\tz}{\widetilde{z}}

\newcommand{\hk}{hyperk\"ahler }
\newcommand{\wt}{\widetilde}

\newcommand{\sh}{\sinh \pi}
\newcommand{\ch}{\cosh \pi}
\newcommand{\s}{\sigma}

\newcommand{\Secref}[1]{Section~\ref{#1}}

\newcommand{\Appref}[1]{Appendix~\ref{#1}}

\newcommand{\Tabref}[1]{Table~\ref{#1}}

\newcommand{\figref}[1]{Fig.~\ref{#1}}
\renewcommand{\eqref}[1]{(\ref{#1})}

\title{Three Dimensional Mirror Symmetry beyond $ADE$ quivers and Argyres-Douglas theories}
\author{Anindya Dey}
\affiliation{Department of Physics and Astronomy, Johns Hopkins University, Baltimore, MD 21218, USA}
\emailAdd{anindya.hepth@gmail.com}
\abstract{Mirror symmetry, a three dimensional $\CN=4$ IR duality, has been studied in detail for quiver gauge theories of the $ADE$-type (as well as their affine versions) with unitary gauge groups. The $A$-type quivers (also known as linear quivers) and the associated mirror dualities have a particularly simple realization in terms of a Type IIB system of D3-D5-NS5-branes. In this paper, we present a systematic field theory prescription for constructing 3d mirror pairs beyond the $ADE$ quiver gauge theories, starting from a dual pair of $A$-type quivers with unitary gauge groups. The construction involves a certain generalization of the $S$ and the $T$ operations, which arise in the context of the $SL(2,\BZ)$ action on a 3d CFT with a $U(1)$ 0-form global symmetry. We implement this construction in terms of two supersymmetric observables -- the round sphere partition function and the superconformal index on $S^2 \times S^1$. We discuss explicit examples of various (non-$ADE$) infinite families of mirror pairs that can be obtained in this fashion. In addition, we use the above construction to conjecture explicit 3d $\CN=4$ Lagrangians for 3d SCFTs, which arise in the deep IR limit of certain Argyres-Douglas theories compactified on a circle.}

\begin{document}
\maketitle

\section{A brief summary of the paper}

\subsection{Background and the basic idea of the paper}
In spite of the impressive success of perturbative QFT, the study of non-perturbative/strongly coupled aspects of a QFT remains a 
challenge for theorists. Probing the physics in this regime requires new computational tools, and there 
has been substantive progress in this area over the last thirty years. In addition to String Dualities, advances in AdS/CFT, as well as
 the more recent developments in the fields of Localization Methods and Conformal Bootstrap, have provided us with a powerful toolbox for 
 studying non-perturbative phenomena in QFTs, particularly ones with supersymmetry.\\

A rather ubiquitous phenomenon in QFTs, living in different space-time dimensions, is the existence of UV/IR dualities. Broadly 
speaking, existence of such a duality implies that a set of theories, which have completely different descriptions (theories with different 
Lagrangians, for example) at a given energy scale, are described by a common physical theory at another energy scale. 
The theory at the latter scale is often a strongly-coupled interacting CFT. 
In addition to High Energy Theory, the study of these dualities play a significant role in other branches of physics like
Condensed Matter Theory.
Given the strongly coupled nature of the problem, some direct or indirect handle on the non-perturbative physics of QFTs is 
necessary to probe these dualities. A special subclass of such dualities in (2+1) space-time dimensions will be our primary focus 
in this paper.\\

There has been significant progress in our understanding of UV/IR dualities for supersymmetric QFTs in the last decade, largely due to the 
development of a set of computational techniques which allows one to calculate certain supersymmetric observables exactly. These are 
collectively referred to as ``Localization Techniques/Methods" (see \cite{Pestun:2016zxk} for a recent review). 
The basic idea involves putting a supersymmetric QFT, with a given Lagrangian in flat space, on a $d$-dimensional Riemannian manifold $\CM_d$, 
such that the theory has an unbroken fermionic generator $Q$ which squares to a bosonic symmetry $\CL$. A systematic prescription for 
obtaining the supersymmetric Lagrangian on $\CM_d$, using the rigid limit of an appropriate supergravity theory with background auxiliary fields, 
was given in \cite{Festuccia:2011ws}. In addition, one can turn on background gauge fields for the non-spacetime global symmetries of the theory. 
The path integral corresponding to a supersymmetric observable on $\CM_d$ can then be deformed in a way such that it 
localizes to a set of ``saddle-point" configurations. For a sufficiently restrictive bosonic symmetry $\CL$, this set may reduce to a set of finite 
number of isolated points on the moduli space. Given the set of saddle-point configurations, one can attempt to compute the functional determinant associated 
with the fluctuations of fields around these configurations and thereby evaluate the path integral exactly. For the purpose of this paper, we will consider 
supersymmetric observables for which the answer can be expressed as a matrix 
integral (with possible sum over flux sectors) over some bosonic zero mode(s) with the integrand being completely determined by the gauge group and matter content of the Lagrangian. Schematically, such a supersymmetric observable computed using localization in a theory $T$, will have 
the following form\footnote{There can also be a sum over flux sectors that we are choosing to ignore in these schematic equations.}:
\begin{align} \label{LOC-Gen}
Z^{(T)}(\wh{A}) = \int \, [d \vec{\varphi}]\, Z^{(T)}_{\rm classical}(\vec{\varphi}, \wh{A})\, Z^{(T)}_{\rm quantum}(\vec{\varphi}, \wh{A}),
\end{align}
where $\vec{\varphi}$ collectively denotes the bosonic zero modes, and $\wh{A}$ denotes a space-time-independent background gauge field\footnote{In supersymmetric theories, $\wh{A}$ should be understood as a bosonic component of a background vector multiplet.} associated with a $0$-form global symmetry. 
$Z_{\rm classical}$ is the part of the matrix integrand that arises from simply evaluating the (possibly regularized) classical action on the saddle point configuration. 
$Z_{\rm quantum}$ captures the rest of the functional determinant and is completely determined by the gauge group as well as the representations of the 
matter multiplets under the gauge and the global symmetries.\\

A very important feature of the localization answer is that it is independent of dimensionful coupling constants and therefore invariant under a renormalization 
group flow. This implies that the observable, although computed using a weakly-coupled description, can be used to extract information about the strongly-coupled
CFT\footnote{For example, the superconformal index in three dimensions counts primary operators of the IR SCFT, the three sphere partition function is related to  
entanglement entropy across a disc, and so on.}. The RG-invariant localization answer is particularly suited for studying the UV/IR dualities. 
For a pair of theories $(X,Y)$ flowing to the same CFT in the UV/IR, 
RG-invariance of the supersymmetric observable $Z$ would imply
\be \label{LOC-Duality1}
Z^{(X)}(\wh{A}) =  Z^{(Y)}(\wh{A}).
\ee
The above equation constitutes a very non-trivial check of the duality statement, since one is comparing a function of the background gauge fields as 
opposed to a number. In particular, the map of the background gauge fields across the duality can be non-trivial.\\

Historically, the discovery and analysis of these UV/IR dualities, beyond simple examples, have relied heavily on String Theory dualities.
Localization techniques, on the other hand, provide an efficient way for studying dualities, from a purely QFT perspective.  It is natural to ask 
whether, within a given class of dualities, one can use these tools to construct a systematic field theory prescription for generating 
new dualities, starting from certain ``basic" ones. A related challenge for any such program would be to find QFT dualities which are 
not realized by the standard String Theory constructions. We will try to answer these issues for a class 
of IR dualities in three dimensions, explaining what we mean by ``basic" dualities along the way. \\

In order to implement such a program for 3d IR dualities, we would first need to discuss a prescription for constructing a new 3d CFT from a given 3d CFT.  
In \cite{Witten:2003ya}, the author studied an $SL(2,\BZ)$ action on 3d CFT $X[\wh{A}]$ with a global $U(1)$ symmetry with a 
background gauge field $\wh{A}$. The action of the $SL(2,\BZ)$ generators $T$ and $S$ are given as
\be
\begin{split} 
& T : \CL(\wh{A}) \to \CL(\wh{ A}) + \frac{1}{4\pi} \wh{ A}\wedge d\wh{ A},\\
& S : \CL(\wh{A}) \to \CL(a) + \frac{1}{2\pi} \wh{ B} \wedge da \qquad (a\,\,{\rm dynamical}),
\end{split}
\ee
where $\wh{B}$ is the background gauge field for the topological $U(1)_J$ symmetry in 3d. The action of the $S$ generator, which amounts to gauging a global symmetry of the theory $X[\wh{ A}]$, generically gives a new 3d CFT $X'[\wh{ B}]$, i.e.
\be
S: X[\wh{A}] \mapsto X'[\wh{B}].
\ee

In this paper, we will present a certain generalization of the $S$-operation on a class of 3d CFTs, which have a weakly coupled description 
with a manifest global symmetry subgroup $G^{\rm sub}_{\rm global}=\prod_\gamma U(M_\gamma)$. 
We will refer to this class of 3d CFTs as class $\CU$. Similar to the $S$-operation above, the generalized operation will allow us to construct new 3d CFTs from a 
given 3d CFT $X$ in the class $\CU$. We will refer to this generalized version as as an ``elementary $S$-type operation", and define it momentarily.\\

To begin with, we introduce a set of four independent operations on a generic 3d Lagrangian theory $\CT$ (not necessarily in $\CU$), where the operations act locally on a given 
global symmetry factor $K_\gamma$ (labelled by $\gamma$):
\begin{enumerate}

\item {\bf{Gauging ($\CG^\gamma$):}} Promotes $K_\gamma$ to a gauge group, adding a $U(1)_J$ background field, if $K_\gamma$ is unitary.

\item {\bf{Flavoring ($\CF^\gamma$):}} Adds matter fields charged under $K_\gamma$, and turns on background gauge fields for an 
additional global symmetry $G^\gamma_{F}$ associated with the added matter.

\item {\bf{Identification ($\CI^{\gamma}_{\gamma'}$):}} Identifies $K_\gamma$ with other factors $K_{\gamma'}$ (Lie group of the same type and same rank as $K_\gamma$) 
in the global symmetry of $\CT$, and turns on appropriate background gauge fields for the new global symmetry that the operation leads to.

\item {\bf{Defects ($\CD^\gamma$):}} Turns on defects for $K_\gamma$ (Wilson lines or vortex lines, for example).
\end{enumerate}

Now consider a Lagrangian theory $X$ in the class $\CU$. 
Let us pick the $\alpha$-th unitary factor $U(M_\alpha)$ in the global symmetry subgroup $G^{\rm sub}_{\rm global}=\prod_\gamma U(M_\gamma)$, and split it as 
$U(M_\alpha) \to U(r_\alpha) \times U(M_\alpha-r_\alpha)$. We will refer to the resultant theory as $(X, \CP)$, where $\CP$ denotes the data which encodes how the $U(r_\alpha)$ background gauge fields are chosen from those of $U(M_\alpha)$\footnote{See \eref{uvdef0} for a concrete realization in the case of $\CN \geq 2$ SUSY theories.}. 
Also, let us label the $U(r_\alpha)$ factor in $(X, \CP)$ as $\alpha'$. 
We will find it convenient to consider a slightly general situation, where we split multiple unitary factors $U(M_\beta) \to U(r_\alpha)_\beta \times U(M_\beta -r_\alpha)$ 
($\beta \neq \alpha$), in addition to the $\alpha$-th unitary factor. We will call the resultant theory $(X, \{\CP_\beta \})$, and label the $U(r_\alpha)_\beta$ factors as $\beta'$. 
An identification operation which identifies all the nodes labelled $\beta'$ with $\alpha'$ will be denoted as $\CI^{\alpha'}_{\beta'}$, following the above notation.\\

We will now define a ``$\CQ$-operation" $\CQ^\alpha_{\vec \CP}$ acting locally at the unitary factor $\alpha$ of the Lagrangian theory $X$ in the following fashion:
\begin{align}
\CQ^\alpha_{\vec \CP}(X) :=& (\CG^{\alpha'})^{n_4} \circ (\CF^{\alpha'})^{n_3} \circ (\CI^{\alpha'}_{\beta'})^{n_2} \circ (\CD^{\alpha'})^{n_1} (X, \{\CP_\beta \}), \quad (n_i=0,1,\,\, \forall i ) \label{QOP-1}\\
 := & (G^\alpha_{\vec \CP})^{n_4} \circ (F^\alpha_{\vec \CP})^{n_3} \circ (I^\alpha_{\vec \CP})^{n_2} \circ (D^\alpha_{\vec \CP})^{n_1}(X).  \label{QOP-2}
\end{align}
The first equality is the definition of $\CQ$-operation on $X$. It states that the action of a given operation $\CQ^\alpha_{\vec \CP}$ at a unitary factor $U(M_\alpha)$ (labelled $\alpha$) on $X$, is defined by acting on the theory $(X, \{\CP_\beta \})$ a certain combination of $\CG,\CF,\CI$ and $\CD$-operations locally at the unitary factor $U(r_\alpha)$ (labelled $\alpha'$). The specific combination is determined by the integers $\{n_i\}$. Note that the composition of the operations $\CF, \CI, \CD$ is commutative, but they 
do not commute with $\CG$.\\

The second equality \eref{QOP-2} defines a set of four basic $\CQ$-operations (discussed in detail in \Secref{GFI-def} for the specific case of 3d $\CN=4$ quivers) in terms of the operations $\CG,\CF,\CI$ and $\CD$, and gives a rule to compose them :
\begin{itemize}

\item {\bf{$\CQ$-Gauging ($G^\alpha_{\vec \CP}$):}} $G^\alpha_{\vec \CP}(X) = \CG^{\alpha'}(X, \{\CP_\beta \})$,

\item {\bf{$\CQ$-Flavoring ($F^\alpha_{\vec \CP}$):}} $F^\alpha_{\vec \CP}(X) = \CF^{\alpha'}(X, \{\CP_\beta \})$,

\item {\bf{$\CQ$-Identification ($I^\alpha_{\vec \CP}$):}} $I^\alpha_{\vec \CP}(X) = \CI^{\alpha'}_{\beta'}(X, \{\CP_\beta \})$,

\item {\bf{$\CQ$-defects ($D^\alpha_{\vec \CP}$):}} $D^\alpha_{\vec \CP}(X) = \CD^{\alpha'}(X, \{\CP_\beta \})$.
\end{itemize} 
In the rest of the paper, we will refer to these basic $\CQ$-operations as gauging, flavoring, identification and defect operations 
respectively, acting on a theory $X$, and the original operations $\CG,\CF,\CI$ and $\CD$ will never appear again.\\

We can now define the ``elementary $S$-type operation" $\CO^\alpha_{\vec \CP}$ on a global symmetry factor $U(M_\alpha)$ of $X$, as a special case of a 
$\CQ^\alpha_{\vec \CP}$ operation which necessarily includes the gauging operation, i.e.
\begin{align}
\boxed{\CO^\alpha_{\vec \CP}(X) := G^\alpha_{\vec \CP} \circ \Big(F^\alpha_{\vec \CP}\Big)^{n_3} \circ \Big(I^\alpha_{\vec \CP}\Big)^{n_2} \circ \Big(D^\alpha_{\vec \CP}\Big)^{n_1} (X),}  \label{SOP-intro}
\end{align}
where the RHS is precisely defined in \eref{QOP-2}. 
The action of $\CO^\alpha_{\vec \CP}$ on $X$ gives a (generically) new 3d CFT $X'$, whose weakly 
coupled description has a different gauge group and matter content from the theory $X$. The elementary $S$-type operation 
$\CO^\alpha_{\vec \CP}$ can therefore be thought of as an operation that acts on a 3d CFT $X[\wh{\vec{A}}]$ and produces 
a new 3d CFT $X'[\wh{\vec B}]$ (with or without defect), i.e.
\be \label{Smap}
\CO^\alpha_{\vec \CP}: X[\vec{\wh{A}}] \mapsto X'[\vec{\wh{B}}],
\ee
where $\wh{\vec{A}}$ denotes the $U(r_\alpha)$ background gauge field. $\wh{\vec B}$ collectively denotes the $U(1)_J$ 
background gauge field, background fields associated with the global symmetries coming from the flavoring/identification operations, 
as well as additional data for the defects, if any. In the trivial case, where $G^{\rm sub}_{\rm global}=U(1)$, and $\CO^\alpha_{\vec \CP}$ 
is constituted of simply a gauging operation, the elementary $S$-type operation coincides with the action of the $S$ generator in \cite{Witten:2003ya}.\\

From \eref{SOP-intro}, ignoring defect operations for the time being (which would be the case for most of this paper), an elementary $S$-type operation can be classified 
into four distinct types:
\begin{enumerate}
\item {\bf{Gauging:}} $\CO^\alpha_{\vec \CP}(X)= G^\alpha_{\vec \CP}(X)$.
\item {\bf{Flavoring-Gauging:}} $\CO^\alpha_{\vec \CP}(X)= G^\alpha_{\vec \CP} \circ F^\alpha_{\vec \CP}(X)$.
\item {\bf{Identification-Gauging:}} $\CO^\alpha_{\vec \CP}(X)= G^\alpha_{\vec \CP} \circ I^\alpha_{\vec \CP}(X)$.
\item {\bf{Identification-Flavoring-Gauging:}} $\CO^\alpha_{\vec \CP}(X)= G^\alpha_{\vec \CP} \circ F^\alpha_{\vec \CP} \circ I^\alpha_{\vec \CP}(X)$.
\end{enumerate}
By definition, a ``generic $S$-type operation" will be understood as a combination of elementary 
$S$-type operations of the four types listed above. Note that the construction \eref{Smap} is completely independent of the 
existence or the amount of supersymmetry of $X[\wh{\vec A}]$. In a similar fashion, one can define an ``elementary $T$-type operation",
 which turns on a Chern-Simons term for $U(r_\alpha)$, and combine it with $S$-type operations to construct new 3d CFTs.\\

Given the map \eref{Smap}, one can now address the issue of constructing new dual pairs starting from 
a given dual pair. Suppose the theory $X[\wh{\vec A}]$ is IR dual to the theory $Y[\wh{\vec A}]$, where both theories 
have a weakly coupled description and $X$ is in class $\CU$. Then, given an $S$-type operation $\CO^\alpha_\CP$ on $X[\wh{\vec A}]$, let us  
define a dual operation $\wt{\CO}^\alpha_\CP$ on $Y[\wh{\vec A}]$,
\be
\wt{\CO}^\alpha_{\CP}: Y[\wh{A}] \mapsto Y'[\wh{B}],
\ee
such that the pair of theories $(X'[\wh{\vec B}],Y'[\wh{\vec B}])$ are IR dual. This is summarized in the figure below.

\begin{figure}[H]
\begin{center}
\begin{tikzpicture}
  \node (X) at (2,1.5) {$X[\wh{\vec A}]$};
  \node (Y) at (6,1.5) {$Y[\wh{\vec A}]$};
  \node (D) at (2,-1) {$X'[\wh{\vec B}]$};
  \node (E) at (6,-1) {$Y'[\wh{\vec B}]$};
   \draw[->] (X) -- (D) node [midway,  right=+3pt] {\footnotesize $\CO^\alpha_{\CP}$};
  \draw[<->] (X) -- (Y)node [midway,above  ] {\footnotesize IR duality};
    \draw[->] (Y) -- (E) node [midway,  right=+3pt] {\footnotesize $\wt{\CO}^\alpha_{\CP}$};
        \draw[<->] (D) -- (E)node  [midway,above  ] {\footnotesize IR duality};
\end{tikzpicture}
\end{center}
\caption{Generating new dual pairs using an elementary $S$-type operation. }
\label{S-OP-duality}
\end{figure}

The challenge then is to give a precise definition for the dual operation $\wt{\CO}^\alpha_{\CP}$, given $(X[\wh{\vec A}],Y[\wh{\vec A}])$ and $\CO^\alpha_{\vec \CP}$, which should also allow one to read off the theory $Y'[\wh{\vec B}]$, if it turns out to be Lagrangian (note that the IR dual of a Lagrangian theory is 
not necessarily Lagrangian). For theories where an RG-invariant observable computable using localization exists, as given in \eref{LOC-Gen},
this problem may be solved explicitly. For three space-time dimensions, this forces us to restrict ourselves to theories with $\CN \geq 2$ 
supersymmetry. Given a pair of dual theories $(X[\wh{\vec A}],Y[\wh{\vec A}])$ and the $S$-type operation $\CO^\alpha_{\vec \CP}$, one can then write explicit formulae for RG-invariant observables of $Y'[\wh{\vec B}]$ using \eref{LOC-Gen} and \eref{LOC-Duality1}. 
To begin with, the operation $\CO^\alpha_{\vec \CP}$ can be implemented at the level of the supersymmetric 
observable in the following fashion:
\be \label{GFI-X-Intro}
\boxed{Z^{(X')}(\wh{\vec B})=\int \, [d \wh{\vec A}] \, \CZ_{{\CO}^\alpha_{\vec \CP}(X)}(\wh{\vec A}, \wh{\vec B}) \, Z^{(X)}(\wh{\vec A}),}
\ee
where $\CZ_{{\CO}^\alpha_{\vec \CP}(X)}$ is an explicitly known operator that depends on the specific elementary $S$-type operation.
The observable for the dual theory $Y'[\wh{\vec B}]$ then assumes the following schematic form (substituting \eref{LOC-Duality1} in \eref{GFI-X-Intro}, and 
changing the order of integration):
\begin{align}\label{LOC-Duality2}
&\boxed{Z^{(Y')}(\wh{\vec B}) = \int \, [d \vec{\varphi}']\, \Big(\int \, [d \wh{\vec A}] \, \CZ_{{\CO}^\alpha_{\vec \CP}(X)}(\wh{\vec A}, \wh{\vec B}) \Big)\, Z^{(Y)}_{\rm classical}(\vec{\varphi}', \wh{A})\, Z^{(Y)}_{\rm quantum}(\vec{\varphi}', \wh{\vec A})} \\
&\text{where } Z^{(Y)}(\wh{\vec A}) = \int \, [d \vec{\varphi}']\,Z^{(Y)}_{\rm classical}(\vec{\varphi}', \wh{\vec A})\, Z^{(Y)}_{\rm quantum}(\vec{\varphi}', \wh{\vec A}). \nn
\end{align}
The relation \eref{LOC-Duality2} is the definition of the dual operation $\wt{\CO}^\alpha_{\CP}$ on the theory $Y$.
If the theory $Y'[\wh{\vec B}]$ is Lagrangian, one should be able to rewrite $Z^{(Y')}(\wh{\vec B})$ in \eref{LOC-Duality2} in the standard form
\be
Z^{(Y')}(\wh{\vec B}) =\int \, [d \vec{\varphi}']\, Z^{(Y')}_{\rm classical}(\vec{\varphi}', \wh{\vec B})\, Z^{(Y')}_{\rm quantum}(\vec{\varphi}', \wh{\vec B}),
\ee
which allows one to read off the gauge group and matter content of $Y'[\wh{\vec B}]$ from the RHS of the above equation. Note that this last step may not be easy to perform 
for a generic duality, and might involve some non-trivial manipulation of the matrix integral.\\

The strategy for generating dualities using $S$-type operations is now clear. For a given type of IR duality, we pick a convenient  
subset of dual theories (with $(X,Y)$ Lagrangian and $X$ in class $\mathcal{U}$) for which the duality is completely understood in terms of RG-invariant observables. We will refer to this as the set of ``basic dualities" for the given IR duality. Picking $(X,Y)$ from this set of basic dualities, one can now 
implement the construction of \figref{S-OP-duality} step-wise to generate new dual pairs.\\ 

In the present paper, we will choose a specific IR duality for 3d $\CN=4$ theories, called mirror symmetry \cite{Intriligator:1996ex, deBoer:1996ck, deBoer:1996mp, Kapustin:1999ha}, 
to illustrate how our construction can be realized in a concrete setting. A discussion of more generic 3d $\CN=2$ dualities will be deferred to a 
future work. Mirror symmetry is a special IR duality for a pair of theories $(X,Y)$ which interchanges the Coulomb branch of $X$ with the Higgs 
branch of $Y$ and vice-versa, in the deep IR regime. In particular, this implies that flavor symmetry (i.e. the Higgs branch global symmetry) on one side 
of the duality gets mapped to topological symmetry (i.e. Coulomb branch global symmetry) on the other side. Non-supersymmetric versions of such dualities 
appear in a wide-range of condensed matter systems \cite{Sachdev:2008wv, Murugan:2016zal, Seiberg:2016gmd}. Recently, it was shown that these supersymmetric dualities can also generate a large 
class of bosonization dualities \cite{Giombi:2011kc, Aharony:2012nh, Aharony:2015mjs, Karch:2016sxi} via soft supersymmetry breaking \cite{Kachru:2016rui, 
Kachru:2016aon}.\\

In the literature, mirror symmetry for quiver gauge theories of the $ADE$ type (and their affine cousins) has been studied in a lot of detail \cite{Dey:2011pt, Dey:2013nf, Dey:2014tka, Dedushenko:2016jxl, Dedushenko:2017avn, Dedushenko:2018icp, Fan:2019jii}. This is essentially due to the fact that these theories have well-known realization in Type IIB String Theory \cite{Hanany:1996ie, Hanany:1999sj, Gaiotto:2008sa, Gaiotto:2008ak} and M-Theory \cite{Porrati:1996xi, Witten:2009xu, Dey:2011pt}. In particular, 
the $A$-type quivers with unitary gauge groups (also known as linear quivers) have a very simple realization in Type IIB, in terms of D3-branes on a 
line segment, along with NS5 and D5-branes. More recently, a large class of non-Lagrangian 3d SCFTs, whose mirror duals are Lagrangian, 
was also obtained using the dimensional reduction of 4d $\CN=2$ theories of class $\CS$ \cite{Benini:2010uu, Xie:2012hs}. In all these String Theory/M-Theory constructions,
the QFT duality follows as a consequence of some String Duality.\\

One of the primary goals of this paper is to construct mirror dual quiver gauge theories beyond these $ADE$ quiver examples, using the $S$-type operations described above. 
The strategy, as outlined before, is to start from a pair of basic dual theories $(X,Y)$, and implement one or more elementary $S$-type operation(s) defined above to obtain a new dual pair $(X',Y')$. As mentioned earlier, the theories $(X,Y)$ must both have Lagrangian descriptions for the construction to work. Also, the precise localization procedure, as we will see later, requires knowing the mirror map, i.e. how flavor symmetries of $X$ map to topological symmetries of $Y$. Given a Lagrangian quiver $X$ in class $\mathcal{U}$, the mirror dual $Y$ is generically not known. Therefore, a convenient choice of basic dualities is necessary. For the purpose of this paper, we will choose the set of basic dualities as 
the set of \textit{good} linear quivers \cite{Gaiotto:2008ak} with unitary gauge groups, where the mirror symmetry (including the mirror map) is completely understood, both from String Theory and QFT. With this choice of the pair $(X,Y)$, the construction of \figref{S-OP-duality} can be seamlessly implemented to generate a new dual pair $(X',Y')$. 
In particular, the expression \eref{LOC-Duality2} greatly simplifies as we will see in \Secref{GFIdual-summary}.\\
 
For the examples treated in this paper, the relevant elementary $S$-type operations are Abelian, i.e. they involve gauging of a single $U(1)$ global symmetry 
combined with flavoring and/or identification operations. Note that the generic $S$-type operations, built out of these elementary ones, can generate dualities for Abelian as well as non-Abelian theories, depending on the basic dual pair $(X,Y)$ one starts from. Examples involving non-Abelian elementary $S$-type operations will be addressed in an upcoming paper.\\

The paper is organized as follows. \Secref{Review} gives a brief review of the various aspects of 3d $\CN=4$ physics relevant for the paper.
\Secref{GFI-main} sets up the general formalism for constructing new dualities, as outlined above.
\Secref{GFI-def} introduces the $S$-type and $T$-type operations on a generic $\CN=4$ quiver in class $\CU$, and discusses their realization in terms of 
a specific RG-invariant observable - the round 3-sphere partition function\cite{Kapustin:2009kz, Kapustin:2010xq}. \Secref{GFI-summary} then discusses the partition function construction of quivers using elementary $S$-type operations from linear quivers (as well as more general quivers), while \Secref{GFIdual-summary} discusses the dual operations. A simple illustrative example, involving an affine $A$-type theory, is presented in \Secref{GFI-exAbPF}, while more involved examples involving (affine) $D$-type quivers are presented in \Appref{GFI-ex}. The $S$-type operations can also be implemented in terms of other RG-invariant observables. We present the analysis in terms of the $S^2 \times S^1$ superconformal index \cite{Imamura:2011su, Kapustin:2011jm, Benini:2011nc} (reviewed in \Appref{SCI}) in \Appref{GFI-SCI}.\\

\Secref{AbMirr} focuses on Abelian $S$-type operations and examples of new dual theories that can be constructed using them.
\Secref{AbS} studies the Abelian version of the four different types of 
elementary $S$-type operations and their duals. \Secref{AbExNew} and \Secref{NAbExNew} 
then apply these operations to construct two infinite families of Abelian mirror pairs and two infinite families non-Abelian mirror pairs respectively, consisting of non-$ADE$-type quiver gauge theories in each case. These two subsections contain the main results of this paper in terms of constructing new mirror dualities.\\

Finally, using results of \Secref{AbExNew} and \Secref{NAbExNew}, we demonstrate in \Secref{ADMirrors} that 3d SCFTs, obtained by compactifying a large family of Argyres-Douglas (AD) theories on a circle and flowing to the IR, turn out to have Lagrangian descriptions. Our strategy in this case involves showing that the 3d Lagrangian mirror \cite{Xie:2012hs, boalch2008irregular} predicted by the class $\CS$ construction of a given AD theory, itself has a Lagrangian mirror. Note that we mean a manifestly $\CN=4$ Lagrangian in this context, 
and not a Lagrangian with less supersymmetry that flows to a theory with $\CN=4$ supersymmetry \cite{Maruyoshi:2016tqk}.

\subsection{Summary of the main results}
In this paper, we introduce a systematic field theory prescription for generating infinite families of 3d $\CN=4$ mirror quiver pairs,
using $S$-type operations, as outlined in \figref{S-OP-duality}. We explicitly realize this program in terms of two RG-invariant 
supersymmetric observables -- the $S^3$ partition function and the $S^2 \times S^1$ superconformal 
index, which can be computed using localization techniques. 
This is then applied to construct examples of dual pairs involving quiver gauge theories beyond the $ADE$-type. 
The main results of the paper can be summarized as follows:

\subsection*{Construction of mirror pairs from $S$-type operations}
Consider a pair of dual quiver gauge theories $(X,Y)$ where $X$ is in class $\CU$, with a Higgs branch global symmetry subgroup 
$G^{\rm sub}_{\rm global}=\prod_\gamma U(M_\gamma)$. 
We first realize the construction of a new pair of dual theories $(X',Y')$ starting from $(X,Y)$ in terms of the $S^3$ partition function. 
In the matrix integral, the background vector multiplets for the Higgs branch global symmetry appear as real masses, while the twisted vector multiplets
 for the Coulomb branch global symmetry appear as real FI parameters. Supersymmetry demands that, in 
 each case, the background multiplets live in the Cartan subalgebra of the respective global symmetry group. 
For a generic elementary $S$-type operation $\CO^\alpha_{\vec \CP}$ on the quiver, the partition function of the theory 
$X'=\CO^\alpha_{\vec \CP}(X)$ is given as (the more detailed form of this equation appears as \eref{PF-OP} in the main text) :
\begin{empheq}[box=\widefbox]{align}\label{PF-OP-Intro}
Z^{\CO^\alpha_{\vec \CP}(X)} (\eta_\alpha, \vec{m}^{\CO^\alpha_{\vec \CP}}, \vec \eta,\ldots)= \int \Big[d\vec{u}^\alpha\Big] \, \CZ_{\CO^\alpha_{\vec \CP}(X)}(\vec u^{\alpha}, \{\vec{u}^\beta\}, \eta_\alpha, \vec{m}^{\CO^\alpha_{\vec \CP}})
\cdot Z^{(X,\{ P_\beta\})} (\{\vec{u}^\beta\},\vec \eta, \ldots),
\end{empheq}
which gives a concrete realization of \eref{GFI-X-Intro}. The function 
$Z^{(X,\{ P_\beta\})}$ (given in \eref{uvZAI0}) is the partition function of the theory $(X,\{ P_\beta\})$, which was introduced 
prior to \eref{QOP-1}. The explicit form of the operator $\CZ_{\CO^\alpha_{\vec \CP}(X)}$ is given in \eref{CZ-OP}.
The parameters ${\vec u}^\beta$ denote the real masses in the Cartan subalgebra of the group $U(r_\alpha)_\beta$, $\forall \beta$, 
while the parameters $\vec{u}^\alpha$ denote the masses for the chosen group $U(r_\alpha)_\alpha$ with which the groups $U(r_\alpha)_\beta$ 
(for $\beta\neq \alpha$) will be identified by the $S$-type operation $\CO^\alpha_{\vec \CP}$. 
The parameters $\eta_\alpha$ and $\vec{m}^{\CO^\alpha_{\vec \CP}}$ denote 
the FI parameter for the $U(r_\alpha)$ gauge group and the masses for the global symmetry introduced by the identification and/or flavoring
operation respectively. Also, $\vec \eta$ collectively denotes the FI parameters of $X$, and the ``$\ldots$" in the argument of $Z^{(X,\{ P_\beta\})}$ 
denote mass parameters of $X$ that remain unaffected by the $S$-type operation. For the special case of $X$ being a linear quiver, the 
corresponding equation is given in \eref{PF-OPLQ}. \\

The partition function of the dual theory $Y'=\wt{\CO}^\alpha_{\vec \CP}(Y)$ can then be written in the following form (appears as 
\eref{PF-wtOPgen} in the main text) :
\begin{empheq}[box=\widefbox]{align}\label{PF-wtOPgen-Intro}
& Z^{\wt{\CO}^\alpha_{\vec \CP}(Y)}(\vec{m}'(\vec{\eta},\eta_{\alpha}); \vec \eta'(\vec{m}^{{\CO}^\alpha_{\vec\CP}},\ldots)) \nn \\
=& \int \prod_{\gamma'} \Big[d\vec{\s}^{\gamma'} \Big]\, \CZ_{\wt{\CO}^\alpha_{\vec \CP}(Y)}(\{\s^{\gamma'}\},\vec{m}^{{\CO}^\alpha_{\vec\CP}}, \eta_{\alpha},\vec \eta)\,\cdot Z^{(Y,\{\CP_\beta\})}_{\rm int}(\{\vec \s^{\gamma'} \}, \vec{m}^Y(\vec{\eta}), \vec{\eta}^Y(\{\vec{u}^\beta =0 \},\ldots)),
\end{empheq}
which gives a concrete realization of \eref{LOC-Duality2}. The function $Z^{(Y,\{\CP_\beta\})}_{\rm int}$ is the full integrand for the partition function 
matrix integral of $(Y,\{ P_\beta\})$, defined in \eref{PfYP-Y}. 
The masses and FI parameters of $Y'$ are denoted as $(\vec{m}', \vec \eta')$, while $(\vec{m}^Y, \vec{\eta}^Y)$ denote the same for the quiver gauge theory $Y$. 
$\CZ_{\wt{\CO}^\alpha_{\vec \CP}(Y)}$ is a function that can be formally written as a Fourier transform of the operator $\CZ_{\CO^\alpha_{\vec \CP}(X)}$:
\begin{empheq}[box=\widefbox]{align} \label{CZ-wtOP-Intro}
\CZ_{\wt{\CO}^\alpha_{\vec \CP}(Y)}
= \int \Big[d\vec{u}^\alpha\Big] \, \CZ_{\CO^\alpha_{\vec \CP}(X)}(\vec u^{\alpha}, \{\vec{u}^\beta\}, \eta_\alpha, \vec{m}^{\CO^\alpha_{\vec \CP}})\, \cdot \prod_{\beta}\,e^{2\pi i \,(g^i_\beta (\{\vec \s^{\gamma'} \}, \CP_\beta) + b^{il}_\beta \eta_l)\,u^{\beta}_i}, 
\end{empheq}
where $g^i_\beta (\{\vec \s^{\gamma'} \}, \CP_\beta)$ is a linear function in the variables $\{\vec \s^{\gamma'} \}$, that can be read off from the mirror map 
relating mass parameters of $X$ and FI parameters of $Y$, while $b^{il}_\beta$ are integer-valued matrices defined in \eref{bil-def}. For the special case of $(X,Y)$ being linear quivers, the functions $g^i_\beta$ are known a priori from the Type IIB construction, and the partition function for the dual theory is given in \eref{PF-wtOPLQ}. If the theory $Y'$ is Lagrangian, then the matrix integral on the RHS of \eref{PF-wtOPgen-Intro} can be recast into the standard form for a Lagrangian theory, such that the gauge group and the matter content can be read off. Note that the computation of the dual partition function essentially boils down to computing the function $\CZ_{\wt{\CO}^\alpha_{\vec \CP}(Y)}$ in \eref{CZ-wtOP-Intro}.\\

Using the general formulae \eref{PF-OP-Intro},\eref{PF-wtOPgen-Intro} and \eref{CZ-wtOP-Intro}, we write down the dual partition functions for the four distinct types of elementary $S$-type operations -- gauging, flavoring-gauging, identification-gauging, and identification-flavoring-gauging.
We refer the reader to \Secref{GFI-summary} and \Secref{GFIdual-summary} for details.\\

The above construction can also be implemented in terms the $S^2 \times S^1$ superconformal index.
 The equations analogous to \eref{PF-OP-Intro},\eref{PF-wtOPgen-Intro} and \eref{CZ-wtOP-Intro} are 
 given by \eref{genSI}, \eref{genSCId1} and \eref{genSCId2} respectively.

\subsection*{Abelian $S$-type operations and their duals}
In \Secref{AbS}, we work out the general rules for the dual operations associated with the four distinct types of elementary Abelian $S$-type operations, 
in terms of the $S^3$ partition function. For the constituent flavoring operations, we restrict ourselves to hypermultiplets
with charge 1 under the new $U(1)$ gauge group. Given any dual pair of quiver gauge theories $(X,Y)$ with $X$ in class $\CU$, we show explicitly that the dual theory $Y'$ 
is a Lagrangian theory for each of the four operations. In each case, we first give the formula for the dual partition function for a generic 
quiver pair $(X,Y)$, and then discuss the special case where $(X,Y)$ are linear quivers. We make the following general observations:

\begin{itemize}

\item For gauging and identification-gauging operations, the function $\CZ_{\wt{\CO}^\alpha_{\vec \CP}(Y)}$ in \eref{PF-wtOPgen-Intro} turns 
out to be a single delta function. This implies that the dual theory $\wt{\CO}^\alpha_{\vec \CP}(Y)$ is given by ungauging a single $U(1)$ factor 
from the gauge group of $Y$. The precise $U(1)$ being ungauged depends on the precise form of the functions $g_\beta$ on the RHS of \eref{CZ-wtOP-Intro}.
The related equations can be found in \eref{GAbgen}
and \eref{dual-AbGIgen} respectively in the main text. 

\item For flavoring-gauging and identification-flavoring-gauging operations, the function $\CZ_{\wt{\CO}^\alpha_{\vec \CP}(Y)}$ in \eref{PF-wtOPgen-Intro}
is made up of Lagrangian building blocks (up to some overall phase factors). This leads to a Lagrangian theory $\wt{\CO}^\alpha_{\vec \CP}(Y)$. 
The related equations can be found in \eref{PFGFgen} and \eref{PFGFIgen} respectively 
in the main text. 

\end{itemize}
 
The above findings lead to the following result. Consider a dual pair of quiver gauge theories $(X,Y)$, and an Abelian $S$-type operation involving 
a sequence of the elementary operations, acting on $X$. Using the results above, one can readily show that the resultant theory $X'$ is guaranteed 
to have a Lagrangian mirror dual $Y'$, which can be worked out explicitly.\\

These results on Abelian $S$-type operations can also be obtained using the superconformal index on $S^2\times S^1$. The relevant discussion 
can be found in \Appref{GFI-SCI-Ab}.

\subsection*{Explicit examples of mirror pairs beyond $ADE$ quivers}
Using the general rules for Abelian $S$-type operations derived in \Secref{AbS}, we construct four infinite families of dual quiver pairs, 
by a sequence of elementary Abelian $S$-type operations starting from a pair of linear quiver gauge theories $(X,Y)$ in each case. 
The quiver gauge theories we consider have the following generic features:
\begin{enumerate}
\item \textbf{Loops attached to a linear quiver tail :} Loops built out of gauge nodes and hypermultiplets in appropriate 
representations, such that one or more of the gauge nodes are attached to linear quiver tail(s).

\item \textbf{Loops with multiple edges :} Loops built out of gauge nodes and hypermultiplets, such that one or more 
pairs of gauge nodes are connected by multiple hypermultiplets transforming in a given representation of 
the associated gauge groups.
\end{enumerate}

\Tabref{Tab:SummaryMain} lists the four infinite families of dual quiver gauge theories. Note that Family I and Family II consist of Abelian 
quiver gauge theories, while Family III and Family IV consist of non-Abelian quiver gauge theories. The details of the partition function 
computation for these dual pairs can be found in \Secref{AbExNew} and \Secref{NAbExNew} respectively. As mentioned earlier, these results can 
be obtained using the superconformal index as well, and the rules for the Abelian $S$-type operations, necessary for constructing the 
duals in this paper, are worked out in \Appref{GFI-SCI-Ab}. We therefore check every proposed duality using two 
RG-invariant supersymmetric observables - the three-sphere partition function and the superconformal 
index. For the sake of brevity, however, we will only include a sample computation in the paper for the index, 
given in \Secref{GFI-exAbSCI}. Other examples can be readily checked using the Abelian $S$-type operations for the index 
in \Appref{GFI-SCI-Ab}, in a fashion analogous to the $S^3$ partition function.

\subsection*{3d $\CN=4$ Lagrangians for Argyres-Douglas theories reduced on a circle}
We propose explicit $\CN=4$ Lagrangians for 3d $\CN=4$ SCFTs which arise by putting certain 4d Argyres-Douglas (AD) theories  
on a circle and flowing to the IR. These 4d SCFTs can be constructed in class $\CS$ \cite{Gaiotto:2009hg, Xie:2012hs} by a twisted compactification of a 6d (2,0) $A_{N-1}$ 
theory on a Riemann sphere with an irregular puncture with/without a single regular puncture. The 3d mirrors of some of these SCFTs  
are known to be Lagrangian, and can be constructed using the 6d description -- we will refer to them as ``class $\CS$ mirrors".
The list of 4d SCFTs with Lagrangian class $\CS$ mirrors includes the $(G,G')$ theories of Cecotti-Neitzke-Vafa \cite{Cecotti:2010fi} of the type 
$(A_s, A_{(s+1)p-1})$ and $(A_s, D_{(s+1)p+2})$ (where $s$ and $p$ are positive integers).  
Our strategy is to show that the class $\CS$ mirrors associated with these 4d SCFTs have Lagrangian mirrors themselves, 
which in turn implies that the original 3d SCFT has a Lagrangian description. \Tabref{Tab:SummaryAD} 
summarizes the class $\CS$ mirror and the Lagrangian that we propose for a given family of AD theories. Given the class $\CS$ mirror,
the proposed Lagrangian can be read off from \Tabref{Tab:SummaryMain} -- Family I, II and IV respectively -- for appropriate choices of the 
integer parameters. A more detailed analysis of the physics of these 3d theories will be presented in a future paper. 

\begin{center}
\begin{table}[h]
\begin{tabular}{|c|c|c|}
\hline
Family & Theory $X$ & Theory $Y$  \\
\hline \hline 
${\rm{I}}_{[n,l,p]}$  & \scalebox{0.6}{\begin{tikzpicture}[node distance=2cm,cnode/.style={circle,draw,thick,minimum size=8mm},snode/.style={rectangle,draw,thick,minimum size=8mm},pnode/.style={rectangle,red,draw,thick,minimum size=8mm}]
\node[cnode] (1) at (-2,0) {$1$};
\node[cnode] (2) at (-1,1) {$1$};
\node[cnode] (3) at (1,1) {$1$};
\node[cnode] (4) at (2,0) {$1$};
\node[cnode] (5) at (1,-1) {$1$};
\node[cnode] (6) at (-1,-1) {$1$};
\node[cnode] (7) at (3,0) {$1$};
\node[cnode] (8) at (4,0) {$1$};
\node[cnode] (9) at (5,0) {$1$};
\node[cnode] (10) at (7,0) {$1$};
\node[snode] (11) at (8,0) {$1$};
\node[snode] (12) at (-2,-2) {$1$};
\draw[thick] (1) to [bend left=40] (2);
\draw[dashed,thick] (2) to [bend left=40] (3);
\draw[thick] (3) to [bend left=40] (4);
\draw[thick] (4) to [bend left=40] (5);
\draw[dashed,thick] (5) to [bend left=40] (6);
\draw[thick] (6) to [bend left=40] (1);
\draw[thick] (1) -- (12);
\draw[thick] (4) -- (7);
\draw[thick] (7) -- (8);
\draw[thick] (8) -- (9);
\draw[thick, dashed] (9) -- (10);
\draw[thick] (10) -- (11);
\node[text width=1.5cm](20) at (-2,0.6) {$n-l+1$};
\node[text width=1.5cm](21) at (-1,1.6) {$n-l+2$};
\node[text width=1cm](22) at (1,1.5) {$n-1$};
\node[text width=1cm](23) at (2,0.5) {$n$};
\node[text width=1cm](24) at (1,-1.5) {$1$};
\node[text width=1cm](25) at (-1,-1.5) {$n-l$};
\node[text width=0.1cm](26) at (3,0.7) {$1$};
\node[text width=0.1cm](27) at (4,0.7) {$2$};
\node[text width=0.1cm](28) at (5,0.7) {$3$};
\node[text width=1 cm](29) at (7,0.7) {$p-2$};
\end{tikzpicture}}
 & \scalebox{0.6}{\begin{tikzpicture}[node distance=2cm,cnode/.style={circle,draw,thick,minimum size=8mm},snode/.style={rectangle,draw,thick,minimum size=8mm},pnode/.style={rectangle,red,draw,thick,minimum size=8mm}]
\node[cnode] (1) at (-2,0) {$1$};
\node[cnode] (2) at (2,0) {$1$};
\node[snode] (3) at (-2,-2) {$n-l+1$};
\node[snode] (4) at (2,-2) {$l-1$};
\draw[thick] (1) -- (3);
\draw[thick] (2) -- (4);
\draw[thick,dotted] (0,0.4) to (0, -0.4);
\draw[thick] (1) to [bend left=20] (2);
\draw[thick] (1) to [bend right=20] (2);
\draw[thick] (1) to [bend left=40] (2);
\draw[thick] (1) to [bend right=40] (2);
\node[text width=0.1cm](15) at (0,1){$p$};
\end{tikzpicture} }\\
\hline
${\rm{II}}_{[n,l,l_1,l_2,p_1,p_2]}$ & \scalebox{0.6}{\begin{tikzpicture}[node distance=2cm,cnode/.style={circle,draw,thick,minimum size=8mm},snode/.style={rectangle,draw,thick,minimum size=8mm},pnode/.style={rectangle,red,draw,thick,minimum size=8mm}]
\node[cnode] (1) at (-3,0) {$1$};
\node[cnode] (2) at (-2,2) {$1$};
\node[cnode] (3) at (0,2.5) {$1$};
\node[cnode] (4) at (2,2) {$1$};
\node[cnode] (5) at (3,0) {$1$};
\node[cnode] (6) at (2,-2) {$1$};
\node[cnode] (7) at (0,-2.5) {$1$};
\node[cnode] (8) at (-2,-2) {$1$};
\node[cnode] (9) at (0,1.5) {$1$};
\node[snode] (10) at (-5,0) {$1$};
\node[cnode] (11) at (4,0) {$1$};
\node[cnode] (12) at (5,0) {$1$};
\node[cnode] (13) at (7.5,0) {$1$};
\node[snode] (14) at (9,0) {$1$};
\node[cnode] (15) at (0,0.5) {$1$};
\node[cnode] (16) at (0,-1.5) {$1$};
\draw[thick] (1) to [bend left=40] (2);
\draw[dashed] (2) to [bend left=40] (3);
\draw[thick] (3) to [bend left=40] (4);
\draw[dashed] (4) to [bend left=40] (5);
\draw[thick] (5) to [bend left=40] (6);
\draw[dashed] (6) to [bend left=40] (7);
\draw[dashed] (7) to [bend left=40] (8);
\draw[thick] (8) to [bend left=40] (1);
\draw[thick] (1) -- (10);
\draw[thick] (5) -- (11);
\draw[thick] (3) -- (9);
\draw[thick] (9) -- (15);
\draw[thick] (11) -- (12);
\draw[dashed] (12) -- (13);
\draw[thick] (13) -- (14);
\draw[dashed] (15) -- (16);
\draw[thick] (16) -- (7);
\node[text width=1cm](10) at (-3,0.5) {$l$};
\node[text width=1cm](11) at (-2,2.5) {$l+1$};
\node[text width=1cm](12) at (0, 3) {$l_2$};
\node[text width=1cm](13) at (2, 2.5) {$l_2+1$};
\node[text width=1cm](14) at (3, 0.5) {$n$};
\node[text width=1cm](15) at (2,-2.5) {$1$};
\node[text width=1.5 cm](16) at (0,-3.2){$l_1$};
\node[text width=1cm](17) at (-2.5,-2.5){$l-1$};
\node[text width=1cm](18) at (4, 0.6){$1$};
\node[text width=1cm](19) at (5, 0.6){$2$};
\node[text width=1cm](20) at (7.5, 0.6){$p_1-2$};
\node[text width=1cm](21) at (1,1.5){$p_2-1$};
\node[text width=1cm](22) at (1,0.5){$p_2-2$};
\node[text width=1cm](23) at (1,-1.5){$1$};
\end{tikzpicture}} 
 & \scalebox{0.65}{\begin{tikzpicture}[node distance=2cm,cnode/.style={circle,draw,thick,minimum size=8mm},snode/.style={rectangle,draw,thick,minimum size=8mm},pnode/.style={rectangle,red,draw,thick,minimum size=8mm}]
\node[cnode] (1) at (-2,0) {$1$};
\node[cnode] (2) at (2,0) {$1$};
\node[cnode] (3) at (0,2) {$1$};
\node[snode] (4) at (-2,-2) {$l_1$};
\node[snode] (5) at (2,-2) {$n-l_2$};
\node[snode] (6) at (0,4) {$p_2$};
\draw[thick] (1) -- (4);
\draw[thick] (2) -- (5);
\draw[thick] (3) -- (6);
\draw[thick] (1) to [bend left=10] (2);
\draw[thick] (1) to [bend right=10] (2);
\draw[thick] (1) to [bend left=20] (2);
\draw[thick] (1) to [bend right=20] (2);
\draw[thick, dotted] (0,0.2) to (0,-0.2);
\node[text width=1cm](15) at (0,-1){$p_1$};
\draw[thick] (1) to [bend left=10] (3);
\draw[thick] (1) to [bend right=10] (3);
\draw[thick] (1) to [bend left=20] (3);
\draw[thick] (1) to [bend right=20] (3);
\draw[thick, dotted] (-1,1.2) to (-0.8, 1.0);
\node[text width=1cm](15) at (-1.5, 1.7){$l-l_1$};
\draw[thick] (2) to [bend left=10] (3);
\draw[thick] (2) to [bend right=10] (3);
\draw[thick] (2) to [bend left=20] (3);
\draw[thick] (2) to [bend right=20] (3);
\draw[thick, dotted] (1,1.2) to (0.8, 1.0);
\node[text width=1cm](15) at (1.5, 1.7){$l_2-l$};
\end{tikzpicture}} \\
\hline
${\rm{III}}_{[p_1,p_2,p_3]}$  & \scalebox{0.6}{\begin{tikzpicture}[node distance=2cm,cnode/.style={circle,draw,thick,minimum size=8mm},snode/.style={rectangle,draw,thick,minimum size=8mm},pnode/.style={rectangle,red,draw,thick,minimum size=8mm}]
\node[cnode] (1) at (-3,0) {$2$};
\node[cnode] (2) at (-2,2) {$1$};
\node[cnode] (3) at (0,2.5) {$1$};
\node[cnode] (4) at (2,2) {$1$};
\node[cnode] (5) at (3,0) {$1$};
\node[cnode] (6) at (2,-2) {$1$};
\node[cnode] (7) at (0,-2.5) {$1$};
\node[cnode] (8) at (-2,-2) {$1$};
\node[snode] (10) at (-5,0) {$3$};
\node[cnode] (11) at (4,0) {$1$};
\node[cnode] (12) at (5,0) {$1$};
\node[cnode] (13) at (7.5,0) {$1$};
\node[snode] (14) at (9,0) {$1$};
\draw[thick] (1) to [bend left=40] (2);
\draw[thick] (2) to [bend left=40] (3);
\draw[thick,dashed] (3) to [bend left=40] (4);
\draw[thick] (4) to [bend left=40] (5);
\draw[thick] (5) to [bend left=40] (6);
\draw[dashed] (6) to [bend left=40] (7);
\draw[thick] (7) to [bend left=40] (8);
\draw[thick] (8) to [bend left=40] (1);
\draw[thick] (1) -- (10);
\draw[thick] (5) -- (11);
\draw[thick] (11) -- (12);
\draw[dashed] (12) -- (13);
\draw[thick] (13) -- (14);
\node[text width=1cm](11) at (-2,2.5) {$1$};
\node[text width=1cm](12) at (0, 3) {$2$};
\node[text width=1cm](13) at (2, 2.5) {$p_1$};
\node[text width=1cm](15) at (2,-2.5) {$p_2$};
\node[text width=1.5 cm](16) at (0,-3.2){$2$};
\node[text width=1cm](17) at (-2.5,-2.5){$1$};
\node[text width=0.1cm](18) at (4, 0.6){$1$};
\node[text width=0.1cm](19) at (5, 0.6){$2$};
\node[text width=1cm](20) at (7.5, 0.6){$p_3-1$};
\end{tikzpicture}} 
 & \scalebox{0.6}{\begin{tikzpicture}[node distance=2cm,cnode/.style={circle,draw,thick,minimum size=8mm},snode/.style={rectangle,draw,thick,minimum size=8mm},pnode/.style={rectangle,red,draw,thick,minimum size=8mm}]
\node[cnode] (1) at (-2,4) {$1$};
\node[cnode] (2) at (-2,0) {$2$};
\node[cnode] (3) at (2,0) {$2$};
\node[cnode] (4) at (2,4) {$1$};
\node[snode] (5) at (-2,6) {$p_2$};
\node[snode] (6) at (-2,-2) {$1$};
\node[snode] (8) at (2,6) {$p_1$};
\node[snode] (7) at (2,-2) {$1$};
\draw[thick] (1) -- (2);
\draw[thick] (2) -- (3);
\draw[thick] (3) -- (4);
\draw[thick] (1) to [bend left=10] (4);
\draw[thick] (1) to [bend right=10] (4);
\draw[thick] (1) to [bend left=20] (4);
\draw[thick] (1) to [bend right=20] (4);
\draw[thick] (1) -- (5);
\draw[thick] (2) -- (6);
\draw[thick] (3) -- (7);
\draw[thick] (4) -- (8);
\draw[thick, dotted] (0,4.2) to (0,3.8);
\node[text width=1cm](15) at (0.2, 3){$p_3$};
\end{tikzpicture}}\\
\hline
${\rm{IV}}_{[p_1,p_2,p_3]}$  & \scalebox{0.6}{\begin{tikzpicture}[node distance=2cm,cnode/.style={circle,draw,thick,minimum size=8mm},snode/.style={rectangle,draw,thick,minimum size=8mm},pnode/.style={rectangle,red,draw,thick,minimum size=8mm}]
\node[cnode] (1) at (-3,0) {$2$};
\node[cnode] (2) at (-2,2) {$1$};
\node[cnode] (3) at (0,2.5) {$1$};
\node[cnode] (4) at (2,2) {$1$};
\node[cnode] (5) at (3,0) {$1$};
\node[cnode] (6) at (2,-2) {$1$};
\node[cnode] (7) at (0,-2.5) {$1$};
\node[cnode] (8) at (-2,-2) {$1$};
\node[snode] (10) at (-5,0) {$3$};
\node[cnode] (11) at (4,0) {$1$};
\node[cnode] (12) at (5,0) {$1$};
\node[cnode] (13) at (7.5,0) {$1$};
\node[snode] (14) at (9,0) {$1$};
\draw[thick] (1) to [bend left=40] (2);
\draw[thick] (2) to [bend left=40] (3);
\draw[thick,dashed] (3) to [bend left=40] (4);
\draw[thick] (4) to [bend left=40] (5);
\draw[thick] (5) to [bend left=40] (6);
\draw[dashed] (6) to [bend left=40] (7);
\draw[thick] (7) to [bend left=40] (8);
\draw[thick] (8) to [bend left=40] (1);
\draw[thick] (1) -- (10);
\draw[thick] (5) -- (11);
\draw[thick] (11) -- (12);
\draw[dashed] (12) -- (13);
\draw[thick] (13) -- (14);
\node[text width=1cm](11) at (-2,2.5) {$1$};
\node[text width=1cm](12) at (0, 3) {$2$};
\node[text width=1cm](13) at (2, 2.5) {$p_2-1$};
\node[text width=1cm](15) at (2,-2.5) {$p_1$};
\node[text width=1.5 cm](16) at (0,-3.2){$2$};
\node[text width=1cm](17) at (-2.5,-2.5){$1$};
\node[text width=0.1cm](18) at (4, 0.6){$1$};
\node[text width=0.1cm](19) at (5, 0.6){$2$};
\node[text width=1cm](20) at (7.5, 0.6){$p_3-1$};
\node[text width=1cm](21) at (-2.5, 1){$\CA$};
\end{tikzpicture}} 
 & \scalebox{0.6}{\begin{tikzpicture}[
cnode/.style={circle,draw,thick, minimum size=1.0cm},snode/.style={rectangle,draw,thick,minimum size=1cm}]
\node[cnode] (9) at (0,1){1};
\node[snode] (10) at (0,-1){1};
\node[cnode] (11) at (2, 0){2};
\node[cnode] (12) at (4, 1){1};
\node[cnode] (13) at (4, -1){1};
\node[snode] (14) at (6, 1){$p_2$};
\node[snode] (15) at (6, -1){$p_3$};
\draw[-] (9) -- (11);
\draw[-] (10) -- (11);
\draw[-] (12) -- (11);
\draw[-] (13) -- (11);
\draw[-] (12) -- (14);
\draw[-] (13) -- (15);
\draw[thick] (12) to [bend left=10] (13);
\draw[thick] (12) to [bend right=10] (13);
\draw[thick] (12) to [bend left=20] (13);
\draw[thick] (12) to [bend right=20] (13);
\draw[thick, dotted] (3.85,0) to (4.2, 0);
\node[text width=0.1cm](20) at (4.5,0){$p_1$};
\node[text width=0.1cm](21)[above=0.2 cm of 9]{1};
\node[text width=0.1cm](22)[above=0.2 cm of 11]{2};
\node[text width=0.1cm](23)[above=0.2 cm of 12]{3};
\node[text width=0.1cm](24)[below=0.05 cm of 13]{4};
\end{tikzpicture}}\\
\hline
\end{tabular}
\caption{Summary table of non-$ADE$ families of quiver gauge theories which are mirror dual to each other. The 3d $\CN=4$ quiver notation, 
used throughout this paper, is explained in \figref{genericquiver}. In particular, the line labelled $\CA$ in quiver $X$ of Family IV denotes 
a hypermultiplet which transforms in the rank-2 antisymmetric representation of the $U(2)$ gauge group (i.e. has charge 2 under the $U(1)\subset U(2)$ and is a singlet 
under the $SU(2)$) and has charge 1 under the $U(1)$.}
\label{Tab:SummaryMain}
\end{table}
\end{center}

\begin{center}
\begin{table}[h]
\begin{tabular}{|c|c|c|}
\hline
AD Theory & 3d mirror from Class $\CS$ & Proposed Lagrangian  \\
\hline \hline 
$(A_2, A_{3p-1})$ 
&\scalebox{0.6}{\begin{tikzpicture}[node distance=2cm,cnode/.style={circle,draw,thick,minimum size=8mm},snode/.style={rectangle,draw,thick,minimum size=8mm},pnode/.style={rectangle,red,draw,thick,minimum size=8mm}]
\node[cnode] (1) at (-2,0) {$1$};
\node[cnode] (2) at (2,0) {$1$};
\node[snode] (3) at (-2,-2) {$p$};
\node[snode] (4) at (2,-2) {$p$};
\draw[thick] (1) -- (3);
\draw[thick] (2) -- (4);
\draw[thick,dotted] (0,0.4) to (0, -0.4);
\draw[thick] (1) to [bend left=20] (2);
\draw[thick] (1) to [bend right=20] (2);
\draw[thick] (1) to [bend left=40] (2);
\draw[thick] (1) to [bend right=40] (2);
\node[text width=0.1cm](15) at (0,1){$p$};
\end{tikzpicture}}
&\scalebox{0.6}{\begin{tikzpicture}[node distance=2cm,cnode/.style={circle,draw,thick,minimum size=8mm},snode/.style={rectangle,draw,thick,minimum size=8mm},pnode/.style={rectangle,red,draw,thick,minimum size=8mm}]
\node[cnode] (1) at (-2,0) {$1$};
\node[cnode] (2) at (-1,1) {$1$};
\node[cnode] (3) at (1,1) {$1$};
\node[cnode] (4) at (2,0) {$1$};
\node[cnode] (5) at (1,-1) {$1$};
\node[cnode] (6) at (-1,-1) {$1$};
\node[cnode] (7) at (3,0) {$1$};
\node[cnode] (8) at (4,0) {$1$};
\node[cnode] (9) at (5,0) {$1$};
\node[cnode] (10) at (7,0) {$1$};
\node[snode] (11) at (8,0) {$1$};
\node[snode] (12) at (-2,-2) {$1$};
\draw[thick] (1) to [bend left=40] (2);
\draw[dashed,thick] (2) to [bend left=40] (3);
\draw[thick] (3) to [bend left=40] (4);
\draw[thick] (4) to [bend left=40] (5);
\draw[dashed,thick] (5) to [bend left=40] (6);
\draw[thick] (6) to [bend left=40] (1);
\draw[thick] (1) -- (12);
\draw[thick] (4) -- (7);
\draw[thick] (7) -- (8);
\draw[thick] (8) -- (9);
\draw[thick, dashed] (9) -- (10);
\draw[thick] (10) -- (11);
\node[text width=1cm](20) at (-2,0.5) {$p$};
\node[text width=1cm](21) at (-1,1.5) {$p+1$};
\node[text width=1cm](22) at (1,1.5) {$2p-1$};
\node[text width=1cm](23) at (2,0.5) {$2p$};
\node[text width=0.1cm](24) at (1,-1.5) {$1$};
\node[text width=1cm](25) at (-1,-1.5) {$p-1$};
\node[text width=0.1cm](26) at (3,0.7) {$1$};
\node[text width=0.1cm](27) at (4,0.7) {$2$};
\node[text width=0.1cm](28) at (5,0.7) {$3$};
\node[text width=1 cm](29) at (7,0.7) {$p-2$};
\end{tikzpicture}} \\
\hline
$(A_3, A_{4p-1})$ & \scalebox{.6}{\begin{tikzpicture}[node distance=2cm,cnode/.style={circle,draw,thick,minimum size=8mm},snode/.style={rectangle,draw,thick,minimum size=8mm},pnode/.style={rectangle,red,draw,thick,minimum size=8mm}]
\node[cnode] (1) at (-2,0) {$1$};
\node[cnode] (2) at (2,0) {$1$};
\node[cnode] (3) at (0,2) {$1$};
\node[snode] (4) at (-2,-2) {$p$};
\node[snode] (5) at (2,-2) {$p$};
\node[snode] (6) at (0,4) {$p$};
\draw[thick] (1) -- (4);
\draw[thick] (2) -- (5);
\draw[thick] (3) -- (6);
\draw[thick] (1) to [bend left=10] (2);
\draw[thick] (1) to [bend right=10] (2);
\draw[thick] (1) to [bend left=20] (2);
\draw[thick] (1) to [bend right=20] (2);
\draw[thick, dotted] (0,0.2) to (0,-0.2);
\node[text width=0.1cm](15) at (0,-1){$p$};
\draw[thick] (1) to [bend left=10] (3);
\draw[thick] (1) to [bend right=10] (3);
\draw[thick] (1) to [bend left=20] (3);
\draw[thick] (1) to [bend right=20] (3);
\draw[thick, dotted] (-1,1.2) to (-0.8, 1.0);
\node[text width=0.1cm](15) at (-1, 1.7){$p$};
\draw[thick] (2) to [bend left=10] (3);
\draw[thick] (2) to [bend right=10] (3);
\draw[thick] (2) to [bend left=20] (3);
\draw[thick] (2) to [bend right=20] (3);
\draw[thick, dotted] (1,1.2) to (0.8, 1.0);
\node[text width=0.1cm](15) at (1, 1.7){$p$};
\end{tikzpicture}}
& \scalebox{0.6}{\begin{tikzpicture}[node distance=2cm,cnode/.style={circle,draw,thick,minimum size=8mm},snode/.style={rectangle,draw,thick,minimum size=8mm},pnode/.style={rectangle,red,draw,thick,minimum size=8mm}]
\node[cnode] (1) at (-3,0) {$1$};
\node[cnode] (2) at (-2,2) {$1$};
\node[cnode] (3) at (0,2.5) {$1$};
\node[cnode] (4) at (2,2) {$1$};
\node[cnode] (5) at (3,0) {$1$};
\node[cnode] (6) at (2,-2) {$1$};
\node[cnode] (7) at (0,-2.5) {$1$};
\node[cnode] (8) at (-2,-2) {$1$};
\node[cnode] (9) at (0,1.5) {$1$};
\node[snode] (10) at (-5,0) {$1$};
\node[cnode] (11) at (4,0) {$1$};
\node[cnode] (12) at (5,0) {$1$};
\node[cnode] (13) at (8,0) {$1$};
\node[snode] (14) at (9,0) {$1$};
\node[cnode] (15) at (0,0.5) {$1$};
\node[cnode] (16) at (0,-1.5) {$1$};
\draw[thick] (1) to [bend left=40] (2);
\draw[thick, dashed] (2) to [bend left=40] (3);
\draw[thick] (3) to [bend left=40] (4);
\draw[thick, dashed] (4) to [bend left=40] (5);
\draw[thick] (5) to [bend left=40] (6);
\draw[thick,dashed] (6) to [bend left=40] (7);
\draw[thick,dashed] (7) to [bend left=40] (8);
\draw[thick] (8) to [bend left=40] (1);
\draw[thick] (1) -- (10);
\draw[thick] (5) -- (11);
\draw[thick] (3) -- (9);
\draw[thick] (9) -- (15);
\draw[thick] (11) -- (12);
\draw[thick,dashed] (12) -- (13);
\draw[thick] (13) -- (14);
\draw[thick, dashed] (15) -- (16);
\draw[thick] (16) -- (7);
\node[text width=1cm](10) at (-3,0.5) {$2p$};
\node[text width=1cm](11) at (-2,2.5) {$2p+1$};
\node[text width=1cm](12) at (0, 3) {$3p$};
\node[text width=1cm](13) at (2, 2.5) {$3p+1$};
\node[text width=1cm](14) at (3, 0.5) {$4p$};
\node[text width=1cm](15) at (2,-2.5) {$1$};
\node[text width=0.1 cm](16) at (0,-3.2){$p$};
\node[text width=1cm](17) at (-2.5,-2.6){$2p-1$};
\node[text width=0.1cm](18) at (4, 0.6){1};
\node[text width=0.1cm](19) at (5, 0.6){2};
\node[text width=1cm](20) at (8, 0.6){$p-2$};
\node[text width=1cm](21) at (1,1.5){$p-1$};
\node[text width=1cm](22) at (1,0.5){$p-2$};
\node[text width=1cm](23) at (1,-1.5){$1$};
\end{tikzpicture}} \\
\hline
$(A_2, D_{3p +2})$ & \scalebox{.7}{\begin{tikzpicture}[node distance=2cm,cnode/.style={circle,draw,thick,minimum size=8mm},snode/.style={rectangle,draw,thick,minimum size=8mm},pnode/.style={rectangle,red,draw,thick,minimum size=8mm}]
\node[cnode] (1) at (-2,0) {$1$};
\node[cnode] (2) at (2,0) {$1$};
\node[cnode] (3) at (0,2) {$1$};
\node[snode] (4) at (-2,-2) {$p$};
\node[snode] (5) at (2,-2) {$p$};
\node[snode] (6) at (0,4) {$1$};
\draw[thick] (1) -- (4);
\draw[thick] (2) -- (5);
\draw[thick] (3) -- (6);
\draw[thick] (1) to [bend left=10] (2);
\draw[thick] (1) to [bend right=10] (2);
\draw[thick] (1) to [bend left=20] (2);
\draw[thick] (1) to [bend right=20] (2);
\draw[thick, dotted] (0,0.2) to (0,-0.2);
\node[text width=0.1cm](15) at (0,-1){$p$};
\draw[thick] (1) -- (3);
\draw[thick] (2) -- (3);
\node[text width=0.1cm](16) at (0,-2){$(X)$};
\end{tikzpicture}}
& \scalebox{0.7}{\begin{tikzpicture}[node distance=2cm,cnode/.style={circle,draw,thick,minimum size=8mm},snode/.style={rectangle,draw,thick,minimum size=8mm},pnode/.style={rectangle,red,draw,thick,minimum size=8mm}]
\node[cnode] (1) at (-2,0) {$1$};
\node[cnode] (2) at (-1,1) {$1$};
\node[cnode] (3) at (1,1) {$1$};
\node[cnode] (4) at (2,0) {$1$};
\node[cnode] (5) at (1,-1) {$1$};
\node[cnode] (6) at (-1,-1) {$1$};
\node[cnode] (7) at (3,0) {$1$};
\node[cnode] (8) at (4,0) {$1$};
\node[cnode] (9) at (5,0) {$1$};
\node[cnode] (10) at (7,0) {$1$};
\node[snode] (11) at (8,0) {$1$};
\node[snode] (12) at (-2,-2) {$1$};
\draw[thick] (1) to [bend left=40] (2);
\draw[dashed,thick] (2) to [bend left=40] (3);
\draw[thick] (3) to [bend left=40] (4);
\draw[thick] (4) to [bend left=40] (5);
\draw[dashed,thick] (5) to [bend left=40] (6);
\draw[thick] (6) to [bend left=40] (1);
\draw[thick] (2) to [bend left=40] (6);
\draw[thick] (1) -- (12);
\draw[thick] (4) -- (7);
\draw[thick] (7) -- (8);
\draw[thick] (8) -- (9);
\draw[thick, dashed] (9) -- (10);
\draw[thick] (10) -- (11);
\node[text width=1cm](20) at (-2,0.5) {$p+1$};
\node[text width=1cm](21) at (-1,1.5) {$p+2$};
\node[text width=1cm](22) at (1,1.5) {$2p+1$};
\node[text width=1cm](23) at (2,0.5) {$2p+2$};
\node[text width=0.1cm](24) at (1,-1.5) {$1$};
\node[text width=1cm](25) at (-1,-1.5) {$p$};
\node[text width=0.1cm](26) at (3,0.7) {$1$};
\node[text width=0.1cm](27) at (4,0.7) {$2$};
\node[text width=0.1cm](28) at (5,0.7) {$3$};
\node[text width=1 cm](29) at (7,0.7) {$p-2$};
\node[text width=0.1cm](30) at (3,-3){$(Y)$};
\end{tikzpicture}}\\
\hline
$A^{{\rm maximal}}_{2,p}$ & \scalebox{0.6}{\begin{tikzpicture}[
cnode/.style={circle,draw,thick, minimum size=1.0cm},snode/.style={rectangle,draw,thick,minimum size=1cm}]
\node[cnode] (9) at (0,1){1};
\node[snode] (10) at (0,-1){1};
\node[cnode] (11) at (2, 0){2};
\node[cnode] (12) at (4, 1){1};
\node[cnode] (13) at (4, -1){1};
\node[snode] (14) at (6, 1){$p$};
\node[snode] (15) at (6, -1){$p$};
\node[] (25) at (6, -3){};
\node[] (26) at (0, -3){};
\draw[-] (9) -- (11);
\draw[-] (10) -- (11);
\draw[-] (12) -- (11);
\draw[-] (13) -- (11);
\draw[-] (12) -- (14);
\draw[-] (13) -- (15);
\draw[thick] (12) to [bend left=10] (13);
\draw[thick] (12) to [bend right=10] (13);
\draw[thick] (12) to [bend left=20] (13);
\draw[thick] (12) to [bend right=20] (13);
\draw[thick, dotted] (3.85,0) to (4.2, 0);
\node[text width=0.1cm](20) at (4.5,0){$p$};
\end{tikzpicture}}
&\scalebox{0.6}{\begin{tikzpicture}[node distance=2cm,cnode/.style={circle,draw,thick,minimum size=8mm},snode/.style={rectangle,draw,thick,minimum size=8mm},pnode/.style={rectangle,red,draw,thick,minimum size=8mm}]
\node[cnode] (1) at (-3,0) {$2$};
\node[cnode] (2) at (-2,2) {$1$};
\node[cnode] (3) at (0,2.5) {$1$};
\node[cnode] (4) at (2,2) {$1$};
\node[cnode] (5) at (3,0) {$1$};
\node[cnode] (6) at (2,-2) {$1$};
\node[cnode] (7) at (0,-2.5) {$1$};
\node[cnode] (8) at (-2,-2) {$1$};
\node[snode] (10) at (-5,0) {$3$};
\node[cnode] (11) at (4,0) {$1$};
\node[cnode] (12) at (5,0) {$1$};
\node[cnode] (13) at (7.5,0) {$1$};
\node[snode] (14) at (9,0) {$1$};
\draw[thick] (1) to [bend left=40] (2);
\draw[thick] (2) to [bend left=40] (3);
\draw[thick,dashed] (3) to [bend left=40] (4);
\draw[thick] (4) to [bend left=40] (5);
\draw[thick] (5) to [bend left=40] (6);
\draw[dashed] (6) to [bend left=40] (7);
\draw[thick] (7) to [bend left=40] (8);
\draw[thick] (8) to [bend left=40] (1);
\draw[thick] (1) -- (10);
\draw[thick] (5) -- (11);
\draw[thick] (11) -- (12);
\draw[dashed] (12) -- (13);
\draw[thick] (13) -- (14);
\node[text width=1cm](11) at (-2,2.5) {$1$};
\node[text width=1cm](12) at (0, 3) {$2$};
\node[text width=1cm](13) at (2, 2.5) {$p-1$};
\node[text width=1cm](15) at (2,-2.5) {$p$};
\node[text width=1.5 cm](16) at (0,-3.2){$2$};
\node[text width=1cm](17) at (-2.5,-2.5){$1$};
\node[text width=0.1cm](18) at (4, 0.6){$1$};
\node[text width=0.1cm](19) at (5, 0.6){$2$};
\node[text width=1cm](20) at (7.5, 0.6){$p-1$};
\node[text width=1cm](21) at (-2.5, 1){$\CA$};
\end{tikzpicture}}\\
\hline
\end{tabular}
\caption{Summary table of 3d mirror pairs associated with certain Argyres-Douglas theories realized in the class $\CS$ construction. The third column tabulates the proposed 
Lagrangians for the 3d SCFTs obtained by putting the Argyres-Douglas theory on a circle and flowing to the deep IR. The theory labelled $A^{{\rm maximal}}_{2,p}$ in the fourth row is an AD theory realized by the twisted compactification of a 6d (2,0) $A_2$ theory on a Riemann sphere with an irregular puncture and a maximal regular puncture.}
\label{Tab:SummaryAD}
\end{table}
\end{center}

\subsection{Future directions}
Before embarking on the main text of the paper, we would like to briefly comment on certain issues that have not been addressed in this 
work, but will be discussed in upcoming papers.
\begin{itemize}
\item {\bf{Examples of dualities from non-Abelian $S$-type operations:}} In this paper, we have restricted ourselves to examples of mirror symmetry 
which can be constructed using Abelian $S$-type operations only. In a paper currently under preparation \cite{Dey:2020xyz}, we show that a generic Non-Abelian $S$-type operation can be  ``abelianized" i.e. written in terms of a set of Abelian $S$-type operations. A much larger class of mirror duals can be 
generated using the construction above, once these Non-Abelian $S$-type operations are incorporated.

\item {\bf{Duality maps for defects beyond linear quivers:}} Even for Abelian $S$-type operations discussed in this paper, we have chosen
not to turn on the defect operations. Incorporating them gives a powerful tool for analyzing the map of defect operators across the duality.
This is especially relevant for dualities beyond linear quivers, where the aforementioned defect operators do not have a known realization 
in Type IIB String Theory. Defects for Non-Abelian $S$-type operations enrich the story further. We explore these directions in a work currently in
progress.

\item {\bf{3d $\CN=2$ dualities:}} Generic dualities in 3d $\CN=2$ gauge theories with Chern-Simons terms can also be generated from a set of basic 
dualities, in a way analogous to the construction presented here. This brings into play the $T$-type operations, in addition to the 
$S$-type operations. In these classes of dualities, there will be additional constraints on the $S$-type operations (in gauging a certain 
flavor symmetry) coming from the parity anomaly. 

\end{itemize}

\section{$\CN=4$ mirror symmetry and linear quivers: Generalities}\label{Review}
In this section, we review some of the important basic concepts and computational tools for three dimensional 
gauge theories with eight real supercharges, which will be useful in the rest of the paper. The reader 
familiar with 3d $\CN=4$ physics can skip to \Secref{GFI-main}.

\subsection{Supermultiplets and Lagrangian description}

\textbf{Supersymmetry and supermultiplets:} $\CN=4$ supersymmetry in three dimensions has 8 real supercharges. We will work in terms of complex supercharges $Q_{\alpha A A'}$ which are doublets of $Spin(2,1)\sim SL(2,\mathbb{R})$ (indexed by $\alpha=1,2$) and transform as $(2,2)$ under the R-symmetry group, $SU(2)_H \times SU(2)_C$ (indexed by $A=1,2$ and $A'=1,2$ respectively). The complex supercharges on $\BR^{1,2}$ (i.e. with signature $(-,+,+)$) generate the following supersymmetry algebra:
\be 
\{ Q_{\alpha A A'}, Q_{\beta B B'}\}= (\gamma_\mu C)_{\alpha \beta} P^\mu \epsilon_{AB} \epsilon_{A'B'},
\ee
where $\gamma_\mu= (\gamma_0, \gamma_1, \gamma_2)= (i\tau_3, \tau_1, \tau_2)$ ($\tau_i$ being the standard Pauli matrices), and the charge conjugation matrix $C=\tau_2$. The complex supercharges obey the reality condition $Q^{\dagger}_{\alpha A A'}= (\tau_1)_\alpha^{\,\,\beta} \epsilon^{AB} \epsilon^{A' B'} Q_{\beta B B'}$. 

A Lagrangian theory with $\CN=4$ supersymmetry consists of a vector multiplet in the adjoint representation of a gauge group $G$, and hypermultiplets in a given quaternionic representation of $G$. The field content of the vector multiplet and the hypermultiplet can be obtained from dimensional reduction of the 4d $\CN=2$ vector multiplet and hypermultiplet respectively, and is given by the second column of Table \ref{Tab:globalsym}\footnote{A 3d gauge field can be dualized to a circle-valued scalar only for an Abelian gauge group. But a non-Abelian gauge group is Higgsed to at most an Abelian subgroup, at a generic point on the moduli space. Therefore, the counting can be carried over to describe the low energy theory.}. The third column of Table \ref{Tab:globalsym} lists the representations of the $SU(2)_H \times SU(2)_C$ R-symmetry in which the constituent fields transform. Note that the supermultiplets are not symmetric with respect to the $SU(2)_H$ and the $SU(2)_C$ representations, and this allows one to define corresponding ``twisted" multiplets, i.e. multiplets where the representations of the two $SU(2)$s are exchanged. In addition, the theory has a global symmetry group $G_H \times G_C$, that we will describe below. 
\begin{table}[htbp]
\begin{center}
\begin{tabular}{|c|c|c|}
\hline
Supermultiplets & Constituent Fields &  $SU(2)_H \times SU(2)_C$  \\
\hline \hline 
Vector & \begin{tabular}{@{}c@{}} Bosons ($\sigma,\Phi, \gamma$) \\  Fermions ($\lambda_{AA'}$) \end{tabular} &  \begin{tabular}{@{}c@{}} $(1, 3\oplus1)$ \\  $(2,2)$  \end{tabular} \\
\hline
Hyper & \begin{tabular}{@{}c@{}} Bosons($\phi, \, \wt{\phi}^\dagger$) \\  Fermions ($\psi, \, \wt{\psi}^\dagger$)\end{tabular} & \begin{tabular}{@{}c@{}} (2,1) \\ (1,2) \end{tabular} \\
\hline
\end{tabular}
\caption{The field content of three dimensional $\CN=4$ vector multiplet and hypermultiplet, along with the representations of R-symmetry in which they transform. Note that for the vector multiplet bosons, $\sigma$ is a real non-compact scalar, $\Phi$ is a complex scalar, and $\gamma$ is a real compact scalar dual to the 3d gauge field. For the hypermultiplet bosons, $\phi$ and $\wt{\phi}$ are both complex scalars.}
\label{Tab:globalsym}
\end{center}
\end{table}

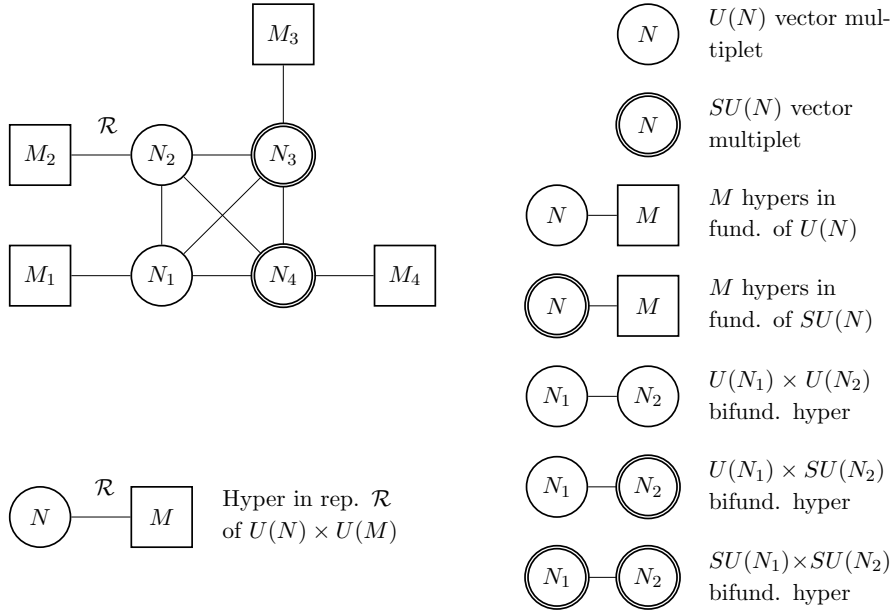
\begin{figure}[htbp]
\begin{center}
\scalebox{0.8}{\begin{tikzpicture}[node distance=2cm,
cnode/.style={circle,draw,thick, minimum size=1.0cm},snode/.style={rectangle,draw,thick,minimum size=1cm}, pnode/.style={circle,double,draw,thick, minimum size=1.0cm}]
\node[cnode] (1) at (-2,0) {$N_1$};
\node[cnode] (2) at (-2,2) {$N_2$};
\node[pnode] (3) at (0,2) {$N_3$};
\node[pnode] (4) at (0,0) {$N_4$};
\node[snode] (5) at (-4,0) {$M_1$};
\node[snode] (6) at (-4,2) {$M_2$};
\node[snode] (7) at (0,4) {$M_3$};
\node[snode] (8) at (2,0) {$M_4$};
\draw[-] (1) -- (2);
\draw[-] (2) -- (3);
\draw[-] (3) -- (4);
\draw[-] (1) -- (4);
\draw[-] (1) -- (5);
\draw[-] (2) -- (6);
\draw[-] (3) -- (7);
\draw[-] (4) -- (8);
\draw[-] (2) -- (4);
\draw[-] (1) -- (3);
\node[cnode] (9) at (6,4) {$N$};
\node[text width=3cm](10) at (8.5, 4){$U(N)$ vector multiplet};
\node[pnode] (11) at (6,2.5) {$N$};
\node[text width=3cm](12) at (8.5, 2.5){$SU(N)$ vector multiplet};
\node[snode] (13) at (6,1) {$M$};
\node[cnode] (14) at (4.5,1) {$N$};
\draw[-] (13)--(14);
\node[text width=3cm](15) at (8.5, 1){$M$ hypers in fund. of $U(N)$};
\node[snode] (16) at (6, -0.5) {$M$};
\node[pnode] (17) at (4.5, -0.5) {$N$};
\draw[-] (16)--(17);
\node[text width=3cm](18) at (8.5, - 0.5){$M$ hypers in fund. of $SU(N)$};
\node[cnode] (19) at (6, -2) {$N_2$};
\node[cnode] (20) at (4.5, -2) {$N_1$};
\draw[-] (19)--(20);
\node[text width=3cm](21) at (8.5, -2){$U(N_1) \times U(N_2)$ bifund. hyper};
\node[pnode] (22) at (6, -3.5) {$N_2$};
\node[cnode] (23) at (4.5, -3.5) {$N_1$};
\draw[-] (22)--(23);
\node[text width=3cm](24) at (8.5, -3.5){$U(N_1) \times SU(N_2)$ bifund. hyper};
\node[pnode] (25) at (6, -5) {$N_2$};
\node[pnode] (26) at (4.5, -5) {$N_1$};
\draw[-] (25)--(26);
\node[text width=3cm](27) at (8.5, -5){$SU(N_1) \times SU(N_2)$ bifund. hyper};
\node[cnode] (31) at (-4,-4) {$N$};
\node[snode] (32) at (-2,-4) {$M$};
\draw[-] (31)--(32);
\node[text width=0.2cm](33) at (-3, -3.5){$\CR$};
\node[text width=3cm](34) at (0.5, -4){Hyper in rep. $\CR$ of $U(N) \times U(M)$};
\node[text width=0.1cm](34) at (-3,2.5){$\CR$};
\end{tikzpicture}}
\caption{LHS: A quiver diagram representing the field content of a 3d $\CN=4$ theory with gauge group $G=U(N_1) \times U(N_2) \times SU(N_3) \times SU(N_4)$, 
and fundamental/bifundamental matter. The various conventions are listed on the RHS. In a quiver diagram, we will refer to the circles as gauge nodes and the 
boxes as flavor nodes.}
\label{genericquiver}
\end{center}
\end{figure}

In this paper, we will only consider theories with unitary (or special unitary) gauge groups with matter in a given representation $\CR$ of the gauge and global symmetries. For a given theory, the field content is most conveniently represented by a 3d $\CN=4$ quiver diagram -- the conventions are explained in terms of an illustrative example 
in \figref{genericquiver}.\\

\noindent \textbf{Lagrangian description:}  The action for the $\mathcal{N}=4$ supersymmetric gauge theories is most conveniently presented in the 3d $\mathcal{N}=2$ superspace language (see \cite{Intriligator:2013lca} for a recent review). An $\mathcal{N}=4$ vector multiplet consists of one $\mathcal{N}=2$ vector multiplet $V$ and one $\mathcal{N}=2$ chiral multiplet $\Phi$ in the adjoint of the gauge group. The action for an $\mathcal{N}=4$ quiver gauge theory on $\BR^{1,2}$ consists of the following terms \cite{deBoer:1996ck, Kapustin:2010xq}:
\begin{itemize}
\item A Super-Yang-Mills term for the gauge group $G$: 
\begin{align}
S_{\rm SYM}[V, \Phi]= & \frac{1}{g_{YM}^2}\int \,d^3x \,d^2\theta \,d^2\wt{\theta}\, \tr \,\Big(- \frac{1}{4} \Sigma^2-\Phi^{\dagger} e^{2V} \Phi \Big),
\end{align}
where $``\tr"$ is an invariant inner product on the Lie algebra of $G$, $V$ is an $\mathcal{N}=2$ vector multiplet and $\Phi$ is an $\mathcal{N}=2$ adjoint chiral 
multiplet inside the $\mathcal{N}=4$ vector multiplet, while $\Sigma$ is an $\mathcal{N}=2$ multiplet (often called a linear multiplet), which 
includes the field strength of the gauge field in $V$. In terms of V, the superfield $\Sigma$ is defined as:
\begin{align}
& \Sigma = -\frac{i}{2} \epsilon^{\alpha \beta} \wt{D}_\alpha D_\beta V\Big(x^\mu, \theta_\alpha, \wt{\theta}_\alpha\Big), \\
& D_\alpha = \frac{\partial}{\partial \theta^\alpha} + i \gamma^\mu_{\alpha \beta} \wt{\theta}^\beta \partial_\mu, \nn \\
& \wt{D}_\alpha= -\frac{\partial}{\partial \wt{\theta}^\alpha} - i \theta^\beta \gamma^\mu_{\beta \alpha}\partial_\mu. \nn
 \end{align}

\item Kinetic terms and minimal gauge couplings for the hypermultiplets transforming in a representation $\CR$ of $G$:

\begin{equation}
S_{\rm matter}[\phi, \wt{\phi}, V]=\int  d^3x d^2\theta d^2\wt{\theta}\, \tr \Big(\phi_i^{\dagger}e^{2V} \phi^i+\wt{\phi}_i^{\dagger}e^{(-2V)} \wt{\phi}^i \Big)
\end{equation}

where $\phi_i,\tilde{\phi}_i$ are $\mathcal{N}=2$ chiral multiplets  constituting a $\mathcal{N}=4$ hypermultiplet with $i$ being the global symmetry index.

\item A holomorphic superpotential term for the $\mathcal{N}=2$ chiral multiplets, which preserves $\mathcal{N}=4$ supersymmetry.

\begin{equation}
S_{\rm sup}[\phi, \wt{\phi}, \Phi]= i \sqrt{2} \int  d^3x d^2\theta \tr \Big(\tilde{\phi}_i\Phi\phi^i \Big) + h.c.\, .
\end{equation}

\end{itemize}

In addition, there are two possible $\CN=4$-preserving deformations of the theory, which correspond to turning on mass terms for the hypermultiplets and Fayet-Iliopoulos (FI) terms for the $U(1)$ factors in the gauge group $G$.

\begin{itemize}
\item The hypermultiplet masses transform as triplets under $SU(2)_C$ and can be interpreted as the scalar components of a background $\mathcal{N}=4$ vector multiplet 
associated with the Cartan subalgebra of the global symmetry group $G_{H}$. They couple to the flavor symmetry current in the standard way:

\begin{align}
S_{\rm mass}[\phi, \wt{\phi}, V_m, \Phi_m]=&\int  d^3x d^2\theta d^2\wt{\theta}\, \tr \Big(\phi_i^{\dagger} e^{2V^{ij}_m} \phi_j+\tilde{\phi}_i^{\dagger} e^{(-2V^{ij}_m)} \tilde{\phi}_j \Big) \nn \\
& + i\sqrt{2} \int  d^3x d^2\theta \tr \Big(\tilde{\phi}_i\Phi^{ij}_m\phi_j \Big) + h.c.
\end{align}

where $V^{ij}_m = V^a_m T_a^{ij}$ and $\Phi^{ij}_m= \Phi^a_m T_a^{ij}$ are respectively the $\mathcal{N}=2$ vector and chiral multiplet which make up the $\mathcal{N}=4$ background vector multiplet, with $T_a^{ij}$ being a Cartan generator of the Lie algebra of $G_{H}$. The $\tr$ is an invariant inner product on the Lie algebra of the gauge group $G$, as before.

\item The FI parameters transform as triplets under $SU(2)_H$ and can be thought of as the scalar components of a \emph{twisted} $\mathcal{N}=4$ background vector multiplet 
associated with the Cartan subalgebra of the global symmetry group $G_{C}$. They couple to the 3d topological currents for the $U(1)$ factors in the gauge group $G$ by a BF term:
\begin{equation}
S_{\rm FI}[V, \Phi,  \hat{V}_{\rm FI}, \hat{\Phi}_{\rm FI}]=\int  d^3x d^2\theta d^2\wt{\theta}\, \tr \,\Sigma \hat{V}_{\rm FI} + \int  d^3x d^2\theta \, \tr \, \Phi\hat{\Phi}_{\rm FI}+ h.c.
\end{equation}
where $\hat{V}_{\rm FI}, \hat{\Phi}_{\rm FI}$ denote the twisted $\CN=2$ vector and chiral multiplet that constitute an $\CN=4$ twisted vector multiplet.
\end{itemize}

The complete 3d $\CN=4$ action is a sum of the terms listed above:
\begin{align}
S_{\CN=4}[V, \Phi, \phi, \wt{\phi},V_m, \Phi_m,  \hat{V}_{\rm FI}, \hat{\Phi}_{\rm FI}] = & S_{\rm SYM}[V, \Phi]+ S_{\rm matter}[\phi, \wt{\phi}, V]+ S_{\rm sup}[\phi, \wt{\phi}, \Phi] \nn\\
& + S_{\rm mass}[\phi, \wt{\phi}, V_m, \Phi_m]+ S_{\rm FI}[V, \Phi,  \hat{V}_{\rm FI}, \hat{\Phi}_{\rm FI}].
\end{align}

\subsection{IR physics of 3d $\CN=4$ theories and mirror symmetry}\label{GBU-GW}

\noindent \textbf{Moduli spaces and LEET:} A 3d $\CN=4$ theory has a dimensionful gauge coupling constant -- the theory is asymptotically free in the UV 
and generically flows to a strongly-coupled interacting SCFT in the deep IR. The moduli space of vacua of these theories is a \hk manifold with certain 
distinguished branches, which we will describe momentarily.  The low energy effective theory (LEET) around any point on the moduli space gives the
physics at the associated energy scale. In particular, the low energy theory at a generic smooth point on the moduli space is a free theory, while LEETs living 
on the singular loci of the moduli space can involve interesting SCFTs.\\

The moduli space of vacua has the following distinguished branches:
\begin{itemize}
\item \textbf{Higgs branch:} The Higgs branch $\CM_{\rm H}$ corresponds to the branch of the moduli space where all hypermultiplet scalars have non-zero vacuum expectation value (vev), all the adjoint scalars have zero vev, and generically, the gauge group is completely Higgsed. It is a \hk manifold and additionally admits a realization as a \hk quotient. 
The quaternionic dimension of the manifold is given in terms of the dimensions of the gauge group and matter representation $\CR$ (of the gauge group and the global symmetry group $G_{H}$) by the following formula:
\be
{\rm dim}\, \CM_{\rm H} = {\rm dim}_{\mathbb{C}}\,(\CR) -  {\rm dim}_{\mathbb{R}}\,(G).
\ee
The \hk metric on the Higgs branch is protected against quantum corrections by supersymmetry. There is a natural action of $SU(2)_H \times G_H$ on the Higgs branch, 
where $G_H$ is a manifest symmetry of the Lagrangian and commutes with the R-symmetry.\\
Viewed as an algebraic variety in a chosen complex structure, the associated chiral ring of holomorphic functions is generated by the expectation values of half-BPS local operators, which transform in irreps of $SU(2)_H \times G_H$ (and are singlets under $SU(2)_C$). These are built out of gauge-invariant polynomials of the hypermultiplet scalars.

\item \textbf{Coulomb branch:} The Coulomb branch $\CM_{\rm C}$ corresponds to the branch of the moduli space where the triplet of real adjoint scalars have non-zero vev, all hypermultiplet scalars have zero vev, and the gauge group $G$, at a generic point on the branch, is broken to its maximal torus. The resultant Abelian gauge fields can be dualized to scalar fields, which together with the adjoint scalars give a \hk manifold (but not necessarily a \hk quotient). The \hk metric on the manifold is \textit{not} protected against quantum corrections, and the full quantum corrected metric can be directly computed only for very special cases. The quaternionic dimension of the manifold is:
\be
{\rm dim}\, \CM_{\rm C} = {\rm rank}\, (G).
\ee
There is a natural action of $SU(2)_C \times G_C$ on the Coulomb branch, where $G_C$ is a global symmetry which commutes with the R-symmetry. In the UV, the manifest form of this global symmetry is $G_C=U(1)^k$, where $k$ is the number of U(1) factors in the gauge group $G$, while in the IR, $G_C$ can be enhanced to a nonabelian group with rank $k$.\\
An alternative and modern way to describe the Coulomb branch is given in terms of BPS monopole operators \cite{Borokhov:2002cg, Borokhov:2002ib}, which are local disorder operators very similar to the 't Hooft defects in four dimensions. In a 3d $\CN=4$ theory, a monopole operator is defined by introducing a Dirac monopole singularity, labelled by a cocharacter $\vec{B}$, for the gauge fields at an insertion point. Preserving half of the supersymmetry requires that one of the three real adjoint scalars should be singular at the insertion point. This implies introducing the following boundary condition in the 3d QFT at the insertion point $\vec x$:
\be
A_{\pm} \sim \frac{B}{2}(\pm 1 - \cos{\theta})\,d\phi, \qquad \s \sim \frac{B}{2r}, 
\ee
where $(r, \theta, \phi)$ are spherical coordinates with $\vec x$ as origin, $A_{\pm}$ are the gauge fields on the northern/southern patches of the $S^2$ with $\vec x$ as the 
center, and $\s$ is a real adjoint scalar.
The remaining two real scalar fields, combined into a single complex scalar $\Phi$, are regular and must transform in the adjoint of the subgroup of $G$ left unbroken by the monopole singularity. A monopole operator with $\Phi=0$ is referred to as a ``bare monopole operator" while those with $\Phi \neq 0$ is referred to as a ``dressed monopole operator". The monopole operators transform in irreps of $SU(2)_C \times G_C$ and are singlets under $SU(2)_H$.\\
In a chosen complex structure, the Coulomb branch can be viewed as an algebraic variety and the associated chiral ring of holomorphic functions is generated by the expectation values of the bare and dressed monopole operators described above. The choice of the complex structure, in particular, determines which of the real scalars 
is picked to be singular at the insertion point.

\item \textbf{Mixed branches:} A mixed branch $\CM_{\rm M}$ corresponds to a branch of the moduli space, where some of the hypermultiplet scalars as well as some of the adjoint scalars have non-zero vevs. The Higgsing of the gauge group on a given mixed branch depends on the precise scalar vevs turned on. \\

\end{itemize}

\noindent \textbf{Good, bad, and ugly classification:} A 3d $\CN=4$ SCFT has an $Spin(4)_{\rm IR}$ R-symmetry. Generally speaking, 
the IR R-symmetry may not coincides with the $SU(2)_H \times SU(2)_C$ R-symmetry manifest in the UV Lagrangian. Assuming that
$Spin(4)_{\rm IR} \cong SU(2)_H \times SU(2)_C $, one can check \cite{Gaiotto:2008ak} whether the various 1/2-BPS chiral primary 
operators of the SCFT satisfy the unitarity bound $\Delta \geq \frac{1}{2}$, where $\Delta$ is the conformal dimension of the operator in 
the IR. The Higgs branch chiral operators, which are built out of hypermultiplet scalars, trivially satisfy the bound. 
The Coulomb branch monopole operators, however, receive non-trivial corrections to their conformal dimension. For a  theory 
with gauge group $G$, Higgs branch global symmetry $G_H$, and hypermultiplets transforming in a representation $\CR$ of $G \times G_H$, 
the conformal dimension of a monopole operator is given as \cite{Bashkirov:2010kz, Benna:2009xd}:
\be
\Delta(B) = q_R = -  \sum_{\alpha \in \Delta^+} |\alpha(B)| + \frac{1}{2} \sum_{\rho\in \CR} | \rho (B)|,
\ee
where $B \in \Lambda_{\rm cochar}(G)$ labels the monopole operator, $q_R$ is the charge of the monopole 
operator under the $U(1)_R$ R-symmetry of the $\CN=2$ subalgebra  that it preserves, $\alpha$ is a positive root of the 
Lie algebra of $G$, and $\rho$ is a weight of the $G \times G_H$ Lie algebra in the representation $\CR$.\\

Consistency with unitarity leads to the following classification of 3d $\CN=4$ theories: 
\begin{itemize}
\item \textbf{Good theory:} A given theory is good if $\Delta(B) > \frac{1}{2}$ for all $B \in \Lambda_{\rm cochar}$. For such a 
theory, the most singular locus (i.e. singular locus of highest codimension) on the Coulomb branch is a point where the Coulomb 
and the Higgs branches intersect. The local geometry around this point is that of a conical variety and represents the moduli space 
of an $\CN=4$ SCFT for which $Spin(4)_{\rm IR} \cong SU(2)_H \times SU(2)_C$. In the deep IR, the good theory flows to this SCFT. 

\item \textbf{Ugly theory:} A given theory is ugly if there exists at least one monopole operator for which $\Delta(B)= \frac{1}{2}$, 
but none with $\Delta(B) < \frac{1}{2}$. The most singular locus on the Coulomb branch is generically not a point. However, the local geometry 
around a generic point in the most singular locus is a product of some flat directions and a conical singularity, such that the former represents 
the moduli space of certain free twisted hypermultiplets, while the latter represents the moduli space of an $\CN=4$ SCFT for which 
$Spin(4)_{\rm IR} \cong SU(2)_H \times SU(2)_C$. In the deep IR, the ugly theory therefore flows to an SCFT along with some free decoupled twisted
hypermultiplets.

\item \textbf{Bad theory:} A given theory is bad if there exits at least one unitarity violating monopole operator, i.e. for which $\Delta(B) < \frac{1}{2}$. 
Similar to ugly theories, the most singular locus on the Coulomb branch is not a point, and the local geometry around a generic point in this singular 
locus is a product of some flat directions and a conical singularity, with the former representing the moduli space of certain free twisted hypermultiplets.
However, the conical singularity represents the moduli space of an $\CN=4$ SCFT for which $Spin(4)_{\rm IR}$ does not coincide with the UV R-symmetry 
of the original theory, and is realized as an embedding inside the product of the UV R-symmetry and certain accidental flavor symmetry that appear in the IR 
\cite{Assel:2017jgo, Dey:2017fqs, Assel:2018exy}. 

\end{itemize}

\noindent \textbf{Three dimensional mirror symmetry:} An IR duality implies that a set of theories with different UV Lagrangians flow to the same IR 
SCFT. Mirror symmetry is a special case of an IR duality in three dimensions with the following properties:
\begin{itemize}

\item Given a pair of dual theories $X$ and $Y$, mirror symmetry exchanges the Coulomb and the Higgs branches in the deep IR, 
i.e. as $g^2_{YM} \to \infty$:
\be
\mathcal{M}^{(X)}_{\rm C}= \mathcal{M}^{(Y)}_{\rm H}, \qquad \mathcal{M}^{(X)}_{\rm H}= \mathcal{M}^{(Y)}_{\rm C}.
\ee

\item The duality exchanges $SU(2)_C$ and $SU(2)_H$, and therefore exchanges background vector and twisted vector multiplets.
This implies that hypermultiplet masses and FI parameters are exchanged under mirror symmetry. 

\end{itemize}
Mirror symmetry relates observables in theory $X$ with observables in theory $Y$, and the precise map is referred to as the 
``mirror map". The simplest mirror map is the one which relates hypermultiplet masses on one side of the duality with FI 
parameters on the other (or vice-versa).

\subsection{Checking mirror symmetry using RG flow invariant observables}\label{S3-review}
Localization techniques \cite{Pestun:2007rz,Kapustin:2009kz} can be used to compute exact expressions for various RG flow-invariant supersymmetric 
observables in a theory using the UV Lagrangian. A non-trivial check of an IR duality is to show that such supersymmetric observables 
for a given pair of dual theories agree. In this subsection, we present a brief review of a supersymmetric observable - the partition function of a 
3d $\CN=4$ theory on a round three sphere, which will be one of the main tools of analysis in this paper.
Consider an $\CN = 4$ quiver gauge theory with gauge group $G$ and global symmetry group $G_{H}$, 
such that the matter hypermultiplets transform in a representation $\CR$ of $G \times G_{H}$. 
We turn on background vector multiplets in the Cartan subalgebra of the global symmetry group $G_{H}$ 
(hypermultiplet masses) as well as twisted background vector multiplets in the Cartan subalgebra of the topological symmetry 
group $G_C$ (FI parameters). However, instead of the full triplet, we can only turn on a single real parameter in each case to preserve 
supersymmetry.
In addition, one can turn on various supersymmetric defects in the theory. For a generic 3d $\CN=2$ theory, one 
can also turn on Chern-Simons interaction for the gauge group. In this paper, we will focus exclusively on $\CN=4$ theories,  
for which we set the Chern-Simons level $\kappa=0$.\\

The rules for writing down the $S^3$ partition function for a generic 3d $\CN=4$ theory \cite{Kapustin:2009kz, Kapustin:2010xq} may be summarized as follows. 
Localization on $S^3$ ensures that the partition function can be written as a matrix integral in terms of a single real scalar $s$  which lives in the 
Cartan of the gauge group, where $s$ is the zero-mode associated with the real adjoint scalar that sits inside a 
3d $\mathcal{N}=2$ vector multiplet (which in turn sits inside a 3d $\CN=4$ vector multiplet). 
The partition function is a function of the real masses $\vec m$ and real FI parameters $\vec \eta$, which should be thought of as real adjoint 
scalars inside the respective background $\CN=2$ vector multiplets (which in turn sits inside a 3d $\CN=4$ vector multiplet). In presence of defects, the partition 
function will also depend on some additional data $\CD$. For example, Wilson line defects, wrapping a great circle on $S^3$, will be labelled by a representation 
$\CR_{W}$ of the gauge group.\\

Since $S^3$ does not have any instantons (unlike the case of $S^4$), the partition function may be written as a matrix integral where the integrand is built out of 
classical (FI and Chern-Simons) and one-loop contributions, as well possible supersymmetric defects : 
\begin{align}\label{PF-main}
Z(\vec m; \vec \eta; \vec k; \CD)= & \int  \Big[d\vec s\Big] \, Z_{\rm int}(\vec s, \vec m, \vec \eta,\CD) \nn \\ = & \int  \Big[d\vec s\Big] \, Z_{\rm CS}(\vec s, \vec k) \, Z_{\rm FI}(\vec s, \vec \eta)\, Z_{\rm defect}(\vec s, \CD)\, Z^{\rm vector}_{\rm{1-loop}}(\vec s)\,Z^{\rm hyper}_{\rm{1-loop}}(\vec s, \vec m),
\end{align}
where $\Big[d\vec s\Big]=\frac{d^k \vec{s}}{|{W}(G)|}$, and $|{W}(G)|$ is the order of the Weyl group of $G$. The individual terms in the integrand on the RHS are given as follows. 
\begin{itemize}

\item The Chern-Simons interactions for various factors in the gauge group gives the classical contribution:
\be \label{PF-main-CS}
Z_{\rm CS}(\vec s, \vec k) =  \prod_{\gamma} e^{2\pi i k_\gamma \tr (\vec s^{\gamma})^2},
\ee
where $\gamma$ runs over all the factors in the gauge group. 

\item The $l$ $U(1)$ factors in the gauge group gives the following classical contribution
\begin{equation} \label{PF-main-FI}
Z_{\rm FI}(\vec s, \vec \eta) = \prod^l_{\gamma=1} e^{2\pi i \eta_\gamma \, \text{Tr}(\vec{s}^\gamma)}\,,
\end{equation}
where $\gamma$ runs over the $l$ gauge nodes with a $U(1)$ factor and $\eta_\gamma$ is the associated FI parameter. 

\item The contribution of a Wilson line defect in a representation $\CR_W$ of the gauge group, is given as
\be \label{PF-main-defect}
Z_{\rm defect}(\vec s, \CD) := Z_{\rm Wilson}(\vec s, \CR_W)= \frac{1}{{\rm{dim} (\CR_W)}}\text{Tr}_{\CR_W} \Big(e^{2\pi \vec s}\Big).
\ee

\item The  $\mathcal{N}=4$ vector multiplet contributes a one-loop term:
\begin{equation}\label{PF-main-vec}
Z^{\rm vector}_{\text{1-loop}}(\vec s)=\prod_{\alpha} \sinh{\pi \alpha(\vec s)}\,,
\end{equation}
where the product extends over the roots of the Lie algebra of $G$. In fact, this is precisely the contribution of an $\mathcal{N}=2$ vector multiplet since contribution of the adjoint chiral which is part of the $\mathcal{N}=4$ vector multiplet is trivial \cite{Kapustin:2010xq}. \footnote{Strictly speaking, the vector one-loop contribution contains a factor of the Vandermonte determinant $\prod_{\alpha} \alpha(\vec s)$ in the denominator. However, this factor exactly cancels with another factor of Vandermonte determinant coming from the measure of the integration over the Cartan of the gauge group. We will also ignore certain factors of 2 that appear in the 1-loop contributions for the vector multiplet and the hypermultiplet.}

\item The one-loop contribution from $\mathcal{N}=4$ hypermultiplets transforming in a representation  $\CR$ of $G \times G_{H}$ :
\begin{equation} \label{PF-main-hyper}
Z^{\rm hyper}_{\text{1-loop}}(\vec s, \vec m)=\prod_{\rho(\CR)} \frac{1}{\cosh{\pi \rho(\vec s, \vec m)}}\,,
\end{equation}
where the product extends over the weights of the representation $\CR$.
\end{itemize}

Consider a pair of theories $X$ and $Y$ which are mirror dual to each other. Assuming no line defects and Chern-Simons terms are turned on, the IR duality 
implies that the partition functions of the theories\footnote{We will denote the partition function of a quiver $Q$ with real mass parameters $\vec x$ and FI parameters $\vec y$ as
$Z^{(Q)}(\vec x ; \vec y)$.} are related as:
\be
Z^{(X)}(\vec m ; \vec \eta) = e^{2\pi i a^{kl} m_k \eta_l}\, Z^{(Y)}(\vec m'(\vec \eta) ; \vec \eta' (\vec m)), 
\ee
where $(\vec m, \vec \eta)$, $(\vec m', \vec \eta' )$ are the masses and FI parameters for $X$ and $Y$ respectively. The mass parameters $\vec m'$ of $Y$ are linear functions of the FI parameters $\vec \eta$ of $X$, while FI parameters $\vec \eta'$ of $Y$ are linear functions of the FI parameters $\vec m$ of $X$, as expected under mirror symmetry. The overall phase factor $e^{2\pi i a^{kl} m_k \eta_l}$, whose exponents are linear in $\vec m$ and $\vec \eta$ with integer $a^{kl}$, correspond to three-dimensional contact terms as discussed in \cite{Closset:2012vg}.\\

The localization computation relies on the assumption that the IR conformal dimensions of fields can be read off from their transformation properties under the R-symmetry visible in the UV Lagrangian,
which in turn implies that the UV R-symmetry is assumed to be the same as the R-symmetry of the IR SCFT. In terms of the classification of 3d $\CN=4$ theories presented in \Secref{GBU-GW}, the formula for 
the sphere partition function presented in \eqref{PF-main} is only valid for the good and ugly theories, and not for the bad theories. In fact, one can show that the condition 
-- $\Delta(B) \geq \frac{1}{2}$ for all cocharacters $B$ -- for a good/ugly theory derived by Gaiotto and Witten \cite{Gaiotto:2008ak} is equivalent to the condition for the above partition function to be absolutely convergent in the absence of defects.
This can be seen in the following fashion \cite{Kapustin:2010mh}. First note that the integrand is regular at all finite real values of $\vec s$ (the poles coming from hyperbolic cosine functions are all located on the imaginary axis), 
and therefore any divergence in the integral must arise from very large values of the integration variable $\vec s$. Consider a ray in the Cartan subalgebra of $G$ in the direction specified by a cocharacter $B$, and let $r \in \BR_{+}$ be a coordinate along the ray. The subset of such rays is dense in the set of all rays in the Cartan subalgebra of $G$. 
Now, at large values of the coordinate $r$, the integrand behaves as $e^{- r q(B)}$, where $q(B) \in \BZ$ is
\be
q(B) = - \sum_{\alpha \in \Delta^+} |\alpha(B)| +  \frac{1}{2} \sum_{\rho\in \CR} | \rho (B)|. 
\ee
The Gaiotto-Witten condition  $\Delta(B) \geq \frac{1}{2}$ for good/ugly theories guarantees that $q(B) \geq 1$ for all cocharacters $B$, which implies that the integrand falls off sufficiently fast at large $\vec s$ along any ray in the Cartan subalgebra and therefore gives a finite integral. For a bad theory, one can have $q(B) =0$ for some cocharacter(s) $B$ which will lead to a divergent integral.\\

In this paper, we will use another RG-invariant supersymmetric observable, i.e. superconformal index on $S^2 \times S^1$, which we review in 
\Secref{SCI}.

\subsection{Mirror symmetry as S-duality in Type IIB construction : Linear quivers}\label{LQ-IIB}
\begin{figure}[htbp]
\begin{center}
\scalebox{.8}{\begin{tikzpicture}[
cnode/.style={circle,draw,thick,minimum size=4mm},snode/.style={rectangle,draw,thick,minimum size=8mm},pnode/.style={rectangle,red,draw,thick,minimum size=1.0cm}]
\node[cnode] (1) {$N_1$};
\node[cnode] (2) [right=.5cm  of 1]{$N_2$};
\node[cnode] (3) [right=.5cm of 2]{$N_3$};
\node[cnode] (4) [right=1cm of 3]{$N_{\alpha-1}$};
\node[cnode] (5) [right=0.5cm of 4]{$N_{\alpha}$};
\node[cnode] (6) [right=0.5cm of 5]{$N_{\alpha + 1}$};
\node[cnode] (7) [right=1cm of 6]{{$N_{L-2}$}};
\node[cnode] (8) [right=0.5cm of 7]{$N_{L-1}$};
\node[cnode] (9) [right=0.5cm of 8]{$N_L$};
\node[snode] (10) [below=0.5cm of 1]{$M_1$};
\node[snode] (11) [below=0.5cm of 2]{$M_2$};
\node[snode] (12) [below=0.5cm of 3]{$M_3$};
\node[snode] (13) [below=0.5cm of 4]{$M_{\alpha-1}$};
\node[snode] (14) [below=0.5cm of 5]{$M_{\alpha}$};
\node[snode] (15) [below=0.5cm of 6]{$M_{\alpha+1}$};
\node[snode] (16) [below=0.5cm of 7]{$M_{L-2}$};
\node[snode] (17) [below=0.5cm of 8]{$M_{L-1}$};
\node[snode] (18) [below=0.5cm of 9]{$M_{L}$};
\draw[-] (1) -- (2);
\draw[-] (2)-- (3);
\draw[dashed] (3) -- (4);
\draw[-] (4) --(5);
\draw[-] (5) --(6);
\draw[dashed] (6) -- (7);
\draw[-] (7) -- (8);
\draw[-] (8) --(9);
\draw[-] (1) -- (10);
\draw[-] (2) -- (11);
\draw[-] (3) -- (12);
\draw[-] (4) -- (13);
\draw[-] (5) -- (14);
\draw[-] (6) -- (15);
\draw[-] (7) -- (16);
\draw[-] (8) -- (17);
\draw[-] (9) -- (18);
\end{tikzpicture}}
\caption{A generic linear quiver with $L$ gauge nodes.}
\label{fig: LQGen}
\end{center}
\end{figure}
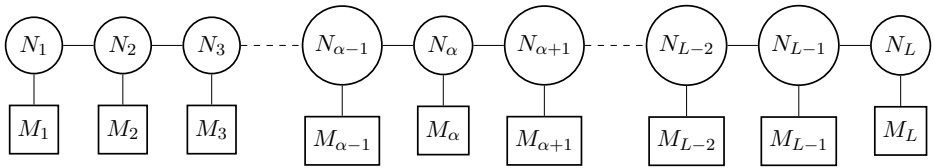

In this section, we present a brief review of a very special class of 3d $\CN=4$ quiver gauge theory -- linear quivers with unitary gauge groups. A generic 
example with $L$ gauge nodes is shown in \figref{fig: LQGen}. For a more detailed account on the physics and geometry of linear quivers, we refer the 
reader to the papers \cite{Gaiotto:2008ak,Gaiotto:2008sa}.\\

The linear quivers have a very simple realization in terms of a Type IIB brane construction of the Hanany-Witten type \cite{Hanany:1996ie}. 
A large class of 3d $\CN=4$ Lagrangian theories can be obtained by considering D3 branes extending along a compact direction $L$, with 1/2-BPS 
boundary conditions at the two ends \cite{Gaiotto:2008ak}. For linear quivers, the set-up involves D3, D5 and NS5 branes, with their respective world-volumes 
specified in Table \ref{tab:HWbranes}. The gauge theory data can be read off from a configuration where all D3 branes end on NS5 branes using the following set of 
rules:
\begin{itemize}
\item D3-D3 open strings in the $\gamma$-th NS5 chamber containing $N_\gamma$ D3 branes give a $U(N_\gamma)$ vector multiplet.

\item D3-D5 open strings in the $\gamma$-th NS5 chamber, containing $M_\gamma$ D5 branes, give $M_\gamma$ hypermultiplets in the 
fundamental representation of $U(N_\gamma)$. 

\item D3-D3 open strings running between the $\gamma$-th and the $\gamma+1$-th NS5 chambers give hypermultiplets in the 
bifundamental of $U(N_\gamma) \times U(N_{\gamma+1})$. 

\item The triplet of mass parameters $m^Z_\beta$, with $\beta=1,\ldots, \sum^L_{\gamma=1} M_\gamma$, correspond to the position of the D5 branes in $\BR^{3}_{7,8,9}$,
while for the triplet of FI parameters $\eta^Y_{\gamma}= t^Y_{\gamma} - t^Y_{\gamma +1}$, with $\gamma=1,\ldots,L$, the parameters $\vec{t}^Y$ correspond to the position of the NS5 branes in $\BR^{3}_{4,5,6}$. Given the translational symmetry on $\BR^3$, both sets of moduli should be counted up to an overall shift.

\end{itemize}

\begin{table}[htbp]
\begin{center}
\begin{tabular}{|c|c|c|c|c|c|c|c|c|c|c|}
\hline
         & 0 & 1 &  2 & 3 & 4 & 5 & 6 & 7 & 8 & 9 \\
\hline \hline
NS5 & x &  x &  x & $\cdot$  & $\cdot$  & $\cdot$ & $\cdot$   &  x     &  x & x  \\
\hline
D5    & x &  x & x  &  $\cdot$ & x  & x  & x &  $\cdot$  &  $\cdot$   &  $\cdot$  \\
\hline
D3    & x &  x & x & x &  $\cdot$  &  $\cdot$   & $ \cdot$  & $\cdot$  &   $\cdot$   &  $\cdot$  \\
\hline
\end{tabular}
\caption{Basic Type IIB brane construction.} 
\label{tab:HWbranes}
\end{center}
\end{table}

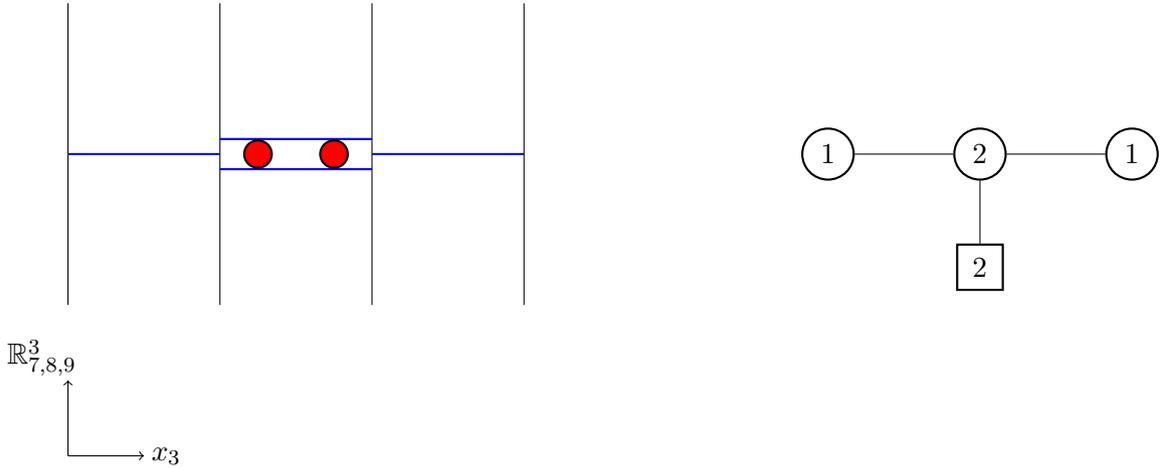
\begin{figure}[h]
\begin{center}
\begin{tikzpicture}[
nnode/.style={circle,draw,thick, fill=red},cnode/.style={circle,draw,thick,minimum size=4mm},snode/.style={rectangle,draw,thick,minimum size=6mm}]
\draw[-] (0,-2) -- (0,2);
\draw[-] (2,-2) -- (2,2);
\draw[-] (4,-2) -- (4,2);
\draw[-] (6,-2) -- (6,2);
\draw[blue,thick,-] (0,0) -- (2,0);
\draw[blue,thick,-] (2,0.2) -- (4,0.2);
\draw[blue,thick,-] (2,-0.2) -- (4,-0.2);
\draw[blue,thick,-] (4,0) -- (6,0);
\node[nnode] (1) at (2.5,0) {} ;
\node[nnode] (2) at (3.5,0) {};
\node[cnode] (5) at (10,0) {1};
\node[cnode] (6) at (12,0) {2};
\node[cnode] (7) at (14,0) {1};
\node[snode] (8) at (12, -1.5) {2};
\draw[-] (5)--(6);
\draw[-] (6)--(7);
\draw[-] (6)--(8);
\draw[->] (0,-4) -- (1, -4);
\draw[->] (0,-4) -- (0,-3);
\node[text width=.2cm](9) at (1.2, -4){$x_3$};
\node[text width=1cm](10) at (-0.3, -2.7){$\BR^3_{7,8,9}$};
\end{tikzpicture}
\caption{The figure on the left shows the Type IIB brane construction for the linear quiver on the right. 
The red nodes represent D5 branes, the horizontal blue lines are D3 branes, and the vertical black lines 
represent NS5 branes.}
\label{Fig:HWEx1}
\end{center}
\end{figure}

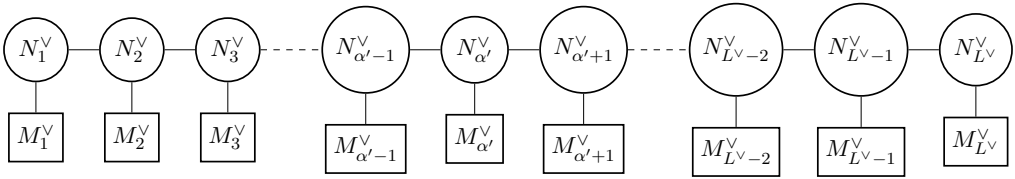
\begin{figure}[htbp]
\begin{center}
\scalebox{.8}{\begin{tikzpicture}[
cnode/.style={circle,draw,thick,minimum size=4mm},snode/.style={rectangle,draw,thick,minimum size=8mm},pnode/.style={rectangle,red,draw,thick,minimum size=1.0cm}]
\node[cnode] (1) {$N^{\vee}_1$};
\node[cnode] (2) [right=.5cm  of 1]{$N^{\vee}_2$};
\node[cnode] (3) [right=.5cm of 2]{$N^{\vee}_3$};
\node[cnode] (4) [right=1cm of 3]{$N^{\vee}_{\alpha'-1}$};
\node[cnode] (5) [right=0.5cm of 4]{$N^{\vee}_{\alpha'}$};
\node[cnode] (6) [right=0.5cm of 5]{$N^{\vee}_{\alpha' + 1}$};
\node[cnode] (7) [right=1cm of 6]{{$N^{\vee}_{L^{\vee}-2}$}};
\node[cnode] (8) [right=0.5cm of 7]{$N^{\vee}_{L^{\vee}-1}$};
\node[cnode] (9) [right=0.5cm of 8]{$N^{\vee}_{L^{\vee}}$};
\node[snode] (10) [below=0.5cm of 1]{$M^{\vee}_1$};
\node[snode] (11) [below=0.5cm of 2]{$M^{\vee}_2$};
\node[snode] (12) [below=0.5cm of 3]{$M^{\vee}_3$};
\node[snode] (13) [below=0.5cm of 4]{$M^{\vee}_{\alpha'-1}$};
\node[snode] (14) [below=0.5cm of 5]{$M^{\vee}_{\alpha'}$};
\node[snode] (15) [below=0.5cm of 6]{$M^{\vee}_{\alpha'+1}$};
\node[snode] (16) [below=0.5cm of 7]{$M^{\vee}_{L^{\vee}-2}$};
\node[snode] (17) [below=0.5cm of 8]{$M^{\vee}_{L^{\vee}-1}$};
\node[snode] (18) [below=0.5cm of 9]{$M^{\vee}_{L^{\vee}}$};
\draw[-] (1) -- (2);
\draw[-] (2)-- (3);
\draw[dashed] (3) -- (4);
\draw[-] (4) --(5);
\draw[-] (5) --(6);
\draw[dashed] (6) -- (7);
\draw[-] (7) -- (8);
\draw[-] (8) --(9);
\draw[-] (1) -- (10);
\draw[-] (2) -- (11);
\draw[-] (3) -- (12);
\draw[-] (4) -- (13);
\draw[-] (5) -- (14);
\draw[-] (6) -- (15);
\draw[-] (7) -- (16);
\draw[-] (8) -- (17);
\draw[-] (9) -- (18);
\end{tikzpicture}}
\caption{The linear quiver which is mirror dual to the generic linear quiver in \figref{fig: LQGen}. The total number of gauge 
nodes is $L^{\vee}$.}
\label{fig: LQGenDual}
\end{center}
\end{figure}

\figref{Fig:HWEx1} gives an illustrative example of how one can read off the gauge theory content from the brane set up. 
Mirror symmetry in three dimensions can be understood as an S-duality of the above brane construction, followed by a 
rotation $R: \vec x^{7,8,9} \to - \vec {x}^{4,5,6}, \, \vec x^{4,5,6} \to \vec x^{7,8,9}$. NS5 and D5 branes are exchanged 
under S-duality, while D3 branes are self-dual. To read off the dual gauge theory from the rotated S-dual brane system,
one has to move to a configuration where the all D3 branes end on NS5 branes. This can be done by performing a series 
of Hanany-Witten moves, where NS5 and D5 branes are moved past each other along the compact direction $x^3$. 
This generically results in creation/annihilation of D3 branes which are required to keep the {\it linking numbers} of the 
individual 5-branes invariant \cite{Hanany:1996ie,Gaiotto:2008ak}.
The linking numbers of the 5-branes $(l^{\rm NS5}_{\gamma}, l^{\rm D5}_\beta)$ in a generic brane configuration are given as:
\begin{align}
& l^{\rm NS5}_{\gamma} = n_{\rm left}({\rm D5}) - \wt{n}_{\rm left}({\rm D3}) + \wt{n}_{\rm right}({\rm D3}),\qquad \gamma =1,\ldots, L+1,  \\
& l^{\rm D5}_\beta = n_{\rm left}({\rm NS5}) - \wt{n}_{\rm left}({\rm D3}) + \wt{n}_{\rm right}({\rm D3}), \qquad \beta=1, \ldots, L^\vee +1,
\end{align}
where $L^\vee=\sum^L_{\gamma=1} M_\gamma -1$, $n_{\rm left,right}({\rm D5/NS5})$ denotes the number of D5/NS5-branes to the left or right of the 5-brane in question, 
while $\wt{n}_{\rm left,right}({\rm D3})$ denotes the number of D3 branes ending on the 5-brane from the left and the right 
respectively. The mirror dual of the generic quiver in \figref{fig: LQGen} is then given by the linear quiver in \figref{fig: LQGenDual}, 
where the ranks of the gauge group factors and the flavor symmetry factors can be computed from
\begin{align}
& M^{\vee}_{\gamma'}= \sum^{L+1}_{\gamma=1}  \delta_{\gamma'\, l^{\rm NS5}_{\gamma}}, \qquad \gamma'=1,\ldots,L^\vee, \label{rankdf}\\ 
& N^{\vee}_{\beta+1}+N^{\vee}_{\beta-1}  - 2N^{\vee}_{\beta} + M^{\vee}_{\beta}= l^{\rm D5}_{\beta +1} -  l^{\rm D5}_{\beta}, \qquad \beta=1,\ldots,L^\vee ,  \label{rankdg}
\end{align}
where the latter equation should be solved subject to the boundary conditions $M^{\vee}_{0}=M^\vee_{L^\vee +1}=0$, and $N^{\vee}_{0}=N^{\vee}_{L^\vee +1}=0$.\\

\begin{figure}[h]
\begin{center}
\begin{tikzpicture}[
nnode/.style={circle,draw,thick, fill=red},cnode/.style={circle,draw,thick,minimum size=4mm},snode/.style={rectangle,draw,thick,minimum size=6mm}]
\draw[-] (0,-2) -- (0,2);
\draw[-] (4,-2) -- (4,2);
\draw[blue,thick,-] (0,-0.2) -- (4,-0.2);
\draw[blue,thick,-] (0,0.2) -- (4,0.2);
\node[nnode] (1) at (0.5,0) {} ;
\node[nnode] (2) at (1.5,0) {};
\node[nnode] (3) at (2.5,0) {};
\node[nnode] (4) at (3.5,0) {};
\node[cnode] (5) at (10,0) {2};
\node[snode] (6) at (10,-2) {4};
\draw[-] (5)--(6);
\draw[->] (0,-4) -- (1, -4);
\draw[->] (0,-4) -- (0,-3);
\node[text width=.2cm](9) at (1.2, -4){$x_3$};
\node[text width=1cm](10) at (-0.3, -2.7){$\BR^3_{7,8,9}$};
\end{tikzpicture}
\caption{The figure on the left shows the Type IIB brane construction for the linear quiver on the right. 
The red nodes represent D5 branes, the horizontal blue lines are D3 branes, and the vertical black lines 
represent NS5 branes.}
\label{Fig:HWEx2}
\end{center}
\end{figure}
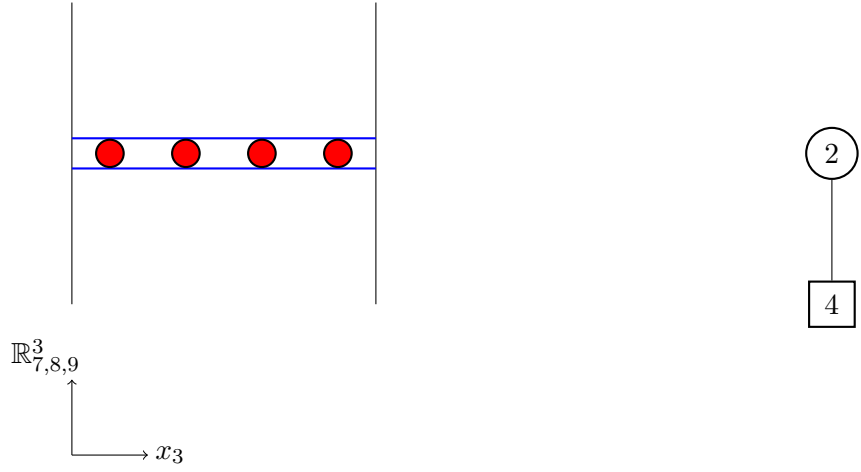

The S-duality followed by a rotation exchanges the positions of NS5 and D5 branes up to a sign, which explains the form of the mirror map in 
\eqref{mirrormap}. As a concrete example, consider the linear quiver and its Type IIB realization in \figref{Fig:HWEx1}. The rotated S-dual configuration, after appropriate Hanany-Witten moves, is shown in the LHS of \figref{Fig:HWEx2} and the corresponding 3d quiver is shown on the RHS. The latter is the mirror dual of the linear quiver in 
\figref{Fig:HWEx1}.

Note that we can easily generalize this Type IIB description to include affine $\hat{A}_N$ quivers with unitary gauge groups. In this case, the compact direction $x^3$
wrapped by D3 branes is a circle.\\

\begin{figure}[htbp]
\begin{center}
\begin{tikzpicture}[
cnode/.style={circle,draw,thick, minimum size=1.0cm},snode/.style={rectangle,draw,thick,minimum size=1cm}]
\node[cnode] (1) {1};
\node[cnode] (2) [right=1cm  of 1]{2};
\node[cnode] (3) [right=1cm  of 2]{1};
\node[snode] (4) [below=1cm of 2]{2};
\node[text width=0.1cm](20)[below=0.5 cm of 4]{$(X)$};
\draw[-] (1) -- (2);
\draw[-] (2)-- (3);
\draw[-] (2)-- (4);
\end{tikzpicture}
\qquad \qquad \qquad \qquad 
\begin{tikzpicture}[cnode/.style={circle,draw,thick, minimum size=1.0cm},snode/.style={rectangle,draw,thick,minimum size=1cm}]
\node[cnode] (1) {2};
\node[snode] (2) [below=1cm of 1]{4};
\draw[-] (1) -- (2);
\node[text width=0.1cm](20)[below=0.5 cm of 2]{$(Y)$};
\end{tikzpicture}
\caption{An example of a pair of linear quivers with unitary gauge groups which are 3d mirrors.}
\label{fig: LQEx1}
\end{center}
\end{figure}
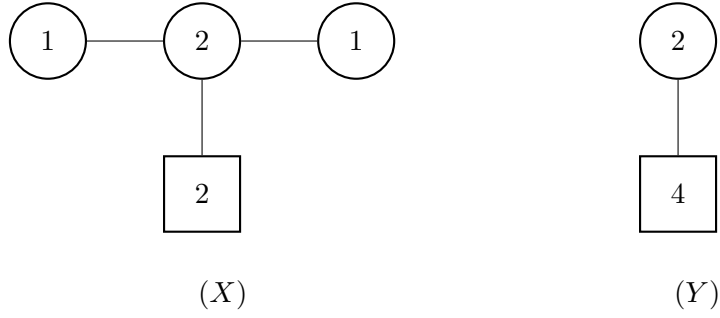

We will now summarize a set of properties of linear quivers that will be useful in the rest of the paper:
\begin{itemize}

\item \textbf{Global symmetries:} The global symmetry on the Higgs branch of a linear quiver is $G_{H}= (\prod^L_{\gamma=1} U(M_\gamma))/U(1)$, while 
the global symmetry on the Coulomb branch manifest in the Lagrangian is $G_{\rm C}= U(1)^L$. The Coulomb branch symmetry can be enhanced 
if one or more gauge nodes are balanced, i.e. $N_{\gamma-1} + N_{\gamma+1} + M_\gamma =2N_\gamma$ for a balanced gauge node $\gamma$. For an array 
of $n$ consecutive balanced nodes, the corresponding global symmetry is enhanced as $U(1)^n \to SU(n+1)$. 

\item \textbf{``Goodness'':}  A linear quiver is a good theory in the Gaiotto-Witten classification if every gauge node in the quiver is individually 
good, i.e. $\Delta_\gamma = N_{\gamma-1} + N_{\gamma+1} + M_\gamma - 2N_\gamma \geq 0$, for $\gamma=1,\ldots,L$. Note that this is a special property of linear quivers, and 
a similar statement is not true for quivers of arbitrary shape.

\item \textbf{Mirror symmetry:} The mirror dual of a good linear quiver is another good linear quiver. In addition, the mirror map between the masses 
and FI parameters is extremely simple. Let the FI parameters of the linear quiver in \figref{fig: LQGen} be parametrized as $\eta^Y_\gamma= t^Y_\gamma - t^Y_{\gamma+1}$, with 
$\gamma=1,\ldots,L$ and $Y=1,2,3$, and the hypermultiplet masses be $m^Z_\beta=\{\vec m^{Z\,\gamma}\}=\{\vec{m}^{Z\,1}, \ldots, \vec{m}^{Z\,L}\}$, with $\beta=1,\ldots, L^\vee +1$, $\gamma=1,\ldots,L$, and $Z=1,2,3$. Recall that the mass parameters are triplets of $SU(2)_C$ (indexed by $Z$), and the FI parameters are triplets of $SU(2)_H$ (indexed by $Y$). 
The dual theory, where $SU(2)_C$ and $SU(2)_H$ are exchanged, is given by the linear quiver in \figref{fig: LQGenDual}, with the 
ranks of the gauge group and flavor symmetry group given in \eref{rankdf}-\eref{rankdg}. 
The dual theory has mass parameters ${m}^{\vee\,Y}_{\beta'}=\{{m}^{\vee\,\gamma'}_{Y}\}=\{\vec{m}^{\vee\,1}_Y, \ldots, \vec{m}^{\vee \, L^\vee}_Y\}$, with $\beta'=1,\ldots,L+1$, and FI parameters ${\eta}^{\vee\,Z}_{\gamma'} ={t}^{\vee\,Z}_{\gamma'}  - {t}^{\vee\,Z}_{\gamma'+1} $, with $\gamma'=1,\ldots, L^{\vee} $. The mirror map in this case is simply given by:
\be \label{mirrormap}
{m}^{\vee\,Y}_{\beta'} = -  t^Y_{\beta'},\, \forall Y, \forall \beta'=1,\ldots,L+1, \qquad {t}^{\vee\,Z}_\beta=m^Z_\beta, \,\forall Z, \forall \beta=1,\ldots, L^\vee +1.
\ee

\item \textbf{Partition function and its dual:} As discussed in \Secref{S3-review}, one can directly check mirror symmetry for a pair of theories using RG-invariant 
supersymmetric observables computed using localization. The partition function on round $S^3$ (where only a real deformation parameter is turned on instead of a triplet)  
of the quiver $X$ is given as 
\begin{align}
&Z^{(X)}(\vec m; \vec t) = \int \prod^L_{\gamma=1} \Big[ d \vec s^\gamma\Big]\, Z^{(X)}_{\rm int} (\{\vec s^\gamma\},\vec m, \vec t) = \int \prod^L_{\gamma=1} \Big[ d \vec s^\gamma\Big]\, Z^{(X)}_{\rm FI}(\{\vec s^\gamma \}, \vec t)\,Z^{(X)}_{\rm{1-loop}}(\{\vec s^\gamma \}, \{\vec m^\gamma \} ) \nn \\
&= \int \prod^L_{\gamma=1}\Big[ d \vec s^\gamma\Big]\,Z^{(X)}_{\rm FI}(\{\vec s^\gamma \}, \vec t)\, \prod^L_{\gamma=1} Z^{\rm vector}_{\rm{1-loop}}(\vec s^\gamma)\,Z^{\rm fund}_{\rm{1-loop}}(\vec s^\gamma, \vec m^{\gamma})\,\prod^{L-1}_{\gamma=1} Z^{\rm bif}_{\rm{1-loop}}(\vec s^\gamma, \vec s^{\gamma +1},0) \label{PF-Agen}.
\end{align}
The classical and the 1-loop contributions to the integrand are given as
\begin{align}\label{PF-Agen1}
& Z^{(X)}_{\rm FI}(\{\vec s^\gamma \}, \vec t)=\prod^L_{\gamma=1} e^{2\pi i \sum^{N_\gamma}_{i_\gamma=1} s^\gamma_{i_\gamma} (t_\gamma - t_{\gamma+1})}, \quad Z^{\rm vector}_{\rm{1-loop}}(\vec s^\gamma)= \prod_{i_\gamma \neq j_\gamma} \sh{(s^\gamma_{i_\gamma} - s^\gamma_{j_\gamma})},\nn \\
& Z^{\rm fund}_{\rm{1-loop}}(\vec s^\gamma, \vec m^{\gamma})=\frac{1}{\prod^{N_\gamma}_{i_\gamma=1} \prod^{M_\gamma}_{l_\gamma=1} \ch{(s^\gamma_{i_\gamma} - m^{\gamma}_{l_\gamma})}}, \nn \\ 
&Z^{\rm bif}_{\rm{1-loop}}(\vec s^\gamma, \vec s^{\gamma +1}, m^{\rm bif}_\gamma)=\frac{1}{\prod^{N_\gamma}_{i_\gamma=1} \prod^{N_{\gamma +1}}_{i_{\gamma +1}=1} \ch{(s^\gamma_{i_\gamma} - s^{\gamma+1}_{i_{\gamma+1}} - m^{\rm bif}_\gamma)}},
\end{align}
where the fundamental masses and the bifundamental masses are labelled as $\vec {m}^{\gamma}$ and $\vec{m}^{\rm bif}_\gamma$ respectively. 
The partition function of the dual quiver $Y$ is similarly given as
\begin{align}
&Z^{(Y)}(\vec{m}^\vee; \vec{t}^\vee)= \int \prod^{L^\vee}_{\gamma'=1} \Big[d\vec \s^{\gamma'}\Big]\, Z^{(Y)}_{\rm int}(\{\vec \s^{\gamma'} \},\vec{m}^\vee, \vec{t}^\vee)\nn \\ & =\int \prod^{L^\vee}_{\gamma'=1}\,\Big[d\vec \s^{\gamma'}\Big] \,Z^{(Y)}_{\rm FI}(\{\vec \s^{\gamma'} \}, \vec t^\vee)\, Z^{(Y)}_{\rm{1-loop}}(\{\vec \s^{\gamma'} \}, \{\vec m^{\vee\,\gamma'} \} ) \nn \\
&= \int \prod^{L^\vee}_{\gamma'=1} \,\Big[d\vec \s^{\gamma'}\Big]\, Z^{(Y)}_{\rm FI}(\{\vec \s^{\gamma'} \}, \vec t^\vee) \prod^{L^\vee}_{\gamma'=1} Z^{\rm vector}_{\rm{1-loop}}(\vec \s^{\gamma'})\, Z^{\rm fund}_{\rm{1-loop}}(\vec \s^{\gamma'}, \vec m^{\vee\,\gamma'})\, \prod^{L^\vee-1}_{\gamma'=1} Z^{\rm bif}_{\rm{1-loop}}(\vec \s^{\gamma'}, \vec \s^{\gamma' +1},0),  \label{PF-Bgen}
\end{align}
where the classical and the 1-loop contributions to the integrand are given as follows:
\begin{align}\label{PF-Bgen1}
& Z^{(Y)}_{\rm FI}(\{\vec \s^{\gamma'} \}, \vec t^\vee) =\prod^{L^\vee}_{\gamma'=1} e^{2\pi i \sum^{N^\vee_{\gamma'}}_{i_{\gamma'}=1} \s^{\gamma'}_{i_{\gamma'}} (t^\vee_{\gamma'} - t^\vee_{\gamma'+1})}, \quad Z^{\rm vector}_{\rm{1-loop}}(\vec \s^{\gamma'})=\prod_{i_{\gamma'} \neq j_{\gamma'}}\sh{(\s^{\gamma'}_{i_{\gamma'}} - \s^{\gamma'}_{j_{\gamma'}})},\nn \\
& Z^{\rm fund}_{\rm{1-loop}}(\vec \s^{\gamma'}, \vec m^{\vee\,\gamma'})=\frac{1}{\prod^{N^\vee_{\gamma'}}_{i_{\gamma'}=1} \prod^{M^\vee_{\gamma'}}_{l_{\gamma'}=1} \ch{(\s^{\gamma'}_{i_{\gamma'}} - m^{\vee\,\gamma'}_{l_{\gamma'}})}}, \nn \\
& Z^{\rm bif}_{\rm{1-loop}}(\vec \s^{\gamma'}, \vec \s^{\gamma' +1}, m^{\vee\,\gamma'}_{\rm bif})=\frac{1}{\prod^{N^\vee_{\gamma'}}_{i_{\gamma'}=1} \prod^{N^\vee_{\gamma' +1}}_{i_{\gamma'+1}=1} \ch{(\s^{\gamma'}_{i_{\gamma'}} - \s^{\gamma+1}_{i_{\gamma' +1}} - m^{\vee\,\gamma'}_{\rm bif})}} .
\end{align}

Mirror symmetry implies that the partition functions of $X$ and $Y$ 
are related as 
\be \label{pf-LQmirror}
Z^{(X)}(\vec m; \vec t)= e^{2\pi i a^{kl} m_k t_l} \, Z^{(Y)}(- \vec t; \vec m),
\ee
where the overall phase factors can interpreted as three-dimensional contact terms. Here $a^{kl}$ is an $M \times (L+1)$ matrix with integer entries, $M=\sum^L_{\alpha=1} M_\alpha$ and $L$ is the number of nodes in quiver $A$. The general proof of the above equality is non-trivial, and we refer the reader to \cite{Kapustin:2010xq} for details.\\

We would like to emphasize an important feature of this partition function equality. For quiver $X$, the integrand is manifestly invariant under a $S_{M_\gamma}$-transformation of the mass parameters $\vec{m}^\gamma$ (this is the Weyl symmetry of the flavor group $U(M_\gamma)$), 
which in turn makes the partition function of $X$ invariant under these transformations. However, for the dual theory $Y$, where $\vec{m}^\gamma$ 
appear as FI parameters, the partition function is not manifestly invariant under these transformations. In fact, one can show that the matrix integral for $(Y)$, 
generically changes by some overall phase factor under these transformations. The contact term contribution also changes by a phase factor. It turns out that these two phase factors exactly cancel each other. There is an analogous argument for permutation of the parameters $\vec t$, which appear as mass parameters of $Y$, and FI parameters for $X$.

\end{itemize}

Finally, let us show the equality \eref{pf-LQmirror} explicitly for the simple example of the pair of theories in \figref{fig: LQEx1}.
The partition function of the quiver $X$ is
\begin{equation}
\begin{split}
& Z^{(X)}(\vec{m}; \vec{t})=\int \prod^2_{\alpha=1} d s_{\alpha}  \frac{d^2 s_0}{2!} \frac{e^{2\pi i s_{1} (t_1-t_2)} e^{2\pi i \sum_i s^i_0 (t_2 -t_3)} e^{2\pi i s_{2} (t_3-t_4)} \sinh^2{\pi(s^1_0-s^2_0)}}{\prod^2_{i=1}\cosh{\pi(s_1-s^i_0)}\prod^2_{a=1} \cosh{\pi(s^i_0 - m_a)}\cosh{\pi(s_2-s^i_0)}},
\end{split}
\label{eq:ZAex}
\end{equation}
where $m_1$ and $m_2$ are the masses of the fundamental hypermultiplets in the middle node, and the FI parameters of the three gauge nodes are $\eta_1=t_1-t_2, \eta_0=t_2-t_3, \eta_2=t_3-t_4$.\\

Similarly, the partition function of the quiver $Y$ is
\begin{equation}
\begin{split}
& Z^{(Y)}(\vec{\wt{m}}; \vec{\widetilde{t}})=\int \frac{d^2\s}{2!} \frac{\prod^2_{i=1} e^{2\pi i \s^i (\widetilde{t}_1-\widetilde{t}_2)}  \sinh^2{\pi(\s^1-\s^2)}}{\prod^2_{i=1}\prod^4_{a=1} \cosh{\pi(\s^i - \wt{m}_a)}},
\end{split}
\label{eq:ZBex}
\end{equation}
where $\wt{m}_1, \wt{m}_2, \wt{m}_3, \wt{m}_4$ are hypermultiplet masses, while the FI parameter of the gauge group is $\eta= \widetilde{t}_1-\widetilde{t}_2$. 
Evaluating the two matrix integrals explicitly, one can check that
\begin{align}
Z^{(X)}[\vec m ; \vec t] = e^{2\pi i m_1(t_1+t_2)} e^{-2\pi i m_2(t_3+t_4)}\, Z^{(Y)}[-\vec t; \vec m].
\end{align}
The expressions agree exactly (i.e. the phase factor vanishes) when one imposes the constraints $m_1+m_2=0, \; t_1+t_2+t_3+t_4=0$.

\section{Generating mirrors from linear quiver pairs using $S$-type operations}\label{GFI-main}
In this section, we define the $S$-type operations on a generic  3d $\CN=4$ quiver gauge theory $X$ in class $\CU$, 
i.e. the Higgs branch global symmetry of $X$ has a subgroup of the form $\prod_{\gamma} U(M_\gamma)$.
We then discuss the field theory machinery for generating new pairs of mirror dual theories starting from a pair of linear quivers, 
in terms of the $S^3$ partition function.\\

In \Secref{GFI-def}, we discuss the four basic $\CQ$-operations 
(defined in \eref{QOP-1}-\eref{QOP-2}) on $X$, and then define an elementary $S$-type operation in terms 
of these operations. We discuss the realization of these operations at the level of the $\CN=4$ quiver diagram as well as the $S^3$ partition function. 
In \Secref{GFI-summary}, we discuss how quivers of arbitrary shapes can be 
constructed in steps by the action of elementary $S$-type operations starting from a linear quiver. In particular, we 
discuss the four distinct types of elementary $S$-type operations separately.
We give a partition function prescription 
for writing down the duals of these elementary $S$-type operations in \Secref{GFIdual-summary}. 
\Secref{GFI-summary}-\Secref{GFIdual-summary} therefore gives the complete partition function recipe for constructing 
a new pair of dual theories from the original dual pair of linear quivers. 
In \Secref{GFI-exAbPF}, we present a simple illustrative example involving a dual pair of Abelian quiver gauge theories that 
can be constructed following the recipe of \Secref{GFI-summary} and \Secref{GFIdual-summary}. Another set of examples
involving (affine) $D$-type quivers can be found in \Appref{GFI-ex}. \\

In \Appref{GFI-SCI}, we discuss how the elementary
$S$-type operations can be implemented in terms of the superconformal index on $S^2 \times S^1$. 
Analogous to the case of the partition function on $S^3$, this leads to a natural prescription for the dual operations.

\subsection{Elementary $S$-type and $T$-type operations on a generic quiver}\label{GFI-def}
Consider a quiver gauge theory $X$ in class $\CU$. 
We will assume that the theory $X$ is good in the Gaiotto-Witten sense \cite{Gaiotto:2008ak}, which in turn implies that the round three
sphere partition function of the theory is convergent \cite{Kapustin:2010mh}, as discussed in \Secref{S3-review}. Given the quiver $X$, one can describe the 
set of four basic $\CQ$-operations in the concrete case of 3d $\CN=4$ quiver gauge theories as follows:

\begin{itemize}
\item \textbf{Gauging operation}:  A \textit{gauging operation} $G^\alpha_\CP$ at a flavor node $\alpha$ of the theory $X$ (shown schematically in \figref{fig: GenQuivG}) 
involves the following two steps :
\begin{enumerate}
\item Given a flavor node $U(M_\alpha)$, we split it into two flavor nodes, corresponding to a $U(r_\alpha) \times U(M_\alpha - r_\alpha)$ global symmetry. 
At the level of the Lagrangian, this simply implies identifying the $U(1)^{M_\alpha}$ background vector multiplets as 
$U(1)^{r_\alpha} \times U(1)^{M_\alpha-r_\alpha}$ background vector multiplets. In particular, this implies identifying 
the $U(1)^{M_\alpha}$ mass parameters $\overrightarrow{\vec{m}^\alpha}$ with the $U(1)^{r_\alpha} \times U(1)^{M_\alpha-r_\alpha}$ 
mass parameters $(\overrightarrow{\vec{u}^\alpha},\overrightarrow{\vec{v}^\alpha})$. 
There is an $S_{M_\alpha}$ (permutation group of $M_\alpha$ objects) worth of freedom in this identification, and 
a specific choice is part of the data for this procedure. We will choose to parametrize this identification as:
\be \label{uvdef0}
\overrightarrow{m^\alpha}_{i_\alpha} = \CP_{i_\alpha i} \, \overrightarrow{u}^\alpha_i + \CP_{i_\alpha \, r_\alpha + j} \, \overrightarrow{v}^\alpha_j, \quad i_\alpha=1,\ldots, M_\alpha,\quad i=1,\ldots, r_\alpha, \quad j=1,\ldots, M_\alpha - r_\alpha,
\ee
where $\CP$ is a permutation matrix of order $M_\alpha$. A choice of $\CP$ therefore encodes the additional data of how the $U(1)^{r_\alpha} \times U(1)^{M_\alpha-r_\alpha}$ background fields are chosen from the original $U(1)^{M_\alpha}$ ones. We will denote the theory deformed by the $U(r_\alpha) \times U(M_\alpha - r_\alpha)$ mass parameters as $(X, \CP)$.

\item Given the theory $(X, \CP)$, we promote the flavor symmetry node $U(r_\alpha) $ to a gauge node, as shown on the RHS of \figref{fig: GenQuivG}, i.e. make the vector multiplets for $U(r_\alpha)$ dynamical, as well as turning on a background twisted vector multiplet for the $U(1)_J$ topological symmetry.
\end{enumerate}

The operation can be implemented in terms of the $S^3$ partition function for the quiver $X$. Recall that preserving supersymmetry on $S^3$ permits turning on a single real mass parameter as opposed to the triplet in the Cartan of $G_H$. The partition function of $(X,\CP)$ is given as
\be \label{PF-XP}
Z^{(X, \CP)} (\vec{u}^\alpha, \vec{v}^\alpha,\ldots;\vec{\eta}) :=  Z^{(X)} (\vec m^\alpha(\CP, \vec{u}^\alpha, \vec{v}^\alpha),\ldots;\vec{\eta}),
\ee
where the $\ldots$ denote the mass parameters of theory $X$ aside from $\vec{m}^\alpha$, and $\vec \eta$ collectively denotes the FI parameters.
Let $G^\alpha_\CP(X)$ denote the quiver gauge theory obtained from the theory $X$ via the gauging operation. 
The partition function of the theory $G^\alpha_\CP(X)$ is therefore given by:
 \begin{empheq}[box=\widefbox]{align}
Z^{G^\alpha_{\CP}(X)} (\vec{v}^\alpha, \ldots;\vec{\eta},{\eta}_\alpha) =  \int \Big[d\vec{u}^\alpha\Big] \, \CZ_{G^\alpha_\CP(X)}(\vec{u}^\alpha,\eta_\alpha) \,Z^{(X,\CP)} (\vec{u}^\alpha, \vec{v}^\alpha,\ldots;\vec{t}), \label{G-basic0}
\end{empheq}
where $\vec{u}^\alpha$ live in the Cartan subalgebra of the group $U(r_\alpha)$, $\Big[d\vec{u}^\alpha\Big]=\frac{\prod^{r_\alpha}_{i=1} \de {u^\alpha_i}}{|{W}_\alpha|}$ with $|{W}_\alpha|= r_\alpha!$ being the order of the Weyl group for the new gauge node. The function $\CZ_{G^\alpha_\CP(X)}$ is given as
\be \label{defZG}
\boxed{\CZ_{G^\alpha_\CP(X)}(\vec{u}^\alpha,\eta_\alpha) = Z_{\rm FI} (\vec{u}^\alpha,\eta_\alpha) \,Z_{\rm 1-loop} ^{\rm vector} (\vec{u}^\alpha),}
 \ee
where $Z_{\rm FI} (\vec{u}^\alpha,\eta_\alpha)$ and $Z_{\rm 1-loop} ^{\rm vector} (\vec{u}^\alpha)$ are the appropriate classical and one-loop contributions respectively of a $U(r_\alpha)$ vector multiplet.

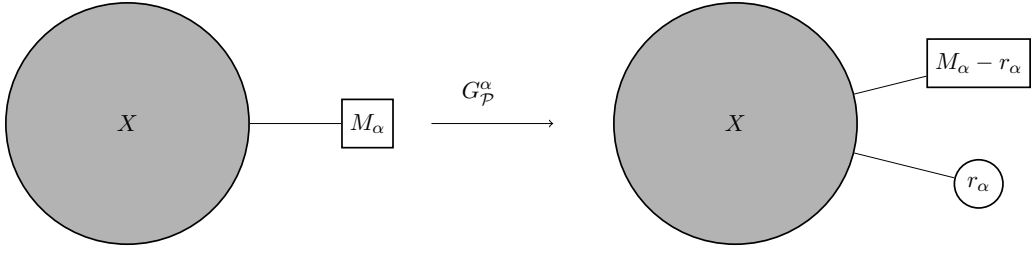
\begin{figure}[htbp]
\begin{center}
\scalebox{0.8}{\begin{tikzpicture}[
cnode/.style={circle,draw,thick,minimum size=4mm},snode/.style={rectangle,draw,thick,minimum size=8mm},pnode/.style={rectangle,red,draw,thick,minimum size=1.0cm}, bnode/.style={circle,draw, thick, fill=black!30,minimum size=4cm}]
\node[bnode] (1) at (0,0){$X$} ;
\node[snode] (2) [right=1.5cm  of 1]{$M_\alpha$} ;
\draw[-] (1)--(2);
\draw[->] (5,0) -- (7,0); 
\node[bnode] (3) at (10,0){$X$} ;
\node[snode] (4) at (14, 1){$M_\alpha- r_\alpha$} ;
\node[cnode] (5) at (14,-1){$r_\alpha$} ;
\draw[-] (3)--(4);
\draw[-] (3)--(5);
\node[text width=1cm](6) at (6,0.5){$G^\alpha_\CP$};
\end{tikzpicture}}
\caption{A gauging operation $G^\alpha_\CP$ on a $U(M_\alpha)$ subgroup of the global symmetry group of a generic quiver gauge theory (represented by the grey circle).}
\label{fig: GenQuivG}
\end{center}
\end{figure}

\item  \textbf{Flavoring Operation}: A \textit{flavoring operation} $F^\alpha_\CP$ at a flavor node $\alpha$ of the theory $X$ (shown schematically in \figref{fig: GenQuivF}) involves the following two steps:
\begin{enumerate}
\item Given the flavor node $U(M_\alpha)$, we split it into two flavor nodes, corresponding to a $U(r_\alpha) \times U(M_\alpha - r_\alpha)$ global symmetry. This requires identifying the $U(1)^{M_\alpha}$ background vector multiplets as $U(1)^{r_\alpha} \times U(1)^{M_\alpha-r_\alpha}$ background vector multiplets which is parametrized by a permutation matrix $\CP$, as given in \eref{uvdef0}. The resultant theory is denoted as $(X, \CP)$.  

\item Given the theory $(X, \CP)$, we attach a flavor node denoted by ${G}^\alpha_{ F}$ to the flavor node $U(r_\alpha)$, as shown on the RHS of \figref{fig: GenQuivF}. This amounts to introducing some free hypermultiplets in the theory which transform under some representation of the global symmetry group $U(r_\alpha) \times {G}^\alpha_{ F}$.
\end{enumerate}

The operation can be implemented in terms of the $S^3$ partition function as follows. Let $F^\alpha_\CP(X)$ denote the theory obtained by implementing the 
flavoring operation on $X$. The partition function of the theory $F^\alpha_\CP(X)$ is then given as (with $Z^{(X,\CP)}$ defined in \eref{PF-XP})
\begin{empheq}[box=\widefbox]{align}\label{F-basic0}
Z^{F^\alpha_\CP(X)}(\vec{u}^\alpha,\vec{v}^\alpha, \ldots;\vec{\eta})= & \CZ_{F^\alpha_\CP(X)}(\vec{u}^\alpha, \vec{m}^\alpha_F) \, Z^{(X,\CP)} (\vec{u}^\alpha, \vec{v}^\alpha, \ldots;\vec{\eta}).
\end{empheq}
The function $\CZ_{F^\alpha_\CP(X)}$ is given as
\begin{empheq}[box=\widefbox]{align}\label{defZF}
\CZ_{F^\alpha_\CP(X)}(\vec{u}^\alpha, \vec{m}^\alpha_F)= Z_{\rm 1-loop} ^{\rm hyper} (\vec{u}^\alpha, \vec{m}^\alpha_F),
\end{empheq} 
where the factor $Z_{\rm 1-loop} ^{\rm hyper} (\vec{u}^\alpha, \vec{m}^\alpha_F)$ denotes the contribution of the free hypermutiplets which are charged under the symmetry $U(r_\alpha) \times {G}^\alpha_{ F}$, with the parameters $\vec{m}^\alpha_F$ in the Cartan subalgebra of the group ${G}^\alpha_{ F}$.

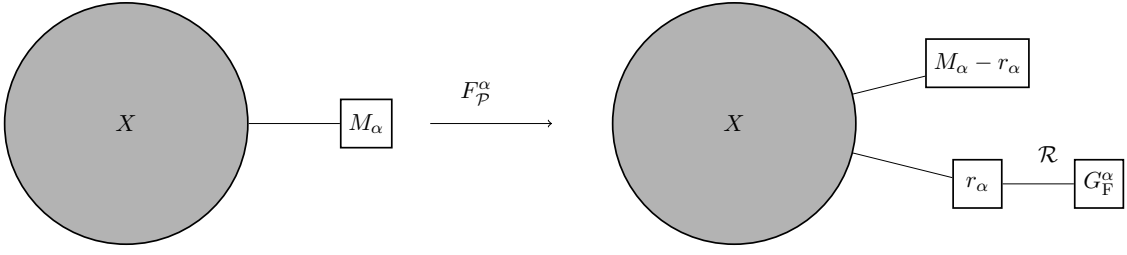
\begin{figure}[htbp]
\begin{center}
\scalebox{0.8}{\begin{tikzpicture}[
cnode/.style={circle,draw,thick,minimum size=4mm},snode/.style={rectangle,draw,thick,minimum size=8mm},pnode/.style={rectangle,red,draw,thick,minimum size=1.0cm}, bnode/.style={circle,draw, thick, fill=black!30,minimum size=4cm}]
\node[bnode] (1) at (0,0){$X$} ;
\node[snode] (2) [right=1.5cm  of 1]{$M_\alpha$} ;
\draw[-] (1)--(2);
\draw[->] (5,0) -- (7,0); 
\node[bnode] (3) at (10,0){$X$} ;
\node[snode] (4) at (14, 1){$M_\alpha- r_\alpha$} ;
\node[snode] (5) at (14,-1){$r_\alpha$} ;
\node[snode] (6) at (16,-1){$G^\alpha_{\rm F}$} ;
\draw[-] (3)--(4);
\draw[-] (3)--(5);
\draw[-] (5)--(6);
\node[text width=1cm](10) at (6,0.5){$F^\alpha_\CP$};
\node[text width=1cm](11) at (15.5, - 0.5){$\CR$};
\end{tikzpicture}}
\caption{A flavoring operation $F^\alpha_\CP$ on a $U(M_\alpha)$ subgroup of the global symmetry group of a generic quiver gauge theory (represented by the grey circle).}
\label{fig: GenQuivF}
\end{center}
\end{figure}

\item \textbf{Identification operation:}  Given the quiver gauge theory $X$ with a global symmetry subgroup $\prod^L_{\gamma=1}U(M_\gamma)$,
let  $N^{\{\gamma_j\}}_{p,r_\alpha}$ denote a set of (not necessarily consecutive) $p \leq L$ flavor nodes - $\gamma_1, \ldots, \gamma_p$, with $r_\alpha$ 
being a positive integer such that $r_\alpha \leq {\rm{Min}}(\{M_\beta\}| \beta \in N^{\{\gamma_j\}}_{p,r_\alpha})$. Let $\alpha$ be a chosen node in 
$N^{\{\gamma_j\}}_{p,r_\alpha}$. An identification operation $I^\alpha_\CP$ (shown schematically in \figref{fig: GenQuivI}) is then performed in two steps: 
\begin{enumerate}
\item For all $\beta \in N^{\{\gamma_j\}}_{p,r_\alpha}$, we split the corresponding flavor node $U(M_\beta)$ into two flavor nodes, associated to a $U(r_\alpha)_\beta \times U(M_\beta -r_\alpha)$ global symmetry. The special case of $p=2$ nodes $\beta = \alpha-1, \alpha,$ is shown in \figref{fig: GenQuivI}. For a given $\beta$, the choice of $U(1)^{r_\alpha}_\beta \times U(1)^{M_\beta-r_\alpha}$ background vector multiplets from the original $U(1)^{M_\beta}$ background vector multiplets is parametrized by a permutation matrix $\CP_\beta$ of order $M_\beta$, i.e.
\be
\overrightarrow{m^\beta}_{i_\beta} = (\CP_\beta)_{i_\beta i} \, \overrightarrow{u}^\beta_i + (\CP_\beta)_{i_\beta \, r_\alpha + j} \, \overrightarrow{v}^\beta_j,
\ee
where $i_\beta=1,\ldots, M_\beta$, $i=1,\ldots, r_\alpha$, and $j=1,\ldots, M_\beta - r_\alpha$. We denote the resultant theory as $(X, \{\CP_\beta \})$.

\item Given the theory $(X, \{\CP_\beta \})$, we identify the flavor nodes $U(r_\alpha)_\beta$ for all $\beta \neq \alpha$ to the flavor node $U(r_\alpha)_\alpha$, as shown on the RHS of \figref{fig: GenQuivI}. 
\end{enumerate}

The identification operation can be implemented in terms of the $S^3$ partition function in the following fashion. First, the partition function of the theory
$(X, \{ P_\beta \})$ is given as:
\be \label{uvZAI0}
Z^{(X, \{ P_\beta\})} (\{u^\beta\},\{v^\beta\}, \ldots; \vec \eta)=: Z^{(X)} (\{\vec m^\beta(\CP_\beta, u^\beta, v^\beta) \}, \ldots; \vec \eta).
\ee

The identification operation on $(X, \{ P_\beta\})$ then implies imposing the following constraints on the mass parameters $\{\vec u^\beta \}$:
\be \label{cartan-id}
u^{\gamma_1}_{i} - \mu^{\gamma_1}= u^{\gamma_2}_{i} - \mu^{\gamma_2} = \ldots = u^{\gamma_p}_{i} - \mu^{\gamma_p}=u^{\alpha}_{i}, \quad {\rm with} \quad 
i=1,\ldots,r_\alpha\, ,
\ee
where $\{\mu^{\gamma_i}\}$ are constant parameters. The choice of $\alpha=\gamma_k$ for a certain $\gamma_k \in N^{\{\gamma_j\}}_{p,r_\alpha}$ corresponds to 
setting $\mu^{\gamma_k}=0$ in the above equation. We will, however, prefer to keep the parameters $\{\mu^{\gamma_i}\}$ arbitrary in our computation and 
express the final answer in terms of independent linear combinations of these parameters, instead of using the constraint $\mu^{\gamma_k}=0$ upfront. 

Let $I^\alpha_\CP(X)$ denote the quiver gauge theory obtained by an identification operation on the quiver $X$. 
The partition function of $I^\alpha_\CP(X)$ is then given as
\begin{empheq}[box=\widefbox]{align}\label{I-basic0}
 Z^{I^\alpha_\CP(X)}(\vec u^{\alpha}, \{ \vec{v}^\beta \},\ldots,\vec{\mu};\vec{\eta}) = \CZ_{I^\alpha_\CP(X)}(\vec u^{\alpha}, \{\vec{u}^\beta\}, \vec \mu) \cdot Z^{(X,\{ P_\beta\})} (\{\vec{u}^\beta\}, \{\vec{v}^\beta\},\ldots;\vec{\eta}),
\end{empheq} 
where $\CZ_{I^\alpha_\CP(X)}$ should be thought of as an operator acting on the function $Z^{(X,\{ P_\beta\})}$, which is explicitly given as
\begin{empheq}[box=\widefbox]{align}\label{defZI}
\CZ_{I^\alpha_\CP(X)}(\vec u^{\alpha}, \{\vec{u}^\beta\}, \vec \mu) = \int \prod^{p}_{j=1} \prod^{r_\alpha}_{i=1} d {u_i^{\gamma_j}} \, \prod^{p}_{j=1} \delta^{(r_\alpha)}\Big(\vec{u}^{\alpha} - \vec{u}^{\gamma_{j}} + {\mu}^{\gamma_j} \Big),
\end{empheq}

\begin{figure}[htbp]
\begin{center}
\scalebox{0.8}{\begin{tikzpicture}[
cnode/.style={circle,draw,thick,minimum size=4mm},snode/.style={rectangle,draw,thick,minimum size=8mm},pnode/.style={rectangle,red,draw,thick,minimum size=1.0cm}, bnode/.style={circle,draw, thick, fill=black!30,minimum size=4cm}]
\node[bnode] (1) at (0,0){$X$} ;
\node[snode] (2) at (4,-2) {$M_\alpha$} ;
\node[snode] (3) at (4, 2) {$M_{\alpha -1}$} ;
\draw[-] (1)--(2);
\draw[-] (1)--(3);
\draw[->] (5,0) -- (7,0); 
\node[bnode] (3) at (10,0){$X$} ;
\node[snode] (4) at (12, 3){$M_{\alpha-1}- r_\alpha$} ;
\node[snode] (5) at (14,0){$r_\alpha$} ;
\node[snode] (6) at (12,-3){$M_\alpha- r_\alpha$} ;
\draw[-] (3)--(4);
\draw[-] (3.north east) to  (5);
\draw[-] (3.south east) to (5);
\draw[-] (3)--(6);
\node[text width=1cm](6) at (6,0.5){$I^\alpha_\CP$};
\end{tikzpicture}}
\caption{An identification operation $I^\alpha_{\vec \CP}$ involving two flavors nodes $U(M_\alpha)$ and $U(M_{\alpha-1})$ in a generic quiver gauge theory.}
\label{fig: GenQuivI}
\end{center}
\end{figure}
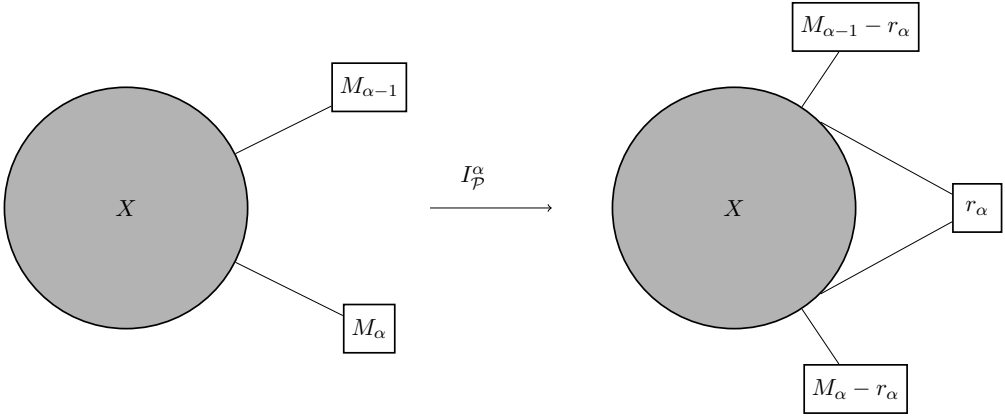

\item  \textbf{Defect operation:} Given a $U(M_\alpha)$ flavor node of a quiver gauge theory $X$, the operation $D^\alpha_\CP$ can be defined in the following 
fashion. One first constructs the theory $(X, \CP)$ deformed by the $U(r_\alpha) \times U(M_\alpha - r_\alpha)$ masses, as before. One can then turn on a
defect for the flavor node $U(r_\alpha)$, labelled by some data $\CD$. In terms of the $S^3$ partition function, the operation is implemented as
\begin{empheq}[box=\widefbox]{align}\label{D-basic0}
Z^{D^\alpha_\CP(X)}(\vec{u}^\alpha,\vec{v}^\alpha, \ldots;\vec{\eta}, \CD)=  \CZ_{D^\alpha_\CP(X)}(\vec{u}^\alpha, \CD) \, Z^{(X,\CP)} (\vec{u}^\alpha, \vec{v}^\alpha, \ldots;\vec{\eta}),
\end{empheq} 
where $D^\alpha_\CP(X)$ is the quiver obtained by implementing the operation $D^\alpha_\CP$ on $X$, and $\CD$ is the data associated with the defect. 
The function $\CZ_{D^\alpha_\CP(X)}$ is given as
\begin{empheq}[box=\widefbox]{align}\label{defZD}
\CZ_{D^\alpha_\CP(X)}(\vec{u}^\alpha, \CD)= Z_{\rm defect} (\vec{u}^\alpha, \CD).
\end{empheq} 
The simplest example of such a defect will be a Wilson line in a representation $\CR$ of $U(r_\alpha)$. In this case, we have
\be
Z_{\rm defect} (\vec{u}^\alpha, \CD):= Z_{\rm Wilson}(\vec{u}^\alpha, \CR) = \sum_{\rho \in \CR} e^{2\pi \rho(\vec u^\alpha)},
\ee
where $\rho$ is a weight of the representation $\CR$ of $U(r_\alpha)$.\\
\end{itemize}

\textbf{Definition.} An elementary $S$-type operation $\CO^\alpha_{\vec \CP}$ on $X$ at a flavor node $\alpha$, is defined by the action of any possible combination of the $F^\alpha_{\CP}$, $I^\alpha_{\vec \CP}$, and $D^\alpha_{\CP}$ operations followed by a \textit{single} gauging operation $G^\alpha_{\CP}$. 
\be \label{Sbasic-def}
\CO^\alpha_{\vec \CP}(X) := (G^\alpha_{\vec \CP}) \circ (F^\alpha_{\vec \CP})^{n_3} \circ (I^\alpha_{\vec \CP})^{n_2} \circ (D^\alpha_{\vec \CP})^{n_1}(X), 
\quad (n_i=0,1, \,\, \forall i).
\ee

The operation $\CO^\alpha_{\vec \CP}$ can be implemented in terms of the partition function as follows. 
Let $\beta$ label the flavor nodes of the theory $X$ on which the given  $S$-type operation
 $\CO^\alpha_{\vec \CP}$ acts, via identification/gauging operations. The partition function of the theory $\CO^\alpha_{\vec \CP}(X)$ is then given by
\begin{empheq}[box=\widefbox]{align}\label{PF-OP}
Z^{\CO^\alpha_{\vec \CP}(X)} = \int \Big[d\vec{u}^\alpha\Big] \, \CZ_{\CO^\alpha_{\vec \CP}(X)}(\vec u^{\alpha}, \{\vec{u}^\beta\}, \eta_\alpha, \vec{m}^{\CO^\alpha_{\vec \CP}}, \CD)
\cdot Z^{(X,\{ P_\beta\})} (\{\vec{u}^\beta\}, \{\vec{v}^\beta\},\ldots;\vec{\eta}),
\end{empheq}
where $\CZ_{\CO^\alpha_{\vec \CP}(X)}$ should be understood as an operator acting on the function $Z^{(X,\{ P_\beta\})}$. The explicit operator can be constructed using 
the expressions of $ \CZ_{G^\alpha_\CP(X)}$, $ \CZ_{F^\alpha_\CP(X)}$, $\CZ_{I^\alpha_\CP(X)}$, and $\CZ_{D^\alpha_\CP(X)}$, given in \eref{defZG}, \eref{defZF}, \eref{defZI} and \eref{defZD} 
respectively, and following the definition \eref{Sbasic-def} of $\CO^\alpha_{\vec \CP}$ in terms of the gauging, flavoring, identification, and defect operations:
\be \label{CZ-OP}
\boxed{\CZ_{\CO^\alpha_{\vec \CP}(X)} =\CZ_{G^\alpha_{\vec \CP}(X)} \cdot \Big(\CZ_{F^\alpha_{\vec \CP}(X)}\Big)^{n_3} \cdot \Big(\CZ_{I^\alpha_{\vec \CP}(X)}\Big)^{n_2} \cdot \Big(\CZ_{D^\alpha_{\vec \CP}(X)}\Big)^{n_1},}
\ee
where the dependence on the mass and FI parameters is implicit. A generic $S$-type operation on the quiver gauge theory $X$ is defined simply as the action of successive elementary $S$-type operations:
\be \label{Comp-OP}
\CO^{(\alpha_1, \ldots, \alpha_l)}_{({\vec \CP_1},\ldots,{\vec \CP_l})}(X) :={\CO^{\alpha_l}_{\vec \CP_l}} \circ {\CO^{\alpha_{l-1}}_{\vec \CP_{l-1}}} \circ \ldots \circ {\CO^{\alpha_2}_{\vec \CP_2}} \circ  {\CO^{\alpha_1}_{\vec \CP_1}}(X).
\ee
Note that the gauging operation in a given constituent elementary $S$-type operation can either involve flavor symmetries present in the theory $X$, 
or flavor symmetries introduced by a previous $\CO^\alpha_{\vec \CP}$. The partition function of the theory $\CO^{(\alpha_1, \ldots, \alpha_l)}_{({\vec \CP_1},\ldots,{\vec \CP_l})}(X)$ can be obtained by using \eref{PF-OP} iteratively.\\

Finally, let us define an elementary $T$-type operation $T^\alpha_\CP$, analogous to Witten's $T$ operation in \cite{Witten:2003ya}), at a 
flavor node $\alpha$ of the quiver gauge theory $X$. Given a flavor node $U(M_\alpha)$ of $X$, one first constructs 
the theory $(X, \CP)$ deformed by the $U(r_\alpha) \times U(M_\alpha - r_\alpha)$ masses. Then, one turns on a Chern-Simons term for 
the flavor symmetry group $U(r_\alpha)$. In terms of the $S^3$ partition function, the operation is implemented as
\begin{empheq}[box=\widefbox]{align}\label{CS-basic0}
Z^{T^\alpha_{\CP}(X)}(\vec{u}^\alpha,\vec{v}^\alpha, \ldots;\vec{\eta}, k)= & \CZ_{T^\alpha_{\CP}}(\vec{u}^\alpha, k) \, Z^{(X,\CP)} (\vec{u}^\alpha, \vec{v}^\alpha, \ldots;\vec{\eta}) \nn \\
=& Z_{\rm CS} (\vec{u}^\alpha, k) \, Z^{(X,\CP)} (\vec{u}^\alpha, \vec{v}^\alpha, \ldots;\vec{\eta}),
\end{empheq} 
where $T^\alpha_{\CP}(X)$ is the quiver obtained by implementing the operation $T^\alpha_\CP$ on $X$, $k$ is the level of the Chern-Simons term, 
and $Z_{\rm CS} (\vec{u}^\alpha, k)$ is the partition function contribution of the Chern-Simons term given in \eref{PF-main-CS}. 
Obviously, one can consider the action of generic operations built out of elementary $S$-type and $T$-type operations on the quiver $X$. The partition 
function of the resultant theory can be written down by combining \eref{PF-OP} and \eref{CS-basic0} appropriately.\\

\subsection{Construction of generic quivers from linear quivers using $S$-type operations}\label{GFI-summary}
Given a quiver gauge theory $X$ in class $\CU$ and an elementary $S$-type operation ${\CO}^\alpha_{\vec \CP}$, finding the mirror dual of the theory 
${\CO}^\alpha_{\vec \CP}(X)$ requires knowing the mirror dual of the theory $X$. For a generic $X$, the mirror dual $Y$ is obviously unknown, and 
there is no guarantee that $Y$ will be a Lagrangian theory. Therefore, one needs a convenient starting point where both the theories 
$X$ and $Y$ are good Lagrangian theories, and the map of masses and FI parameters across the duality is explicitly known.\\

Our strategy in this paper will be to construct a quiver gauge theory $X'$ from a good linear quiver $X$ using a sequence of elementary 
$S$-type operations with no defects. 
The dual of the theory $X'$ can then be read off from the associated dual operations on the good linear quiver $Y$
\footnote{The mirror dual of a good linear quiver is guaranteed to be a good linear quiver  \cite{Gaiotto:2008ak}. Note that this is only true for a linear quiver, and not for quivers of arbitrary shape.}. In this subsection, we will present the formula for the partition 
function realization of an elementary $S$-type operation on a linear quiver $X$, discussing the four distinct types of $S$-type operations separetely. In
\Secref{GFIdual-summary}, we will present the corresponding dual operations.\\

Consider a generic linear quiver theory $X$ with $L$ nodes, as shown in \figref{fig: LQGen}. The mass parameters of $X$ can be parametrized in the following fashion:
\be \label{massconv}
m_i = \{m^1_{i_1}, m^2_{i_2}, \ldots, m^\alpha_{i_\alpha},\ldots, m^{L-1}_{i_{L-1}}, m^L_{i_L} \}, \qquad i=1,\ldots, L^\vee +1,
\ee
with $i_1=1,\ldots,M_1$, $i_2=1,\ldots,M_2$,$\ldots$, $i_\alpha=1,\ldots, M_\alpha$, $i_L=1,\ldots, M_L$, and $L^\vee=\sum^L_{\gamma=1} M_\gamma -1$.
Recall that the Higgs branch global symmetry of $X$ is $G^{(X)}_{H} = (\prod^L_{\gamma=1} U(M_\gamma))/ U(1)$. If $\beta$ labels the set of flavor nodes on which an ${\CO}^\alpha_{\vec \CP}$ acts, then we will assume that the $U(1)$ quotient has been implemented by 
constraining some mass parameters $\{m^{\alpha'}_{i_{\alpha'}}\}$ with $\alpha' \neq \beta$, so that the parameters $\{m^\beta_{i_\beta}\}$ for all $\beta$ are completely unconstrained. Also, let the FI parameters of the linear quiver in \figref{fig: LQGen} be parametrized as 
\be
\eta_\gamma= t_\gamma - t_{\gamma+1}, \qquad \gamma=1,\ldots,L.
\ee
Now, proceeding in the same fashion as in \Secref{GFI-def}, we can write down the partition function of the theory ${\CO}^\alpha_{\vec \CP}(X)$ 
with $X$ being a linear quiver:
\begin{empheq}[box=\widefbox]{align}\label{PF-OPLQ}
Z^{\CO^\alpha_{\vec \CP}(X)} = \int \Big[d\vec{u}^\alpha\Big] \, \CZ_{\CO^\alpha_{\vec \CP}(X)}(\vec u^{\alpha}, \{\vec{u}^\beta\}, \eta_\alpha, \vec{m}^{\CO^\alpha_{\vec \CP}})
\cdot Z^{(X,\{ P_\beta\})} (\{\vec{u}^\beta\}, \{\vec{v}^\beta\},\{\vec m^\gamma\}_{\gamma \neq \alpha};\vec{t}),
\end{empheq}
where the operator $\CZ_{\CO^\alpha_{\vec \CP}(X)}$ can be constructed from the gauging, flavoring and identification operators, as described in \eref{CZ-OP},
and the function $ Z^{(X,\{ P_\beta\})}$ is defined as
\be \label{PF-XPbeta}
Z^{(X,\{ P_\beta\})} (\{\vec{u}^\beta\}, \{\vec{v}^\beta\},\{\vec m^\gamma\}_{\gamma \neq \alpha};\vec{t}):= Z^{(X)} (\{\vec m^\beta(\CP_\beta, u^\beta, v^\beta) \}, \{\vec m^\gamma\}_{\gamma \neq \alpha}; \vec t).
\ee
One can then continue building more general quivers by implementing another elementary $S$-type operation on the quiver $\CO^\alpha_{\vec \CP}(X)$ 
(which is generically not a linear quiver), and so on, following the general recipe given in \eref{PF-OP} for implementing $S$-type operations on 
a generic quiver. \\

We would like to emphasize that the Lagrangian of the theory $\CO^\alpha_{\vec \CP}(X)$ is manifestly independent of the permutation matrices $\vec \CP$.
However, as we will see in the discussion of the dual operations, the Lagrangian of the theory dual to $\CO^\alpha_{\vec \CP}(X)$ manifestly depends on the data $\vec \CP$. This is related to the fact that while the integrand for the partition function of $X$ is manifestly invariant under a permutation of the $M_\beta$ masses $\vec{m}^\beta$ for a given $\beta$, the integrand for the partition function of $Y$ (where $\vec{m}^\beta$ appear as FI parameters) is not.\\

The four distinct types of elementary $S$-type operations can be read off from \eref{Sbasic-def}, when no defect is turned on.
We will now present the partition function realization of these operations on linear quivers. 
Extending these operations to generic quivers is straightforward, and the corresponding partition functions can be written down from the general formula \eref{PF-OP}.

\begin{itemize}
\item \textbf{Gauging operation}:  The simplest example of an elementary $S$-type operation is a gauging operation acting on a flavor node $\alpha$ 
of the quiver $X$, as shown in \figref{fig: GaugingGen}. 
The partition function of the theory $G^\alpha_{\CP}(X)$ is given by:
 \begin{align}
Z^{G^\alpha_{\CP}(X)} (\vec{v}^\alpha,\{\vec m^\gamma\}_{\gamma \neq \alpha};\vec{t},{\eta}_\alpha) =  \int \Big[d\vec{u}^\alpha\Big] \, \CZ_{G^\alpha_\CP(X)}(\vec{u}^\alpha,\eta_\alpha) \,Z^{(X,\CP)} (\vec{u}^\alpha, \vec{v}^\alpha,\{\vec m^\gamma\}_{\gamma \neq \alpha};\vec{t}), \label{G-basic}
\end{align}
where the function $\CZ_{G_{\vec \CP}(X)}$ is given as
\be \label{defZGL}
\CZ_{G^\alpha_\CP(X)}(\vec{u}^\alpha,\eta_\alpha) = Z_{\rm FI} (\vec{u}^\alpha,\eta_\alpha) \,Z_{\rm 1-loop} ^{\rm vector} (\vec{u}^\alpha).
\ee

Using the results \eref{PF-XP}, \eref{defZGL} and \eref{PF-Agen}, the partition function can be put in the standard form of \eref{PF-main} (with no defects):
\begin{align} \label{G-basic1}
Z^{G^\alpha_{\CP}(X)} (\vec{v}^\alpha,\{\vec m^\gamma\}_{\gamma \neq \alpha};\vec{t},{\eta}_\alpha)  = : \int \, \Big[d\vec{u}^\alpha\Big] \, \prod^L_{\gamma=1} \Big[d\vec{s}^\gamma\Big] \, Z^{G^\alpha_\CP(X)}_{\rm FI}\, \cdot Z^{G^\alpha_{\CP}(X)}_{\rm{1-loop}},
\end{align}
where the functions $Z^{G^\alpha_\CP(X)}_{\rm FI}$ and $Z^{G^\alpha_{\CP}(X)}_{\rm{1-loop}}$ are given as 
 \begin{align}
& Z^{G^\alpha_\CP(X)}_{\rm FI}(\{\vec s^\gamma \}, u^\alpha, \vec t, \eta_\alpha)= Z_{\rm FI} (\vec{u}^\alpha,\eta_\alpha) \, Z^{(X)}_{\rm FI}(\{\vec s^\gamma \}, \vec t), \\
 & Z^{G^\alpha_{\CP}(X)}_{\rm{1-loop}}(\{\vec s^\gamma\}, \vec u^\alpha, \{\vec m^\gamma\}_{\gamma \neq \alpha}, \vec v^\alpha)=  \Big(Z_{\rm 1-loop} ^{\rm vector} (\vec{u}^\alpha) \,\prod^L_{\gamma=1} Z^{\rm vector}_{\rm{1-loop}}(\vec s^\gamma) \Big) \nn \\
& \times \Big(Z^{\rm fund}_{\rm{1-loop}}(\vec s^\alpha, \vec v^{\alpha})\, \prod_{\gamma \neq \alpha}Z^{\rm fund}_{\rm{1-loop}}(\vec s^\gamma, \vec m^{\gamma})\Big) \,\Big(Z^{\rm bif}_{\rm{1-loop}}(\vec s^\alpha, \vec u^{\alpha},0) \, \prod^{L-1}_{\gamma=1} Z^{\rm bif}_{\rm{1-loop}}(\vec s^\gamma, \vec s^{\gamma +1},0) \Big).
 \end{align}

The Lagrangian of the theory $G^\alpha_{\CP}(X)$ can be read off from the integrand of the matrix integral on the RHS of \eref{G-basic1}, and reproduces the quiver gauge theory in the third line of \figref{fig: GaugingGen}.

\begin{figure}[htbp]
\begin{center}
\scalebox{.6}{\begin{tikzpicture}[
cnode/.style={circle,draw,thick,minimum size=4mm},snode/.style={rectangle,draw,thick,minimum size=8mm},pnode/.style={rectangle,red,draw,thick,minimum size=1.0cm}]
\node[cnode] (1) {$N_1$};
\node[cnode] (2) [right=.5cm  of 1]{$N_2$};
\node[cnode] (3) [right=.5cm of 2]{$N_3$};
\node[cnode] (4) [right=1cm of 3]{$N_{\alpha-1}$};
\node[cnode] (5) [right=0.5cm of 4]{$N_{\alpha}$};
\node[cnode] (6) [right=0.5cm of 5]{$N_{\alpha + 1}$};
\node[cnode] (7) [right=1cm of 6]{{$N_{l-2}$}};
\node[cnode] (8) [right=0.5cm of 7]{$N_{l-1}$};
\node[cnode] (9) [right=0.5cm of 8]{$N_l$};
\node[snode] (10) [below=0.5cm of 1]{$M_1$};
\node[snode] (11) [below=0.5cm of 2]{$M_2$};
\node[snode] (12) [below=0.5cm of 3]{$M_3$};
\node[snode] (13) [above=0.5cm of 4]{$M_{\alpha-1}$};
\node[pnode] (14) [below=0.5cm of 5]{$M_{\alpha}$};
\node[snode] (15) [above=0.5cm of 6]{$M_{\alpha+1}$};
\node[snode] (16) [below=0.5cm of 7]{$M_{l-2}$};
\node[snode] (17) [below=0.5cm of 8]{$M_{l-1}$};
\node[snode] (18) [below=0.5cm of 9]{$M_{l}$};
\draw[-] (1) -- (2);
\draw[-] (2)-- (3);
\draw[dashed] (3) -- (4);
\draw[-] (4) --(5);
\draw[-] (5) --(6);
\draw[dashed] (6) -- (7);
\draw[-] (7) -- (8);
\draw[-] (8) --(9);
\draw[-] (1) -- (10);
\draw[-] (2) -- (11);
\draw[-] (3) -- (12);
\draw[-] (4) -- (13);
\draw[-] (5) -- (14);
\draw[-] (6) -- (15);
\draw[-] (7) -- (16);
\draw[-] (8) -- (17);
\draw[-] (9) -- (18);
\draw[->] (6.5,-2.5) -- (6.5,-4);
\end{tikzpicture}}
\qquad
\scalebox{.6}{\begin{tikzpicture}[
cnode/.style={circle,draw,thick,minimum size=4mm},snode/.style={rectangle,draw,thick,minimum size=8mm}, pnode/.style={rectangle,red,draw,thick,minimum size=1.0cm}]
\node[cnode] (1) {$N_1$};
\node[cnode] (2) [right=.5cm  of 1]{$N_2$};
\node[cnode] (3) [right=.5cm of 2]{$N_3$};
\node[cnode] (4) [right=1cm of 3]{$N_{\alpha-1}$};
\node[cnode] (5) [right=0.5cm of 4]{$N_{\alpha}$};
\node[cnode] (6) [right=0.5cm of 5]{$N_{\alpha + 1}$};
\node[cnode] (7) [right=1cm of 6]{{$N_{l-2}$}};
\node[cnode] (8) [right=0.5cm of 7]{$N_{l-1}$};
\node[cnode] (9) [right=0.5cm of 8]{$N_l$};
\node[snode] (10) [below=0.5cm of 1]{$M_1$};
\node[snode] (11) [below=0.5cm of 2]{$M_2$};
\node[snode] (12) [below=0.5cm of 3]{$M_3$};
\node[snode] (13) [above=0.5cm of 4]{$M_{\alpha-1}$};
\node[pnode] (14) at (6,-2){$r_\alpha$};
\node[pnode] (19) at (7.5,-2){$M_\alpha - r_\alpha$};
\node[snode] (15) [above=0.5cm of 6]{$M_{\alpha+1}$};
\node[snode] (16) [below=0.5cm of 7]{$M_{l-2}$};
\node[snode] (17) [below=0.5cm of 8]{$M_{l-1}$};
\node[snode] (18) [below=0.5cm of 9]{$M_{l}$};
\draw[-] (1) -- (2);
\draw[-] (2)-- (3);
\draw[dashed] (3) -- (4);
\draw[-] (4) --(5);
\draw[-] (5) --(6);
\draw[dashed] (6) -- (7);
\draw[-] (7) -- (8);
\draw[-] (8) --(9);
\draw[-] (1) -- (10);
\draw[-] (2) -- (11);
\draw[-] (3) -- (12);
\draw[-] (4) -- (13);
\draw[-] (5) -- (14);
\draw[-] (6) -- (15);
\draw[-] (7) -- (16);
\draw[-] (8) -- (17);
\draw[-] (9) -- (18);
\draw[-] (5) -- (19);
\draw[->] (6.5,-2.75) -- (6.5,-4);
\end{tikzpicture}}
\qquad
\scalebox{.6}{\begin{tikzpicture}[
cnode/.style={circle,draw,thick,minimum size=4mm},snode/.style={rectangle,draw,thick,minimum size=8mm}, nnode/.style={circle,red,draw,thick,minimum size=1.0cm}, pnode/.style={rectangle,red,draw,thick,minimum size=1.0cm}]
\node[cnode] (1) {$N_1$};
\node[cnode] (2) [right=.5cm  of 1]{$N_2$};
\node[cnode] (3) [right=.5cm of 2]{$N_3$};
\node[cnode] (4) [right=1cm of 3]{$N_{\alpha-1}$};
\node[cnode] (5) [right=0.5cm of 4]{$N_{\alpha}$};
\node[cnode] (6) [right=0.5cm of 5]{$N_{\alpha + 1}$};
\node[cnode] (7) [right=1cm of 6]{{$N_{l-2}$}};
\node[cnode] (8) [right=0.5cm of 7]{$N_{l-1}$};
\node[cnode] (9) [right=0.5cm of 8]{$N_l$};
\node[snode] (10) [below=0.5cm of 1]{$M_1$};
\node[snode] (11) [below=0.5cm of 2]{$M_2$};
\node[snode] (12) [below=0.5cm of 3]{$M_3$};
\node[snode] (13) [above=0.5cm of 4]{$M_{\alpha-1}$};
\node[nnode] (14) at (6,-2){$r_\alpha$};
\node[pnode] (19) at (7.5,-2){$M_\alpha - r_\alpha$};
\node[snode] (15) [above=0.5cm of 6]{$M_{\alpha+1}$};
\node[snode] (16) [below=0.5cm of 7]{$M_{l-2}$};
\node[snode] (17) [below=0.5cm of 8]{$M_{l-1}$};
\node[snode] (18) [below=0.5cm of 9]{$M_{l}$};
\draw[-] (1) -- (2);
\draw[-] (2)-- (3);
\draw[dashed] (3) -- (4);
\draw[-] (4) --(5);
\draw[-] (5) --(6);
\draw[dashed] (6) -- (7);
\draw[-] (7) -- (8);
\draw[-] (8) --(9);
\draw[-] (1) -- (10);
\draw[-] (2) -- (11);
\draw[-] (3) -- (12);
\draw[-] (4) -- (13);
\draw[-] (5) -- (14);
\draw[-] (5) -- (19);
\draw[-] (6) -- (15);
\draw[-] (7) -- (16);
\draw[-] (8) -- (17);
\draw[-] (9) -- (18);
\end{tikzpicture}}
\caption{This figure illustrates the gauging operation $G^\alpha_\CP$ on a generic linear quiver $X$ at a flavor node $U(M_\alpha)$.}
\label{fig: GaugingGen}
\end{center}
\end{figure}
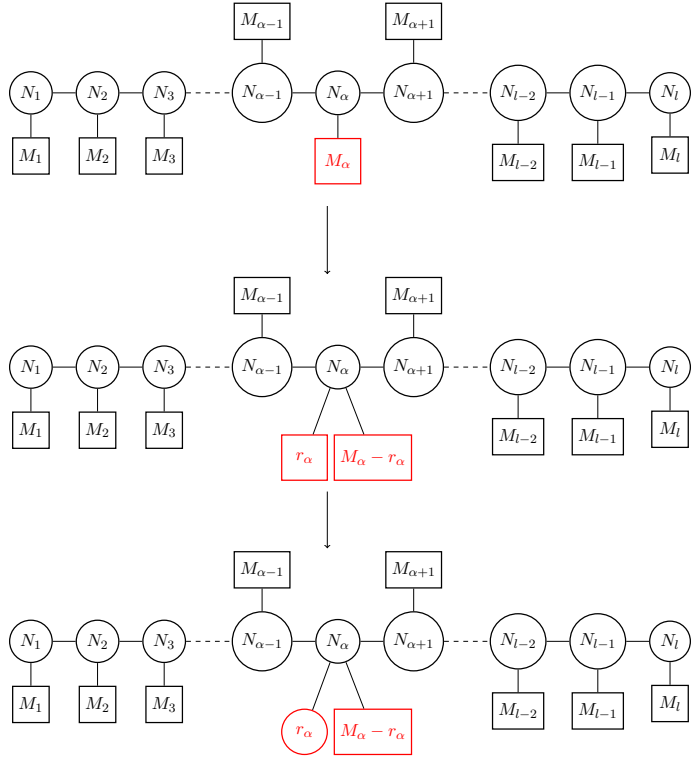

\item  \textbf{Flavoring-Gauging Operation}: The second elementary $S$-type operation involves a flavoring operation 
combined with a gauging operation, as shown in \figref{fig: GaugingFlavGen}. In the notation of \eref{Sbasic-def}, the 
combined operation can be denoted as 
\be 
\CO^\alpha_\CP (X) =  G^\alpha_{\CP} \circ F^\alpha_\CP (X),
\ee
where $\CP$ is a permutation matrix of order $M_\alpha$. We will refer to $\CO^\alpha_\CP$ as the \textit{flavoring-gauging} operation.
Following \eref{PF-OPLQ}, the partition function of the theory $G^\alpha_{\CP} \circ F^\alpha_\CP(X)$ is given as: 
\begin{align} \label{GF-basic}
&Z^{G^\alpha_{\CP} \circ F^\alpha_\CP(X)}(\vec{v}^\alpha, \vec{m}^\alpha_F, \{\vec m^\gamma\}_{\gamma \neq \alpha};\vec{t}, \eta_\alpha) \nn \\
& =  \int \, \Big[\de {u^\alpha} \Big]\, \CZ_{G^\alpha_{\CP} \circ F^\alpha_\CP(X)}(\vec{u}^\alpha, \vec m^\alpha_F, \eta_\alpha) \, Z^{(X,\CP)}(\vec{u}^\alpha,\vec{v}^\alpha, \{\vec m^\gamma\}_{\gamma \neq \alpha};\vec{t}),
\end{align}
where the function $\CZ_{G^\alpha_{\CP} \circ F^\alpha_\CP(X)}$ can be constructed using \eref{CZ-OP}:
\be \label{defZGF}
\CZ_{G^\alpha_{\CP} \circ F^\alpha_\CP(X)}(\vec{u}^\alpha, \vec m^\alpha_F, \eta_\alpha)= Z_{\rm FI} (\vec{u}^\alpha,\eta_\alpha)\,Z_{\rm 1-loop}^{\rm vector} (\vec{u}^\alpha)\,Z_{\rm 1-loop}^{\rm hyper} (\vec{u}^\alpha, \vec m^\alpha_F).
\ee

Using the results \eref{PF-XP}, \eref{defZGF} and  \eref{PF-Agen}, the partition function can be put in the standard form of \eref{PF-main} (with no defects): 
\begin{align}\label{GF-basic2}
 Z^{G^\alpha_{\CP} \circ F^\alpha_\CP(X)} (\vec{v}^\alpha,\vec{m}^\alpha_F, \{\vec m^\gamma\}_{\gamma \neq \alpha};\vec{t},{\eta}_\alpha) 
= : \int  \Big[\de {u^\alpha} \Big]  \, \prod^L_{\gamma=1} \Big[d\vec{s}^\gamma\Big]  \, Z^{G^\alpha_{\CP} \circ F^\alpha_\CP(X)}_{\rm FI} \cdot Z^{G^\alpha_{\CP} \circ F^\alpha_\CP(X)}_{\rm{1-loop}},
\end{align}
where the functions $Z^{G^\alpha_{\CP} \circ F^\alpha_\CP(X)}_{\rm FI}$ and $Z^{G^\alpha_{\CP} \circ F^\alpha_\CP(X)}_{\rm{1-loop}}$ are given as
\begin{align}
& Z^{G^\alpha_{\CP} \circ F^\alpha_\CP(X)}_{\rm FI}(\{\vec s^\gamma \}, u^\alpha, \vec t, \eta_\alpha)= Z_{\rm FI} (\vec{u}^\alpha,\eta_\alpha) \, Z^{(X)}_{\rm FI}(\{\vec s^\gamma \}, \vec t), \label{GF-basic2a}\\
& Z^{G^\alpha_{\CP} \circ F^\alpha_\CP(X)}_{\rm{1-loop}}(\{\vec s^\gamma\}, \vec u^\alpha, \{\vec m^\gamma\}_{\gamma \neq \alpha}, \vec v^\alpha, \vec{m}^\alpha_F)= \Big(Z_{\rm 1-loop} ^{\rm vector} (\vec{u}^\alpha) \,\prod^L_{\gamma=1} Z^{\rm vector}_{\rm{1-loop}}(\vec s^\gamma) \Big)\nn \\
& \qquad \qquad \qquad \qquad \qquad \qquad \times \,\Big(Z^{\rm fund}_{\rm{1-loop}}(\vec u^\alpha, \vec m^{\alpha}_F)\,Z^{\rm fund}_{\rm{1-loop}}(\vec s^\alpha, \vec v^{\alpha})\, \prod_{\gamma \neq \alpha}Z^{\rm fund}_{\rm{1-loop}}(\vec s^\gamma, \vec m^{\gamma})\Big) \nn \\
& \qquad \qquad \qquad \qquad \qquad \qquad \times \Big(Z^{\rm bif}_{\rm{1-loop}}(\vec s^\alpha, \vec u^{\alpha},0) \, \prod^{L-1}_{\gamma=1} Z^{\rm bif}_{\rm{1-loop}}(\vec s^\gamma, \vec s^{\gamma +1},0) \Big). \label{GF-basic2b}
\end{align} 
The Lagrangian of the theory $G^\alpha_{\CP} \circ F^\alpha_\CP(X)$ can be read off from the integrand of the matrix integral on the RHS of \eref{GF-basic2}, and reproduces the quiver gauge theory in the third line of \figref{fig: GaugingFlavGen}. The parameters $\vec m^{\alpha}_F$, which live in the Cartan subalgebra of $G^\alpha_F$, can be identified as masses of the added hypermultiplets. 

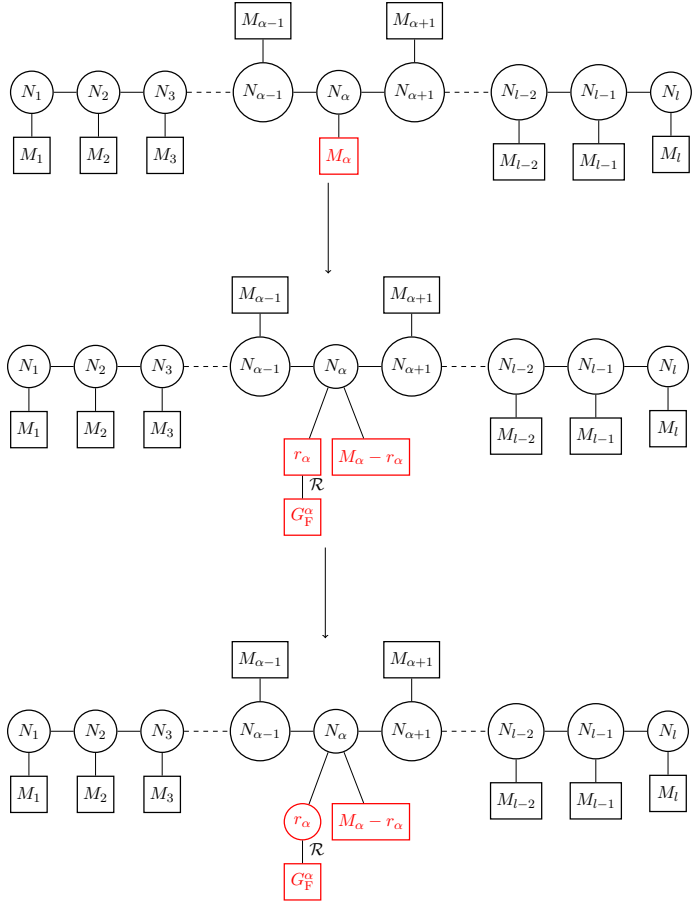
\begin{figure}[htbp]
\begin{center}
\scalebox{0.6}{
\begin{tikzpicture}[
cnode/.style={circle,draw,thick,minimum size=4mm},snode/.style={rectangle,draw,thick,minimum size=8mm},pnode/.style={rectangle,red,draw,thick,minimum size=8mm}]
\node[cnode] (1) {$N_1$};
\node[cnode] (2) [right=.5cm  of 1]{$N_2$};
\node[cnode] (3) [right=.5cm of 2]{$N_3$};
\node[cnode] (4) [right=1cm of 3]{$N_{\alpha-1}$};
\node[cnode] (5) [right=0.5cm of 4]{$N_{\alpha}$};
\node[cnode] (6) [right=0.5cm of 5]{$N_{\alpha + 1}$};
\node[cnode] (7) [right=1cm of 6]{{$N_{l-2}$}};
\node[cnode] (8) [right=0.5cm of 7]{$N_{l-1}$};
\node[cnode] (9) [right=0.5cm of 8]{$N_l$};
\node[snode] (10) [below=0.5cm of 1]{$M_1$};
\node[snode] (11) [below=0.5cm of 2]{$M_2$};
\node[snode] (12) [below=0.5cm of 3]{$M_3$};
\node[snode] (13) [above=0.5cm of 4]{$M_{\alpha-1}$};
\node[pnode] (14) [below=0.5cm of 5]{$M_{\alpha}$};
\node[snode] (15) [above=0.5cm of 6]{$M_{\alpha+1}$};
\node[snode] (16) [below=0.5cm of 7]{$M_{l-2}$};
\node[snode] (17) [below=0.5cm of 8]{$M_{l-1}$};
\node[snode] (18) [below=0.5cm of 9]{$M_{l}$};
\draw[-] (1) -- (2);
\draw[-] (2)-- (3);
\draw[dashed] (3) -- (4);
\draw[-] (4) --(5);
\draw[-] (5) --(6);
\draw[dashed] (6) -- (7);
\draw[-] (7) -- (8);
\draw[-] (8) --(9);
\draw[-] (1) -- (10);
\draw[-] (2) -- (11);
\draw[-] (3) -- (12);
\draw[-] (4) -- (13);
\draw[-] (5) -- (14);
\draw[-] (6) -- (15);
\draw[-] (7) -- (16);
\draw[-] (8) -- (17);
\draw[-] (9) -- (18);
\draw[->] (6.5,-2) -- (6.5,-4);
\end{tikzpicture}}
\qquad
\scalebox{0.6}{\begin{tikzpicture}[
cnode/.style={circle,draw,thick,minimum size=4mm},snode/.style={rectangle,draw,thick,minimum size=8mm},pnode/.style={rectangle,red,draw,thick,minimum size=8mm}]
\node[cnode] (1) {$N_1$};
\node[cnode] (2) [right=.5cm  of 1]{$N_2$};
\node[cnode] (3) [right=.5cm of 2]{$N_3$};
\node[cnode] (4) [right=1cm of 3]{$N_{\alpha-1}$};
\node[cnode] (5) [right=0.5cm of 4]{$N_{\alpha}$};
\node[cnode] (6) [right=0.5cm of 5]{$N_{\alpha + 1}$};
\node[cnode] (7) [right=1cm of 6]{{$N_{l-2}$}};
\node[cnode] (8) [right=0.5cm of 7]{$N_{l-1}$};
\node[cnode] (9) [right=0.5cm of 8]{$N_l$};
\node[snode] (10) [below=0.5cm of 1]{$M_1$};
\node[snode] (11) [below=0.5cm of 2]{$M_2$};
\node[snode] (12) [below=0.5cm of 3]{$M_3$};
\node[snode] (13) [above=0.5cm of 4]{$M_{\alpha-1}$};
\node[pnode] (14) at (6,-2){$r_\alpha$};
\node[pnode] (19) at (7.5,-2){$M_\alpha - r_\alpha$};
\node[pnode] (20) [below=0.5cm of 14]{$G^\alpha_{\rm F}$};
\node[text width=0.1cm](30) at (6.2,-2.6) {$\CR$};
\node[snode] (15) [above=0.5cm of 6]{$M_{\alpha+1}$};
\node[snode] (16) [below=0.5cm of 7]{$M_{l-2}$};
\node[snode] (17) [below=0.5cm of 8]{$M_{l-1}$};
\node[snode] (18) [below=0.5cm of 9]{$M_{l}$};
\draw[-] (1) -- (2);
\draw[-] (2)-- (3);
\draw[dashed] (3) -- (4);
\draw[-] (4) --(5);
\draw[-] (5) --(6);
\draw[dashed] (6) -- (7);
\draw[-] (7) -- (8);
\draw[-] (8) --(9);
\draw[-] (1) -- (10);
\draw[-] (2) -- (11);
\draw[-] (3) -- (12);
\draw[-] (4) -- (13);
\draw[-] (5) -- (14);
\draw[-] (5) -- (19);
\draw[-] (20) -- (14);
\draw[-] (6) -- (15);
\draw[-] (7) -- (16);
\draw[-] (8) -- (17);
\draw[-] (9) -- (18);
\draw[->] (6.5,-4) -- (6.5,-6);
\end{tikzpicture}}
\qquad
\scalebox{0.6}{\begin{tikzpicture}[
cnode/.style={circle,draw,thick,minimum size=4mm},snode/.style={rectangle,draw,thick,minimum size=8mm}, pnode/.style={rectangle,red,draw,thick,minimum size=8mm},nnode/.style={circle,red,draw,thick,minimum size=8mm}]
\node[cnode] (1) {$N_1$};
\node[cnode] (2) [right=.5cm  of 1]{$N_2$};
\node[cnode] (3) [right=.5cm of 2]{$N_3$};
\node[cnode] (4) [right=1cm of 3]{$N_{\alpha-1}$};
\node[cnode] (5) [right=0.5cm of 4]{$N_{\alpha}$};
\node[cnode] (6) [right=0.5cm of 5]{$N_{\alpha + 1}$};
\node[cnode] (7) [right=1cm of 6]{{$N_{l-2}$}};
\node[cnode] (8) [right=0.5cm of 7]{$N_{l-1}$};
\node[cnode] (9) [right=0.5cm of 8]{$N_l$};
\node[snode] (10) [below=0.5cm of 1]{$M_1$};
\node[snode] (11) [below=0.5cm of 2]{$M_2$};
\node[snode] (12) [below=0.5cm of 3]{$M_3$};
\node[snode] (13) [above=0.5cm of 4]{$M_{\alpha-1}$};
\node[nnode] (14) at (6,-2){$r_\alpha$};
\node[pnode] (19) at (7.5,-2){$M_\alpha - r_\alpha$};
\node[pnode] (20) [below=0.5cm of 14]{$G^\alpha_{\rm F}$};
\node[text width=0.1cm](30) at (6.2,-2.6) {$\CR$};
\node[snode] (15) [above=0.5cm of 6]{$M_{\alpha+1}$};
\node[snode] (16) [below=0.5cm of 7]{$M_{l-2}$};
\node[snode] (17) [below=0.5cm of 8]{$M_{l-1}$};
\node[snode] (18) [below=0.5cm of 9]{$M_{l}$};
\draw[-] (1) -- (2);
\draw[-] (2)-- (3);
\draw[dashed] (3) -- (4);
\draw[-] (4) --(5);
\draw[-] (5) --(6);
\draw[dashed] (6) -- (7);
\draw[-] (7) -- (8);
\draw[-] (8) --(9);
\draw[-] (1) -- (10);
\draw[-] (2) -- (11);
\draw[-] (3) -- (12);
\draw[-] (4) -- (13);
\draw[-] (5) -- (14);
\draw[-] (5) -- (19);
\draw[-] (20) -- (14);
\draw[-] (6) -- (15);
\draw[-] (7) -- (16);
\draw[-] (8) -- (17);
\draw[-] (9) -- (18);
\end{tikzpicture}}
\caption{This figure illustrates the flavoring-gauging operation $G^\alpha_{\CP} \circ F^\alpha_\CP$ on a generic linear quiver $X$ at the flavor node $U(M_\alpha)$.}
\label{fig: GaugingFlavGen}
\end{center}
\end{figure}

\item \textbf{Identification-gauging operation} : The third elementary $S$-type operation involves an identification operation 
combined with a gauging operation, as shown in \figref{fig: IdGaugingGen} for $p=3$ nodes. In the notation of \eref{Sbasic-def}, 
the combined operation can be denoted as 
\be 
\CO^\alpha_{\vec \CP}(X)= G^\alpha_{\vec \CP} \circ I^\alpha_{\vec \CP}(X)
\ee
where $\vec \CP=\{ \CP_\beta\} $ collectively denotes the permutation matrices of order $M_\beta$. We will refer to as $\CO^\alpha_{\vec \CP}$ as the \textit{identification-gauging} operation. The quiver gauge theory  $G^\alpha_{\vec \CP} \circ I^\alpha_{\vec \CP} (X)$ has the partition function:
\begin{align}\label{GI-basic}
& Z^{G^\alpha_{\vec \CP} \circ I^\alpha_{\vec \CP} (X)}( \{ \vec{v}^\beta \},\{\vec m^\gamma\}_{\gamma \neq \beta},\vec{\mu};\vec{t},\eta_{\alpha}) \nn \\
&= \int \Big[ d{\vec{u}^{\alpha}} \Big]\,\CZ_{G^\alpha_{\vec \CP} \circ I^\alpha_{\vec \CP}(X)}(\vec{u}^{\alpha},\{\vec{u}^\beta\}, \eta_{\alpha}, \vec \mu) \,
 Z^{(X,\{ P_\beta\})} (\{\vec{u}^\beta\}, \{\vec{v}^\beta\},\{\vec m^\gamma\}_{\gamma \neq \beta};\vec{t}),
\end{align} 
where $\CZ_{G^\alpha_{\vec \CP} \circ I^\alpha_{\vec \CP}(X)}$ is an operator of the following form: 
\be \label{defZGI}
\CZ_{G^\alpha_{\vec \CP} \circ I^\alpha_{\vec \CP}(X)}= Z_{\rm FI} (\vec{u}^{\alpha},\eta_{\alpha})\,Z_{\rm 1-loop} ^{\rm vector} (\vec{u}^{\alpha}) \,\int \prod^{p}_{j=1} \prod^{r_\alpha}_{i=1} d u_i^{\gamma_j} \, \prod^{p}_{j=1} \delta^{(r_\alpha)}\Big(\vec{u}^{\alpha} - \vec{u}^{\gamma_{j}} + {\mu}^{\gamma_j} \Big).
\ee

Using the results \eref{PF-XP}, \eref{defZGI} and \eref{PF-Agen}, the partition function can be put in the standard form of \eref{PF-main} (with no defects):
\begin{align}\label{GI-basic1}
Z^{G^\alpha_{\vec \CP} \circ I^\alpha_{\vec \CP} (X)} (\{ \vec{v}^\beta \},\{\vec m^\gamma\}_{\gamma \neq \beta},\vec{\mu};\vec{t}, {\eta}_{\alpha}) 
= : \int \,\Big[ d {\vec{u}^{\alpha}} \Big]  \, \prod^L_{\gamma=1} \Big[ d\vec{s}^\gamma \Big] \, Z^{G^\alpha_{\vec \CP} \circ I^\alpha_{\vec \CP} (X)}_{\rm FI}
\cdot Z^{G^\alpha_{\vec \CP} \circ I^\alpha_{\vec \CP} (X)}_{\rm{1-loop}},
\end{align}
where the functions $Z^{G^\alpha_{\vec \CP} \circ I^\alpha_{\vec \CP} (X)}_{\rm FI}$ and $Z^{G^\alpha_{\vec \CP} \circ I^\alpha_{\vec \CP} (X)}_{\rm{1-loop}}$ are given as
\begin{align}
& Z^{G^\alpha_{\vec \CP} \circ I^\alpha_{\vec \CP} (X)}_{\rm FI}(\{\vec s^{\gamma} \}, u^{\alpha}, \vec t, \eta_{\alpha})= Z_{\rm FI} (\vec{u}^{\alpha},\eta_{\alpha}) \, Z^{(X)}_{\rm FI}(\{\vec s^\gamma \}, \vec t), \label{GI-basic2a}\\
& Z^{G^\alpha_{\vec \CP} \circ I^\alpha_{\vec \CP} (X)}_{\rm{1-loop}}(\{\vec s^\gamma\}, \vec u^{\alpha}, \{\vec m^\gamma\}_{\gamma \neq \beta}, \{\vec v^\beta\},\vec \mu)=\Big(Z_{\rm 1-loop} ^{\rm vector} (\vec{u}^{\alpha}) \,\prod^L_{\gamma=1} Z^{\rm vector}_{\rm{1-loop}}(\vec s^\gamma) \Big) \nn \\
& \times \Big(\prod^p_{j=1} Z^{\rm fund}_{\rm{1-loop}}(\vec s^{\gamma_j}, \vec v^{\gamma_j})\, \prod_{\gamma \neq \beta}Z^{\rm fund}_{\rm{1-loop}}(\vec s^\gamma, \vec m^{\gamma})\Big)\,\Big(\prod^p_{j=1} Z^{\rm bif}_{\rm{1-loop}}(\vec s^{\gamma_j}, \vec u^{\alpha}, - \mu^{\gamma_j}) \, \prod^{L-1}_{\gamma=1} Z^{\rm bif}_{\rm{1-loop}}(\vec s^\gamma, \vec s^{\gamma +1},0) \Big). \label{GI-basic2b}
\end{align}
The Lagrangian can now be read off from the integrand of the matrix model on the RHS of \eref{GI-basic1}, and agrees with the quiver gauge theory in the third line of \figref{fig: IdGaugingGen}. Note that the parameters $\vec \mu$ appear as masses of the hypermultiplets in the bifundamental of $U(M_{\gamma_j}) \times U(r)$.

\begin{figure}[htbp]
\begin{center}
\scalebox{0.6}{\begin{tikzpicture}[
cnode/.style={circle,draw,thick,minimum size=4mm},snode/.style={rectangle,draw,thick,minimum size=8mm},pnode/.style={rectangle,red,draw,thick,minimum size=8mm} ]
\node[cnode] (1) {$N_1$};
\node[cnode] (2) [right=.5cm  of 1]{$N_2$};
\node[cnode] (3) [right=.5cm of 2]{$N_3$};
\node[cnode] (4) [right=1cm of 3]{$N_{\alpha-1}$};
\node[cnode] (5) [right=0.5cm of 4]{$N_{\alpha}$};
\node[cnode] (6) [right=0.5cm of 5]{$N_{\alpha + 1}$};
\node[cnode] (7) [right=1cm of 6]{{$N_{l-2}$}};
\node[cnode] (8) [right=0.5cm of 7]{$N_{l-1}$};
\node[cnode] (9) [right=0.5cm of 8]{$N_l$};
\node[snode] (10) [below=0.5cm of 1]{$M_1$};
\node[snode] (11) [below=0.5cm of 2]{$M_2$};
\node[snode] (12) [below=0.5cm of 3]{$M_3$};
\node[pnode] (13) [above=0.5cm of 4]{$M_{\alpha-1}$};
\node[pnode] (14) [below=0.5cm of 5]{$M_{\alpha}$};
\node[pnode] (15) [above=0.5cm of 6]{$M_{\alpha+1}$};
\node[snode] (16) [below=0.5cm of 7]{$M_{l-2}$};
\node[snode] (17) [below=0.5cm of 8]{$M_{l-1}$};
\node[snode] (18) [below=0.5cm of 9]{$M_{l}$};
\draw[-] (1) -- (2);
\draw[-] (2)-- (3);
\draw[dashed] (3) -- (4);
\draw[-] (4) --(5);
\draw[-] (5) --(6);
\draw[dashed] (6) -- (7);
\draw[-] (7) -- (8);
\draw[-] (8) --(9);
\draw[-] (1) -- (10);
\draw[-] (2) -- (11);
\draw[-] (3) -- (12);
\draw[-] (4) -- (13);
\draw[-] (5) -- (14);
\draw[-] (6) -- (15);
\draw[-] (7) -- (16);
\draw[-] (8) -- (17);
\draw[-] (9) -- (18);
\draw[->] (6.5,-2) -- (6.5,-4);
\end{tikzpicture}}
\qquad
\scalebox{0.6}{\begin{tikzpicture}[
cnode/.style={circle,draw,thick,minimum size=4mm},snode/.style={rectangle,draw,thick,minimum size=8mm}, pnode/.style={rectangle,red,draw,thick,minimum size=8mm}]
\node[cnode] (1) {$N_1$};
\node[cnode] (2) [right=.5cm  of 1]{$N_2$};
\node[cnode] (3) [right=.5cm of 2]{$N_3$};
\node[cnode] (4) [right=1cm of 3]{$N_{\alpha-1}$};
\node[cnode] (5) [right=0.5cm of 4]{$N_{\alpha}$};
\node[cnode] (6) [right=0.5cm of 5]{$N_{\alpha + 1}$};
\node[cnode] (7) [right=1cm of 6]{{$N_{l-2}$}};
\node[cnode] (8) [right=0.5cm of 7]{$N_{l-1}$};
\node[cnode] (9) [right=0.5cm of 8]{$N_l$};
\node[snode] (10) [below=0.5cm of 1]{$M_1$};
\node[snode] (11) [below=0.5cm of 2]{$M_2$};
\node[snode] (12) [below=0.5cm of 3]{$M_3$};
\node[pnode] (13) [above=1.5cm of 4]{$M_{\alpha-1}-r$};
\node[pnode] (20) [below=0.5cm of 4]{$r$};
\node[pnode] (14) [below=0.5cm of 5]{$r$};
\node[pnode] (21) [above=0.5cm of 5]{$M_{\alpha}-r$};
\node[pnode] (15) [above=1.5cm of 6]{$M_{\alpha+1}-r$};
\node[pnode] (22) [below=0.5cm of 6]{$r$};
\node[snode] (16) [below=0.5cm of 7]{$M_{l-2}$};
\node[snode] (17) [below=0.5cm of 8]{$M_{l-1}$};
\node[snode] (18) [below=0.5cm of 9]{$M_{l}$};
\draw[-] (1) -- (2);
\draw[-] (2)-- (3);
\draw[dashed] (3) -- (4);
\draw[-] (4) --(5);
\draw[-] (5) --(6);
\draw[-] (5) --(21);
\draw[dashed] (6) -- (7);
\draw[-] (7) -- (8);
\draw[-] (8) --(9);
\draw[-] (1) -- (10);
\draw[-] (2) -- (11);
\draw[-] (3) -- (12);
\draw[-] (4) -- (13);
\draw[-] (5) -- (14);
\draw[-] (6) -- (15);
\draw[-] (7) -- (16);
\draw[-] (8) -- (17);
\draw[-] (9) -- (18);
\draw[-] (4) -- (20);
\draw[-] (6) -- (22);
\draw[->] (6.5,-2) -- (6.5,-4);
\end{tikzpicture}}
\scalebox{0.6}{\begin{tikzpicture}[
cnode/.style={circle,draw,thick,minimum size=4mm},snode/.style={rectangle,draw,thick,minimum size=8mm},pnode/.style={rectangle,red,draw,thick,minimum size=8mm},nnode/.style={circle,red,draw,thick,minimum size=8mm}]
\node[cnode] (1) {$N_1$};
\node[cnode] (2) [right=.5cm  of 1]{$N_2$};
\node[cnode] (3) [right=.5cm of 2]{$N_3$};
\node[cnode] (4) [right=1cm of 3]{$N_{\alpha-1}$};
\node[cnode] (5) [right=0.5cm of 4]{$N_{\alpha}$};
\node[cnode] (6) [right=0.5cm of 5]{$N_{\alpha + 1}$};
\node[cnode] (7) [right=1cm of 6]{{$N_{l-2}$}};
\node[cnode] (8) [right=0.5cm of 7]{$N_{l-1}$};
\node[cnode] (9) [right=0.5cm of 8]{$N_l$};
\node[snode] (10) [below=0.5cm of 1]{$M_1$};
\node[snode] (11) [below=0.5cm of 2]{$M_2$};
\node[snode] (12) [below=0.5cm of 3]{$M_3$};
\node[pnode] (13) [above=1.5cm of 4]{$M_{\alpha-1}-r$};
\node[pnode] (14) [below=0.5cm of 5]{$r$};
\node[pnode] (19) [above=0.5cm of 5]{$M_\alpha -r$};
\node[pnode] (15) [above=1.5cm of 6]{$M_{\alpha+1} -r$};
\node[snode] (16) [below=0.5cm of 7]{$M_{l-2}$};
\node[snode] (17) [below=0.5cm of 8]{$M_{l-1}$};
\node[snode] (18) [below=0.5cm of 9]{$M_{l}$};
\draw[-] (1) -- (2);
\draw[-] (2)-- (3);
\draw[dashed] (3) -- (4);
\draw[-] (4) --(5);
\draw[-] (5) --(6);
\draw[dashed] (6) -- (7);
\draw[-] (7) -- (8);
\draw[-] (8) --(9);
\draw[-] (1) -- (10);
\draw[-] (2) -- (11);
\draw[-] (3) -- (12);
\draw[-] (4) -- (13);
\draw[-] (5) -- (14);
\draw[-] (5) -- (19);
\draw[-] (4) -- (14);
\draw[-] (6) -- (14);
\draw[-] (6) -- (15);
\draw[-] (7) -- (16);
\draw[-] (8) -- (17);
\draw[-] (9) -- (18);
\draw[->] (6.5,-2) -- (6.5,-4);
\end{tikzpicture}}
\scalebox{0.6}{\begin{tikzpicture}[
cnode/.style={circle,draw,thick,minimum size=4mm},snode/.style={rectangle,draw,thick,minimum size=8mm},pnode/.style={rectangle,red,draw,thick,minimum size=8mm},nnode/.style={circle,red,draw,thick,minimum size=8mm}]
\node[cnode] (1) {$N_1$};
\node[cnode] (2) [right=.5cm  of 1]{$N_2$};
\node[cnode] (3) [right=.5cm of 2]{$N_3$};
\node[cnode] (4) [right=1cm of 3]{$N_{\alpha-1}$};
\node[cnode] (5) [right=0.5cm of 4]{$N_{\alpha}$};
\node[cnode] (6) [right=0.5cm of 5]{$N_{\alpha + 1}$};
\node[cnode] (7) [right=1cm of 6]{{$N_{l-2}$}};
\node[cnode] (8) [right=0.5cm of 7]{$N_{l-1}$};
\node[cnode] (9) [right=0.5cm of 8]{$N_l$};
\node[snode] (10) [below=0.5cm of 1]{$M_1$};
\node[snode] (11) [below=0.5cm of 2]{$M_2$};
\node[snode] (12) [below=0.5cm of 3]{$M_3$};
\node[pnode] (13) [above=1.5cm of 4]{$M_{\alpha-1}-r$};
\node[nnode] (14) [below=0.5cm of 5]{$r$};
\node[pnode] (19) [above=0.5cm of 5]{$M_\alpha -r$};
\node[pnode] (15) [above=1.5cm of 6]{$M_{\alpha+1} -r$};
\node[snode] (16) [below=0.5cm of 7]{$M_{l-2}$};
\node[snode] (17) [below=0.5cm of 8]{$M_{l-1}$};
\node[snode] (18) [below=0.5cm of 9]{$M_{l}$};
\draw[-] (1) -- (2);
\draw[-] (2)-- (3);
\draw[dashed] (3) -- (4);
\draw[-] (4) --(5);
\draw[-] (5) --(6);
\draw[dashed] (6) -- (7);
\draw[-] (7) -- (8);
\draw[-] (8) --(9);
\draw[-] (1) -- (10);
\draw[-] (2) -- (11);
\draw[-] (3) -- (12);
\draw[-] (4) -- (13);
\draw[-] (5) -- (14);
\draw[-] (5) -- (19);
\draw[-] (4) -- (14);
\draw[-] (6) -- (14);
\draw[-] (6) -- (15);
\draw[-] (7) -- (16);
\draw[-] (8) -- (17);
\draw[-] (9) -- (18);
\end{tikzpicture}}
\caption{This figure illustrates the identification-gauging $G^\alpha_{\vec \CP} \circ I^\alpha_{\vec \CP}$ operation on a generic linear quiver $X$ involving $p=3$ nodes labelled by $\beta=\alpha-1, \alpha, \alpha+1$.}
\label{fig: IdGaugingGen}
\end{center}
\end{figure}
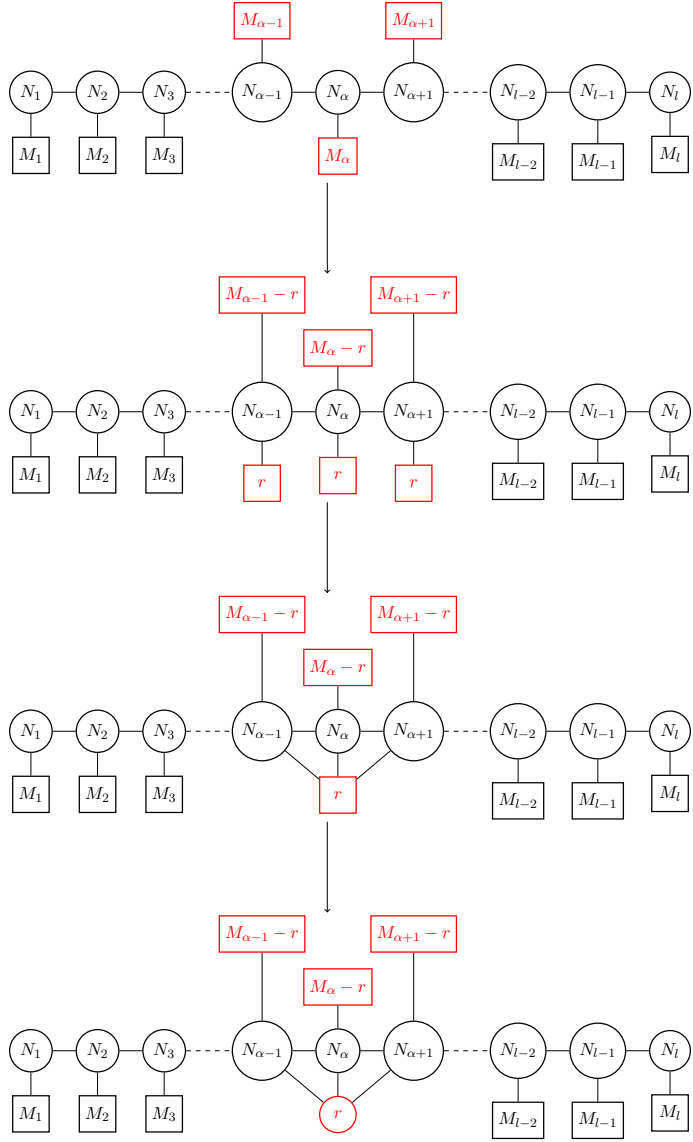

\item \textbf{Identification-flavoring-gauging operation} : The fourth elementary $S$-type operation involves a combination of an identification operation,  
a flavoring operation, and a gauging operation, as shown in \figref{fig: IdFGaugingGen} for $p=3$ nodes. 
In the notation of \eref{Sbasic-def}, therefore, the combined operation can be denoted as 
\be 
\CO^\alpha_{\vec \CP}(X)= G^\alpha_{\vec \CP} \circ F^\alpha_{\vec \CP} \circ I^\alpha_{\vec \CP}(X)
\ee
where $\vec \CP=\{ \CP_\beta\} $ collectively denotes the permutation matrices of order $M_\beta$. We will refer to as $\CO^\alpha_{\vec \CP}$ as the \textit{identification-flavoring-gauging} operation.
The quiver gauge theory  $G^\alpha_{\vec \CP} \circ F^\alpha_{\vec \CP} \circ I^\alpha_{\vec \CP} (X)$ has the partition function:
\begin{align}\label{GFI-basic}
& Z^{G^\alpha_{\vec \CP} \circ F^\alpha_{\vec \CP} \circ I^\alpha_{\vec \CP} (X)}( \{ \vec{v}^\beta \},\{\vec m^\gamma\}_{\gamma \neq \beta},\vec{\mu}, \vec m^\alpha_F; \vec{t},\eta_{\alpha}) \nn \\
&= \int \Big[ d{\vec{u}^{\alpha}} \Big]\,\CZ_{G^\alpha_{\vec \CP} \circ F^\alpha_{\vec \CP} \circ I^\alpha_{\vec \CP}}(\vec{u}^{\alpha},\{\vec{u}^\beta\}, \eta_{\alpha}, \vec \mu, \vec m^\alpha_F) \,
 Z^{(X,\{ P_\beta\})} (\{\vec{u}^\beta\}, \{\vec{v}^\beta\},\{\vec m^\gamma\}_{\gamma \neq \beta};\vec{t}),
\end{align} 
where $\CZ_{G^\alpha_{\vec \CP} \circ F^\alpha_{\vec \CP} \circ I^\alpha_{\vec \CP}(X)}$ is an operator of the following form: 
\begin{align} \label{defZGFI}
\CZ_{G^\alpha_{\vec \CP} \circ F^\alpha_{\vec \CP} \circ I^\alpha_{\vec \CP}(X)}=&Z_{\rm FI} (\vec{u}^{\alpha},\eta_{\alpha})\,Z_{\rm 1-loop} ^{\rm vector} (\vec{u}^{\alpha}) \,Z_{\rm 1-loop}^{\rm hyper} (\vec{u}^\alpha, \vec m^\alpha_F) \nn \\
& \times \, \int \prod^{p}_{j=1} \prod^{r_\alpha}_{i=1} d u_i^{\gamma_j} \,\prod^{p}_{j=1} \delta^{(r_\alpha)}\Big(\vec{u}^{\alpha} - \vec{u}^{\gamma_{j}} + {\mu}^{\gamma_j} \Big).
\end{align}

By using the result \eref{PF-XP} and the formula \eref{PF-Agen}, the partition function can be put in the standard form of \eref{PF-main} (with no defects):
\begin{align}\label{GFI-basic1}
Z^{G^\alpha_{\vec \CP} \circ F^\alpha_{\vec \CP} \circ I^\alpha_{\vec \CP} (X)} 
= : \int \,\Big[ d {\vec{u}^{\alpha}} \Big]  \, \prod^L_{\gamma=1} \Big[ d\vec{s}^\gamma \Big] \, Z^{G^\alpha_{\vec \CP} \circ F^\alpha_{\vec \CP} \circ I^\alpha_{\vec \CP} (X)}_{\rm FI}
\cdot Z^{G^\alpha_{\vec \CP} \circ F^\alpha_{\vec \CP} \circ I^\alpha_{\vec \CP} (X)}_{\rm{1-loop}},
\end{align}
where the functions $Z^{G^\alpha_{\vec \CP} \circ F^\alpha_{\vec \CP} \circ I^\alpha_{\vec \CP} (X)}_{\rm FI}$ and $Z^{G^\alpha_{\vec \CP} \circ F^\alpha_{\vec \CP} \circ I^\alpha_{\vec \CP} (X)}_{\rm{1-loop}}$ are given as
\begin{align}
& Z^{G^\alpha_{\vec \CP} \circ F^\alpha_{\vec \CP} \circ I^\alpha_{\vec \CP} (X)}_{\rm FI}(\{\vec s^{\gamma} \}, u^{\alpha}, \vec t, \eta_{\alpha})= Z_{\rm FI} (\vec{u}^{\alpha},\eta_{\alpha}) \, Z^{(X)}_{\rm FI}(\{\vec s^\gamma \}, \vec t), \label{GFI-basic2a}\\
& Z^{G^\alpha_{\vec \CP} \circ F^\alpha_{\vec \CP} \circ I^\alpha_{\vec \CP} (X)}_{\rm{1-loop}}(\{\vec s^\gamma\}, \vec u^{\alpha}, \{\vec m^\gamma\}_{\gamma \neq \beta}, \{\vec v^\beta\},\vec \mu, \vec m^\alpha_F)=\Big(Z_{\rm 1-loop} ^{\rm vector} (\vec{u}^{\alpha}) \,\prod^L_{\gamma=1} Z^{\rm vector}_{\rm{1-loop}}(\vec s^\gamma) \Big) \nn \\
& \qquad \qquad \qquad \qquad \qquad \qquad \times \Big(\prod^p_{j=1} Z^{\rm fund}_{\rm{1-loop}}(\vec s^{\gamma_j}, \vec v^{\gamma_j})\, \prod_{\gamma \neq \beta}Z^{\rm fund}_{\rm{1-loop}}(\vec s^\gamma, \vec m^{\gamma})\, Z_{\rm 1-loop}^{\rm hyper} (\vec{u}^\alpha, \vec m^\alpha_F)\Big)\nn \\ 
&  \qquad \qquad \qquad \qquad \qquad \qquad \times \,\Big(\prod^p_{j=1} Z^{\rm bif}_{\rm{1-loop}}(\vec s^{\gamma_j}, \vec u^{\alpha}, - \mu^{\gamma_j}) \, \prod^{L-1}_{\gamma=1} Z^{\rm bif}_{\rm{1-loop}}(\vec s^\gamma, \vec s^{\gamma +1},0) \Big). \label{GFI-basic2b}
\end{align}
Similar to the previous examples, the Lagrangian for the theory $G^\alpha_{\vec \CP} \circ F^\alpha_{\vec \CP} \circ I^\alpha_{\vec \CP} (X)$ can now be read off from the integrand of the matrix integral on the RHS of \eref{GFI-basic1}, and agrees with the quiver gauge theory in the fourth line of \figref{fig: IdFGaugingGen}. The parameters $\vec \mu$ appear as masses of the hypermultiplets in the bifundamental of $U(M_{\gamma_j}) \times U(r)$, while
$\vec m^\alpha_F$ are the masses of the hypers transforming in a representation $\CR$ of $U(r) \times G^\alpha_F$.

\begin{figure}[htbp]
\begin{center}
\scalebox{0.6}{\begin{tikzpicture}[
cnode/.style={circle,draw,thick,minimum size=4mm},snode/.style={rectangle,draw,thick,minimum size=8mm},pnode/.style={rectangle,red,draw,thick,minimum size=8mm} ]
\node[cnode] (1) {$N_1$};
\node[cnode] (2) [right=.5cm  of 1]{$N_2$};
\node[cnode] (3) [right=.5cm of 2]{$N_3$};
\node[cnode] (4) [right=1cm of 3]{$N_{\alpha-1}$};
\node[cnode] (5) [right=0.5cm of 4]{$N_{\alpha}$};
\node[cnode] (6) [right=0.5cm of 5]{$N_{\alpha + 1}$};
\node[cnode] (7) [right=1cm of 6]{{$N_{l-2}$}};
\node[cnode] (8) [right=0.5cm of 7]{$N_{l-1}$};
\node[cnode] (9) [right=0.5cm of 8]{$N_l$};
\node[snode] (10) [below=0.5cm of 1]{$M_1$};
\node[snode] (11) [below=0.5cm of 2]{$M_2$};
\node[snode] (12) [below=0.5cm of 3]{$M_3$};
\node[pnode] (13) [above=0.5cm of 4]{$M_{\alpha-1}$};
\node[pnode] (14) [below=0.5cm of 5]{$M_{\alpha}$};
\node[pnode] (15) [above=0.5cm of 6]{$M_{\alpha+1}$};
\node[snode] (16) [below=0.5cm of 7]{$M_{l-2}$};
\node[snode] (17) [below=0.5cm of 8]{$M_{l-1}$};
\node[snode] (18) [below=0.5cm of 9]{$M_{l}$};
\draw[-] (1) -- (2);
\draw[-] (2)-- (3);
\draw[dashed] (3) -- (4);
\draw[-] (4) --(5);
\draw[-] (5) --(6);
\draw[dashed] (6) -- (7);
\draw[-] (7) -- (8);
\draw[-] (8) --(9);
\draw[-] (1) -- (10);
\draw[-] (2) -- (11);
\draw[-] (3) -- (12);
\draw[-] (4) -- (13);
\draw[-] (5) -- (14);
\draw[-] (6) -- (15);
\draw[-] (7) -- (16);
\draw[-] (8) -- (17);
\draw[-] (9) -- (18);
\draw[->] (6.5,-2) -- (6.5,-4); 
\end{tikzpicture}}
\qquad
\scalebox{0.6}{\begin{tikzpicture}[
cnode/.style={circle,draw,thick,minimum size=4mm},snode/.style={rectangle,draw,thick,minimum size=8mm},pnode/.style={rectangle,red,draw,thick,minimum size=8mm},nnode/.style={circle,red,draw,thick,minimum size=8mm}]
\node[cnode] (1) {$N_1$};
\node[cnode] (2) [right=.5cm  of 1]{$N_2$};
\node[cnode] (3) [right=.5cm of 2]{$N_3$};
\node[cnode] (4) [right=1cm of 3]{$N_{\alpha-1}$};
\node[cnode] (5) [right=0.5cm of 4]{$N_{\alpha}$};
\node[cnode] (6) [right=0.5cm of 5]{$N_{\alpha + 1}$};
\node[cnode] (7) [right=1cm of 6]{{$N_{l-2}$}};
\node[cnode] (8) [right=0.5cm of 7]{$N_{l-1}$};
\node[cnode] (9) [right=0.5cm of 8]{$N_l$};
\node[snode] (10) [below=0.5cm of 1]{$M_1$};
\node[snode] (11) [below=0.5cm of 2]{$M_2$};
\node[snode] (12) [below=0.5cm of 3]{$M_3$};
\node[pnode] (13) [above=1.5cm of 4]{$M_{\alpha-1}-r$};
\node[pnode] (14) [below=0.5cm of 5]{$r$};
\node[pnode] (19) [above=0.5cm of 5]{$M_\alpha -r$};
\node[pnode] (15) [above=1.5cm of 6]{$M_{\alpha+1} -r$};
\node[snode] (16) [below=0.5cm of 7]{$M_{l-2}$};
\node[snode] (17) [below=0.5cm of 8]{$M_{l-1}$};
\node[snode] (18) [below=0.5cm of 9]{$M_{l}$};
\draw[-] (1) -- (2);
\draw[-] (2)-- (3);
\draw[dashed] (3) -- (4);
\draw[-] (4) --(5);
\draw[-] (5) --(6);
\draw[dashed] (6) -- (7);
\draw[-] (7) -- (8);
\draw[-] (8) --(9);
\draw[-] (1) -- (10);
\draw[-] (2) -- (11);
\draw[-] (3) -- (12);
\draw[-] (4) -- (13);
\draw[-] (5) -- (14);
\draw[-] (5) -- (19);
\draw[-] (4) -- (14);
\draw[-] (6) -- (14);
\draw[-] (6) -- (15);
\draw[-] (7) -- (16);
\draw[-] (8) -- (17);
\draw[-] (9) -- (18);
\draw[->] (6.5,-2) -- (6.5,-4);
\end{tikzpicture}}
\scalebox{0.6}{\begin{tikzpicture}[
cnode/.style={circle,draw,thick,minimum size=4mm},snode/.style={rectangle,draw,thick,minimum size=8mm},pnode/.style={rectangle,red,draw,thick,minimum size=8mm},nnode/.style={circle,red,draw,thick,minimum size=8mm}]
\node[cnode] (1) {$N_1$};
\node[cnode] (2) [right=.5cm  of 1]{$N_2$};
\node[cnode] (3) [right=.5cm of 2]{$N_3$};
\node[cnode] (4) [right=1cm of 3]{$N_{\alpha-1}$};
\node[cnode] (5) [right=0.5cm of 4]{$N_{\alpha}$};
\node[cnode] (6) [right=0.5cm of 5]{$N_{\alpha + 1}$};
\node[cnode] (7) [right=1cm of 6]{{$N_{l-2}$}};
\node[cnode] (8) [right=0.5cm of 7]{$N_{l-1}$};
\node[cnode] (9) [right=0.5cm of 8]{$N_l$};
\node[snode] (10) [below=0.5cm of 1]{$M_1$};
\node[snode] (11) [below=0.5cm of 2]{$M_2$};
\node[snode] (12) [below=0.5cm of 3]{$M_3$};
\node[pnode] (13) [above=1.5cm of 4]{$M_{\alpha-1}-r$};
\node[pnode] (14) [below=0.5cm of 5]{$r$};
\node[pnode] (20) [below=0.5cm of 14]{$G^\alpha_F$};
\node[text width=0.1cm](30) at (7,-2) {$\CR$};
\node[pnode] (19) [above=0.5cm of 5]{$M_\alpha -r$};
\node[pnode] (15) [above=1.5cm of 6]{$M_{\alpha+1} -r$};
\node[snode] (16) [below=0.5cm of 7]{$M_{l-2}$};
\node[snode] (17) [below=0.5cm of 8]{$M_{l-1}$};
\node[snode] (18) [below=0.5cm of 9]{$M_{l}$};
\draw[-] (1) -- (2);
\draw[-] (2)-- (3);
\draw[dashed] (3) -- (4);
\draw[-] (4) --(5);
\draw[-] (5) --(6);
\draw[dashed] (6) -- (7);
\draw[-] (7) -- (8);
\draw[-] (8) --(9);
\draw[-] (1) -- (10);
\draw[-] (2) -- (11);
\draw[-] (3) -- (12);
\draw[-] (4) -- (13);
\draw[-] (5) -- (14);
\draw[-] (5) -- (19);
\draw[-] (4) -- (14);
\draw[-] (6) -- (14);
\draw[-] (6) -- (15);
\draw[-] (7) -- (16);
\draw[-] (8) -- (17);
\draw[-] (9) -- (18);
\draw[-] (14) -- (20);
\draw[->] (6.5,-3.5) -- (6.5,-5);
\end{tikzpicture}}
\scalebox{0.6}{\begin{tikzpicture}[
cnode/.style={circle,draw,thick,minimum size=4mm},snode/.style={rectangle,draw,thick,minimum size=8mm},pnode/.style={rectangle,red,draw,thick,minimum size=8mm},nnode/.style={circle,red,draw,thick,minimum size=8mm}]
\node[cnode] (1) {$N_1$};
\node[cnode] (2) [right=.5cm  of 1]{$N_2$};
\node[cnode] (3) [right=.5cm of 2]{$N_3$};
\node[cnode] (4) [right=1cm of 3]{$N_{\alpha-1}$};
\node[cnode] (5) [right=0.5cm of 4]{$N_{\alpha}$};
\node[cnode] (6) [right=0.5cm of 5]{$N_{\alpha + 1}$};
\node[cnode] (7) [right=1cm of 6]{{$N_{l-2}$}};
\node[cnode] (8) [right=0.5cm of 7]{$N_{l-1}$};
\node[cnode] (9) [right=0.5cm of 8]{$N_l$};
\node[snode] (10) [below=0.5cm of 1]{$M_1$};
\node[snode] (11) [below=0.5cm of 2]{$M_2$};
\node[snode] (12) [below=0.5cm of 3]{$M_3$};
\node[pnode] (13) [above=1.5cm of 4]{$M_{\alpha-1}-r$};
\node[nnode] (14) [below=0.5cm of 5]{$r$};
\node[pnode] (20) [below=0.5cm of 14]{$G^\alpha_F$};
\node[text width=0.1cm](30) at (7,-2) {$\CR$};
\node[pnode] (19) [above=0.5cm of 5]{$M_\alpha -r$};
\node[pnode] (15) [above=1.5cm of 6]{$M_{\alpha+1} -r$};
\node[snode] (16) [below=0.5cm of 7]{$M_{l-2}$};
\node[snode] (17) [below=0.5cm of 8]{$M_{l-1}$};
\node[snode] (18) [below=0.5cm of 9]{$M_{l}$};
\draw[-] (1) -- (2);
\draw[-] (2)-- (3);
\draw[dashed] (3) -- (4);
\draw[-] (4) --(5);
\draw[-] (5) --(6);
\draw[dashed] (6) -- (7);
\draw[-] (7) -- (8);
\draw[-] (8) --(9);
\draw[-] (1) -- (10);
\draw[-] (2) -- (11);
\draw[-] (3) -- (12);
\draw[-] (4) -- (13);
\draw[-] (5) -- (14);
\draw[-] (5) -- (19);
\draw[-] (4) -- (14);
\draw[-] (6) -- (14);
\draw[-] (6) -- (15);
\draw[-] (7) -- (16);
\draw[-] (8) -- (17);
\draw[-] (9) -- (18);
\draw[-] (14) -- (20);
\end{tikzpicture}}
\caption{This figure illustrates the identification-gauging $G^\alpha_{\vec \CP} \circ F^\alpha_{\vec \CP} \circ I^\alpha_{\vec \CP}$ operation on a generic linear quiver $X$ involving $p=3$ nodes labelled by $\beta=\alpha-1, \alpha, \alpha+1$.}
\label{fig: IdFGaugingGen}
\end{center}
\end{figure}
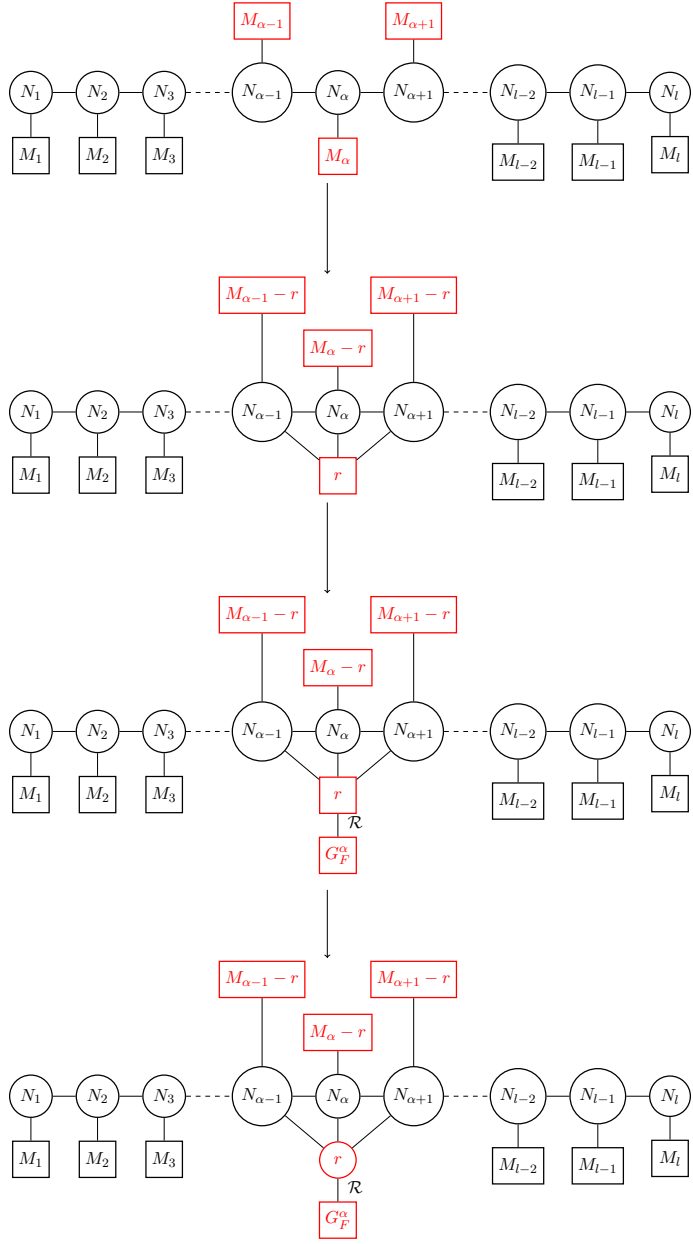

\end{itemize}

\subsection{Reading off the dual gauge theory}\label{GFIdual-summary}

In the previous section, we have described how new quivers can be constructed from a given linear quiver gauge theory $X$ using the elementary $S$-type operations. To determine the IR dual of the new quiver, one needs to understand the dual operations acting on the mirror quiver gauge theory $Y$, as one performs the $S$-type operation on quiver $X$. In this section, we will write down a general formula for the partition function of the mirror dual of the theory 
$\CO^\alpha_{\vec \CP}(X)$ (as defined in \eref{PF-OPLQ}) in terms of the partition function of $Y$. We will denote the mirror dual as $\wt{\CO}^\alpha_{\vec \CP}(Y)$. We will give explicit expressions for the dual partition functions of the four elementary $S$-type operations discussed in 
\Secref{GFI-summary}. Finally, we will also give a formula for the dual partition function, when the theory $X$ is a generic quiver gauge theory 
(i.e. not a linear quiver) and has a Lagrangian mirror dual $Y$.\\

The properties of linear quivers imply that the partition functions of the mirror pair $X$ and $Y$ are related follows:
\be \label{PfMirrorLQ1}
Z^{(X)}(\vec{m};\vec{t})=e^{2\pi i a^{kl} m_k t_l}Z^{(Y)}(-\vec{t};\vec{m}),
\ee
where $a^{kl}$ is an $M \times (L+1)$ matrix with integer entries, $M=\sum^L_{\alpha=1} M_\alpha$ and $L$ is the number of nodes in quiver $X$. Note that $Z^{(X)}$ is convergent since $X$ is a good quiver. Given that $X$ is a good linear quiver, the mirror $Y$ is guaranteed to be a good linear quiver 
\cite{Gaiotto:2008ak}, which implies that $Z^{(Y)}$ is also convergent. Therefore, the above equation is well-defined.\\ 

The partition function of the theory $(X, \{\CP_\beta\})$ is given in \eref{PF-XPbeta}. Similarly, the partition function of the mirror dual, which we denote as $(Y, \{\CP_\beta\})$, is given as
\be\label{PfYP-Y}
Z^{(Y, \{\CP_\beta\})} (\vec{t}; \{\vec{u}^\beta\}, \{\vec{v}^\beta\},\{\vec m^\gamma\}_{\gamma \neq \beta}) := Z^{(Y)} (\vec{t}; \{\vec m^\beta(\CP_\beta, \vec{u}^\beta, \vec{v}^\beta)\},\{\vec m^\gamma\}_{\gamma \neq \beta}).
\ee
The mirror symmetry statement \eref{PfMirrorLQ1} can be rewritten for the theory $(X, \{\CP_\beta\})$ and its mirror dual 
$(Y, \{\CP_\beta\})$ as follows:
\begin{align} \label{PfMirrorLQ3}
Z^{(X,\{ P_\beta\})} (\{\vec{u}^\beta\}, \{\vec{v}^\beta\},\{\vec m^\gamma\}_{\gamma 
 \neq \beta};\vec{t}) 
 = e^{2\pi i b^{il}_\beta u^\beta_i t_l} \, Z^{(Y, \{\CP_\beta\})}(-\vec{t}; \{\vec{u}^\beta\}, \{\vec{v}^\beta\},\{\vec m^\gamma\}_{\gamma \neq \beta}),
\end{align}
where for a given $\beta=\gamma_1,\ldots,\gamma_p$, $b^{il}_\beta$ is an $r_\alpha \times (L+1)$ matrix with integer entries for a given $\beta$. 
In writing the above equality, we have suppressed  a phase factor independent of the parameters $\vec{u}^\alpha$.\\

Since $Y$ is a good linear quiver, the function $Z^{(Y,\{\CP_\beta\})}$ can be written as a matrix integral as follows:
\be \label{PF-Bgenu}
\begin{split}
& Z^{(Y, \{\CP_\beta\})} (\vec{t}; \{\vec{u}^\beta\}, \{\vec{v}^\beta\},\{\vec m^\gamma\}_{\gamma \neq \beta}) =: \int \prod^{L^\vee}_{\gamma'=1}  \Big[d\vec{\s}^{\gamma'} \Big]\, Z^{(Y, \{\CP_\beta\})}_{\rm int}(\{\vec \s^{\gamma'} \}, \vec t, \{\vec{u}^\beta\}, \{\vec{v}^\beta\}, \{\vec m^\gamma\}_{\gamma \neq \beta} ),\\
& Z^{(Y, \{\CP_\beta\})}_{\rm int} =  Z^{(Y,\{\CP_\beta\})}_{\rm FI}(\{\vec \s^{\gamma'} \},\{\vec{u}^\beta\}, \{\vec{v}^\beta\}, \{\vec m^\gamma\}_{\gamma \neq \beta} )\cdot Z^{(Y)}_{\rm{1-loop}}(\{\vec \s^{\gamma'} \}, \vec t ), 
\end{split}
\ee
where the function $Z^{(Y)}_{\rm{1-loop}}$ is independent of $\{\vec{u}^\beta\}$ and $\{\CP_\beta\}$, and explicitly given as
\begin{align}\label{Z1LB}
Z^{(Y)}_{\rm{1-loop}}(\{\vec \s^{\gamma'} \}, \vec t )= \prod^{L^\vee}_{\gamma'=1} Z^{\rm vector}_{\rm{1-loop}}(\vec \s^{\gamma'})\, Z^{\rm fund}_{\rm{1-loop}}(\vec \s^{\gamma'}, \vec t)\, \prod^{L^\vee-1}_{\gamma'=1} Z^{\rm bif}_{\rm{1-loop}}(\vec \s^{\gamma'}, \vec \s^{\gamma' +1},0).
\end{align}
The $\{\vec{u}^\beta\}$-dependent FI term, which also depends on $\{\CP_\beta\}$, can be written in the following fashion:
\begin{align} \label{ZFIB}
& Z^{(Y,\{\CP_\beta\})}_{\rm FI}(\{\vec \s^{\gamma'} \},\{\vec{u}^\beta\}, \{\vec{v}^\beta\}, \{\vec m^\gamma\}_{\gamma \neq \beta}) = Z^{(Y)}_{\rm FI}(\{\vec \s^{\gamma'} \}, \{\vec m^\beta(\CP_\beta, \vec{u}^\beta, \vec{v}^\beta)\},\{\vec m^\gamma\}_{\gamma \neq \beta}) \nn \\
=& \Big(\prod_{\beta}\,e^{2\pi i \,g^i_\beta (\{\vec \s^{\gamma'} \}, \CP_\beta)\,u^{\beta}_i}\Big) \cdot Z^{(Y,\{\CP_\beta\})}_{\rm FI}(\{\vec \s^{\gamma'} \},\{\vec{u}^\beta =0\}, \{\vec{v}^\beta\}, \{\vec m^\gamma\}_{\gamma \neq \beta}),
\end{align}
where for the second equality, we have isolated the $\{\vec{u}^\beta\}$-dependent part. 
Given that the function $Z^{(Y)}_{\rm FI}(\{\vec \s^{\gamma'} \}, \{\vec m^\gamma\})$ is precisely known for any linear quiver $Y$
(and given in \eref{PF-Bgen1}), we can explicitly write down the functions $g^i_\beta (\{\vec \s^{\gamma'} \}, \CP_\beta)$:
\begin{align}
g^i_\beta \Big(\{\vec \s^{\gamma'} \}, \CP_\beta \Big) = & -\tr{\s^{M_1+ \ldots+ M_{\beta-1}}}\, \CP_{\beta\,1i} + 
\sum^{M_\beta -1}_{i_\beta=1} \tr{\s^{M_1+ \ldots+ M_{\beta-1} + i_\beta}}\, (\CP_{\beta \, i_\beta i} - \CP_{\beta\, (i_\beta +1) i}) \nn \\ & + \tr{\s^{M_1+ \ldots+ M_{\beta-1} + M_\beta}}\, \CP_{\beta \, M_\beta i} \nn \\
= & \sum^{M_\beta }_{i_\beta=1} \CP_{\beta\, i_\beta i} (- \tr{\s^{M_1+ \ldots+ M_{\beta-1} + i_\beta -1}} +  \tr{\s^{M_1+ \ldots+ M_{\beta-1} + i_\beta}}) \nn \\
= & (- \tr{\s^{M_1+ \ldots+ M_{\beta-1} + j -1}} +  \tr{\s^{M_1+ \ldots+ M_{\beta-1} + j}}) \label{giu}
\end{align}
where $\CP_{\beta\, i_\beta i} =1$ for some $i_\beta=j$ and a fixed $i$, and vanishes otherwise. The relation is subject to the boundary conditions 
$\tr{\s^{M_0}}=\tr{\s^{M_1+ \ldots+ M_{\alpha} }}=0$.
Now, let $\wt{\CO}^\alpha_{\vec \CP}(Y)$ denote the mirror dual of the theory $\CO^\alpha_{\vec \CP}(X)$. The IR duality, along with the fact that both 
$\CO^\alpha_{\vec \CP}(X)$ and $\wt{\CO}^\alpha_{\vec \CP}(Y)$ are assumed to be good theories, will imply that their partition functions are related as
\begin{align}\label{IR-OP}
Z^{\CO^\alpha_{\vec \CP}(X)} (\vec m; \vec \eta ) = Z^{\wt{\CO}^\alpha_{\vec \CP}(Y)}(\vec m'(\vec \eta) ; \vec \eta'(\vec m)),
\end{align}
up to some phase factor, where $(\vec m,\vec \eta)$ and $(\vec m',\vec \eta')$ collectively denote the $\CN=4$ preserving masses and FI parameters of 
$\CO^\alpha_{\vec \CP}(X)$ and $\wt{\CO}^\alpha_{\vec \CP}(Y)$ respectively.\\

Using \eref{PF-OPLQ}, the mirror symmetry relation \eref{PfMirrorLQ3}, and the equations \eref{PF-Bgenu}-\eref{ZFIB}, in \eref{IR-OP} above, the partition function of the theory $\wt{\CO}^\alpha_{\vec \CP}(Y)$ can be written in the following fashion:
\begin{empheq}[box=\widefbox]{align}\label{PF-wtOPLQ}
& Z^{\wt{\CO}^\alpha_{\vec \CP}(Y)}(\vec{m}'(\vec{t},\eta_{\alpha}); \vec \eta'(\{ \vec{v}^\beta \},\{\vec m^\gamma\}_{\gamma \neq \beta},\vec{m}^{{\CO}_\CP})) \nn \\
=& \int \prod^{L^\vee}_{\gamma'=1} \Big[d\vec{\s}^{\gamma'} \Big]\, \CZ_{\wt{\CO}^\alpha_{\vec \CP}(Y)}(\{\s^{\gamma'}\},\vec{m}^{{\CO}_\CP}, \eta_{\alpha},\vec t)\,\cdot Z^{(Y,\{\CP_\beta\})}_{\rm int}(\{\vec \s^{\gamma'} \}, -\vec t,\{\vec{u}^\beta =0\}, \{\vec{v}^\beta\}, \{\vec m^\gamma\}_{\gamma \neq \beta}),
\end{empheq}
where the function $Z^{(Y,\{\CP_\beta\})}_{\rm int}$ is given in \eref{PF-Bgenu}. The function $\CZ_{\wt{\CO}^\alpha_{\vec \CP}(Y)}$ can be explicitly written (as a formal Fourier transform) in terms of the operator $\CZ_{{\CO}_{\vec \CP}(X)}$ that appears in \eref{PF-OPLQ}:
\begin{empheq}[box=\widefbox]{align} \label{CZ-wtOPLQ}
\CZ_{\wt{\CO}^\alpha_{\vec \CP}(Y)}
= \int \Big[d\vec{u}^\alpha\Big] \, \CZ_{\CO^\alpha_{\vec \CP}(X)}(\vec u^{\alpha}, \{\vec{u}^\beta\}, \eta_\alpha, \vec{m}^{\CO^\alpha_{\vec \CP}})\,\prod_{\beta}\,e^{2\pi i \,(g^i_\beta (\{\vec \s^{\gamma'} \}, \CP_\beta) + b^{il}_\beta t_l)\,u^{\beta}_i}.
\end{empheq}

We shall take \eref{PF-wtOPLQ}-\eref{CZ-wtOPLQ} as the working definition of the dual operation $\wt{\CO}^\alpha_{\vec \CP}$ on quiver $Y$. 
Given the expression for $Z^{(Y,\{\CP_\beta\})}_{\rm int}$ in \eref{PF-Bgenu}, and the expression for $\CZ_{\wt{\CO}^\alpha_{\vec \CP}(Y)}$ 
computed above, the goal is to rewrite the RHS of \eref{PF-wtOPLQ} in the standard form of \eref{PF-main}. Of course, this is only 
possible if the theory $\wt{\CO}^\alpha_{\vec \CP}(Y)$ is Lagrangian. From the standard form, the gauge group and the matter content of the 
theory $\wt{\CO}^\alpha_{\vec \CP}(Y)$ can then be simply read off. In addition, the precise form of the linear functions $\vec{m}'(\vec{t},\eta_\alpha)$ and 
$\vec{\eta}'(\{ \vec{v}^\beta \},\{\vec m^\gamma\}_{\gamma \neq \beta},\vec{m}^{{\CO}_\CP}))$ can also be read off from the standard form of the partition function.\\
Note that the integrand of the partition function of the dual theory $\wt{\CO}_{\vec\CP}(Y)$ explicitly depends on the permutation matrix 
$\vec \CP$, even though the partition function itself is independent of it. Therefore, the Lagrangian description of the theory $\wt{\CO}_{\vec\CP}(Y)$ 
depends on $\vec \CP$. If the theory $\wt{\CO}_{\vec\CP}(Y)$ can be written as a Lagrangian theory for more than one $\CP$ (and they are not related by some trivial change of variables), then all such Lagrangians are conjectured to be IR dual among themselves.\\
Given the general expression \eref{PF-wtOPLQ}, we can write down explicitly the dual partition functions for  the four types of elementary $S$-type operations 
discussed in \Secref{GFI-summary} - gauging, flavoring-gauging, identification-gauging, and flavoring-identification-gauging. The appropriate expressions for 
$\CZ_{\CO^\alpha_{\vec \CP}(X)}(\vec u^{\alpha}, \{\vec{u}^\beta\}, \eta_\alpha, \vec{m}^{\CO^\alpha_{\vec \CP}})$ are given 
in \eref{defZG}, \eref{defZGF}, \eref{defZGI} and \eref{defZGFI} respectively.\\

We end this subsection by writing a formula for the dual partition function when an elementary $S$-type operation $\CO^\alpha_{\vec \CP}$ acts on a 
generic quiver $X$ in class $\CU$, as discussed in \eref{PF-OP}, when no defect is turned on.
This is only possible if the quiver $X$ has a Lagrangian mirror dual $Y$. The mirror symmetry relation between $X$ and $Y$ can be written as:
\be \label{bil-def}
Z^{(X,\{ P_\beta\})} (\{\vec{u}^\beta\}, \{\vec{v}^\beta\},\ldots;\vec{\eta}) 
 = e^{2\pi i b^{il}_\beta u^\beta_i \eta_l} \, Z^{(Y, \{\CP_\beta\})}(\vec{m}^Y(\vec{\eta}) ; \vec{\eta}^Y(\{\vec{u}^\beta\}, \{\vec{v}^\beta\},\ldots)),
\ee
where $\vec{\eta}$ collectively denotes the FI parameters of $X$, and the $\ldots$ in the argument of $Z^{(X,\{ P_\beta\})}$ denote the other masses 
of $X$. The masses and the FI parameters of $Y$ are collectively denoted as $(\vec{m}^Y, \vec{\eta}^Y)$.
The partition function of $(Y, \{\CP_\beta\})$ (being a Lagrangian theory) can again be written as
\begin{align}\label{PF-GenBu}
& Z^{(Y, \{\CP_\beta\})}(\vec{m}^Y(\vec{\eta}) ; \vec{\eta}^Y(\{\vec{u}^\beta\}, \{\vec{v}^\beta\},\ldots)) \nn \\
= & \int \prod_{\gamma'} \Big[d\vec{\s}^{\gamma'} \Big]\,\prod_{\beta}\,e^{2\pi i \, g^i_\beta (\{\vec \s^{\gamma'} \}, \CP_\beta)\, u^\beta_i}\,
Z^{(Y, \{\CP_\beta\})}_{\rm int}(\{\vec \s^{\gamma'} \}, \vec{m}^Y(\vec{\eta}), \vec{\eta}^Y(\{\vec{u}^\beta =0 \}, \{\vec{v}^\beta\},\ldots)),
\end{align}
where $\gamma'$ labels the gauge nodes of the theory $Y$. Note that the functions $g^i_\beta (\{\vec \s^{\gamma'} \}, \CP_\beta)$ 
are not known a priori since $X,Y$ are not linear quivers, but have to be provided as additional data about the duality. 
In cases where $X,Y$ appear as intermediate dual pairs in our construction of a given dual pair starting from linear quivers,
these functions are known by construction.\\

Proceeding in the same fashion as we did for the linear quiver case, the formula for the partition function of the theory 
$\wt{\CO}^\alpha_{\vec \CP}(Y)$ can be written as:
\begin{empheq}[box=\widefbox]{align}\label{PF-wtOPgen}
& Z^{\wt{\CO}^\alpha_{\vec \CP}(Y)}(\vec{m}'(\vec{\eta},\eta_{\alpha}); \vec \eta'(\{ \vec{v}^\beta \},\ldots,\vec{m}^{{\CO}_{\vec\CP}})) \nn \\
=& \int \prod_{\gamma'} \Big[d\vec{\s}^{\gamma'} \Big]\, \CZ_{\wt{\CO}^\alpha_{\vec \CP}(Y)}(\{\s^{\gamma'}\},\vec{m}^{{\CO}_{\vec \CP}}, \eta_{\alpha},\vec \eta)\,\cdot Z^{(Y,\{\CP_\beta\})}_{\rm int}(\{\vec \s^{\gamma'} \}, \vec{m}^Y(\vec{\eta}), \vec{\eta}^Y(\{\vec{u}^\beta =0 \}, \{\vec{v}^\beta\},\ldots)),
\end{empheq}
where the function $Z^{(Y,\{\CP_\beta\})}_{\rm int}$ is given in \eref{PF-GenBu}, and $(\vec{m}', \vec{\eta}')$ collectively denote the masses and FI 
parameters of the theory $\wt{\CO}^\alpha_{\vec \CP}(Y)$.
The operator $\CZ_{\wt{\CO}^\alpha_{\vec \CP}(Y)}$ can be explicitly written (as a formal Fourier transform) in terms the operator $\CZ_{{\CO}_{\vec \CP}(X)}$ that appears in \eref{PF-OP}:
\begin{empheq}[box=\widefbox]{align} \label{CZ-wtOP}
&\CZ_{\wt{\CO}^\alpha_{\vec \CP}(Y)}
= \int \Big[d\vec{u}^\alpha\Big] \, \CZ_{\CO^\alpha_{\vec \CP}(X)}(\vec u^{\alpha}, \{\vec{u}^\beta\}, \eta_\alpha, \vec{m}^{\CO^\alpha_{\vec \CP}})\, \cdot \prod_{\beta}\,e^{2\pi i \,(g^i_\beta (\{\vec \s^{\gamma'} \}, \CP_\beta) + b^{il}_\beta \eta_l)\,u^{\beta}_i}, 
\end{empheq}
where $g^i_\beta (\{\vec \s^{\gamma'} \}, \CP_\beta)$ is a function linear in the variables $\{\vec \s^{\gamma'} \}$, and can be read off from the mirror map 
relating mass parameters of $X$ and FI parameters of $Y$.

\subsection{Simple illustrative example: Flavored $\wh{A}_{n-1}$ quiver}\label{GFI-exAbPF}

\begin{figure}[htbp]
\begin{center}
\scalebox{.7}{\begin{tikzpicture}[node distance=2cm,cnode/.style={circle,draw,thick,minimum size=1.0cm},snode/.style={rectangle,draw,thick,minimum size=1.0cm},pnode/.style={rectangle,red,draw,thick,minimum size=1.0cm}, nnode/.style={circle, red, draw,thick,minimum size=1.0cm}]
\node[cnode] (1) at (2,0) {$1$};
\node[cnode] (2) at (3,1) {$1$};
\node[cnode] (3) at (5,1) {$1$};
\node[cnode] (4) at (6,0) {$1$};
\node[cnode] (5) at (5,-1) {$1$};
\node[cnode] (6) at (3,-1) {$1$};
\node[snode] (7) at (8,0) {$1$};
\node[snode] (9) at (2,-2) {$1$};
\draw[thick] (1) to [bend left=40] (2);
\draw[dashed,thick] (2) to [bend left=40] (3);
\draw[thick] (3) to [bend left=40] (4);
\draw[thick] (4) to [bend left=40] (5);
\draw[dashed,thick] (5) to [bend left=40] (6);
\draw[thick] (6) to [bend left=40] (1);
\draw[thick] (1) -- (9);
\draw[thick] (4) -- (7);
\node[text width=1.5cm](10) at (1,0.75) {$n-l+1$};
\node[text width=1.5cm](11) at (3,2) {$n-l+2$};
\node[text width=1cm](12) at (5,1.75) {$n-1$};
\node[text width=1cm](13) at (7,0.5) {$n$};
\node[text width=1cm](14) at (6,-1.5) {$1$};
\node[text width=1cm](15) at (3.25,-1.75) {$n-l$};
\node[text width=2cm](20) at (6, -3) {$(X')$};
\end{tikzpicture}}
\qquad \qquad \qquad 
\scalebox{.8}{\begin{tikzpicture}[node distance=2cm,cnode/.style={circle,draw,thick,minimum size=1.0cm},snode/.style={rectangle,draw,thick,minimum size=1.0cm},pnode/.style={rectangle,red,draw,thick,minimum size=1.0cm}, nnode/.style={circle, red, draw,thick,minimum size=1.0cm}]
\node[cnode] (25) at (16,0){1};
\node[snode] (26) [below=1cm of 25]{$l-1$};
\node[cnode] (27) [left=1.5 cm of 25]{1};
\node[snode] (28) [below=1cm of 27]{$n-l+1$};
\draw[-] (25) -- (26);
\draw[thick] (25) to [bend left=40] (27);
\draw[thick] (25) to [bend right=40] (27);
\draw[-] (27) -- (28);
\node[text width=6cm](29) at (17, -3) {$(Y')$};
\end{tikzpicture}}
\caption{$\wh{A}_{n-1}$ quiver $X'$ with two fundamental hypermultiplets, and its mirror $Y'$ which is an $\wh{A}_{1}$ quiver with a total of $n$ fundamental hypers.}
\label{SimpAbEx1}
\end{center}
\end{figure}
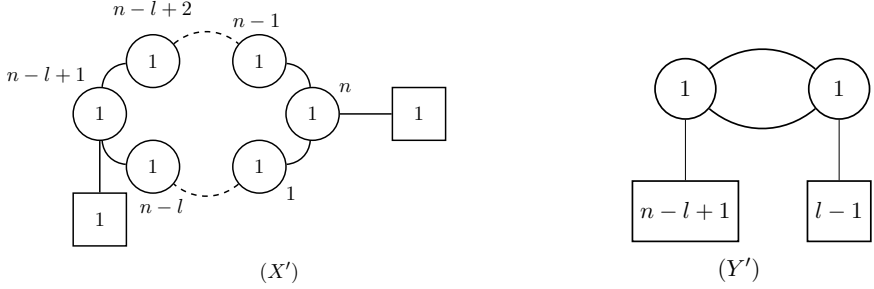

In this section, we present a simple illustrative example of an Abelian mirror dual pair that can be constructed 
from a dual pair of Abelian linear quivers, following the general recipe given in \Secref{GFI-summary} and \Secref{GFIdual-summary}.
The former dual pair, shown in \figref{SimpAbEx1}, are affine $A$-type quiver gauge theories (circular quivers) with some flavors, and are known 
to have simple Type IIB descriptions.\\

\begin{figure}[htbp]
\begin{center}
\begin{tabular}{ccc}
\scalebox{.7}{\begin{tikzpicture}[node distance=2cm,
cnode/.style={circle,draw,thick, minimum size=1.0cm},snode/.style={rectangle,draw,thick,minimum size=1cm}, pnode/.style={red,rectangle,draw,thick, minimum size=1.0cm},]
\node[cnode] (1) {1};
\node[cnode] (2) [right=1.5cm  of 1]{1};
\node[cnode] (3) [right= 1 cm  of 2]{1};
\node[snode] (4) [above=1 cm of 3]{1};
\node[cnode] (5) [right=1 cm  of 3]{1};
\node[cnode] (6) [right=1.5cm  of 5]{1};
\node[pnode] (7) [below=1 cm of 1]{1};
\node[pnode] (8) [below=1 cm of 6]{1};
\node[text width=1.5cm](9)[above=0.2cm of 3]{$n-l+1$};
\node[text width=1.5cm](10)[above=0.2cm of 5]{$n-l+2$};
\node[text width=1 cm](11)[above=0.2cm of 6]{$n-1$};
\node[text width=1cm](12)[above=0.2cm of 2]{$n-l$};
\node[text width=0.1 cm](13)[above=0.2cm of 1]{$1$};
\draw[thick, dashed] (1) -- (2);
\draw[-] (2)-- (3);
\draw[-] (3)-- (4);
\draw[-] (3)-- (5);
\draw[thick, dashed] (5)-- (6);
\draw[-] (1)-- (7);
\draw[-] (8)-- (6);
\node[text width=6cm](20) at (7, -3) {$(X)$};
\end{tikzpicture}}
& \qquad 
& \scalebox{.7}{\begin{tikzpicture}[node distance=2cm,
cnode/.style={circle,draw,thick, minimum size=1.0cm},snode/.style={rectangle,draw,thick,minimum size=1cm}, pnode/.style={red,rectangle,draw,thick, minimum size=1.0cm},]\node[cnode] (25) at (16,0){1};
\node[snode] (26) [below=1cm of 25]{$l-1$};
\node[cnode] (27) [left=1.5 cm of 25]{1};
\node[snode] (28) [below=1cm of 27]{$n-l+1$};
\draw[-] (25) -- (26);
\draw[-] (25) -- (27);
\draw[-] (27) -- (28);
\node[text width=1cm](30) at (15, -3) {$(Y)$};
\end{tikzpicture}}\\
 \scalebox{.7}{\begin{tikzpicture}
\draw[->] (15,-3) -- (15,-5);
\node[text width=0.1cm](20) at (14.5, -4) {$\CO$};
\end{tikzpicture}}
&\qquad
& \scalebox{.7}{\begin{tikzpicture}
\draw[->] (15,-3) -- (15,-5);
\node[text width=0.1cm](29) at (15.5, -4) {$\wt{\CO}$};
\end{tikzpicture}}\\
\scalebox{.7}{\begin{tikzpicture}[node distance=2cm,cnode/.style={circle,draw,thick,minimum size=1.0cm},snode/.style={rectangle,draw,thick,minimum size=1.0cm},pnode/.style={rectangle,red,draw,thick,minimum size=1.0cm}, nnode/.style={circle, red, draw,thick,minimum size=1.0cm}]
\node[cnode] (1) at (2,0) {$1$};
\node[cnode] (2) at (3,1) {$1$};
\node[cnode] (3) at (5,1) {$1$};
\node[cnode] (4) at (6,0) {$1$};
\node[cnode] (5) at (5,-1) {$1$};
\node[cnode] (6) at (3,-1) {$1$};
\node[snode] (7) at (8,0) {$1$};
\node[snode] (9) at (2,-2) {$1$};
\draw[thick] (1) to [bend left=40] (2);
\draw[dashed,thick] (2) to [bend left=40] (3);
\draw[thick] (3) to [bend left=40] (4);
\draw[thick] (4) to [bend left=40] (5);
\draw[dashed,thick] (5) to [bend left=40] (6);
\draw[thick] (6) to [bend left=40] (1);
\draw[thick] (1) -- (9);
\draw[thick] (4) -- (7);
\node[text width=1.5cm](10) at (1,0.75) {$n-l+1$};
\node[text width=1.5cm](11) at (3,2) {$n-l+2$};
\node[text width=1cm](12) at (5,1.75) {$n-1$};
\node[text width=1cm](13) at (7,0.5) {$n$};
\node[text width=1cm](14) at (6,-1.5) {$1$};
\node[text width=1cm](15) at (3.25,-1.75) {$n-l$};
\node[text width=2cm](20) at (6, -3) {$(X')$};
\end{tikzpicture}}
&\qquad 
& \scalebox{.8}{\begin{tikzpicture}[node distance=2cm,cnode/.style={circle,draw,thick,minimum size=1.0cm},snode/.style={rectangle,draw,thick,minimum size=1.0cm},pnode/.style={rectangle,red,draw,thick,minimum size=1.0cm}, nnode/.style={circle, red, draw,thick,minimum size=1.0cm}]
\node[] (24) at (10,0){};
\node[cnode] (25) at (16,0){1};
\node[snode] (26) [below=1cm of 25]{$l-1$};
\node[cnode] (27) [left=1.5 cm of 25]{1};
\node[snode] (28) [below=1cm of 27]{$n-l+1$};
\draw[-] (25) -- (26);
\draw[thick] (25) to [bend left=40] (27);
\draw[thick] (25) to [bend right=40] (27);
\draw[-] (27) -- (28);
\node[text width=6cm](29) at (17, -3) {$(Y')$};
\end{tikzpicture}}
\end{tabular}
\caption{The quiver $(X')$ and its mirror dual $(Y')$ are generated by an elementary S-type operation $\CO=G \circ F \circ I$ of the identification-flavoring-gauging type on the linear quiver $X$ and the dual operation $\wt{\CO}$ on the linear quiver $Y$. The flavor nodes on which $\CO$ acts are shown in red 
on the first line. The dual operation simply involves adding a single bifundamental hyper in the quiver $Y$. }
\label{SimpAbEx1GFI}
\end{center}
\end{figure}
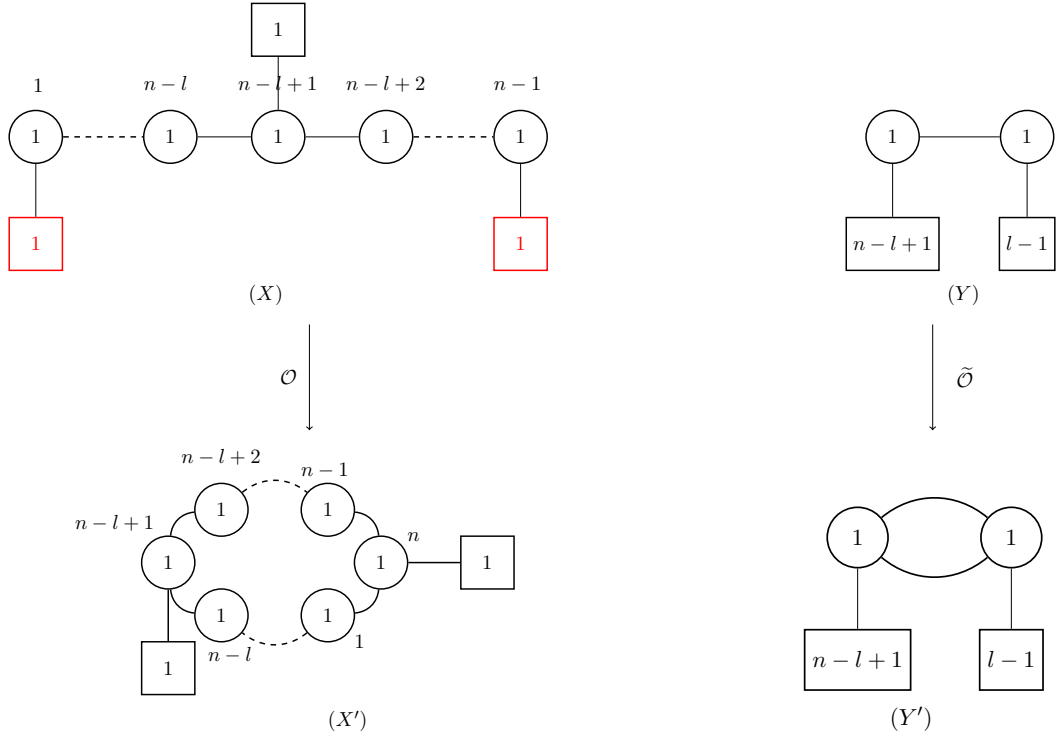

The quiver pair $(X',Y')$ in \figref{SimpAbEx1} can be obtained from a linear dual pair $(X,Y)$ by a single elementary $S$-type 
operation $\CO=(G\circ F\circ I)$ on $X$ and the dual operation $\wt{\CO}$ on $Y$, as shown in \figref{SimpAbEx1GFI} 
\footnote{In this example and subsequent ones, we drop the superscript $\alpha$ from the notation $\CO^\alpha_\CP$, 
when there is no ambiguity regarding the flavor node on which the $S$-type operation acts. In cases where $\CP$ is 
trivial, we drop the subscript as well.}. 
The flavor nodes of $X$ (marked in red) on which $\CO$ acts, correspond to $U(1)$ flavor symmetries, and therefore the permutation matrix $\CP$ is trivial in 
this case. The dual operation in this case is particularly simple -- it amounts to adding a single bifundamental hyper to the linear quiver $Y$. 
We will derive this fact using the $S^3$ partition function\footnote{The same fact will be derived using the superconformal index
in \Appref{GFI-exAbSCI}.}, following the general formulae for the dual partition function 
derived in \Secref{GFIdual-summary}. The purpose of this exercise is to demonstrate how the computation 
works for the simple example under consideration. Therefore, we will be very detailed in our presentation.

The $S^3$ partition function for linear quivers and their mirror dual was discussed in \Secref{LQ-IIB}.
Following the notation in \eref{PF-Agen}, the partition function of the linear quiver $X$ is given as
\begin{align}
& Z^{(X)}(\vec{m}; \vec{t})= \int \prod^{n-1}_{\gamma=1}\Big[d{s}^\gamma \Big] Z^{(X)}_{\rm int} (\{{s}^{\gamma}\}, \vec{m}, \vec{t})
=\int \prod^{n-1}_{\gamma=1}\Big[d{s}^\gamma \Big] Z^{(X)}_{\rm FI} (\{{s}^\gamma\}, \vec t)\, Z^{(X)}_{\rm{1-loop}}(\{{s}^\gamma\}, \vec m) \nn \\
= & \int  \prod^{n-1}_{\gamma=1} \Big[d{s}^\gamma \Big] Z^{(X)}_{\rm FI} (\{{s}^\gamma\}, \vec t)\, Z^{\rm fund}_{\rm{1-loop}}(s^1, m_1) \, Z^{\rm fund}_{\rm{1-loop}}(s^{n-l+1}, m_2)\,Z^{\rm fund}_{\rm{1-loop}}(s^{n-1}, m_3)\prod^{n-2}_{\gamma=1} Z^{\rm bif}_{\rm{1-loop}}(s^\gamma, s^{\gamma +1},0),
\end{align}
where $Z^{(X)}_{\rm FI}$ and the one-loop factors are given in \eref{PF-Agen1}. The Higgs branch global symmetry of $X$ is given by $G^{X}_{H} = U(1)^3/ U(1) = U(1) \times U(1)$, where the quotient by the overall $U(1)$ factor in the flavor symmetry of quiver $X$ can be implemented by the 
constraint $m_2=0$. The $U(1) \times U(1)$ flavor symmetry of $X$ can then be identified with the two 
terminal $U(1)$ flavor nodes, parametrized by the masses $(m_1, m_3)$. Following the general recipe 
in \Secref{GFI-summary}, let us define:
\be \label{defuAbEx1GFI}
u^1=m_1, u^2=m_3.
\ee
From the general formula in \eref{GFI-basic}, the partition function of the theory $\CO(X)$ is 
\begin{align} \label{PF-X1}
Z^{\CO(X)}(\vec{\mu}, \vec m^{\alpha}_F; \vec{t},\eta_\alpha)
= \int d{u}^{\alpha}\,\CZ_{\CO(X)}({u}^{\alpha}, {u}^{1}, {u}^{2}, \eta_\alpha, \vec \mu,  m^{\alpha}_F) \, Z^{(X)} ({u}^1,u^2;\vec{t}),
\end{align} 
where $\CO=G \circ F \circ I$ is the $S$-type operation shown in \figref{SimpAbEx1GFI}. 
The operator $\CZ_{\CO(X)}$ assumes the following form in this case:
\begin{align}\label{PF-ZX}
\CZ_{\CO(X)}({u}^{\alpha}, {u}^{1}, {u}^{2}, \eta_\alpha, \vec\mu,  m^{\alpha}_F)= & Z_{\rm FI} ({u}^\alpha,\eta_\alpha)\, Z_{\rm 1-loop}^{\rm hyper} ({u}^\alpha, m^{\alpha}_F)
\, \int  d {u^1} d {u^2} \,\prod^2_{\beta=1}\delta \Big(u^\alpha - u^\beta + {\mu}^\beta \Big) \nn \\
= & \frac{e^{2\pi i \eta_\alpha u^\alpha}}{\ch{({u}^\alpha - m^{\alpha}_F)}} \, \int  d {u^1} d {u^2} \, \prod^2_{\beta=1}\delta \Big(u^\alpha - u^\beta + {\mu}^\beta \Big).
\end{align} 
Using \eref{PF-X1}-\eref{PF-ZX}, and performing the integration over $u^\beta$, it trivially follows that:
\be 
Z^{\CO(X)}(\vec{\mu}, \vec m^{\alpha}_F; \vec{t},\eta_\alpha)= Z^{(X')}(\vec{\mu}, \vec m^{\alpha}_F; \vec{t},\eta_\alpha),
\ee
where the quiver $X'$ is given in \figref{SimpAbEx1}. Shifting the integration variable $u^\alpha \to u^\alpha - \mu^1$, and redefining the mass parameter $m^{\alpha}_F \to m^{\alpha}_F - \mu^1$, one can show that $Z^{(X')}$ only depends on the parameters $\mu=\mu^2 - \mu^1$ and $m^{\alpha}_F$.
The parameter $\mu$ is the mass associated with one of the bifundamental hypers in the loop, while $m^{\alpha}_F$ is the 
mass of the new fundamental hyper. The parameters $(\mu, m^{\alpha}_F)$ parametrize the Cartan subalgebra of the $U(1) \times U(1)$ 
flavor symmetry of $X'$. \\

Now, let us compute the dual partition function. Following \eref{PF-Bgen}, the partition function of the quiver $Y$ is given as
\begin{align}\label{PF-ZY}
& Z^{(Y)}(\vec{t}; \vec{m})=  \int \prod^2_{\gamma'=1}\Big[d{\s}^{\gamma'} \Big] Z^{(Y)}_{\rm int} (\{{\s}^\gamma\}, \vec{t}, \vec{m})
= \int \prod^2_{\gamma'=1} \Big[d{\s}^{\gamma'} \Big] Z^{(Y)}_{\rm FI}(\{ \s^{\gamma'} \}, \vec m) \,Z^{(Y)}_{\rm{1-loop}}(\{\s^{\gamma'} \}, \vec t)
\nn \\
=& \int \prod^2_{\gamma'=1} \Big[d{\s}^{\gamma'} \Big] Z^{(Y)}_{\rm FI}(\{ \s^{\gamma'} \}, \vec m) \, \prod^{n-l+1}_{i=1}Z^{\rm fund}_{\rm{1-loop}}(\s^{1}, t_i)\, 
\prod^{l-1}_{j=1}Z^{\rm fund}_{\rm{1-loop}}(\s^{2}, t_{n-l+1+j})\, Z^{\rm bif}_{\rm{1-loop}}(\s^1, \s^2,0), 
\end{align}
where $Z^{(Y)}_{\rm FI}$ and the one-loop factors are given in \eref{PF-Bgen1}. In particular, the FI term $Z^{(Y)}_{\rm FI}$ can 
be explicitly written as
\begin{align}
Z^{(Y)}_{\rm FI}(\{ \s^{\gamma'} \}, \vec m)= e^{2\pi i ({m}_1- {m}_2) \s^1}\, e^{2\pi i  ({m}_2 -m_3) \s^2}, \quad
\implies Z^{(Y)}_{\rm FI}(\{ \s^{\gamma'} \}, \vec u)= e^{2\pi i {u}^1 \s^1}\, e^{-2\pi i u^2 \s^2}. 
\end{align}
Mirror symmetry implies that the partition functions of $X$ and $Y$ are related in the following fashion:
\begin{align}
& Z^{(X)}(\vec m; \vec t) = e^{2\pi i t_1 m_1} \, e^{-2\pi i t_n m_3} \, Z^{(Y)} (-\vec t; \vec m), \nn \\
\implies & Z^{(X)}(\vec u; \vec t) = e^{2\pi i b^{l}_\beta u^\beta t_l}\, Z^{(Y)} (-\vec t; \vec u) =e^{2\pi i t_1 u^1} \, e^{-2\pi i t_n u^2} \, Z^{(Y)} (-\vec t; \vec u).
\end{align}

Using the general formula \eref{CZ-wtOPLQ}, the function $\CZ_{\wt\CO(Y)}$ is given by the expression:
\begin{align}
\CZ_{\wt\CO(Y)}(\s^1, \s^2, \eta, \vec\mu,  m^{\alpha}_F)= & \int du^\alpha\, \CZ_{\CO(X)}(u^\alpha, {u}^{1}, {u}^{2}, \eta, \vec \mu,  m^{\alpha}_F) \, e^{2\pi i {u}^1 \s^1}\, e^{-2\pi i u^2 \s^2}\,e^{2\pi i u^\beta b^{l}_\beta t_l},
\end{align} 
where the operator $\CZ_{\CO(X)}$ is explicitly given in \eref{PF-ZX}.
To simply the expression on the RHS, we first shift the integration variable $u^\alpha \to u^\alpha - \mu^1$, redefine the mass parameter 
$m^{\alpha}_F \to m^{\alpha}_F - \mu^1$, and use the identity $\frac{1}{\ch{(u^\alpha - m^\alpha_F)}} = \int d\tau \frac{e^{2\pi i \tau (u^\alpha - m^\alpha_F)}}{\ch{\tau}}$. 
Finally, integrating over $u^\alpha$ we obtain (up to some overall phase factor):
\begin{align}
\CZ_{\wt\CO(Y)}(\s^1, \s^2, \eta_\alpha, \vec\mu,  m^{\alpha}_F)
= &  \int d\tau \, \frac{e^{-2\pi i \mu \s^2} \,e^{-2\pi i \tau m^{\alpha}_F} }{\ch{\tau}} \,\delta(\tau + \eta_\alpha + \s^1 -\s^2 + \sum_\beta b^{l}_\beta t_l) \nn \\
=& \frac{e^{-2\pi i \mu \s^2} \,e^{2\pi i m^{\alpha}_F (\s^1 -\s^2 + \eta_\alpha+\sum_\beta b^{l}_\beta t_l)} }{\ch{( \s^1 -\s^2 + \eta_\alpha+\sum_\beta b^{l}_\beta t_l)}} \nn \\
= & e^{-2\pi i \mu \s^2} \,e^{2\pi i m^{\alpha}_F (\s^1 -\s^2 + \eta_\alpha +\sum_\beta b^{l}_\beta t_l)}\, Z^{\rm bif}_{\rm{1-loop}}(\s^1,\s^2, -\eta_\alpha -\sum_\beta b^{l}_\beta t_l ),
\end{align} 
where $\mu=\mu^2-\mu^1$. Given the expression of the dual partition function in \eref{PF-wtOPLQ}, the above form of $\CZ_{\wt\CO(Y)}$ implies that dual operation $\wt{\CO}$ amounts to adding a bifundamental hyper to the theory $Y$, along with some phase factors which contribute to the FI terms. The dual partition function can be written as,
\begin{align}
&Z^{\wt{\CO}(Y)}(\vec{m}'(\vec{t},\eta_\alpha); \vec{\eta}'(\mu, m^{\alpha}_F))= \int \prod^2_{\gamma'=1}\Big[d{\s}^{\gamma'} \Big] \, Z^{\rm bif}_{\rm{1-loop}}(\s^1,\s^2, -\eta_\alpha -\sum_i b^{il} t_l )\, Z^{(Y)}_{\rm int} (\{{\s}^{\gamma'}\}, -\vec{t}, m^{\alpha}_F, 0, m^{\alpha}_F + \mu) \nn \\
&=  \int Z^{\wt{\CO}(Y)}_{\rm FI}(\{\s^{\gamma'} \}, \vec \eta') \, \prod^{n-l+1}_{i=1}Z^{\rm fund}_{\rm{1-loop}}(\s^{1}, m'^{1}_{{\rm fund}\,i})\, 
\prod^{l-1}_{j=1}Z^{\rm fund}_{\rm{1-loop}}(\s^{2}, m'^{2}_{{\rm fund}\,j})\, \prod^2_{k=1} Z^{\rm bif}_{\rm{1-loop}}(\s^1, \s^2, m'^{\rm bif}_k),
\end{align} 
where, for the second equality, we have used the expression for $Z^{(Y)}_{\rm int}$ from \eref{PF-ZY}, 
followed by shifting the integration variables and ignoring some overall phase factors. From the second equality, one 
can manifestly see that 
\begin{align}
Z^{\wt{\CO}(Y)}(\vec{m}'(\vec{t},\eta_\alpha); \vec{\eta}'(\mu, m^{\alpha}_F))=& \int \prod^2_{\gamma'=1}\Big[d{\s}^{\gamma'} \Big] \, Z^{(Y')}_{\rm FI}(\{\s^{\gamma'} \}, \vec \eta') \, Z^{(Y')}_{\rm 1-loop}(\{ \s^{\gamma'} \}, \vec{m}') \\
=& Z^{(Y')}(\vec{m}'(\vec{t},\eta); \vec{\eta}'(\mu, m^{\alpha}_F)),
\end{align}
where $Y'$ is the quiver in \figref{SimpAbEx1}. The fundamental and bifundamental masses of $Y'$ are given as linear functions 
of the FI parameters of $X'$ as:
\be \label{AbEx1MM1}
\begin{split}
&  m'^{1}_{{\rm fund}\,i}= - t_{i} + \delta_1 , \quad (i=1,\ldots, n-l+1), \quad \delta_1= \frac{1}{n-l+1} \sum^{n-l+1}_{i=1} t_i,\\
&  m'^{2}_{{\rm fund}\,j}= - t_{n-l+1+j} + \delta_2, \quad (j=1,\ldots, l-1), \quad \delta_2= \frac{1}{l-1} \sum^{l-1}_{j=1} t_{n-l+1+j}, \\
&m'_{{\rm bif}\,1}=\delta_1-\delta_2, \, m'_{{\rm bif}\,2}=(\delta_1-\delta_2) - (\eta_\alpha + \sum_\beta b^{l}_\beta t_l).
\end{split}
\ee
These masses parametrize the Cartan subalgebra of the Higgs branch global symmetry $SU(n-l+1) \times SU(l-1) \times U(2)$ of $Y'$.
The FI parameters of $Y'$ are similarly given in terms of the mass parameters of $X'$ as:
\begin{align} 
\eta'_1=m^{\alpha}_F, \qquad \eta'_2= -m^{\alpha}_F -\mu  \label{AbEx1MM2},
\end{align}
which parametrize the Cartan of the $U(1) \times U(1)$ Coulomb branch global symmetry for $Y'$.\\

Note that reading off the Coulomb branch global symmetry of the quiver $X'$ can be a bit counter-intuitive. Naively, 
one would guess the $SU(n-l+1) \times SU(l-1)$ factor from the presence of the two unbalanced linear subquivers 
in $X'$, consisting of $n-l$ and $l-2$ gauge nodes respectively. However, in this case, the remaining $U(1) \times U(1)$ 
also gets enhanced to an $U(1) \times SU(2)$. This can be directly seen by computing the character expansion of the Coulomb 
branch Hilbert Series of $X'$, as we do in \eref{SCI-AbEx1X'} in course of our discussion of the superconformal index realization of 
the $S$-type operation.

\section{Non-ADE mirror duals from Abelian $S$-type operations}\label{AbMirr}
In this section, we study 3d mirror symmetry for several infinite families of quiver gauge theories, starting from linear quivers, using 
Abelian $S$-type operations. A Hanany-Witten construction for these quiver gauge theories is not known, and therefore the standard
 Type IIB description for mirror symmetry is not understood for these cases. Some special cases of these theories have realization as 3d mirrors of certain
4d $\CN=2$ theories of class $\CS$ on a circle in the deep IR. We will comment on some of these special cases in \Secref{ADMirrors}. \\

The quivers studied in this section have the following generic features:
\begin{enumerate}
\item \textbf{Loops attached to a linear quiver tail :} Loops built out of gauge nodes and matter in appropriate 
representations, such that one or more of the gauge nodes are attached to linear quiver tails.

\item \textbf{Loops with multiple edges :} Loops built out of gauge nodes and matter, such that one or more 
pairs of gauge nodes are connected by multiple hypermultiplets transforming in a given representation of the associated gauge groups.

\end{enumerate}

In \Secref{AbS}, we present a general discussion of elementary Abelian $S$-type operations and their duals. In 
particular, we discuss Abelian versions of the four distinct types of elementary $S$-type operations 
discussed in \Secref{GFI-summary}. 
We show that the dual operations for the Abelian case lead to Lagrangian theories, which can be explicitly represented as 
quiver gauge theories.
Using these Abelian operations, we construct families of dual pairs involving Abelian quiver gauge theories with the features 
outlined above, in \Secref{AbExNew}. Families of dual quiver pairs with non-Abelian gauge groups are 
discussed next in \Secref{NAbExNew}. As outlined in \Secref{GFI-summary}, our strategy will be to construct 
these quiver pairs from a pair of linear quivers by a sequence of several Abelian $S$-type operations.

\subsection{Abelian $S$-type operations: General discussion}\label{AbS}
From the general definition \eref{Sbasic-def}, an elementary Abelian $S$-type operation $\CO^\alpha_{\vec \CP}$ on a generic quiver $X$ is defined 
as a combination of flavoring, identification and defect operations at a flavor node $\alpha$, followed by a \textit{single} Abelian 
gauging operation, i.e. 
\be \label{AbSbasic-def}
\CO^\alpha_{\vec \CP}(X) := G^{\rm Ab}_{\vec \CP}\circ (F^\alpha_{\vec \CP})^{n_3} \circ (I^\alpha_{\vec \CP})^{n_2} \circ (D^\alpha_{\vec \CP})^{n_1}(X), 
\quad (n_i=0,1, \,\, \forall i),
\ee
where $G^{\rm Ab}_{\vec \CP}$ is an Abelian gauging operation. The partition function of the theory $\CO^\alpha_{\vec \CP}(X)$ can be obtained from the general formula 
\eref{PF-OP}, while the partition function of the dual theory $\wt{\CO}^\alpha_{\vec \CP}(Y)$ (assuming that $X$ has a Lagrangian mirror $Y$)
can be obtained from \eref{PF-wtOPgen}-\eref{CZ-wtOP}. \\

In this section, we will write down explicit formulae for the dual partition functions, 
for the Abelian version of each distinct type of elementary $S$-type operations\footnote{We will restrict ourselves to flavoring by hypermultiplets with gauge charge 1.} (without defects) studied in \Secref{GFI-summary} - \Secref{GFIdual-summary}. In each case, we will first discuss the action of $\CO^\alpha_{\vec \CP}$ on a generic quiver $X$, followed by the special case where $X$ is a linear quiver. Using these explicit formulae, we will show that, for any dual pair of quiver gauge theories $(X,Y)$, the theory $X'$ (obtained via an 
elementary Abelian $S$-type operation on $X$) always has a Lagrangian dual $Y'$. This naturally leads to the following general result. For any dual pair of quiver gauge theories $(X,Y)$,
and an Abelian $S$-type operation acting on $X$, i.e.
\be
\CO^{(\alpha_1, \ldots, \alpha_l)}_{({\vec \CP_1},\ldots,{\vec \CP_l})}(X) :={\CO^{\alpha_l}_{\vec \CP_l}} \circ {\CO^{\alpha_{l-1}}_{\vec \CP_{l-1}}} \circ \ldots \circ {\CO^{\alpha_2}_{\vec \CP_2}} \circ  {\CO^{\alpha_1}_{\vec \CP_1}}(X),
\ee
where ${\CO^{\alpha_i}_{\vec \CP_i}}$ are elementary Abelian operations, the theory $\CO^{(\alpha_1, \ldots, \alpha_l)}_{({\vec \CP_1},\ldots,{\vec \CP_l})}(X)$ is guaranteed to 
have a Lagrangian dual. In addition, our construction will allow one to write down the dual Lagrangian explicitly as a quiver gauge theory.

\subsubsection{Gauging}\label{AbSG}
For an Abelian gauging operation $G^\alpha_{\CP}$, the functions $\CZ_{G^\alpha_\CP(X)}$ and $\CZ_{\wt{G}^\alpha_{\CP}(Y)}$, as defined in \eref{defZG} and \eref{CZ-wtOP} respectively, 
are given as
\begin{align}
& \CZ_{G^\alpha_\CP(X)}(u^\alpha, \eta_\alpha)= Z_{\rm FI} ({u}^\alpha, \eta_\alpha)= e^{2\pi i \eta_\alpha u^\alpha},\\
& \CZ_{\wt{G}^\alpha_{\CP}(Y)}=  \delta \Big(\eta_\alpha + b^{l}_\alpha \eta_l + g_\alpha \Big(\{\vec \s^{\gamma'} \}, \CP \Big) \Big),
\end{align}
where the function $g_\alpha\Big(\{\vec \s^{\gamma'} \}, \CP \Big)$ can be read off from the mirror map between masses of $X$ 
and FI parameters of $Y$. The dual partition function, following \eref{PF-wtOPgen}, is then given as
\begin{empheq}[box=\widefbox]{align}\label{GAbgen}
 Z^{\wt{G}^\alpha_{\CP}(Y)}(\vec{m}'(\vec{\eta},\eta_{\alpha}); \vec \eta'(\vec{v}^\alpha,\ldots))
=\int \prod_{\gamma'} \Big[d\vec{\s}^{\gamma'} \Big]\, & \delta \Big(\eta_\alpha + b^{l}_\alpha \eta_l + g_\alpha \Big(\{\vec \s^{\gamma'} \}, \CP \Big) \Big)\nn \\
\times & \, Z^{(Y, \CP)}_{\rm int}(\{\vec \s^{\gamma'} \}, \vec{m}^Y(\vec{\eta}), \vec{\eta}^Y({u}^\alpha =0, v^\alpha, \ldots)).
\end{empheq}

\noindent \textbf{Linear quivers:} If $X$ is a linear quiver, the function $\CZ_{G^\alpha_\CP(X)}$ is still given by the above formula. The formula for $\CZ_{\wt{G}^\alpha_{\CP}(Y)}$ is 
modified by taking $\eta_l \to t_l$, and writing the function  $g_\alpha\Big(\{\vec \s^{\gamma'} \}, \CP \Big)$ explicitly (following 
 \eref{giu}):
\begin{align} \label{galphaL}
g^L_\alpha \Big(\{\vec \s^{\gamma'} \}, \CP \Big)= & \sum^{M_\alpha }_{i_\alpha=1} \CP_{i_\alpha 1} (- \tr{\vec\s^{M_1+ \ldots+ M_{\alpha-1} + i_\alpha -1}} + 
 \tr{\vec\s^{M_1+ \ldots+ M_{\alpha-1} + i_\alpha}}) \nn \\
= & (- \tr{\vec\s^{M_1+ \ldots+ M_{\alpha-1} + j -1}} +  \tr{\vec\s^{M_1+ \ldots+ M_{\alpha-1} + j}}), \quad 1 \leq j \leq M_\alpha, \nn \\
= & -\tr{\vec\s^{\alpha' -1}} + \tr {\vec\s^{\alpha'}}, \quad \alpha'= M_1+ \ldots+ M_{\alpha-1} + j,
\end{align}
where $\CP_{i_\alpha 1} =1$ for a fixed $i_\alpha=j$, and vanishes otherwise. The relation is subject to the boundary conditions 
$\tr{\vec\s^{M_0}}=\tr{\vec\s^{M_1+ \ldots+ M_{\alpha} }}=0$.
From the expression \eref{PF-wtOPLQ} for a linear quiver (or the general prescription \eref{PF-wtOPgen}), the partition function of the theory 
$\wt{G}^\alpha_{\CP}(Y)$ is given as:
\begin{align} \label{G0Ab}
Z^{\wt{G}^\alpha_{\CP}(Y)}(\vec{m}'(\vec{t},\eta_\alpha); \vec{\eta}'(\vec{v}^\alpha,\{\vec{m}^\gamma\}_{\gamma \neq \alpha})) =\int \prod^{L^\vee}_{\gamma'=1} \Big[d\s^{\gamma'}\Big]\, & \delta \Big(\eta_\alpha + b^{l}_\alpha t_l -\tr {\vec\s^{\alpha' -1}} + \tr{\vec\s^{\alpha'}}\Big)\nn \\
\times & \, Z^{(Y,\CP)}_{\rm int}(\{\vec \s^{\gamma'} \}, -\vec t, {u}^\alpha=0, \vec{v}^\alpha, \{\vec{m}^\gamma\}_{\gamma \neq \alpha}).
\end{align}

There are a couple of observations from the form of the dual partition function $\wt{G}^\alpha_{\CP}(Y)$ in \eref{GAbgen} and \eref{G0Ab}:
\begin{enumerate}

\item The equations give a very clear prescription for the dual of a $U(1)$ gauging operation. The action of the dual operation 
$\wt{G}^\alpha_{\CP}$ on the quiver $Y$ amounts to introducing a delta function in the integrand for the partition function of $Y$, which removes a single  
$U(1)$ factor (the function $g_\alpha$ specifies which one) from the gauge group. The theory $\wt{G}^\alpha_{\CP}(Y)$ is therefore manifestly Lagrangian, and the rank of the gauge group of the theory is less than that of the theory $Y$ by $\Delta_{\rm rank}=1$. This is a way of seeing that the dual of the gauging operation amounts to ungauging, i.e. 
reduction in the rank of the gauge group of $Y$.

\item The delta function and therefore the integrand of the partition function of the dual theory $\wt{G}^\alpha_{\CP}(Y)$ in principle depends on the permutation matrix 
$\CP$. In other words, the Lagrangian description of the theory $\wt{G}^\alpha_{\CP}(Y)$ depends on $\CP$, even though the partition function itself is independent of 
it. In certain cases, the Lagrangians obtained for different choices of $\CP$ may be related by field redefinitions.
The distinct Lagrangians obtained for different choices of $\CP$ are expected to be IR dual among themselves\footnote{There might be additional 
subtleties associated with discrete symmetries in establishing these IR dualities which are not captured by the partition function analysis.}. 
We will present explicit examples of this phenomenon in \Secref{GFI-ex}, where we study mirror duals of $D_4$ quiver gauge theories and their affine cousins.

\end{enumerate}

\subsubsection{Flavoring-gauging}\label{AbSGF}
Consider an elementary $S$-type operation ${\CO}^\alpha_{\vec \CP}  = G^\alpha_{\CP} \circ F^\alpha_\CP$ on 
a generic quiver $X$, where the global symmetry group $U(M_\alpha)$ associated with the flavor node $\alpha$  
is split into $U(1) \times U(M_\alpha -1)$, and the resulting $U(1)$ node is flavored by $N^{\alpha}_F$ hypermultiplets of charge 1, 
followed by a gauging operation. The case for hypermultiplets with a generic charge $Q \neq 1$, or with different charges, can be worked 
out in an analogous fashion.\\

The flavor symmetry introduced by the operation ${\CO}^\alpha_{\vec \CP}$ is $G^{\alpha}_{\rm F} = U(N^{\alpha}_F)$, 
and the masses in the Cartan of $U(N^{\alpha}_F)$ are labelled as $\{m^{\alpha}_{k\, F}\}$,
where $k=1,\ldots,N^{\alpha}_F$. The function $\CZ_{G^\alpha_{\CP} \circ F^\alpha_\CP(X)}$, which can be read off from \eref{defZG}, \eref{defZF} 
and \eref{CZ-OP}, is given as
\begin{align}
\CZ_{G^\alpha_{\CP} \circ F^\alpha_\CP(X)}({u}^\alpha, \vec m^{\alpha}_F, \eta_\alpha)= Z_{\rm FI} ({u}^\alpha,\eta_\alpha)\,Z_{\rm 1-loop}^{\rm hyper} ({u}^\alpha, \vec m^{\alpha}_F)
= \frac{e^{2\pi i \eta_\alpha u^\alpha}}{\prod_k \ch{({u}^\alpha - m^{\alpha}_{k\, F})}}. 
\end{align}
In addition, the function $Z_{\rm 1-loop}^{\rm hyper} ({u}^\alpha, \vec m^{\alpha}_F)$ satisfies a useful integral identity:
\begin{align}\label{AbHMId}
& Z_{\rm 1-loop}^{\rm hyper} ({u}^\alpha, \vec m^{\alpha}_F) =  \frac{1}{\prod^{N^\alpha_F}_{k=1} \ch{({u}^\alpha - m^{\alpha}_{k\, F})}}
=\int \prod^{N^{\alpha }_F}_{k=1} d\tau^{\alpha}_{k} \,
\frac{e^{2\pi i (u^\alpha - m^{\alpha}_{1\, F}) \tau^{\alpha}_{1}}\,\prod^{N^{\alpha}_F}_{k=2} e^{2\pi i \tau^{\alpha}_{k}(m^{\alpha}_{k-1\, F} - m^{\alpha}_{k\, F})}}{\prod^{N^{\alpha}_F-1}_{k=1} \ch{(\tau^{\alpha}_{k}-\tau^{\alpha}_{k+1})}\, \ch{\tau^{\alpha}_{N^{\alpha}_F }}} \nn \\
=& \int \prod^{N^{\alpha}_F}_{k=1} d\tau^{\alpha}_{k} \, Z_{\rm FI} (\tau^{\alpha}_{1},(u^\alpha - m^{\alpha}_{1\,F}))\,\prod^{N^{\alpha}_F}_{k=2} Z_{\rm FI}(\tau^{\alpha}_{k}, (m^{\alpha}_{k-1\, F} - m^{\alpha}_{k\, F}))
\, Z^{\rm fund}(\tau^{\alpha}_{N^{\alpha}_F}, 0)\, \prod^{N^{\alpha}_{F} -1}_{k=1} Z^{\rm bifund}(\tau^{\alpha}_{k}, \tau^{\alpha}_{k+1},0) \nn\\
=: & \int \prod^{N^{\alpha}_F}_{k=1} d\tau^{\alpha}_{k} \, Z^{(\CT[N^{\alpha}_F])}_{\rm int}(\vec{\tau}^{\alpha}, \vec t'_\alpha,0),
\end{align}
where $\CT[N^{\alpha}_F]$ is an Abelian linear quiver gauge theory with $N^{\alpha}_F$ gauge nodes and a single flavor attached to the $N^{\alpha}_F$-th gauge node. The FI parameters are parametrized by $\vec t'^\alpha$, with $t'_{\alpha\,1}=u^\alpha$, $t'_{\alpha\,k}= m^{\alpha}_{k-1\, F}$ for $k=2,\ldots, N^{\alpha}_F$, and the single mass parameter $m^\alpha=0$.

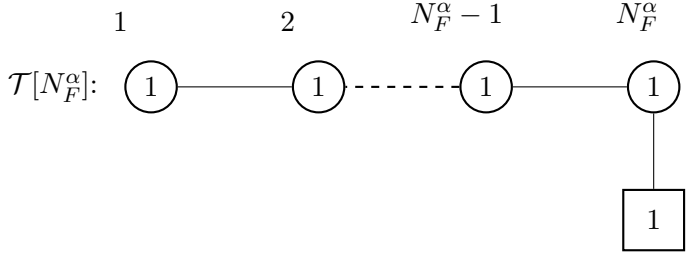
\begin{tikzpicture}[
cnode/.style={circle,draw,thick,minimum size=4mm},snode/.style={rectangle,draw,thick,minimum size=8mm},pnode/.style={rectangle,red,draw,thick,minimum size=8mm},nnode/.style={circle,red,draw,thick,minimum size=8mm}]
\node[text width=1cm](1) at (0,0) {$\CT[N^{\alpha}_F]$:};
\node[cnode] (25) at (8,0){1};
\node[snode] (26) [below=1cm of 25]{1};
\node[cnode] (27) [left=1.5 cm of 25]{1};
\node[cnode] (28) [left=1.5 cm of 27]{1};
\node[cnode] (29) [left=1.5 cm of 28]{1};
\draw[-] (25) -- (26);
\draw[-] (25) -- (27);
\draw[thick, dashed] (27) -- (28);
\draw[-] (28) -- (29);
\node[text width=1cm](31) [above=0.25 of 25] {$N^{\alpha}_F$};
\node[text width=2 cm](32) [above=0.25 of 27] {$N^{\alpha}_F-1$};
\node[text width=1cm](33) [above=0.25 of 28] {$2$};
\node[text width=1cm](34) [above=0.25 of 29] {$1$};
\end{tikzpicture}

Using the general formula given in \eref{CZ-wtOP}, the function $\CZ_{\wt{(G^\alpha_{\CP} \circ F^\alpha_\CP)}(Y)}$ can be written as
\begin{align}
& \CZ_{\wt{(G^\alpha_{\CP} \circ F^\alpha_\CP)}(Y)}(\{\vec\s^{\gamma'}\}, \eta_\alpha, \vec m^{\alpha}_F, \vec \eta) 
=  \int \de u^\alpha \,\CZ_{G^\alpha_{\CP} \circ F^\alpha_\CP(X)}({u}^\alpha, \vec m^{\alpha}_F, \eta_\alpha)\,e^{2\pi i \,(g_\alpha (\{\vec \s^{\gamma'} \}, \CP) + b^{l}_\alpha \eta_l)\,u^\alpha} \nn \\
=& \int \prod^{N^{\alpha}_F}_{k=1} d\tau^{\alpha}_{k} \,Z^{(\CT[N^{\alpha}_F])}_{\rm int}(\vec{\tau}^{\alpha}, \{ t'_{\alpha\,1}=0, t'_{\alpha \, {k \neq 1}} \},0)
\,\delta \Big(\tau^{\alpha}_{1}+\eta_\alpha + b^{l}_\alpha \eta_l + g_\alpha \Big(\{\vec \s^{\gamma'} \}, \CP \Big) \Big).
\end{align}
Implementing the delta function and relabelling the integration variables $\tau^\alpha_k \to \tau^\alpha_{k-1}$, 
$t'_{\alpha \, {k}} \to t'_{\alpha \, {k -1}}$, we can rewrite the above function as
\begin{align}
& \CZ_{\wt{(G^\alpha_{\CP} \circ F^\alpha_\CP)}(Y)}
= \int \prod^{N^{\alpha}_F -1}_{k=1} d\tau^{\alpha}_{k} \,e^{2\pi i m^\alpha_{1\,F}g_\alpha (\{\vec \s^{\gamma'} \}, \CP)}\,
Z^{\rm hyper}_{\wt{(G^\alpha_{\CP} \circ F^\alpha_\CP)}(Y)}\Big(\{\vec \s^{\gamma'} \}, \tau^\alpha_1, \eta_\alpha, \vec \eta \Big)\,Z^{(\CT[N^{\alpha}_F -1])}_{\rm int}(\vec{\tau}^{\alpha},\vec m^{\alpha}_F,0), \label{CZGFgen}\\
& Z^{\rm hyper}_{\wt{(G^\alpha_{\CP} \circ F^\alpha_\CP)}(Y)}\Big(\{\vec \s^{\gamma'} \}, \tau^\alpha_1, \eta_\alpha, \vec \eta \Big)= \frac{1}{\ch{(g_\alpha (\{\vec \s^{\gamma'} \}, \CP)+ \tau^\alpha_1+ \eta_\alpha + b^{l}_\alpha \eta_l)}}. \label{HyperGFgen}
\end{align}
Note that $Z^{\rm hyper}_{\wt{(G^\alpha_{\CP} \circ F^\alpha_\CP)}(Y)}$ can be interpreted as the 1-loop contribution of a single hypermultiplet charged under $U(1)_{1}$ 
of $\CT[N^{\alpha}_F -1]$ and the various $U(1)$ factors in the gauge group of quiver $Y$ -- the precise gauge nodes and the charges can be read off from the function $g_\alpha(\{\vec \s^{\gamma'} \}, \CP)$. 
Using the prescription \eref{PF-wtOPgen}, and the simplified expression for $\CZ_{\wt{(G^\alpha_{\CP} \circ F^\alpha_\CP)}(Y)}$ in \eref{CZGFgen}-\eref{HyperGFgen}, 
the dual partition function can be written as (up to some phase factor)
\begin{empheq}[box=\widefbox]{align} \label{PFGFgen}
 & Z^{\wt{(G^\alpha_{\CP} \circ F^\alpha_\CP)}(Y)}(\vec{m}'(\vec{\eta},\eta_{\alpha}); \vec \eta'( \vec{v}^\alpha,\ldots,\vec{m}^{\alpha}_F))
= \int \prod_{\gamma'} \Big[d\vec{\s}^{\gamma'} \Big]\, \prod^{N^{\alpha}_F-1}_{k=1} d\tau^{\alpha}_{k} \, Z^{(\CT[N^{\alpha}_F -1])}_{\rm int}(\vec{\tau}^{\alpha},\vec m^{\alpha}_F,0)\nn \\ & \times Z^{\rm hyper}_{\wt{(G^\alpha_{\CP} \circ F^\alpha_\CP)}(Y)}\Big(\{\vec \s^{\gamma'} \}, \tau^\alpha_1, \eta_\alpha, \vec \eta \Big) \cdot Z^{(Y, \CP)}_{\rm int}(\{\vec \s^{\gamma'} \}, \vec{m}^Y(\vec{\eta}), \vec{\eta}^Y(\vec{u}^\alpha =m^\alpha_{1\,F}, \vec{v}^\alpha,\ldots)).
\end{empheq}
The dual theory is manifestly Lagrangian, which can be read off from the integrand in the RHS of the above equation. The theory consists of a 
$\CT[N^{\alpha}_F -1]$ tail attached to the quiver $Y$ by a single hypermultiplet, which is also charged under some of the $U(1)$ factors in the 
gauge group of $Y$. As a quiver diagram, the theory can be represented as \\

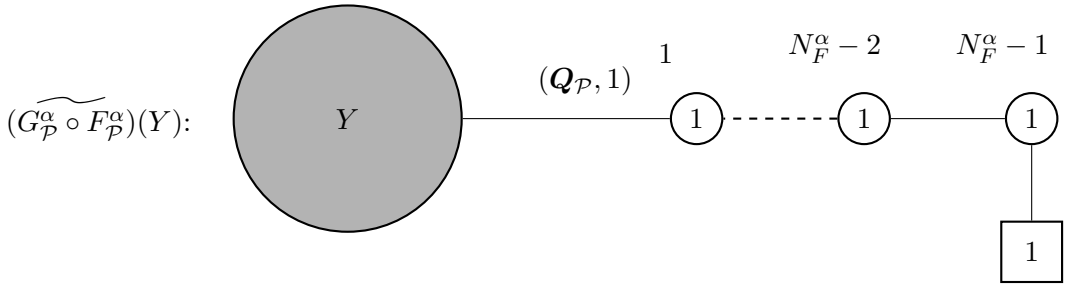
\begin{tikzpicture}[
cnode/.style={circle,draw,thick,minimum size=4mm},snode/.style={rectangle,draw,thick,minimum size=8mm},pnode/.style={rectangle,red,draw,thick,minimum size=1.0cm}, bnode/.style={circle,draw, thick, fill=black!30,minimum size=3cm}]
\node[text width=1cm](2) at (-4,0) {$\wt{(G^\alpha_{\CP} \circ F^\alpha_\CP)}(Y)$:};
\node[bnode] (1) at (0,0){$Y$} ;
\node[cnode] (25) at (9,0){1};
\node[snode] (26) [below=1cm of 25]{1};
\node[cnode] (27) [left= 1.5 cm of 25]{1};
\node[cnode] (28) [left=1.5 cm of 27]{1};
\draw[-] (25) -- (26);
\draw[-] (25) -- (27);
\draw[thick, dashed] (27) -- (28);
\draw[-] (28) -- (1);
\node[text width=2cm](31) [above=0.25 of 25] {$N^{\alpha}_F-1$};
\node[text width=2 cm](32) [above=0.25 of 27] {$N^{\alpha}_F-2$};
\node[text width=1cm](33) [above=0.25 of 28] {$1$};
\node[text width=1 cm](34) at (3,0.5) {$(\vec Q_\CP, 1)$};
\end{tikzpicture}

In the above quiver gauge theory, $(\vec Q_\CP,1)$ collectively denotes the charges of the hypermultiplet connecting $Y$ and $\CT[N^{\alpha}_F -1]$, under various $U(1)$ factors of $Y$ and $U(1)_1$ respectively. There are a couple of observations about the dual theory:
\begin{enumerate}
\item The charge vector $\vec Q_\CP$, and therefore the precise shape of the quiver, depends on the choice of the permutation matrix $\CP$.
\item In the special case, where $N^{\alpha}_F=1$, the function $\CZ_{\wt{(G^\alpha_{\CP} \circ F^\alpha_\CP)}(Y)}$ is given as
\begin{align}
& \CZ_{\wt{(G^\alpha_{\CP} \circ F^\alpha_\CP)}(Y)} =  e^{2\pi i m^\alpha_{\,F}g_\alpha (\{\vec \s^{\gamma'} \}, \CP)}\,Z^{\rm hyper}_{\wt{(G^\alpha_{\CP} \circ F^\alpha_\CP)}(Y)}\Big(\{\vec \s^{\gamma'} \}, \eta_\alpha, \vec \eta \Big), \nn \\ 
& Z^{\rm hyper}_{\wt{(G^\alpha_{\CP} \circ F^\alpha_\CP)}(Y)}\Big(\{\vec \s^{\gamma'} \}, \eta_\alpha, \vec \eta \Big) = \frac{1}{\ch{(g_\alpha (\{\vec \s^{\gamma'} \}, \CP) + \eta_\alpha + b^{l}_\alpha \eta_l)}}. \label{HyperGFgenNf1}
\end{align}
The dual paritition function is then given as
\begin{empheq}[box=\widefbox]{align}\label{PFGFgenNf1}
 & Z^{\wt{(G^\alpha_{\CP} \circ F^\alpha_\CP)}(Y)}(\vec{m}'(\vec{\eta},\eta_{\alpha}); \vec \eta'( \vec{v}^\alpha,\ldots, {m}^{\alpha}_F))  \nn \\
& = \int  \prod_{\gamma'} \Big[d\vec{\s}^{\gamma'} \Big]\, Z^{\rm hyper}_{\wt{(G^\alpha_{\CP} \circ F^\alpha_\CP)}(Y)}\Big(\{\vec \s^{\gamma'} \}, \eta_\alpha, \vec \eta \Big) \cdot 
Z^{(Y, \CP)}_{\rm int}(\{\vec \s^{\gamma'} \}, \vec{m}^Y(\vec{\eta}), \vec{\eta}^Y(\vec{u}^\alpha =m^\alpha_{\,F}, \vec{v}^\alpha,\ldots)).
\end{empheq}
The dual theory $\wt{(G^\alpha_{\CP} \circ F^\alpha_\CP)}(Y)$ then is even simpler. It merely involves introducing an extra hypermultiplet in the 
quiver gauge theory $(Y)$, such that the hypermultiplet is charged under a certain $U(1)^r$ subgroup ($r$ is an integer) of the gauge group.
Note that the mass parameter $m^\alpha_F$ enters the RHS of the above equation as an FI parameter.\\
\end{enumerate}

\noindent \textbf{Linear quivers:} For a linear quiver pair $(X,Y)$, the function $g_\alpha=g^L_\alpha$ has the simple form \eref{galphaL}, while $\CZ_{\wt{(G^\alpha_{\CP} \circ F^\alpha_\CP)}(Y)}$ 
is given by the formulae \eref{CZGFgen}-\eref{HyperGFgen}.
The function $\CZ_{\wt{(G^\alpha_{\CP} \circ F^\alpha_\CP)}(Y)}$ can be further simplified after a change of variables, 
$\tau^{\alpha}_{k} \to \tau^{\alpha}_{k} - \tr {\vec\s^{\alpha'}}$, and can be written as follows:
\begin{align}
 \CZ_{\wt{(G^\alpha_{\CP} \circ F^\alpha_\CP)}(Y)}
= \int \prod^{N^{\alpha}_F -1}_{k=1} d\tau^{\alpha}_{k} \,e^{-2\pi i m^\alpha_{1\,F}\,\tr \s^{\alpha' -1}}\,e^{2\pi i m^\alpha_{N^\alpha_F\,F}\,\tr \s^{\alpha'}}\,
& Z^{\rm hyper}_{(N^{\vee}_{\alpha'-1},1)}\Big(\vec\s^{\alpha' -1}, \tau^\alpha_1, \eta_\alpha + b^{l}_\alpha t_l \Big)\nn \\
& \times Z^{(\CT[N^{\alpha}_F -1])}_{\rm int}(\vec{\tau}^{\alpha},\vec m^{\alpha}_F, \tr{\vec\s^{\alpha'}}),
\end{align}
where the function $Z^{\rm hyper}_{(N^{\vee}_{\alpha'-1},1)}$ has the explicit form:
\be
Z^{\rm hyper}_{(N^{\vee}_{\alpha'-1},1)}\Big(\vec\s^{\alpha' -1}, \tau^\alpha_1, \eta_\alpha + b^{l}_\alpha t_l \Big)=\frac{1}{\ch{(-\tr{\vec\s^{\alpha' -1}} 
+\tau^\alpha_1 + \eta_\alpha + b^{l}_\alpha t_l)}}.
\ee

The subscript in $Z^{\rm hyper}_{(N^{\vee}_{\alpha'-1},1)}$ denotes the charges of the hypermultiplet under the
$U(1)$ subgroup of $U(N^{\vee}_{\alpha'-1})$ and $U(1)_1$ of $\CT[N^{\alpha}_F -1]$ respectively.
Using the general prescription in \eref{PF-wtOPgen}), the dual partition function can be written as
\begin{empheq}[box=\vwidefbox]{align} 
 & Z^{\wt{(G^\alpha_{\CP} \circ F^\alpha_\CP)}(Y)} (\vec{m}'(\vec{t}, \eta_\alpha); \vec{\eta}'(\vec{v}^\alpha, \vec{m}^\alpha_F,\{\vec m^\gamma\}_{\gamma \neq \alpha}))
 =\int \prod^{L^\vee}_{\gamma'=1}  \Big[d\s^{\gamma'}\Big]\,\prod^{N^{\alpha}_F-1}_{k=1} d\tau^{\alpha}_{k}\, Z^{(\CT[N^{\alpha}_F -1])}_{\rm int}(\vec{\tau}^{\alpha},\vec m^{\alpha}_F, \tr \s^{\alpha'})\nn \\ & \times Z^{\rm hyper}_{(N^{\vee}_{\alpha'-1},1)}\Big(\vec\s^{\alpha' -1}, \tau^\alpha_1, \eta_\alpha + b^{l}_\alpha t_l \Big) \,Z_{\rm FI} (\vec\s^{\alpha'}, m^\alpha_{N^{\alpha}_F\,F} - m^\alpha_{1\,F})\, Z^{(Y,\CP)}_{\rm int}(\{\vec \s^{\gamma'} \}, -\vec t, \vec{u}^\alpha=m^\alpha_{1\,F}, \vec{v}^\alpha, \{\vec{m}^\gamma\}_{\gamma \neq \alpha}). \label{GF0Ab}
\end{empheq}

Graphically, the dual theory $\wt{(G^\alpha_{\CP} \circ F^\alpha_\CP)}(Y)$ for a linear quiver $Y$ is given as:\\

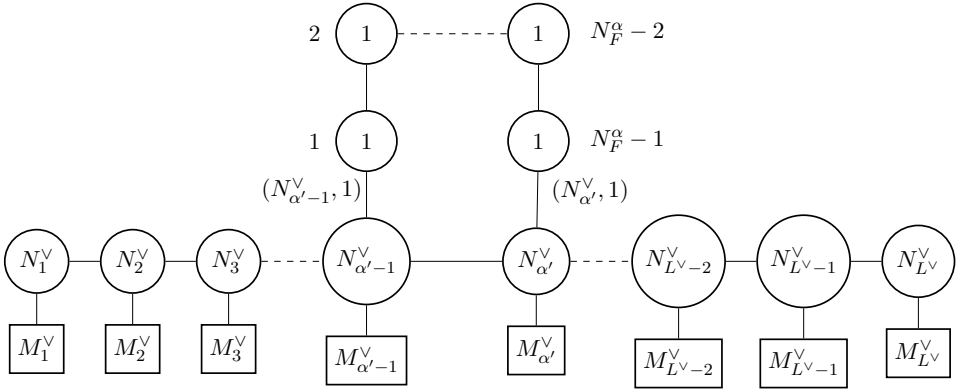
\scalebox{.8}{\begin{tikzpicture}[
cnode/.style={circle,draw,thick,minimum size=1cm},snode/.style={rectangle,draw,thick,minimum size=8mm},pnode/.style={rectangle,red,draw,thick,minimum size=1.0cm}]
\node[cnode] (1) {$N^{\vee}_1$};
\node[cnode] (2) [right=.5cm  of 1]{$N^{\vee}_2$};
\node[cnode] (3) [right=.5cm of 2]{$N^{\vee}_3$};
\node[cnode] (4) [right=1cm of 3]{$N^{\vee}_{\alpha'-1}$};
\node[cnode] (6) [right=1.5 cm of 4]{$N^{\vee}_{\alpha'}$};
\node[cnode] (7) [right=1cm of 6]{{$N^{\vee}_{L^{\vee}-2}$}};
\node[cnode] (8) [right=0.5cm of 7]{$N^{\vee}_{L^{\vee}-1}$};
\node[cnode] (9) [right=0.5cm of 8]{$N^{\vee}_{L^{\vee}}$};
\node[snode] (10) [below=0.5cm of 1]{$M^{\vee}_1$};
\node[snode] (11) [below=0.5cm of 2]{$M^{\vee}_2$};
\node[snode] (12) [below=0.5cm of 3]{$M^{\vee}_3$};
\node[snode] (13) [below=0.5cm of 4]{$M^{\vee}_{\alpha'-1}$};
\node[snode] (15) [below=0.5cm of 6]{$M^{\vee}_{\alpha'}$};
\node[snode] (16) [below=0.5cm of 7]{$M^{\vee}_{L^{\vee}-2}$};
\node[snode] (17) [below=0.5cm of 8]{$M^{\vee}_{L^{\vee}-1}$};
\node[snode] (18) [below=0.5cm of 9]{$M^{\vee}_{L^{\vee}}$};
\node[cnode] (19) [above=0.75cm of 4]{1};
\node[cnode] (20) [above=0.75cm of 19]{1};
\node[cnode] (21) [right= 1.8 cm of 20]{1};
\node[cnode] (22) [below=0.75cm of 21]{1};
\draw[-] (1) -- (2);
\draw[-] (2)-- (3);
\draw[dashed] (3) -- (4);
\draw[-] (4) --(6);
\draw[dashed] (6) -- (7);
\draw[-] (7) -- (8);
\draw[-] (8) --(9);
\draw[-] (1) -- (10);
\draw[-] (2) -- (11);
\draw[-] (3) -- (12);
\draw[-] (4) -- (13);
\draw[-] (6) -- (15);
\draw[-] (7) -- (16);
\draw[-] (8) -- (17);
\draw[-] (9) -- (18);
\draw[-] (4) -- (19);
\draw[-] (19) -- (20);
\draw[dashed] (20) -- (21);
\draw[-] (21) -- (22);
\draw[-] (6) -- (22);
\node[text width=0.1cm](23) [left=0.2cm of 19] {$1$};
\node[text width=0.1cm](24) [left=0.2cm of  20] {$2$};
\node[text width=2cm](25) [right=0.2cm of 21] {$N^\alpha_F -2$};
\node[text width=2cm](26) [right=0.2cm of 22] {$N^\alpha_F -1$};
\node[text width=2cm](27) [above=0.8 cm of 4.west] {$(N^{\vee}_{\alpha'-1},1)$};
\node[text width=2cm](28) [right=2.5 cm of 27] {$(N^{\vee}_{\alpha'},1)$};
\end{tikzpicture}}
\\

There are a few observations that we can make about this dual theory:
\begin{enumerate}

\item The location of the two consecutive nodes at which the Abelian loop is attached depends on the choice of $\CP$.

\item For $N^\alpha_F =1$, the dual theory has a single hypermultiplet (as opposed to the loop) charged under the 
$U(1)$ subgroups of $U(N^{\vee}_{\alpha'-1}) \times U(N^{\vee}_{\alpha'})$. The dual partition function can be computed 
as before:
\begin{align}\label{PFGFLinNf1}
Z^{\wt{(G^\alpha_{\CP} \circ F^\alpha_\CP)}(Y)} (\vec{m}'(\vec{t}, \eta_\alpha); \vec{\eta}'(\vec{v}^\alpha, \vec{m}^\alpha_F,\{\vec m^\gamma\}_{\gamma \neq \alpha}))
 =\int & \prod^{L^\vee}_{\gamma'=1}  \Big[d\vec\s^{\gamma'}\Big]\,Z^{\rm hyper}_{(N^{\vee}_{\alpha'-1}, N^{\vee}_{\alpha'})}\Big(\vec\s^{\alpha' -1}, \vec\s^{\alpha'}, \eta_\alpha, \vec t \Big) \nn \\ \times & Z^{(Y,\CP)}_{\rm int}(\{\vec \s^{\gamma'} \}, -\vec t, \vec{u}^\alpha=m^\alpha_{\,F}, \vec{v}^\alpha, \{\vec{m}^\gamma\}_{\gamma \neq \alpha}),
\end{align} 
\begin{align}\label{HyperGFLinNf1}
Z^{\rm hyper}_{(N^{\vee}_{\alpha'-1}, N^{\vee}_{\alpha'})}\Big(\{\vec \s^{\gamma'} \}, \eta_\alpha, \vec t \Big)
= \frac{1}{\ch{( -\tr{\vec\s^{\alpha' -1}} + \tr{\vec\s^{\alpha'}} + \eta_\alpha + b^{l}_\alpha t_l)}}.
\end{align}
In the special case $N^{\vee}_{\alpha'-1} = N^{\vee}_{\alpha'}=1$, \eref{HyperGFLinNf1} simply corresponds to a hypermultiplet in the bifundamental representation of 
$U(1)_{\alpha'-1} \times U(1)_{\alpha'}$.
\end{enumerate}

The special cases of flavoring-gauging operations with $N^\alpha=1$, given by \eref{HyperGFgenNf1}-\eref{PFGFgenNf1} and 
\eref{PFGFLinNf1}-\eref{HyperGFLinNf1} for a generic quiver and a linear quiver respectively, will be used frequently for the construction 
of non-ADE mirror duals later in this section.

\subsubsection{Identification-gauging}\label{AbSGI}
The Abelian identification-gauging operation on a generic quiver $X$ can be worked out in a similar fashion. The operator 
$\CZ_{(G^\alpha_{\vec \CP} \circ  I^\alpha_{\vec\CP})(X)}$ can be read off from \eref{defZG} and \eref{defZI}:
\be
\CZ_{(G^\alpha_{\vec \CP} \circ  I^\alpha_{\vec\CP})(X)}(u^\alpha, \{{u}^\beta\}, \eta_{\alpha}, \vec \mu)=  Z_{\rm FI} ({u}^{\alpha},\eta_{\alpha})\, \int \prod^{p}_{j=1} \, d {u^{\gamma_j}} \, \delta \Big({u}^{\alpha} - {u}^{\gamma_j} + {\mu}^{\gamma_j} \Big),
\ee
where $\beta=\gamma_1,\ldots, \gamma_p$. The function $\CZ_{\wt{(G^\alpha_{\vec\CP} \circ I^\alpha_{\vec\CP})}(Y)}$, as defined in \eref{CZ-wtOP}, is given as
\begin{align}
\CZ_{\wt{(G^\alpha_{\vec\CP} \circ I^\alpha_{\vec\CP})}(Y)}(\{\vec\s^{\gamma'}\}, \eta_{\alpha}, \vec \mu, \vec \eta)= \int\,d{u}^{\alpha}  \,\CZ_{(G^\alpha_{\vec \CP} \circ  I^\alpha_{\vec\CP})(X)}(u^\alpha, \{{u}^\beta\}, \eta_{\alpha}, \vec \mu)\,\prod_{\beta}\,e^{2\pi i \,(g_\beta (\{\vec \s^{\gamma'} \}, \CP_\beta) + b^{l}_\beta \eta_l)\,u^{\beta}}.
\end{align}
Note that we have dropped the index $i$, since the operation is Abelian. Implementing the $p$ delta function integrals, and integrating over ${u}^{\alpha}$,
we have the following form:
\begin{align}\label{CZ-GI}
\CZ_{\wt{(G^\alpha_{\vec\CP} \circ I^\alpha_{\vec\CP})}(Y)}(\{\vec\s^{\gamma'}\}, \eta_{\alpha}, \vec \mu, \vec \eta)
= \delta \Big(\eta_{\alpha} + \sum_{\beta} b^{l}_\beta \eta_l + \sum_{\beta} g_\beta \Big(\{\vec \s^{\gamma'} \}, \CP_\beta \Big) \Big)\, \prod_{\beta}e^{2\pi i ( g_\beta (\{\vec \s^{\gamma'} \}, \CP_\beta) + b^{l}_\beta \eta_l)\,\mu^{\beta}}. 
\end{align}
From the general prescription \eref{PF-wtOPgen}, the partition function of the theory $\wt{(G^\alpha_{\vec\CP} \circ I^\alpha_{\vec\CP})}(Y)$ is given as
\begin{align} 
&Z^{\wt{(G^\alpha_{\vec\CP} \circ I^\alpha_{\vec\CP})}(Y)}(\vec{m}'(\vec{\eta},\eta_{\alpha}); \vec \eta'(\{ \vec{v}^\beta \},\ldots) 
= \int \prod_{\gamma'} \Big[d \vec\s^{\gamma'}\Big]\,  \delta \Big(\eta_{\alpha} + \sum_{\beta} b^{l}_\beta \eta_l + \sum_{\beta} g_\beta \Big(\{\vec \s^{\gamma'} \}, \CP_\beta \Big) \Big) \nn \\
& \times \, \prod_{\beta}e^{2\pi i ( g_\beta (\{\vec \s^{\gamma'} \}, \CP_\beta) + b^{l}_\beta \eta_l)\,\mu^{\beta}} \cdot Z^{(Y,\{\CP_\beta\})}_{\rm int}(\{\vec \s^{\gamma'} \},\vec{m}^Y(\vec{\eta}), \vec{\eta}^Y(\{\vec{u}^\beta =0 \}, \{\vec{v}^\beta\},\ldots)),
\end{align}
which can be massaged into the following expression:
\begin{empheq}[box=\widefbox]{align} \label{dual-AbGIgen}
Z^{\wt{(G^\alpha_{\vec\CP} \circ I^\alpha_{\vec\CP})}(Y)}(\vec{m}'(\vec{\eta},\eta_{\alpha}); \vec \eta'(\{ \vec{v}^\beta \},\ldots) 
=&  \int \prod_{\gamma'} \Big[d \vec\s^{\gamma'}\Big]\, \delta \Big(\eta_{\alpha} + \sum_{\beta} b^{l}_\beta \eta_l + \sum_{\beta} g_\beta \Big(\{\vec \s^{\gamma'} \}, \CP_\beta \Big) \Big) \nn \\
& \times \, Z^{(Y,\{\CP_\beta\})}_{\rm int}(\{\vec \s^{\gamma'} \},\vec{m}^Y(\vec{\eta}), \vec{\eta}^Y(\{\vec{u}^\beta = \mu^\beta \}, \{\vec{v}^\beta\},\ldots)).
\end{empheq}

\noindent \textbf{Linear quivers:} For a linear quiver $X$, the function $\CZ_{\wt{(G^\alpha_{\vec\CP} \circ I^\alpha_{\vec\CP})}(Y)}$ is obtained from \eref{CZ-GI} by 
replacing $\eta_l \to t_l$, and the function $g_\beta \to g^L_\beta$, where $g^L_\beta\Big(\{\vec \s^{\gamma'} \}, \CP_\beta \Big)$ 
is given as
\begin{align}\label{def-gLbeta}
g^L_\beta \Big(\{\vec \s^{\gamma'} \}, \CP_\beta \Big)  
= & \sum^{M_\beta }_{i_\beta=1} \CP_{\beta\, i_\beta 1} (- \tr{\vec\s^{M_1+ \ldots+ M_{\beta-1} + i_\beta -1}} +  \tr{\vec\s^{M_1+ \ldots+ M_{\beta-1} + i_\beta}}) \nn \\
=& - \tr{\vec\s^{M_1+ \ldots+ M_{\beta-1} + j_\beta -1}} +  \tr{\vec\s^{M_1+ \ldots+ M_{\beta-1} + j_\beta}}, \qquad 1 \leq j_\beta \leq M_\beta \nn \\
= & -\tr {\vec\s^{\alpha'_\beta -1} }+ \tr {\vec\s^{\alpha'_\beta}}, \quad \alpha'_\beta= M_1+ \ldots+ M_{\beta-1} + j_\beta.
\end{align}

From the expression \eref{PF-wtOPLQ} for a linear quiver (or the more general prescription \eref{PF-wtOPgen}), 
the partition function of the theory $\wt{(G^\alpha_{\vec\CP} \circ I^\alpha_{\vec\CP})}(Y)$ is then given as
\begin{empheq}[box=\vwidefbox]{align}\label{dual-AbGIL}
Z^{\wt{(G^\alpha_{\vec\CP} \circ I^\alpha_{\vec\CP})}(Y)}(\vec{m}'(\vec{t},\eta_{\alpha}), \, \vec \eta'(\{ \vec{v}^\beta \},\{\vec m^\gamma\}_{\gamma \neq \beta},\vec{\mu})) &
 = \int \prod^{L^\vee}_{\gamma'=1} \Big[d \vec\s^{\gamma'}\Big]\, \delta \Big(\eta_{\alpha} + \sum_{\beta} b^{l}_\beta t_l + \sum_{\beta} (-\tr \vec\s^{\alpha'_\beta -1} + \tr \vec\s^{\alpha'_\beta}) \Big) \nn \\
& \times  Z^{(Y,\{\CP_\beta\})}_{\rm int}(\{\vec \s^{\gamma'} \},
-\vec t,\{\vec{u}^\beta =\mu^\beta \}, \{\vec{v}^\beta\}, \{\vec m^\gamma\}_{\gamma \neq \beta}).
\end{empheq}
The form of the expressions on the RHS of \eref{dual-AbGIgen} and \eref{dual-AbGIL} imply that the dual of the Abelian $S$-type operation 
$\CO^\alpha_{\vec \CP}=(G^\alpha_{\vec \CP} \circ  I^\alpha_{\vec\CP})$ is an ungauging operation which removes a single $U(1)$ factor from the gauge group. The 
precise $U(1)$ factor to be removed is specified by the function $ \sum_{\beta} g_\beta$ in the delta function. Similar to the simple gauging 
case, the ungauging operation corresponding to the dual of $(G^\alpha_{\vec \CP} \circ  I^\alpha_{\vec\CP})$ also results in a manifestly Lagrangian theory.

\subsubsection{Identification-flavoring-gauging}\label{AbSGFI}
Consider an Abelian elementary $S$-type operation $\CO^\alpha_{\vec \CP}$ which involves identifying $U(1)$ flavor symmetries from 
$p$ distinct nodes in a generic quiver $X$, followed by flavoring the identified node with $N^\alpha_F$ hypers of charge 1, 
and finally by gauging the identified $U(1)$ flavor node. Following the notation of \Secref{GFI-summary}, we denote this operation 
as $\CO^\alpha_{\vec \CP}={G^\alpha_{\vec \CP} \circ F^\alpha_{\vec \CP} \circ I^\alpha_{\vec \CP}}$.\\

The flavor symmetry introduced by the operation $\CO_{\vec\CP}$ is $G^{\alpha}_{\rm F} = U(p) \times U(N^{\alpha}_F)$, modulo 
an overall $U(1)$ factor. The $U(p)$-valued masses are labelled as $\vec \mu = (\mu^{\gamma_1}, \ldots, \mu^{\gamma_p})$, 
and the $U(N^{\alpha}_F)$-valued masses are labelled as $\{m^{\alpha}_{k\, F}\}$,
where $k=1,\ldots,N^{\alpha}_F$. The function $\CZ_{G^\alpha_{\vec \CP} \circ F^\alpha_{\vec \CP} \circ I^\alpha_{\vec \CP}(X)}$, which can be read off from \eref{defZG}, \eref{defZF},\eref{defZI}, and \eref{CZ-OP}, is given as
\be \label{AbZGFI}
\CZ_{G^\alpha_{\vec \CP} \circ F^\alpha_{\vec \CP} \circ I^\alpha_{\vec \CP}(X)}(u^\alpha, \{{u}^\beta\}, \eta_{\alpha}, \vec m^{\alpha}_F, \vec \mu)=  Z_{\rm FI} ({u}^{\alpha},\eta_{\alpha})\, Z_{\rm 1-loop}^{\rm hyper} ({u}^{\alpha}, \vec m^{\alpha}_F)\, \int \prod^{p}_{j=1} \, d {u^{\gamma_j}} \,\prod^{p}_{j=1} \delta \Big({u}^{\alpha} - {u}^{\gamma_j} + {\mu}^{\gamma_j} \Big).
\ee
Using the general formula \eref{CZ-wtOP}, and the identity \eref{AbHMId} for $Z_{\rm 1-loop}^{\rm hyper}$, 
the function $\CZ_{\wt{(G^\alpha_{\vec\CP} \circ F^\alpha_{\vec\CP} \circ I^\alpha_{\vec\CP})}(Y)}$ is given as
\begin{align}
\CZ_{\wt{(G^\alpha_{\vec\CP} \circ F^\alpha_{\vec\CP} \circ I^\alpha_{\vec\CP})}(Y)}
= &\int\,du^\alpha  \,\CZ_{G^\alpha_{\vec \CP} \circ F^\alpha_{\vec \CP} \circ I^\alpha_{\vec \CP}(X)}(\{{u}^\beta\}, \eta_\alpha,  \vec m^{\alpha}_F, \vec \mu)\,\prod_{\beta}\,e^{2\pi i \,(g_\beta (\{\vec \s^{\gamma'} \}, \CP) + b^{l}_\beta \eta_l)\,u^{\beta}} \nn \\
=& \int \prod^{N^\alpha_F}_{k=1} d\tau^\alpha_k \, Z^{(\CT[N^{\alpha}_F])}_{\rm int}(\vec{\tau}^{\alpha}, \{{u}^\alpha=0, \vec m^{\alpha}_F\},0)\,
\prod_{\beta}\,e^{2\pi i \,(g_\beta (\{\vec \s^{\gamma'} \}, \CP_\beta) + b^{l}_\beta \eta_l)\,\mu^{\beta}} \nn \\
&\times \delta \Big(\tau^\alpha_1+\eta_\alpha + \sum_{\beta} b^{l}_\beta \eta_l + \sum_{\beta} g_\beta\Big(\{\vec \s^{\gamma'} \}, \CP_\beta \Big) \Big),
\end{align}
where, in obtaining the second equality, we have used the delta functions in \eref{AbZGFI} and integrated over $u^\alpha$. Proceeding in the 
same way as in the flavoring-gauging case and relabelling the integration variables $\tau^\alpha_k \to \tau^\alpha_{k-1}$, one can recast $\CZ_{\wt{(G^\alpha_{\vec\CP} \circ F^\alpha_{\vec\CP} \circ I^\alpha_{\vec\CP})}(Y)}$ in the following form (up to a $\s^{\gamma'}$-independent phase factor):
\begin{align}
\CZ_{\wt{(G^\alpha_{\vec\CP} \circ F^\alpha_{\vec\CP} \circ I^\alpha_{\vec\CP})}(Y)}= \int \prod^{N^{\alpha}_F -1}_{k=1} & d\tau^{\alpha}_{k}\,e^{2\pi i \sum_\beta g_\beta (\{\vec \s^{\gamma'} \}, \CP_\beta)\,(\mu^{\beta} + m^\alpha_{1\,F})} \,Z^{(\CT[N^{\alpha}_F -1])}_{\rm int}(\vec{\tau}^{\alpha},\vec m^{\alpha}_F,0) \\ \nn
& \times Z^{\rm hyper}_{\wt{(G^\alpha_{\vec\CP} \circ F^\alpha_{\vec\CP} \circ I^\alpha_{\vec\CP})}(Y)}\Big(\{\vec \s^{\gamma'} \}, \tau^\alpha_1, \eta_\alpha, \vec\eta \Big),
\end{align}
\begin{align}\label{HyperGFIgen}
& Z^{\rm hyper}_{\wt{(G^\alpha_{\vec\CP} \circ F^\alpha_{\vec\CP} \circ I^\alpha_{\vec\CP})}(Y)}\Big(\{\vec \s^{\gamma'} \}, \tau^\alpha_1, \eta_\alpha, \vec\eta \Big)= \frac{1}{\ch{(\sum_\beta g_\beta (\{\vec \s^{\gamma'} \}, \CP_\beta)+ \tau^\alpha_1+ \eta_\alpha + \sum_\beta b^{l}_\beta \eta_l)}}.
\end{align}
The function $Z^{\wt{(G^\alpha_{\vec\CP} \circ F^\alpha_{\vec\CP} \circ I^\alpha_{\vec\CP})}(Y)}$ can be identified as the 1-loop contribution of a single hypermultiplet charged under certain $U(1)$ subgroups of 
the gauge group of $Y$ and the group $U(1)_1$ of $\CT[N^{\alpha}_F -1]$. 
Following the general prescription $\eref{PF-wtOPgen}$, the partition function of the theory $\wt{(G^\alpha_{\vec\CP} \circ F^\alpha_{\vec\CP} \circ I^\alpha_{\vec\CP})}(Y)$ is given as
\begin{empheq}[box=\vwidefbox]{align} \label{PFGFIgen}
& Z^{\wt{(G^\alpha_{\vec\CP} \circ F^\alpha_{\vec\CP} \circ I^\alpha_{\vec\CP})}(Y)}(\vec{m}'(\vec{\eta},\eta_\alpha), \vec \eta'(\{ \vec v^\beta\}, \ldots, \vec{\mu}, \vec m^\alpha_F))= \int \prod_{\gamma'} \Big[d\vec{\s}^{\gamma'} \Big]\, \int \prod^{N^{\alpha}_F-1}_{k=1} d\tau^{\alpha}_{k} \, Z^{(\CT[N^{\alpha}_F -1])}_{\rm int}(\vec{\tau}^{\alpha},\vec m^{\alpha}_F,0)\nn \\ & \times Z^{\rm hyper}_{\wt{(G^\alpha_{\vec\CP} \circ F^\alpha_{\vec\CP} \circ I^\alpha_{\vec\CP})}(Y)}\Big(\{\vec \s^{\gamma'} \}, \tau^\alpha_1, \eta_\alpha, \vec\eta \Big) \cdot Z^{(Y,\{\CP_\beta\})}_{\rm int}(\{\vec \s^{\gamma'} \}, \vec{m}^Y(\vec{\eta}), \vec{\eta}^Y(\{\vec{u}^\beta =\mu^\beta + m^\alpha_{1\,F}\}, \{\vec{v}^\beta \},\ldots)).
\end{empheq}
The dual theory $\wt{(G^\alpha_{\vec\CP} \circ F^\alpha_{\vec\CP} \circ I^\alpha_{\vec\CP})}(Y)$ is therefore manifestly Lagrangian, and can be represented by a generic quiver diagram of the following form:\\

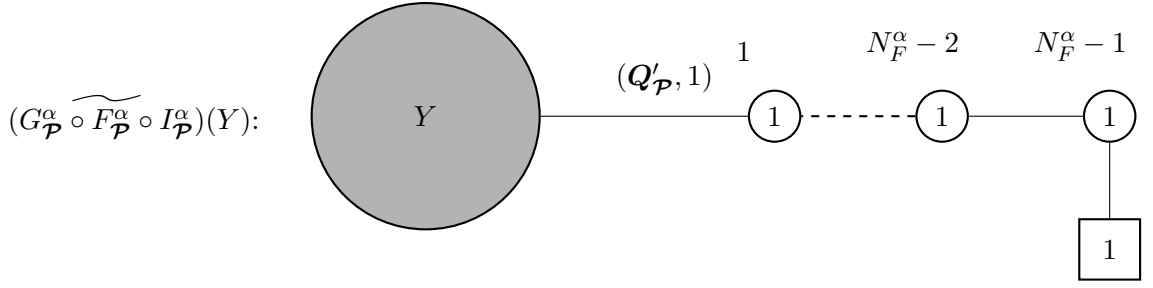
\begin{tikzpicture}[
cnode/.style={circle,draw,thick,minimum size=4mm},snode/.style={rectangle,draw,thick,minimum size=8mm},pnode/.style={rectangle,red,draw,thick,minimum size=1.0cm}, bnode/.style={circle,draw, thick, fill=black!30,minimum size=3cm}]
\node[text width=1cm](2) at (-5,0) {$\wt{(G^\alpha_{\vec\CP} \circ F^\alpha_{\vec\CP} \circ I^\alpha_{\vec\CP})}(Y)$:};
\node[bnode] (1) at (0,0){$Y$} ;
\node[cnode] (25) at (9,0){1};
\node[snode] (26) [below=1cm of 25]{1};
\node[cnode] (27) [left= 1.5 cm of 25]{1};
\node[cnode] (28) [left=1.5 cm of 27]{1};
\draw[-] (25) -- (26);
\draw[-] (25) -- (27);
\draw[thick, dashed] (27) -- (28);
\draw[-] (28) -- (1);
\node[text width=2cm](31) [above=0.25 of 25] {$N^{\alpha}_F-1$};
\node[text width=2 cm](32) [above=0.25 of 27] {$N^{\alpha}_F-2$};
\node[text width=1cm](33) [above=0.25 of 28] {$1$};
\node[text width=1 cm](34) at (3,0.5) {$(\vec Q'_{\vec \CP}, 1)$};
\end{tikzpicture}
The theory consists of a $\CT[N^{\alpha}_F -1]$ tail attached to the quiver $Y$ by a single hypermultiplet, which is also charged under some of the $U(1)$ factors in the gauge group of $Y$. $(\vec Q'_{\vec \CP}, 1)$ collectively denotes the charges of the hypermultiplet connecting $Y$ and $\CT[N^{\alpha}_F -1]$, under various $U(1)$ factors of $Y$ and $U(1)_1$ of $\CT[N^{\alpha}_F -1]$ respectively. Note that
\begin{enumerate}
\item The charge vector $\vec Q'_{\vec \CP}$, and therefore the precise shape of the quiver, depends on the choice of the 
permutation matrices $\{\CP_\beta \}$.

\item In the special case, where $N^{\alpha}_F=1$, the function $\CZ_{(\wt{G\circ F \circ I})_{\{\CP_\beta\}}(Y)}$ is given as
\begin{align}
& \CZ_{\wt{(G^\alpha_{\vec\CP} \circ F^\alpha_{\vec\CP} \circ I^\alpha_{\vec\CP})}(Y)}=  e^{2\pi i \sum_\beta g_\beta (\{\vec \s^{\gamma'} \}, \CP_\beta)\,(\mu^{\beta} + m^\alpha_{F})}\, Z^{\rm hyper}_{\wt{(G^\alpha_{\vec\CP} \circ F^\alpha_{\vec\CP} \circ I^\alpha_{\vec\CP})}(Y)}\Big(\{\vec \s^{\gamma'} \}, \eta_\alpha, \vec\eta \Big), \\
& Z^{\rm hyper}_{\wt{(G^\alpha_{\vec\CP} \circ F^\alpha_{\vec\CP} \circ I^\alpha_{\vec\CP})}(Y)}\Big(\{\vec \s^{\gamma'} \}, \eta_\alpha, \vec\eta \Big) = \frac{1}{\ch{(\sum_\beta g_\beta (\{\vec \s^{\gamma'} \}, \CP_\beta)+ \eta_\alpha + \sum_\beta b^{l}_\beta \eta_l)}}. \label{HyperGFIgenNf1}
\end{align}
The dual paritition function is then given as
\begin{empheq}[box=\vwidefbox]{align}\label{PFGFIgenNf1}
 Z^{\wt{(G^\alpha_{\vec\CP} \circ F^\alpha_{\vec\CP} \circ I^\alpha_{\vec\CP})}(Y)}(\vec{m}'(\vec{\eta},\eta_{\alpha}); \vec \eta'( \vec{v}^\alpha,\ldots, {m}^{\alpha}_F)) &
=  \int  \prod_{\gamma'} \Big[d\vec{\s}^{\gamma'} \Big]\, Z^{\rm hyper}_{\wt{(G^\alpha_{\vec\CP} \circ F^\alpha_{\vec\CP} \circ I^\alpha_{\vec\CP})}(Y)}\Big(\{\vec \s^{\gamma'} \}, \eta_\alpha, \vec \eta \Big)  \nn \\
& \times  Z^{(Y, \CP)}_{\rm int}(\{\vec \s^{\gamma'} \}, \vec{m}^Y(\vec{\eta}), \vec{\eta}^Y(\{\vec{u}^\beta =\mu^\beta + m^\alpha_{F}\}, \{\vec{v}^\beta \},\ldots)).
\end{empheq}

The dual theory ${\wt{(G\circ F\circ I)}_{\{\CP_\beta\}}}(Y)$ in this case is simpler. 
It merely involves introducing an extra hypermultiplet in the quiver gauge theory $Y$, such that the hypermultiplet is charged under a certain $U(1)^r$ subgroup ($r$ is an integer) of the gauge group. The FI parameters $\{u^\beta\}$ in $Z^{(Y, \CP)}_{\rm int}$ is also replaced by the mass parameters $\{\mu^\beta + m^\alpha_F\}$.\\
\end{enumerate}

\noindent \textbf{Linear quivers:} For a linear quiver pair $(X,Y)$, the above partition function can be written as 
\begin{empheq}[box=\vwidefbox]{align} \label{PFGFILin}
&Z^{\wt{(G^\alpha_{\vec\CP} \circ F^\alpha_{\vec\CP} \circ I^\alpha_{\vec\CP})}(Y)}(\vec{m}'(\vec{t},\eta_\alpha), \, \vec \eta'(\{\vec{v}^\beta\},\vec{\mu}, \vec m^\alpha_F, \{\vec m^\gamma\}_{\gamma \neq \beta})) 
= \int \prod^{L^\vee}_{\gamma'=1} \Big[d \vec\s^{\gamma'}\Big]\, \int \prod^{N^\alpha_F-1}_{k=1} d\tau_k \, Z^{(\CT[N^{\alpha}_F -1])}_{\rm int}(\vec{\tau}^{\alpha},\vec m^{\alpha}_F,0)\nn \\ & \times Z^{\rm hyper}_{\wt{(G^\alpha_{\vec\CP} \circ F^\alpha_{\vec\CP} \circ I^\alpha_{\vec\CP})}(Y)}\Big(\{\vec \s^{\gamma'} \}, \tau^\alpha_1, \eta_\alpha, \vec t \Big)
\cdot Z^{(Y,\{\CP_\beta\})}_{\rm int}(\{\vec \s^{\gamma'} \}, -\vec t,\{\vec{u}^\beta =\mu^\beta + m^\alpha_{1\,F}\}, \{\vec{v}^\beta\},\{\vec m^\gamma\}_{\gamma \neq \beta}).
\end{empheq}
In this case, the hypermultiplet contribution can be written explicitly, and the $U(1)$ charges $\vec Q'_{\CP}$ can be read off:
\begin{align}
Z^{\rm hyper}_{\wt{(G^\alpha_{\vec\CP} \circ F^\alpha_{\vec\CP} \circ I^\alpha_{\vec\CP})}(Y)}\Big(\{\vec \s^{\gamma'} \}, \tau^\alpha_1, \eta_\alpha, \vec t \Big)= & \frac{1}{\ch{(\sum_\beta g_\beta (\{\vec \s^{\gamma'} \}, \CP_\beta)+ \tau^\alpha_1+ \eta_\alpha + \sum_\beta b^{l}_\beta \eta_l)}} \nn \\
=  & \frac{1}{\ch{(\sum_\beta (-\tr \vec\s^{\alpha'_\beta -1} + \tr \vec\s^{\alpha'_\beta})+ \tau^\alpha_1+ \eta_\alpha + \sum_\beta b^{l}_\beta t_l)}}, \label{HyperGFILin}
\end{align}
where the functions $\{g_\beta\}$ and the gauge node labels $\{\alpha'_\beta\}$ are defined in \eref{def-gLbeta}.
For the special case $N^{\alpha}_F =1$, the dual partition function again assumes a simplified form:
\begin{align} \label{PFGFILinNf1}
&Z^{\wt{(G^\alpha_{\vec\CP} \circ F^\alpha_{\vec\CP} \circ I^\alpha_{\vec\CP})}(Y)}(\vec{m}'(\vec{t},\eta_\alpha), \, \vec \eta'(\{\vec{v}^\beta\},\vec{\mu}, \vec m^\alpha_F, \{\vec m^\gamma\}_{\gamma \neq \beta})) \\ \nn
=& \int \prod^{L^\vee}_{\gamma'=1} \Big[d \vec\s^{\gamma'}\Big] \, Z^{\rm hyper}_{\wt{(G^\alpha_{\vec\CP} \circ F^\alpha_{\vec\CP} \circ I^\alpha_{\vec\CP})}(Y)}\Big(\{\vec \s^{\gamma'} \}, \eta_\alpha, \vec t \Big)
\cdot Z^{(Y,\{\CP_\beta\})}_{\rm int}(\{\vec \s^{\gamma'} \}, -\vec t,\{\vec{u}^\beta =\mu^\beta + m^\alpha_{F}\}, \{\vec{v}^\beta\},\{\vec m^\gamma\}_{\gamma \neq \beta}),
\end{align}
\begin{align} \label{HyperGFILinNf1}
Z^{\rm hyper}_{\wt{(G^\alpha_{\vec\CP} \circ F^\alpha_{\vec\CP} \circ I^\alpha_{\vec\CP})}(Y)}\Big(\{\vec \s^{\gamma'} \},  \eta_\alpha, \vec t \Big)= 
& \frac{1}{\ch{(\sum_\beta (-\tr \vec\s^{\alpha'_\beta -1} + \tr \vec\s^{\alpha'_\beta})+ \eta_\alpha + \sum_\beta b^{l}_\beta t_l)}}.
\end{align}
The special cases of flavoring-gauging operations with $N^\alpha=1$, given by \eref{HyperGFIgenNf1}-\eref{PFGFIgenNf1} and 
\eref{PFGFILinNf1}-\eref{HyperGFILinNf1} for a generic quiver and a linear quiver respectively, will be used frequently for the construction 
of non-ADE mirror duals later in this section. \\

\subsubsection{A qualitative description of the dual operations} 

While we have determined the precise dual operations at the level of the quiver gauge theories above, their qualitative forms could be guessed 
from the fact that mirror symmetry exchanges Higgs branch global symmetry one one side of the duality with the Coulomb branch global symmetry 
on the other. The dual of the Abelian gauging operation on $X$ will therefore involve gauging of a $U(1)_J$ topological symmetry of $Y$, which is equivalent 
to ungauging a $U(1)$ subgroup of the gauge group of $Y$. The precise $U(1)$ being ungauged can be read off from the matrix model and is realized 
by the function $g_\alpha\Big(\{\vec \s^{\gamma'} \}, \CP \Big)$ in \eref{GAbgen} set to a constant. The same qualitative reasoning also applies to the case of 
Abelian identification-gauging, with the ungauging of the $U(1)$ subgroup 
being realized by setting the function $\sum_{\beta} g_\beta \Big(\{\vec \s^{\gamma'} \}, \CP_\beta \Big)$ 
in \eref{dual-AbGIgen} to a constant.\\

A flavoring-gauging operation can be viewed as identifying the $U(1)$ flavor symmetry of the theory $X$ with a $U(1)$ flavor 
symmetry subgroup of $N^\alpha_F$ free hypermultiplets, and then gauging it. If the $U(1)$ flavor symmetry in $X$ were gauged without this 
identification, then the dual operation would involve ungauging a certain $U(1)$ subgroup of the gauge group of $Y$, as argued above. 
Let us call this $U(1)$ subgroup $U(1)_Y$.
Similarly, if the $U(1)$ flavor symmetry of the free hypers were gauged separately, one would ungauge the $U(1)_1$ node of the quiver 
$\CT[N^{\alpha}_F]$ (see the discussion after \eref{AbHMId}) on the dual side. This would give a linear chain of $N^{\alpha}_F-1$ $U(1)$ nodes with a fundamental hyper at 
each end, which is indeed the correct mirror of a $U(1)$ theory with $N^{\alpha}_F$ flavors. 

In contrast, the identification of the two $U(1)$ flavor symmetries, roughly speaking, leads to an identification of FI parameters associated with the $U(1)_1$ and $U(1)_Y$ on 
the dual side. The gauging operation then, following the logic presented above, leads to the dual operation where a linear combination of $U(1)_1$ and $U(1)_Y$ 
is ungauged. While the precise form of the linear combination can be read off from the matrix model integral, this generically produces a quiver $Y'=\wt{(G^\alpha_{\CP} \circ F^\alpha_\CP)}(Y)$ of the form shown in the paragraph following \eref{PFGFgen}. The same qualitative reasoning also applies to the case of 
Abelian identification-flavoring-gauging.

\subsection{Examples of Abelian Quiver Pairs} \label{AbExNew}
\subsubsection{Family ${\rm I}_{[n,l,p]}$: A single closed loop attached to a linear quiver tail}\label{FamilyI}  

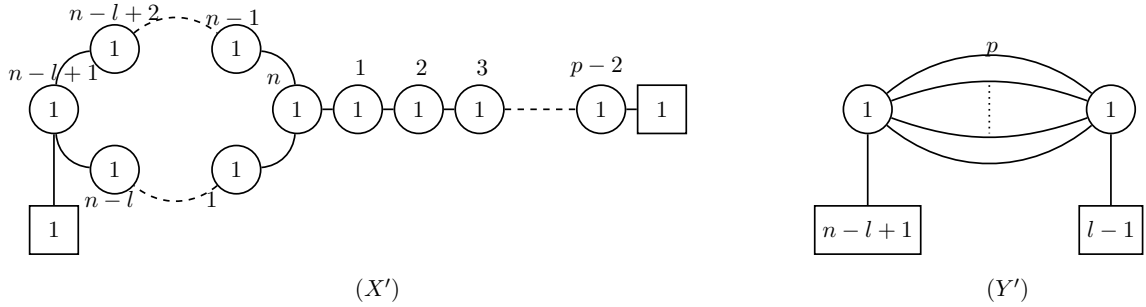
\begin{figure}[htbp]
\begin{center}
\scalebox{0.8}{\begin{tikzpicture}[node distance=2cm,cnode/.style={circle,draw,thick,minimum size=8mm},snode/.style={rectangle,draw,thick,minimum size=8mm},pnode/.style={rectangle,red,draw,thick,minimum size=8mm}]
\node[cnode] (1) at (-2,0) {$1$};
\node[cnode] (2) at (-1,1) {$1$};
\node[cnode] (3) at (1,1) {$1$};
\node[cnode] (4) at (2,0) {$1$};
\node[cnode] (5) at (1,-1) {$1$};
\node[cnode] (6) at (-1,-1) {$1$};
\node[cnode] (7) at (3,0) {$1$};
\node[cnode] (8) at (4,0) {$1$};
\node[cnode] (9) at (5,0) {$1$};
\node[cnode] (10) at (7,0) {$1$};
\node[snode] (11) at (8,0) {$1$};
\node[snode] (12) at (-2,-2) {$1$};
\draw[thick] (1) to [bend left=40] (2);
\draw[dashed,thick] (2) to [bend left=40] (3);
\draw[thick] (3) to [bend left=40] (4);
\draw[thick] (4) to [bend left=40] (5);
\draw[dashed,thick] (5) to [bend left=40] (6);
\draw[thick] (6) to [bend left=40] (1);
\draw[thick] (1) -- (12);
\draw[thick] (4) -- (7);
\draw[thick] (7) -- (8);
\draw[thick] (8) -- (9);
\draw[thick, dashed] (9) -- (10);
\draw[thick] (10) -- (11);
\node[text width=1.5cm](20) at (-2,0.6) {$n-l+1$};
\node[text width=1.5cm](21) at (-1,1.6) {$n-l+2$};
\node[text width=1cm](22) at (1,1.5) {$n-1$};
\node[text width=1cm](23) at (2,0.5) {$n$};
\node[text width=1cm](24) at (1,-1.5) {$1$};
\node[text width=1cm](25) at (-1,-1.5) {$n-l$};
\node[text width=0.1cm](26) at (3,0.7) {$1$};
\node[text width=0.1cm](27) at (4,0.7) {$2$};
\node[text width=0.1cm](28) at (5,0.7) {$3$};
\node[text width=1 cm](29) at (7,0.7) {$p-2$};
\node[text width=0.1cm](30) at (3,-3){$(X')$};
\end{tikzpicture}}
\qquad \qquad
\scalebox{0.8}{\begin{tikzpicture}[node distance=2cm,cnode/.style={circle,draw,thick,minimum size=8mm},snode/.style={rectangle,draw,thick,minimum size=8mm},pnode/.style={rectangle,red,draw,thick,minimum size=8mm}]
\node[cnode] (1) at (-2,0) {$1$};
\node[cnode] (2) at (2,0) {$1$};
\node[snode] (3) at (-2,-2) {$n-l+1$};
\node[snode] (4) at (2,-2) {$l-1$};
\draw[thick] (1) -- (3);
\draw[thick] (2) -- (4);
\draw[thick,dotted] (0,0.4) to (0, -0.4);
\draw[thick] (1) to [bend left=20] (2);
\draw[thick] (1) to [bend right=20] (2);
\draw[thick] (1) to [bend left=40] (2);
\draw[thick] (1) to [bend right=40] (2);
\node[text width=0.1cm](15) at (0,1){$p$};
\node[text width=0.1cm](16) at (0,-3){$(Y')$};
\end{tikzpicture}}
\caption{A three-parameter family of dual quiver gauge theory pairs with $U(1)$ gauge groups labelled by the integers $(n,l,p)$ with $n> l-1>0$, 
and $p \geq 2$. Their construction from linear quivers by identification-gauging operations is shown in \figref{AbEx1GFI}.}
\label{AbEx1gen}
\end{center}
\end{figure}

The first example in the class of Abelian mirrors is the infinite family of mirror duals shown in \figref{AbEx1gen}, labelled by three positive integers $(n,l,p)$ 
with the constraints $n-l+1 > 0$, $l > 1$, and $p \geq 2$. The theory $X'$ has the shape of a single closed loop attached to a linear quiver tail, while the 
theory $Y'$ is a quiver with two nodes connected by multiple edges. The dimensions of the respective Higgs and 
Coulomb branches, and the associated global symmetries are shown in Table \ref{Tab:AbEx1gen}.\\

\begin{center}
\begin{table}[htbp]
\resizebox{\textwidth}{!}{%
\begin{tabular}{|c|c|c|}
\hline
Moduli space data & Theory $X'$ & Theory $Y'$ \\
\hline \hline 
dim\,$\CM_H$ & 2 & $n+p-2$\\
\hline
dim\,$\CM_C$ & $n+p-2$ & 2\\
\hline
$G_H$ & $U(1) \times U(1)$ & $SU(p) \times SU(n-l+1) \times SU(l-1) \times U(1) $\\
\hline
$G_C$ & $SU(p) \times  SU(n-l+1)  \times SU(l-1) \times U(1) $ & $U(1) \times U(1)$\\
\hline
\end{tabular}}
\caption{Summary table for the moduli space dimensions and global symmetries for the mirror pair in \figref{AbEx1gen}.}
\label{Tab:AbEx1gen}
\end{table}
\end{center}

The quiver pair $(X', Y')$ in \figref{AbEx1gen} can be generated from the linear quiver pair $(X, Y)$ in \figref{AbEx1LIN} by a series of 
elementary Abelian operations on $X$ and the dual operations on $Y$, as shown in \figref{AbEx1GFI}. The Higgs branch global symmetry 
of $X$ is given by $G^{X}_{H} = U(1)^3/ U(1) = U(1) \times U(1)$. Using the notation introduced in \Secref{LQ-IIB}, the mass parameters are 
labelled as $\vec m =(m_1, m_2, m_3)$, and the quotient by the overall $U(1)$ factor in the flavor symmetry of quiver $X$ can be 
implemented by the constraint $m_2=0$. The $U(1) \times U(1)$ flavor symmetry of $X$ can then be identified with the two 
terminal $U(1)$ flavor nodes, parametrized by the independent $(m_1, m_3)$.\\

\begin{figure}[htbp]
\begin{center}
\scalebox{0.8}{\begin{tikzpicture}[
cnode/.style={circle,draw,thick, minimum size=1.0cm},snode/.style={rectangle,draw,thick,minimum size=1cm}]
\node[cnode] (1) {1};
\node[cnode] (2) [right=1.5cm  of 1]{1};
\node[cnode] (3) [right= 1 cm  of 2]{1};
\node[snode] (4) [above=1 cm of 3]{1};
\node[cnode] (5) [right= 1 cm  of 3]{1};
\node[cnode] (6) [right=1.5cm  of 5]{1};
\node[snode] (7) [below=1 cm of 1]{1};
\node[snode] (8) [below=1 cm of 6]{1};
\node[text width=1.5cm](9)[above=0.2cm of 3]{$n-l+1$};
\node[text width=1.5cm](10)[above=0.2cm of 5]{$n-l+2$};
\node[text width=1 cm](11)[above=0.2cm of 6]{$n-1$};
\node[text width=1cm](12)[above=0.2cm of 2]{$n-l$};
\node[text width=1.5 cm](13)[above=0.2cm of 1]{1};
\node[text width=0.1cm](20)[below=  1 cm of 3]{$(X)$};
\draw[thick, dashed] (1) -- (2);
\draw[-] (2)-- (3);
\draw[-] (3)-- (4);
\draw[-] (3)-- (5);
\draw[thick, dashed] (5)-- (6);
\draw[-] (1)-- (7);
\draw[-] (8)-- (6);
\end{tikzpicture}}
\qquad \qquad \qquad \qquad \qquad \qquad 
\scalebox{0.8}{\begin{tikzpicture}[
cnode/.style={circle,draw,thick, minimum size=1.0cm},snode/.style={rectangle,draw,thick,minimum size=1cm}]
\node[cnode] (1) {1};
\node[snode] (2) [below=1cm of 1]{$l-1$};
\node[cnode] (3) [left=1.5 cm of 1]{1};
\node[snode] (4) [below=1cm of 3]{$n-l+1$};
\draw[-] (1) -- (2);
\draw[-] (1) -- (3);
\draw[-] (3) -- (4);
\node[text width=0.1cm](20) at (-1,-4){$(Y)$};
\end{tikzpicture}}
\caption{Pair of linear quivers with $U(1)$ gauge groups which generate the 3d mirrors in \figref{AbEx1gen}.}
\label{AbEx1LIN}
\end{center}
\end{figure}

To obtain $X'$ from $X$, we first perform an identification-flavoring-gauging operation\footnote{As mentioned earlier in \Secref{GFI-exAbPF}, we drop the superscript $\alpha$ from the notation $\CO^\alpha_\CP$ when there is no ambiguities regarding the flavor nodes on which the $S$-type operation acts. In cases where $\CP$ is 
trivial, we drop the subscript as well.} $\CO_1$ on the flavor nodes (marked in red) at the two ends of the quiver $X$, with $N^\alpha_F=1$. 
This step is precisely the same as the illustrative example studied in \Secref{GFI-exAbPF}. 
This is followed by a flavoring-gauging operation $\CO_2$ on the quiver $\CO_1(X)$ at the new flavor node (also marked in red), again with 
$N^\alpha_F=1$. The quiver $X'$ can be obtained by implementing a sequence of $p-2$ such flavoring-gauging operations, each acting at the new flavor node obtained in the previous step. Let us adopt the notation:
\begin{align}
& X^{(i)}= \CO_i \circ  \CO_{i-1} \circ \ldots \circ \CO_2 \circ \CO_1(X), \\
& Y^{(i)}= \wt\CO_i \circ  \wt\CO_{i-1} \circ \ldots \circ \wt\CO_2 \circ \wt\CO_1(Y), 
\end{align}
where $\CO_1$ is an identification-flavoring-gauging operation, and $\CO_i$ ($i > 1$) are flavoring-gauging operations. 
We will denote the $\CN=4$-preserving deformations of the quivers $X^{(i)}$ and $Y^{(i)}$ as $(\vec m^{(i)}, \vec \eta^{(i)})$ 
and $(\vec m'^{(i)}, \vec \eta'^{(i)})$ respectively. In this notation, we have
\be
X' = X^{(p-1)}, \, Y' = Y^{(p-1)}.
\ee
We will denote the deformation parameters of the dual theories $X'$ and $Y'$ as $(\vec m, \vec \eta)$ and $(\vec m', \vec \eta')$ respectively, which 
obviously implies
\be
(\vec m, \vec \eta)=(\vec m^{(p-1)}, \vec \eta^{(p-1)}), \quad (\vec m', \vec \eta')=(\vec m'^{(p-1)}, \vec \eta'^{(p-1)}).
\ee
The masses associated with the identification operation are parametrized as $\vec \mu$, while the mass associated with flavoring operation 
at the $i$-th step is $m^\alpha_F=m^{(i)}_F$. The FI parameter associated to the gauging operation at the $i$-th step is $\eta_\alpha=\wt{\eta}^{(i)}$.

\begin{figure}[htbp]
\begin{center}
\begin{tabular}{ccc}
\scalebox{.7}{\begin{tikzpicture}[node distance=2cm,
cnode/.style={circle,draw,thick, minimum size=1.0cm},snode/.style={rectangle,draw,thick,minimum size=1cm}, pnode/.style={red,rectangle,draw,thick, minimum size=1.0cm},]
\node[cnode] (1) {1};
\node[cnode] (2) [right=1.5cm  of 1]{1};
\node[cnode] (3) [right= 1 cm  of 2]{1};
\node[snode] (4) [above=1 cm of 3]{1};
\node[cnode] (5) [right=1 cm  of 3]{1};
\node[cnode] (6) [right=1.5cm  of 5]{1};
\node[pnode] (7) [below=1 cm of 1]{1};
\node[pnode] (8) [below=1 cm of 6]{1};
\node[text width=1.5cm](9)[above=0.2cm of 3]{$n-l+1$};
\node[text width=1.5cm](10)[above=0.2cm of 5]{$n-l+2$};
\node[text width=1 cm](11)[above=0.2cm of 6]{$n-1$};
\node[text width=1cm](12)[above=0.2cm of 2]{$n-l$};
\node[text width=0.1 cm](13)[above=0.2cm of 1]{$1$};
\draw[thick, dashed] (1) -- (2);
\draw[-] (2)-- (3);
\draw[-] (3)-- (4);
\draw[-] (3)-- (5);
\draw[thick, dashed] (5)-- (6);
\draw[-] (1)-- (7);
\draw[-] (8)-- (6);
\node[text width=6cm](20) at (7, -3) {$(X)$};
\end{tikzpicture}}
& \qquad 
& \scalebox{.7}{\begin{tikzpicture}[node distance=2cm,
cnode/.style={circle,draw,thick, minimum size=1.0cm},snode/.style={rectangle,draw,thick,minimum size=1cm}, pnode/.style={red,rectangle,draw,thick, minimum size=1.0cm},]\node[cnode] (25) at (16,0){1};
\node[snode] (26) [below=1cm of 25]{$l-1$};
\node[cnode] (27) [left=1.5 cm of 25]{1};
\node[snode] (28) [below=1cm of 27]{$n-l+1$};
\draw[-] (25) -- (26);
\draw[-] (25) -- (27);
\draw[-] (27) -- (28);
\node[text width=1cm](30) at (15, -3) {$(Y)$};
\end{tikzpicture}}\\
 \scalebox{.7}{\begin{tikzpicture}
\draw[->] (15,-3) -- (15,-5);
\node[text width=0.1cm](20) at (14.5, -4) {$\CO_1$};
\end{tikzpicture}}
&\qquad
& \scalebox{.7}{\begin{tikzpicture}
\draw[->] (15,-3) -- (15,-5);
\node[text width=0.1cm](29) at (15.5, -4) {$\wt{\CO}_1$};
\end{tikzpicture}}\\
\scalebox{.7}{\begin{tikzpicture}[node distance=2cm,cnode/.style={circle,draw,thick,minimum size=1.0cm},snode/.style={rectangle,draw,thick,minimum size=1.0cm},pnode/.style={rectangle,red,draw,thick,minimum size=1.0cm}, nnode/.style={circle, red, draw,thick,minimum size=1.0cm}]
\node[cnode] (1) at (2,0) {$1$};
\node[cnode] (2) at (3,1) {$1$};
\node[cnode] (3) at (5,1) {$1$};
\node[cnode] (4) at (6,0) {$1$};
\node[cnode] (5) at (5,-1) {$1$};
\node[cnode] (6) at (3,-1) {$1$};
\node[pnode] (7) at (8,0) {$1$};
\node[snode] (9) at (2,-2) {$1$};
\draw[thick] (1) to [bend left=40] (2);
\draw[dashed,thick] (2) to [bend left=40] (3);
\draw[thick] (3) to [bend left=40] (4);
\draw[thick] (4) to [bend left=40] (5);
\draw[dashed,thick] (5) to [bend left=40] (6);
\draw[thick] (6) to [bend left=40] (1);
\draw[thick] (1) -- (9);
\draw[thick] (4) -- (7);
\node[text width=1.5cm](10) at (1,0.75) {$n-l+1$};
\node[text width=1.5cm](11) at (3,2) {$n-l+2$};
\node[text width=1cm](12) at (5,1.75) {$n-1$};
\node[text width=1cm](13) at (7,0.5) {$n$};
\node[text width=1cm](14) at (6,-1.5) {$1$};
\node[text width=1cm](15) at (3.25,-1.75) {$n-l$};
\node[text width=2cm](20) at (6, -3) {$(X^{(1)})$};
\end{tikzpicture}}
&\qquad 
& \scalebox{.8}{\begin{tikzpicture}[node distance=2cm,cnode/.style={circle,draw,thick,minimum size=1.0cm},snode/.style={rectangle,draw,thick,minimum size=1.0cm},pnode/.style={rectangle,red,draw,thick,minimum size=1.0cm}, nnode/.style={circle, red, draw,thick,minimum size=1.0cm}]
\node[] (24) at (10,0){};
\node[cnode] (25) at (16,0){1};
\node[snode] (26) [below=1cm of 25]{$l-1$};
\node[cnode] (27) [left=1.5 cm of 25]{1};
\node[snode] (28) [below=1cm of 27]{$n-l+1$};
\draw[-] (25) -- (26);
\draw[thick] (25) to [bend left=40] (27);
\draw[thick] (25) to [bend right=40] (27);
\draw[-] (27) -- (28);
\node[text width=6cm](29) at (17, -3) {$(Y^{(1)})$};
\end{tikzpicture}}\\
\scalebox{.7}{\begin{tikzpicture}
\draw[->] (15,-3) -- (15,-5);
\node[text width=0.1cm](20) at (14.5, -4) {$\CO_2$};
\end{tikzpicture}}
&\qquad
& \scalebox{.7}{\begin{tikzpicture}
\draw[->] (15,-3) -- (15,-5);
\node[text width=0.1cm](29) at (15.5, -4) {$\wt{\CO}_2$};
\end{tikzpicture}}\\
\scalebox{.7}{\begin{tikzpicture}[node distance=2cm,cnode/.style={circle,draw,thick,minimum size=1.0cm},snode/.style={rectangle,draw,thick,minimum size=8mm},pnode/.style={rectangle,red,draw,thick,minimum size=1.0cm}, nnode/.style={circle, red, draw,thick,minimum size=1.0cm}]
\node[cnode] (1) at (-4,0) {$1$};
\node[cnode] (2) at (-3,1) {$1$};
\node[cnode] (3) at (-1,1) {$1$};
\node[cnode] (4) at (0,0) {$1$};
\node[cnode] (5) at (-1,-1) {$1$};
\node[cnode] (6) at (-3,-1) {$1$};
\node[cnode] (7) at (2,0) {$1$};
\node[pnode] (8) at (2,-2) {$1$};
\node[snode] (9) at (-4,-2) {$1$};
\node[text width=1cm](16) at (-1,-3) {$(X^{(2)})$};
\draw[thick] (1) to [bend left=40] (2);
\draw[dashed,thick] (2) to [bend left=40] (3);
\draw[thick] (3) to [bend left=40] (4);
\draw[thick] (4) to [bend left=40] (5);
\draw[dashed,thick] (5) to [bend left=40] (6);
\draw[thick] (6) to [bend left=40] (1);
\draw[thick] (1) -- (9);
\draw[thick] (7) -- (8);
\draw[thick] (4) -- (7);
\node[text width=1.5cm](10) at (-5,0.75) {$n-l+1$};
\node[text width=1.5cm](11) at (-3,1.75) {$n-l+2$};
\node[text width=1cm](12) at (-1,1.75) {$n-1$};
\node[text width=1cm](13) at (1,0.5) {$n$};
\node[text width=1cm](14) at (0,-1.75) {$1$};
\node[text width=1cm](15) at (-2.75,-1.75) {$n-l$};
\end{tikzpicture}}
&\qquad
&\scalebox{.7}{\begin{tikzpicture}[node distance=2cm,cnode/.style={circle,draw,thick,minimum size=1.0cm},snode/.style={rectangle,draw,thick,minimum size=8mm},pnode/.style={rectangle,red,draw,thick,minimum size=1.0cm}, nnode/.style={circle, red, draw,thick,minimum size=1.0cm}]\node[cnode] (21) at (7,0) {$1$};
\node[cnode] (22) at (11,0) {$1$};
\node[snode] (23) at (7,-2) {$n-l+1$};
\node[snode] (24) at (11,-2) {$l-1$};
\draw[thick] (21) -- (23);
\draw[thick] (22) -- (24);
\draw[thick] (21) -- (22);
\draw[thick] (21) to [bend left=40] (22);
\draw[thick] (21) to [bend right=40] (22);
\node[text width=0.1cm](16) at (9,-3) {$(Y^{(2)})$};
\end{tikzpicture}}
\end{tabular}
\caption{The quiver $(X')$ and its mirror dual $(Y')$ in \figref{AbEx1gen} (for $p=3$) is generated by a sequence of 
elementary $S$-type operations on different nodes. In each step, the flavor node(s) on which the $S$-type operation 
acts is shown in red.}
\label{AbEx1GFI}
\end{center}
\end{figure}

Let us begin with the dual operation for $\CO_1(X)$. This was already worked out in \Secref{GFI-exAbPF}. We will now derive that result 
using the general Abelian formula \eref{PFGFILinNf1}-\eref{HyperGFILinNf1} in \Secref{AbSGFI}. Following the notation in \Secref{AbS}
(originally introduced in \Secref{GFI-summary}), let us define the variables $\{ \vec u^\beta \}$, which parametrize the $U(1) \times U(1)$ 
mass parameters of the quiver $X$:
\be \label{defuAbEx1}
u^1=m_1, u^2=m_3,
\ee 
and implement the $S$-type operation $\CO_1$ on the relevant flavor nodes of $X$. Note that the permutation matrices $\{\CP_\beta\}$ 
are trivial in this case. The resultant quiver $\CO_1(X)= X^{(1)}$ is shown in the second line of \figref{AbEx1GFI}.  \\
The dual quiver $\wt{\CO}_1(Y)$ can then be constructed using the general recipe presented in \Secref{AbSGFI} - the expressions
 \eref{PFGFILinNf1}-\eref{HyperGFILinNf1} give the dual partition function. The expression \eref{PFGFILinNf1} implies that the 
 dual quiver $\wt{\CO}_1(Y)$ simply involves adding an extra hypermutiplet to the quiver gauge theory $Y$, i.e.
\begin{align}\label{PFAbEx1gen1}
&Z^{\wt{\CO}_1(Y)}(\vec m'^{(1)}, \vec \eta'^{(1)}) = \int \,\prod^{2}_{\gamma'=1} \Big[d\s^{\gamma'}\Big]\, Z^{\rm hyper}_{\wt{\CO}_1(Y)}(\s^1,\s^2, \wt{\eta}^{(1)}, \vec t)\,Z^{(Y)}_{\rm int}(\s^1,\s^2, -\vec t,\{\vec{u}^\beta =\mu^\beta + m^{(1)}_{F}\}) \nn \\
&=\int \,\prod^{2}_{\gamma'=1} \Big[d\s^{\gamma'}\Big]\, Z^{\rm hyper}_{\wt{\CO}_1(Y)}(\s^1,\s^2, \wt{\eta}^{(1)}, \vec t)\,Z^{(Y)}_{\rm int}(\s^1,\s^2, -\vec t,\{m_1=\mu^1 + m^{(1)}_{F},m_2=0, m_3=\mu^2 + m^{(1)}_{F}\}),
\end{align}
where $\vec m'^{(1)}=\vec m'^{(1)}(\vec t, \wt{\eta}^{(1)}), \vec \eta'^{(1)}= \vec \eta'^{(1)}(m^{(1)}_{F}, \vec \mu)$ are linear functions of their arguments. 
The problem then is reduced to writing down the contribution of the hypermultiplet $Z^{\rm hyper}_{\wt{\CO_1}(Y)}$. From \eref{HyperGFILinNf1}, 
this is given as
\be \label{HyperAbEx1gen1}
Z^{\rm hyper}_{\wt{\CO_1}(Y)}(\s^1,\s^2, \wt{\eta}^{(1)}, \vec t)= \frac{1}{\ch{(\s^{1} -  \s^{2}+ \wt{\eta}^{(1)} + \sum_\beta b^{l}_\beta t_l)}}
=: Z_{\rm 1-loop}^{\rm bif}(\s^1, \s^2, -\wt{\eta}^{(1)} - \sum_\beta b^{l}_\beta t_l).
\ee  
The new hypermultiplet can therefore be identified as a bifundamental hypermultiplet with a mass $m_{\rm bif}= -\wt{\eta}^{(1)} - \sum_\beta b^{l}_\beta t_l$, 
and this leads to the quiver $Y^{(1)}$ in the second line of \figref{AbEx1GFI} \footnote{The matrix $b^l_\beta$ can be read off from the precise mirror symmetry relation between $Z^{(X)}$ and $Z^{(Y)}$, i.e.
\begin{align*}
 Z^{(X)} (\vec u; \vec t) = e^{2 \pi i (t_1 u^1 -t_n u^2)} Z^{(Y)} (-\vec t; \vec u), \,
 \implies b^l_\beta t_l u^\beta = (t_1 u^1 - t_n u^2), \quad (l=1,\ldots, n, \, \beta=1,2).
\end{align*}}.
Finally, the mirror map between the masses and the FI parameters can be read off from \eref{PFAbEx1gen1}-\eref{HyperAbEx1gen1}, 
and can be recast in the form \eref{AbEx1MM1}-\eref{AbEx1MM2} after appropriately redefining the integration variables and mass 
parameters ($m^{(1)}_{F} \to m^{(1)}_{F} - \mu^1$ with $\mu=\mu^2 -\mu^1$).\\

Now consider implementing the flavoring-gauging operation $\CO_2$ on the quiver $X^{(1)}=\CO_1(X)$. 
Following the notation of \Secref{AbMirr}, we label the new flavor node, arising from the operation $\CO_1$ in the previous step, as $\alpha$. 
We define the associated mass parameter as
\be
u^\alpha = m^{(1)}_F.
\ee 
Note that the permutation matrix $\CP$ is trivial in this case. The quiver $\CO_2(X^{(1)})=X^{(2)}$ is shown in \figref{AbEx1GFI}.\\

The dual quiver $\wt{\CO}_2(Y^{(1)})$ can then be constructed using the general recipe presented in \Secref{AbSGF}. The expressions \eref{HyperGFgenNf1}-\eref{PFGFgenNf1} show that the dual quiver $\wt{\CO}_2(Y^{(1)})$ is given by adding one extra hypermultiplet to the quiver gauge theory $Y^{(1)}$, i.e.
\begin{align}\label{PFAbEx1gen2}
Z^{\wt{\CO}_2(Y^{(1)})}(\vec m'^{(2)}, \vec \eta'^{(2)})
= \int \,\prod^{2}_{\gamma'=1} \Big[d\s^{\gamma'}\Big]\, &Z^{(Y^{(1)})}_{\rm int}(\s^1,\s^2, \vec m'^{(1)}(\vec t, \wt{\eta}^{(1)}), \vec \eta'^{(1)}(u^\alpha=m^{(2)}_{F}, \vec \mu)) \nn \\
&\times Z^{\rm hyper}_{\wt{\CO}_2(Y^{(1)})}(\s^1,\s^2, \wt{\eta}^{(2)}, \vec \eta^{(1)}(\vec t, \wt{\eta}^{(1)})), 
\end{align}
where the explicit form of the function $Z^{(Y^{(1)})}_{\rm int}$ can be read off from \eref{PFAbEx1gen1}, and the function $Z^{\rm hyper}_{\wt{\CO}_2(Y^{(1)})}$ is given by \eref{HyperGFgenNf1}. The function $g_\alpha (\s_1,\s_2)$ in \eref{HyperGFgenNf1} can be read off from the FI term appearing in the RHS of \eref{PFAbEx1gen1},
which gives $g_\alpha (\s_1,\s_2)=\s_1-\s_2$. Therefore, the hypermultiplet term reads:
\begin{align}\label{HyperAbEx1gen2}
Z^{\rm hyper}_{\wt{\CO}_2(Y^{(1)})}(\s^1,\s^2, \wt{\eta}^{(2)}, \vec \eta^{(1)}(\vec t, \wt{\eta}^{(1)}))
= Z_{\rm 1-loop}^{\rm bif}(\s^1, \s^2, -\wt{\eta}^{(1)} - \wt{\eta}^{(2)} - \sum_\beta b^{l}_\beta t_l),
\end{align}
The new hypermultiplet can therefore be identified as a bifundamental hyper with a mass parameter 
$m_{\rm bif}= -\wt{\eta}^{(1)} - \wt{\eta}^{(2)} - \sum_\beta b^{l}_\beta t_l$. The resultant quiver then is given by $Y^{(2)}$ 
in the third line of \figref{AbEx1GFI}. The mirror map for the dual theories $X^{(2)}, Y^{(2)}$ can be read off from 
\eref{PFAbEx1gen2}-\eref{HyperAbEx1gen2}.\\

In the next step, we implement the flavoring-gauging operation $\CO_3$ on the quiver $X^{(2)}$ at the flavor node 
arising from the operation $\CO_2$ on $X^{(1)}$, and so on. At the $i$-th step, the dual quiver $Y^{(i)}$ is given by
adding a bifundamental hypermultiplet to the quiver $Y^{(i-1)}$. The dual pair $(X', Y')$ in \figref{AbEx1gen} is therefore 
generated in this fashion for $i=p-1$. The theory $Y'$ can be read off from the partition function:
\begin{align}
\boxed{ Z^{Y'}(\vec m', \vec \eta')
= \int \,\prod^{2}_{\gamma'=1} \Big[d\s^{\gamma'}\Big]\, \prod^{p-1}_{i=1}Z_{\rm 1-loop}^{\rm bif}(\s^1, \s^2, m'_{{\rm bif}\,i})\,Z^{(Y)}_{\rm int}(\s^1,\s^2, -\vec t,\{m^{(p-1)}_{F},0,\mu + m^{(p-1)}_{F}\}),}
\end{align}
where the parameters $m'_{{\rm bif}\,i}$ are given as $m'_{{\rm bif}\,i}= - \sum^{i}_{k=1}\wt{\eta}^{(k)}  -\sum_\beta b^{l}_\beta t_l$.\\

After a change of integration variables, the mirror map for the dual pair $(X', Y')$ can be summarized as follows. 
The mass parameters of theory $Y'$, associated with the bifundamental and fundamental hypermultiplets, can be written as linear 
functions of the FI parameters of $X'$:
\begin{align}
& m'_{{\rm bif }\, j} = \begin{cases}
\delta_1 - \delta_2,& \text{if } j=1\\
-\sum^{j-1}_{k=1} \wt{\eta}^{(k)} + \delta_1 - \delta_2, & \text{if } j=2,\ldots,p,
\end{cases}\\
& m'^{1}_{{\rm fund}\,i_1} = -t_{i_1} + \delta_1, \quad i_1=1,\ldots, n-l+1, \\
& m'^{2}_{{\rm fund}\,i_2} = -t_{i_2+n-l+1} +\delta_2 , \quad i_2=1,\ldots, l-1,
\end{align}
where $\delta_1=\frac{\sum_{i_1}  t_{i_1}}{n-l+1}$ and $\delta_2=\frac{\sum_{i_2}  t_{i_2+n-l+1}}{l-1}$. The FI parameters of the theory $Y'$ are also given in terms of the mass parameters of 
$X'$:
\be
\eta'^1= m^{(p-1)}_F, \quad \eta'^2=-(m^{(p-1)}_F + \mu) .
\ee
The mass parameters of $Y'$ manifestly live in the Cartan subalgebra of the Higgs branch global symmetry 
$G^{Y'}_{H}=U(p) \times SU(n-l+1) \times SU(l-1)$, while the FI parameters live in the Cartan subalgebra
of the Coulomb branch global symmetry $G^{Y'}_{C}=U(1) \times U(1)$.\\

Note that the global symmetries on the Coulomb branch of theory $X'$ looks counter-intuitive. The loop in $X'$  
has two balanced linear subquivers, consisting of $l-2$ and $n-l$ gauge nodes respectively. This leads to the 
factors of $SU(l-1)$ and $SU(n-l+1)$ respectively in the global symmetry as expected. The linear subquiver attached 
to the loop has a set of $p-2$ balanced gauge nodes which naively contribute an $SU(p-1)$ factor to the global symmetry. 
However, it turns out that the naive $U(1) \times U(1) \times SU(p-1)$ is enhanced to $U(1) \times SU(p)$ in the IR. We 
already saw this phenomenon for the example worked out in \Secref{GFI-exAbPF} (which has an explicit Type IIB Hanany-Witten realization), 
corresponding to $p=2$ with generic $n$ and $l$. 
Similar to that example, the enhancement in the global symmetry can be explicitly checked by computing the character expansion for the 
Coulomb branch Hilbert Series of $X'$.

\subsubsection{Family ${\rm II}_{[n,l,l_1, l_2,p_1,p_2]}$: Two closed loops attached to a linear quiver tail}\label{FamilyII}
\begin{figure}[htbp]
\begin{center}
\scalebox{0.7}{\begin{tikzpicture}[node distance=2cm,cnode/.style={circle,draw,thick,minimum size=8mm},snode/.style={rectangle,draw,thick,minimum size=8mm},pnode/.style={rectangle,red,draw,thick,minimum size=8mm}]
\node[cnode] (1) at (-3,0) {$1$};
\node[cnode] (2) at (-2,2) {$1$};
\node[cnode] (3) at (0,2.5) {$1$};
\node[cnode] (4) at (2,2) {$1$};
\node[cnode] (5) at (3,0) {$1$};
\node[cnode] (6) at (2,-2) {$1$};
\node[cnode] (7) at (0,-2.5) {$1$};
\node[cnode] (8) at (-2,-2) {$1$};
\node[cnode] (9) at (0,1.5) {$1$};
\node[snode] (10) at (-5,0) {$1$};
\node[cnode] (11) at (4,0) {$1$};
\node[cnode] (12) at (5,0) {$1$};
\node[cnode] (13) at (7.5,0) {$1$};
\node[snode] (14) at (9,0) {$1$};
\node[cnode] (15) at (0,0.5) {$1$};
\node[cnode] (16) at (0,-1.5) {$1$};
\draw[thick] (1) to [bend left=40] (2);
\draw[dashed] (2) to [bend left=40] (3);
\draw[thick] (3) to [bend left=40] (4);
\draw[dashed] (4) to [bend left=40] (5);
\draw[thick] (5) to [bend left=40] (6);
\draw[dashed] (6) to [bend left=40] (7);
\draw[dashed] (7) to [bend left=40] (8);
\draw[thick] (8) to [bend left=40] (1);
\draw[thick] (1) -- (10);
\draw[thick] (5) -- (11);
\draw[thick] (3) -- (9);
\draw[thick] (9) -- (15);
\draw[thick] (11) -- (12);
\draw[dashed] (12) -- (13);
\draw[thick] (13) -- (14);
\draw[dashed] (15) -- (16);
\draw[thick] (16) -- (7);
\node[text width=1cm](10) at (-3,0.5) {$l$};
\node[text width=1cm](11) at (-2,2.5) {$l+1$};
\node[text width=1cm](12) at (0, 3) {$l_2$};
\node[text width=1cm](13) at (2, 2.5) {$l_2+1$};
\node[text width=1cm](14) at (3, 0.5) {$n$};
\node[text width=1cm](15) at (2,-2.5) {$1$};
\node[text width=1.5 cm](16) at (0,-3.2){$l_1$};
\node[text width=1cm](17) at (-2.5,-2.5){$l-1$};
\node[text width=1cm](18) at (4, 0.6){$1$};
\node[text width=1cm](19) at (5, 0.6){$2$};
\node[text width=1cm](20) at (7.5, 0.6){$p_1-2$};
\node[text width=1cm](21) at (1,1.5){$p_2-1$};
\node[text width=1cm](22) at (1,0.5){$p_2-2$};
\node[text width=1cm](23) at (1,-1.5){$1$};
\node[text width=0.1cm](30) at (0,-4){$(X')$};
\end{tikzpicture}}
\quad
\scalebox{0.8}{\begin{tikzpicture}[node distance=2cm,cnode/.style={circle,draw,thick,minimum size=8mm},snode/.style={rectangle,draw,thick,minimum size=8mm},pnode/.style={rectangle,red,draw,thick,minimum size=8mm}]
\node[cnode] (1) at (-2,0) {$1$};
\node[cnode] (2) at (2,0) {$1$};
\node[cnode] (3) at (0,2) {$1$};
\node[snode] (4) at (-2,-2) {$l_1$};
\node[snode] (5) at (2,-2) {$n-l_2$};
\node[snode] (6) at (0,4) {$p_2$};
\draw[thick] (1) -- (4);
\draw[thick] (2) -- (5);
\draw[thick] (3) -- (6);
\draw[thick] (1) to [bend left=10] (2);
\draw[thick] (1) to [bend right=10] (2);
\draw[thick] (1) to [bend left=20] (2);
\draw[thick] (1) to [bend right=20] (2);
\draw[thick, dotted] (0,0.2) to (0,-0.2);
\node[text width=1cm](15) at (0,-1){$p_1$};
\draw[thick] (1) to [bend left=10] (3);
\draw[thick] (1) to [bend right=10] (3);
\draw[thick] (1) to [bend left=20] (3);
\draw[thick] (1) to [bend right=20] (3);
\draw[thick, dotted] (-1,1.2) to (-0.8, 1.0);
\node[text width=1cm](15) at (-1.5, 1.7){$l-l_1$};
\draw[thick] (2) to [bend left=10] (3);
\draw[thick] (2) to [bend right=10] (3);
\draw[thick] (2) to [bend left=20] (3);
\draw[thick] (2) to [bend right=20] (3);
\draw[thick, dotted] (1,1.2) to (0.8, 1.0);
\node[text width=1cm](15) at (1.5, 1.7){$l_2-l$};
\node[text width=0.1cm](16) at (0,-2){$(Y')$};
\end{tikzpicture}}
\caption{An infinite family of mirror duals labelled by the positive integers $(n,l,l_1, l_2,p_1,p_2)$ subject to the constraints $n > l_2 >l >l_1>0$,
and $p_1>2, p_2>1$.}
\label{AbEx2gen}
\end{center}
\end{figure}
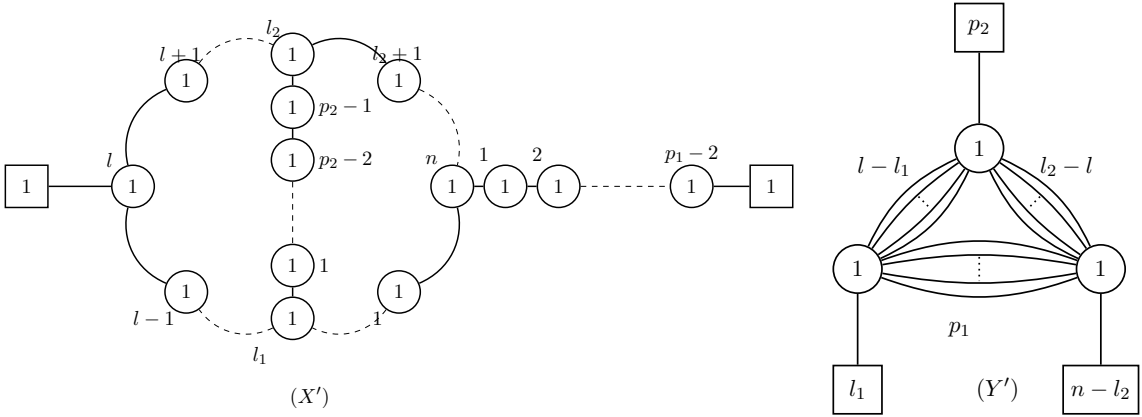

The second example in the class of Abelian mirrors is the infinite family of mirror duals shown in \figref{AbEx2gen}, labelled by the 
positive integers $(n,l,l_1, l_2,p_1,p_2)$ subject to the constraints $n > l_2 >l >l_1>0$, and $p_1>2, p_2>1$. In contrast to Family I 
studied earlier, the quiver $X'$ consists of two closed loops attached to a single linear quiver tail. The proposed mirror dual $Y'$ 
is a quiver with three nodes in a loop connected by multiple edges. The dimensions of the respective Higgs and Coulomb branches, 
and the associated global symmetries are shown in Table \ref{Tab:AbEx2}.\\

\begin{center}
\begin{table}[htbp]
\resizebox{\textwidth}{!}{%
\begin{tabular}{|c|c|c|}
\hline
Moduli space data & Theory $X'$ & Theory $Y'$ \\
\hline \hline 
dim\,$\CM_H$ & $3$ & $n+p_1+p_2-3$\\
\hline
dim\,$\CM_C$ & $n+p_1+p_2-3$ & 3\\
\hline
$G_H$ & $U(1)^3$ & $SU(l_1)\cdot SU(n-l_2) \cdot SU(p_2) \cdot U(l-l_1) \cdot U(l_2-l) \cdot U(p_1)$\\
\hline
$G_C$ & $SU(l_1) \cdot SU(n-l_2) \cdot SU(p_2) \cdot U(l-l_1) \cdot U(l_2-l) \cdot U(p_1)$ & $U(1)^3 $\\
\hline
\end{tabular}}
\caption{Summary table for the moduli space dimensions and global symmetries for the mirror pair in \figref{AbEx2gen}.}
\label{Tab:AbEx2}
\end{table}
\end{center}

The quiver pair $(X', Y')$ in \figref{AbEx2gen} can be generated from the linear quiver pair $(X, Y)$ in \figref{AbEx2LIN} by a series of 
elementary Abelian $S$-type operations on $X$ and the dual operations on $Y$ (shown in \figref{AbEx2GFI}), which we shall describe 
momentarily. The Higgs branch global symmetry of $X$ is given by $G^{X}_{H} = U(1)^5/ U(1) = U(1)^4$. 
Using the notation introduced in \Secref{LQ-IIB}, the mass parameters are labelled as $\vec m =(m_1, m_2, m_3, m_4, m_5)$, and the quotient by the overall $U(1)$ factor in the flavor symmetry of quiver $X$ can be implemented by the constraint $m_3=0$. The $U(1)^4$ flavor symmetry of $X$ can then be identified with the $U(1)_1\times U(1)_{l_1} \times U(1)_{l_2} \times U(1)_{n-1}$ flavor nodes, parametrized by the independent $(m_1, m_2, m_4, m_5)$.\\

\begin{figure}[htbp]
\begin{center}
\begin{tikzpicture}[
cnode/.style={circle,draw,thick,minimum size=4mm},snode/.style={rectangle,draw,thick,minimum size=8mm},pnode/.style={rectangle,red,draw,thick,minimum size=1.0cm}]
\node[cnode] (1) {$1$};
\node[cnode] (2) [right=.5cm  of 1]{$1$};
\node[cnode] (3) [right=1.5cm of 2]{1};
\node[cnode] (4) [right=1.5cm of 3]{1};
\node[cnode] (5) [right=1.5cm of 4]{1};
\node[cnode] (6) [right=1.5cm of 5]{1};
\node[cnode] (7) [right=0.5cm of 6]{1};
\node[snode] (10) [below=0.5cm of 1]{1};
\node[snode] (12) [below=0.5cm of 3]{1};
\node[snode] (13) [below=0.5cm of 4]{1};
\node[snode] (14) [below=0.5cm of 5]{1};
\node[snode] (16) [below=0.5cm of 7]{1};
\draw[-] (1) -- (2);
\draw[dashed] (2) -- (3);
\draw[dashed] (3) -- (4);
\draw[dashed] (4) --(5);
\draw[dashed] (5) --(6);
\draw[-] (6) -- (7);
\draw[-] (1) -- (10);
\draw[-] (3) -- (12);
\draw[-] (4) -- (13);
\draw[-] (5) -- (14);
\draw[-] (7) -- (16);
\node[text width=.5cm](9)[above=0.2cm of 3]{$l_1$};
\node[text width=.5cm](10)[above=0.2cm of 5]{$l_2$};
\node[text width=1 cm](11)[above=0.2cm of 6]{$n-2$};
\node[text width=0.1cm](12)[above=0.2cm of 2]{2};
\node[text width=0.1 cm](13)[above=0.2cm of 1]{1};
\node[text width=1 cm](14)[above=0.2cm of 7]{$n-1$};
\node[text width=.5cm](15)[above=0.2cm of 4]{$l$};
\node[text width=1cm](20)[below=2.5cm of 4]{$(X)$};
\end{tikzpicture}
\qquad \qquad \qquad \qquad \qquad \qquad 
{\begin{tikzpicture}[
cnode/.style={circle,draw,thick, minimum size=4mm},snode/.style={rectangle,draw,thick,minimum size=8mm}]
\node[cnode] (1) {$1$};
\node[cnode] (2) [right=.75cm  of 1]{$1$};
\node[cnode] (3) [right=.75cm of 2]{1};
\node[cnode] (4) [right=.75cm of 3]{1};
\node[snode] (5) [below=0.5cm of 1]{$l_1$};
\node[snode] (6) [below=0.5cm of 2]{$l-l_1$};
\node[snode] (7) [below=0.5cm of 3]{$l_2-l$};
\node[snode] (8) [below=0.5cm of 4]{$n-l_2$};
\draw[-] (1) -- (2);
\draw[-] (2) -- (3);
\draw[-] (3) -- (4);
\draw[-] (1) --(5);
\draw[-] (2) --(6);
\draw[-] (3) -- (7);
\draw[-] (4) -- (8);
\node[text width=0.1cm](10)[above=0.2 cm of 1]{1};
\node[text width=0.1cm](11)[above=0.2 cm of 2]{2};
\node[text width=0.1cm](12)[above=0.2 cm of 3]{3};
\node[text width=0.1cm](13)[above=0.2 cm of 4]{4};
\node[text width=0.1cm](20)[below=2.5cm of 2]{$(Y)$};
\end{tikzpicture}}
\caption{Pair of linear quivers with $U(1)$ gauge groups which generate the 3d mirrors in \figref{AbEx2gen}.}
\label{AbEx2LIN}
\end{center}
\end{figure}

The theory $X'$ can be obtained from the quiver $X$ in three distinct set of steps, each of which involves a sequence of 
elementary Abelian operations:
\begin{enumerate}
\item The first sequence involves an identification-flavoring-gauging operation $\CO_1$ on the $U(1)_1 \times U(1)_{n-1}$ 
flavor nodes (marked in red) at the two ends of the quiver $X$, followed by $p_1-2$ flavoring-gauging operations 
$\CO_i$ (for $i=2,\ldots,p_1-1$), each acting at the new flavor node generated in the previous step. At each step, the flavoring operation 
corresponds to $N^\alpha_F=1$.  This is very similar to the derivation of the quiver $X'$ in Family I, studied earlier. 
We adopt the notation:
\begin{align}
& X^{(i)}= \CO_i \circ  \CO_{i-1} \circ \ldots \circ \CO_2 \circ \CO_1(X), \\
& Y^{(i)}= \wt\CO_i \circ  \wt\CO_{i-1} \circ \ldots \circ \wt\CO_2 \circ \wt\CO_1(Y), 
\end{align}
and denote the $\CN=4$-preserving deformations of the quivers $X^{(i)}$ and $Y^{(i)}$ as $(\vec m^{(i)}, \vec \eta^{(i)})$ 
and $(\vec m'^{(i)}, \vec \eta'^{(i)})$ respectively. The mass associated with the identification operation is $\vec \mu$, 
the mass associated with flavoring at the $i$-th step is labelled as $m^{(i)}_F$, and the FI parameter associated with 
gauging operation at the $i$-th step is $\wt{\eta}^{(i)}$. The quivers $(X^{(p_1-1)},Y^{(p_1-1)})$ are shown in the second line of 
\figref{AbEx2GFI}.

\item The second sequence involves $p_2-2$ $N^\alpha_F=1$ flavoring-gauging operations $\CO'_j$ ($j=1,\ldots,p_2-2$), 
where $\CO'_1$ acts on the $U(1)_{l_1}$ flavor node of the quiver $X^{(p_1-1)}$, while $\CO'_j$ for $j\geq 2$ acts 
on the new flavor node generated by the operation $\CO'_{j-1}$ in the previous step. We will denote these quivers 
as:
\begin{align}
& X^{(p_1-1,j)}= \CO'_j \circ  \CO'_{j-1} \circ \ldots \circ \CO'_2 \circ \CO'_1(X^{(p_1-1)}), \\
& Y^{(p_1-1,j)}= \wt\CO'_i \circ  \wt\CO'_{i-1} \circ \ldots \circ \wt\CO'_2 \circ \wt\CO'_1(Y^{(p_1-1)}), 
\end{align}
and the associated $\CN=4$-preserving deformations will be denoted as $(\vec m^{(p_1-1,j)}, \vec \eta^{(p_1-1,j)})$ 
and $(\vec m'^{(p_1-1,j)}, \vec \eta'^{(p_1-1,j)})$ respectively. The mass associated with the flavoring operation at the 
$j$-th step is $x^{(j)}_F$, and the FI parameter associated with the gauging operation at the $j$-th step is $\xi^{(j)}$.
The quivers $(X^{(p_1-1,p_2-2)}, Y^{(p_1-1,p_2-2)})$ are shown in the third line of \figref{AbEx2GFI}.

\item In the final step, we implement an identification-gauging operation $\CO''$ on the $X^{(p_1-1,p_2-2)}$ involving the 
$U(1)_{l_2}$ flavor node and the $U(1)$ flavor node generated by the operation $\CO'_{p_2-2}$ in the previous step. This 
leads to the quivers $(X', Y')$ in \figref{AbEx2gen}, i.e.
\be
X' = \CO''(X^{(p_1-1,p_2-2)}), \, Y' = \wt{\CO}''(Y^{(p_1-1,p_2-2)}). 
\ee
We will denote the $\CN=4$-preserving deformations of the dual theories as $(\vec m, \vec \eta)$ and 
$(\vec m', \vec \eta')$ respectively. The mass associated with the identification operation is labelled as 
$\wt{\mu}$, and the FI parameter is labelled as $\xi^{(p_2-1)}$.

\end{enumerate}

\begin{figure}[htbp]
\begin{center}
\begin{tabular}{ccc}
\scalebox{.5}{\begin{tikzpicture}[
cnode/.style={circle,draw,thick,minimum size=4mm},snode/.style={rectangle,draw,thick,minimum size=8mm},pnode/.style={rectangle,red,draw,thick,minimum size=1.0cm}]
\node[cnode] (1) {$1$};
\node[cnode] (2) [right=.5cm  of 1]{$1$};
\node[cnode] (3) [right=1.5cm of 2]{1};
\node[cnode] (4) [right=1.5cm of 3]{1};
\node[cnode] (5) [right=1.5cm of 4]{1};
\node[cnode] (6) [right=1.5cm of 5]{1};
\node[cnode] (7) [right=0.5cm of 6]{1};
\node[pnode] (10) [below=0.5cm of 1]{1};
\node[snode] (12) [below=0.5cm of 3]{1};
\node[snode] (13) [below=0.5cm of 4]{1};
\node[snode] (14) [below=0.5cm of 5]{1};
\node[pnode] (16) [below=0.5cm of 7]{1};
\draw[-] (1) -- (2);
\draw[dashed] (2) -- (3);
\draw[dashed] (3) -- (4);
\draw[dashed] (4) --(5);
\draw[dashed] (5) --(6);
\draw[-] (6) -- (7);
\draw[-] (1) -- (10);
\draw[-] (3) -- (12);
\draw[-] (4) -- (13);
\draw[-] (5) -- (14);
\draw[-] (7) -- (16);
\node[text width=.5cm](9)[above=0.2cm of 3]{$l_1$};
\node[text width=.5cm](10)[above=0.2cm of 5]{$l_2$};
\node[text width=1 cm](11)[above=0.2cm of 6]{$n-2$};
\node[text width=0.1cm](12)[above=0.2cm of 2]{2};
\node[text width=0.1 cm](13)[above=0.2cm of 1]{1};
\node[text width=1 cm](14)[above=0.2cm of 7]{$n-1$};
\node[text width=.5cm](15)[above=0.2cm of 4]{$l$};
\node[text width=1cm](20)[below=2.5cm of 4]{$(X)$};
\end{tikzpicture}
}
& \qquad 
& \scalebox{.5}{\begin{tikzpicture}[
cnode/.style={circle,draw,thick, minimum size=4mm},snode/.style={rectangle,draw,thick,minimum size=8mm}]
\node[cnode] (1) {$1$};
\node[cnode] (2) [right=.75cm  of 1]{$1$};
\node[cnode] (3) [right=.75cm of 2]{1};
\node[cnode] (4) [right=.75cm of 3]{1};
\node[snode] (5) [below=0.5cm of 1]{$l_1$};
\node[snode] (6) [below=0.5cm of 2]{$l-l_1$};
\node[snode] (7) [below=0.5cm of 3]{$l_2-l$};
\node[snode] (8) [below=0.5cm of 4]{$n-l_2$};
\draw[-] (1) -- (2);
\draw[-] (2) -- (3);
\draw[-] (3) -- (4);
\draw[-] (1) --(5);
\draw[-] (2) --(6);
\draw[-] (3) -- (7);
\draw[-] (4) -- (8);
\node[text width=0.1cm](10)[above=0.2 cm of 1]{1};
\node[text width=0.1cm](11)[above=0.2 cm of 2]{2};
\node[text width=0.1cm](12)[above=0.2 cm of 3]{3};
\node[text width=0.1cm](13)[above=0.2 cm of 4]{4};
\node[text width=0.1cm](20)[below=2.5cm of 2]{$(Y)$};
\end{tikzpicture}}\\
 \scalebox{.5}{\begin{tikzpicture}
\draw[->] (15,-3) -- (15,-5);
\node[text width=1cm](20) at (14, -4) {$\prod_i\CO_i$};
\end{tikzpicture}}
&\qquad
& \scalebox{.5}{\begin{tikzpicture}
\draw[->] (15,-3) -- (15,-5);
\node[text width=0.1cm](29) at (15.5, -4) {$\prod_i \wt{\CO}_i$};
\end{tikzpicture}}\\
\scalebox{.5}{\begin{tikzpicture}[node distance=2cm,cnode/.style={circle,draw,thick,minimum size=8mm},snode/.style={rectangle,draw,thick,minimum size=8mm},pnode/.style={rectangle,red,draw,thick,minimum size=8mm}]
\node[cnode] (1) at (-3,0) {$1$};
\node[cnode] (2) at (-2,2) {$1$};
\node[cnode] (3) at (0,2.5) {$1$};
\node[cnode] (4) at (2,2) {$1$};
\node[cnode] (5) at (3,0) {$1$};
\node[cnode] (6) at (2,-2) {$1$};
\node[cnode] (7) at (0,-2.5) {$1$};
\node[cnode] (8) at (-2,-2) {$1$};
\node[snode] (9) at (0,1.5) {$1$};
\node[snode] (10) at (-5,0) {$1$};
\node[cnode] (11) at (4,0) {$1$};
\node[cnode] (12) at (5,0) {$1$};
\node[cnode] (13) at (7.5,0) {$1$};
\node[snode] (14) at (9,0) {$1$};
\node[pnode] (16) at (0,-1.5) {$1$};
\draw[thick] (1) to [bend left=40] (2);
\draw[dashed] (2) to [bend left=40] (3);
\draw[thick] (3) to [bend left=40] (4);
\draw[dashed] (4) to [bend left=40] (5);
\draw[thick] (5) to [bend left=40] (6);
\draw[dashed] (6) to [bend left=40] (7);
\draw[dashed] (7) to [bend left=40] (8);
\draw[thick] (8) to [bend left=40] (1);
\draw[thick] (1) -- (10);
\draw[thick] (5) -- (11);
\draw[thick] (3) -- (9);
\draw[thick] (11) -- (12);
\draw[dashed] (12) -- (13);
\draw[thick] (13) -- (14);
\draw[thick] (16) -- (7);
\node[text width=1cm](10) at (-3,0.5) {$l$};
\node[text width=1cm](11) at (-2,2.5) {$l+1$};
\node[text width=1cm](12) at (0, 3) {$l_2$};
\node[text width=1cm](13) at (2, 2.5) {$l_2+1$};
\node[text width=1cm](14) at (3, 0.5) {$n$};
\node[text width=1cm](15) at (2,-2.5) {$1$};
\node[text width=1.5 cm](16) at (0,-3.2){$l_1$};
\node[text width=1cm](17) at (-2.5,-2.5){$l-1$};
\node[text width=1cm](18) at (4, 0.6){$1$};
\node[text width=1cm](19) at (5, 0.6){$2$};
\node[text width=1cm](20) at (7.5, 0.6){$p_1-2$};
\node[text width=0.1cm](30) at (0,-4){$(X^{(p_1-1)})$};
\end{tikzpicture}}
&\qquad 
&{\scalebox{0.5}{\begin{tikzpicture}[node distance=2cm,cnode/.style={circle,draw,thick,minimum size=8mm},snode/.style={rectangle,draw,thick,minimum size=8mm},pnode/.style={rectangle,red,draw,thick,minimum size=8mm}]
\node[cnode] (1) at (-2,4) {$1$};
\node[cnode] (2) at (-2,0) {$1$};
\node[cnode] (3) at (2,0) {$1$};
\node[cnode] (4) at (2,4) {$1$};
\node[snode] (5) at (-2,6) {$l_1$};
\node[snode] (6) at (-2,-2) {$l-l_1$};
\node[snode] (8) at (2,6) {$n-l_2$};
\node[snode] (7) at (2,-2) {$l_2-l$};
\draw[thick] (1) -- (2);
\draw[thick] (2) -- (3);
\draw[thick] (3) -- (4);
\draw[thick] (1) to [bend left=10] (4);
\draw[thick] (1) to [bend right=10] (4);
\draw[thick] (1) to [bend left=20] (4);
\draw[thick] (1) to [bend right=20] (4);
\draw[thick] (1) -- (5);
\draw[thick] (2) -- (6);
\draw[thick] (3) -- (7);
\draw[thick] (4) -- (8);
\node[text width=0.1cm](10)[left=0.2 cm of 1]{1};
\node[text width=0.1cm](11)[left=0.2 cm of 2]{2};
\node[text width=0.1cm](12)[right=0.2 cm of 3]{3};
\node[text width=0.1cm](13)[right=0.2 cm of 4]{4};
\draw[thick, dotted] (0,4.2) to (0,3.8);
\node[text width=1cm](15) at (0.2, 3){$p_1-1$};
\node[text width=1cm](30) at (0,-2){$(Y^{(p_1-1)})$};
\end{tikzpicture}}}\\
\scalebox{.5}{\begin{tikzpicture}
\draw[->] (15,-3) -- (15,-5);
\node[text width=0.1cm](20) at (13.5, -4) {$\prod_j \CO'_j$};
\end{tikzpicture}}
&\qquad
& \scalebox{.5}{\begin{tikzpicture}
\draw[->] (15,-3) -- (15,-5);
\node[text width=0.1cm](29) at (15.5, -4) {$\prod_j \wt{\CO}'_j$};
\end{tikzpicture}}\\
\scalebox{.5}{\begin{tikzpicture}[node distance=2cm,cnode/.style={circle,draw,thick,minimum size=8mm},snode/.style={rectangle,draw,thick,minimum size=8mm},pnode/.style={rectangle,red,draw,thick,minimum size=8mm}]
\node[cnode] (1) at (-3,0) {$1$};
\node[cnode] (2) at (-2,2) {$1$};
\node[cnode] (3) at (0,3) {$1$};
\node[cnode] (4) at (2,2) {$1$};
\node[cnode] (5) at (3,0) {$1$};
\node[cnode] (6) at (2,-2) {$1$};
\node[cnode] (7) at (0,-2.5) {$1$};
\node[cnode] (8) at (-2,-2) {$1$};
\node[pnode] (9) at (0,1.5) {$1$};
\node[snode] (10) at (-5,0) {$1$};
\node[cnode] (11) at (4,0) {$1$};
\node[cnode] (12) at (5,0) {$1$};
\node[cnode] (13) at (7.5,0) {$1$};
\node[snode] (14) at (9,0) {$1$};
\node[cnode] (15) at (0,0.5) {$1$};
\node[cnode] (16) at (0,-1.5) {$1$};
\node[pnode] (25) at (0,4.4) {$1$};
\draw[thick] (1) to [bend left=40] (2);
\draw[dashed] (2) to [bend left=40] (3);
\draw[thick] (3) to [bend left=40] (4);
\draw[dashed] (4) to [bend left=40] (5);
\draw[thick] (5) to [bend left=40] (6);
\draw[dashed] (6) to [bend left=40] (7);
\draw[dashed] (7) to [bend left=40] (8);
\draw[thick] (8) to [bend left=40] (1);
\draw[thick] (1) -- (10);
\draw[thick] (5) -- (11);
\draw[thick] (9) -- (15);
\draw[thick] (11) -- (12);
\draw[dashed] (12) -- (13);
\draw[thick] (13) -- (14);
\draw[dashed] (15) -- (16);
\draw[thick] (16) -- (7);
\draw[thick] (3) -- (25);
\node[text width=1cm](10) at (-3,0.5) {$l$};
\node[text width=1cm](11) at (-2,2.5) {$l+1$};
\node[text width=1cm](12) at (0, 3.5) {$l_2$};
\node[text width=1cm](13) at (2, 2.5) {$l_2+1$};
\node[text width=1cm](14) at (3, 0.5) {$n$};
\node[text width=1cm](15) at (2,-2.5) {$1$};
\node[text width=1.5 cm](16) at (0,-3.2){$l_1$};
\node[text width=1cm](17) at (-2.5,-2.5){$l-1$};
\node[text width=1cm](18) at (4, 0.6){$1$};
\node[text width=1cm](19) at (5, 0.6){$2$};
\node[text width=1cm](20) at (7.5, 0.6){$p_1-2$};
\node[text width=1cm](22) at (1,0.5){$p_2-2$};
\node[text width=1cm](23) at (1,-1.5){$1$};
\node[text width=1cm](30) at (0,-4){$(X^{(p_1-1,p_2-2)})$};
\end{tikzpicture}}
&\qquad
&\scalebox{.5}{\begin{tikzpicture}[node distance=2cm,cnode/.style={circle,draw,thick,minimum size=8mm},snode/.style={rectangle,draw,thick,minimum size=8mm},pnode/.style={rectangle,red,draw,thick,minimum size=8mm}]
\node[cnode] (1) at (-2,4) {$1$};
\node[cnode] (2) at (-2,0) {$1$};
\node[cnode] (3) at (2,0) {$1$};
\node[cnode] (4) at (2,4) {$1$};
\node[snode] (5) at (-2,6) {$l_1$};
\node[snode] (6) at (-2,-2) {$l-l_1$};
\node[snode] (8) at (2,6) {$n-l_2$};
\node[snode] (7) at (2,-2) {$l_2-l$};
\draw[thick] (2) -- (3);
\draw[thick] (3) -- (4);
\draw[thick] (1) to [bend left=10] (4);
\draw[thick] (1) to [bend right=10] (4);
\draw[thick] (1) to [bend left=20] (4);
\draw[thick] (1) to [bend right=20] (4);
\draw[thick] (1) to [bend left=10] (2);
\draw[thick] (1) to [bend right=10] (2);
\draw[thick] (1) to [bend left=20] (2);
\draw[thick] (1) to [bend right=20] (2);
\draw[thick] (1) -- (5);
\draw[thick] (2) -- (6);
\draw[thick] (3) -- (7);
\draw[thick] (4) -- (8);
\node[text width=0.1cm](10)[left=0.2 cm of 1]{1};
\node[text width=0.1cm](11)[left=0.2 cm of 2]{2};
\node[text width=0.1cm](12)[right=0.2 cm of 3]{3};
\node[text width=0.1cm](13)[right=0.2 cm of 4]{4};
\draw[thick, dotted] (0,4.2) to (0,3.8);
\node[text width=1cm](15) at (0.2, 3){$p_1-1$};
\draw[thick, dotted] (-2.2,2) to (-1.8,2);
\node[text width=1cm](16) at (-1, 2){$p_2-1$};
\node[text width=1cm](30) at (-0.5,-2){$(Y^{(p_1-1,p_2-2)})$};
\end{tikzpicture}}\\
\scalebox{.5}{\begin{tikzpicture}
\draw[->] (15,-3) -- (15,-5);
\node[text width=0.1cm](20) at (14.5, -4) {$\CO''$};
\end{tikzpicture}}
&\qquad
& \scalebox{.5}{\begin{tikzpicture}
\draw[->] (15,-3) -- (15,-5);
\node[text width=0.1cm](29) at (15.5, -4) {$\wt{\CO}''$};
\end{tikzpicture}}\\
\scalebox{0.5}{\begin{tikzpicture}[node distance=2cm,cnode/.style={circle,draw,thick,minimum size=8mm},snode/.style={rectangle,draw,thick,minimum size=8mm},pnode/.style={rectangle,red,draw,thick,minimum size=8mm}]
\node[cnode] (1) at (-3,0) {$1$};
\node[cnode] (2) at (-2,2) {$1$};
\node[cnode] (3) at (0,2.5) {$1$};
\node[cnode] (4) at (2,2) {$1$};
\node[cnode] (5) at (3,0) {$1$};
\node[cnode] (6) at (2,-2) {$1$};
\node[cnode] (7) at (0,-2.5) {$1$};
\node[cnode] (8) at (-2,-2) {$1$};
\node[cnode] (9) at (0,1.5) {$1$};
\node[snode] (10) at (-5,0) {$1$};
\node[cnode] (11) at (4,0) {$1$};
\node[cnode] (12) at (5,0) {$1$};
\node[cnode] (13) at (7.5,0) {$1$};
\node[snode] (14) at (9,0) {$1$};
\node[cnode] (15) at (0,0.5) {$1$};
\node[cnode] (16) at (0,-1.5) {$1$};
\draw[thick] (1) to [bend left=40] (2);
\draw[dashed] (2) to [bend left=40] (3);
\draw[thick] (3) to [bend left=40] (4);
\draw[dashed] (4) to [bend left=40] (5);
\draw[thick] (5) to [bend left=40] (6);
\draw[dashed] (6) to [bend left=40] (7);
\draw[dashed] (7) to [bend left=40] (8);
\draw[thick] (8) to [bend left=40] (1);
\draw[thick] (1) -- (10);
\draw[thick] (5) -- (11);
\draw[thick] (3) -- (9);
\draw[thick] (9) -- (15);
\draw[thick] (11) -- (12);
\draw[dashed] (12) -- (13);
\draw[thick] (13) -- (14);
\draw[dashed] (15) -- (16);
\draw[thick] (16) -- (7);
\node[text width=1cm](10) at (-3,0.5) {$l$};
\node[text width=1cm](11) at (-2,2.5) {$l+1$};
\node[text width=1cm](12) at (0, 3) {$l_2$};
\node[text width=1cm](13) at (2, 2.5) {$l_2+1$};
\node[text width=1cm](14) at (3, 0.5) {$n$};
\node[text width=1cm](15) at (2,-2.5) {$1$};
\node[text width=1.5 cm](16) at (0,-3.2){$l_1$};
\node[text width=1cm](17) at (-2.5,-2.5){$l-1$};
\node[text width=1cm](18) at (4, 0.6){$1$};
\node[text width=1cm](19) at (5, 0.6){$2$};
\node[text width=1cm](20) at (7.5, 0.6){$p_1-2$};
\node[text width=1cm](21) at (1,1.5){$p_2-1$};
\node[text width=1cm](22) at (1,0.5){$p_2-2$};
\node[text width=1cm](23) at (1,-1.5){$1$};
\node[text width=0.1cm](30) at (0,-4){$(X')$};
\end{tikzpicture}}
&\qquad
&\scalebox{0.5}{\begin{tikzpicture}[node distance=2cm,cnode/.style={circle,draw,thick,minimum size=8mm},snode/.style={rectangle,draw,thick,minimum size=8mm},pnode/.style={rectangle,red,draw,thick,minimum size=8mm}]
\node[cnode] (1) at (-2,0) {$1$};
\node[cnode] (2) at (2,0) {$1$};
\node[cnode] (3) at (0,-3) {$1$};
\node[snode] (4) at (-2,2) {$l_1$};
\node[snode] (5) at (2,2) {$n-l_2$};
\node[snode] (6) at (0,-5) {$p_2$};
\draw[thick] (1) -- (4);
\draw[thick] (2) -- (5);
\draw[thick] (3) -- (6);
\node[text width=0.1cm](10)[left=0.2 cm of 1]{1};
\node[text width=0.1cm](11)[right=0.2 cm of 2]{3};
\node[text width=0.1cm](12)[right=0.2 cm of 3]{2};
\draw[thick] (1) to [bend left=10] (2);
\draw[thick] (1) to [bend right=10] (2);
\draw[thick] (1) to [bend left=20] (2);
\draw[thick] (1) to [bend right=20] (2);
\draw[thick, dotted] (0,0.2) to (0,-0.2);
\node[text width=1cm](15) at (0.5,1){$p_1$};
\draw[thick] (1) to [bend left=10] (3);
\draw[thick] (1) to [bend right=10] (3);
\draw[thick] (1) to [bend left=20] (3);
\draw[thick] (1) to [bend right=20] (3);
\node[text width=1cm](15) at (-2,-1.5){$l-l_1$};
\draw[thick] (2) to [bend left=10] (3);
\draw[thick] (2) to [bend right=10] (3);
\draw[thick] (2) to [bend left=20] (3);
\draw[thick] (2) to [bend right=20] (3);
\node[text width=1cm](15) at (2.1,-1.5){$l_2-l$};
\node[text width=0.1cm](16) at (1.5,-5){$(Y')$};
\end{tikzpicture}}
\end{tabular}
\caption{The quiver $(X')$ and its mirror dual $(Y')$ in \figref{AbEx2gen} is generated by a sequence of 
elementary $S$-type operations on different nodes. In each step, the flavor node(s) on which the $S$-type operation 
acts is shown in red.}
\label{AbEx2GFI}
\end{center}
\end{figure}

Let us now work out the dual quivers at each step outlined above, using the dual partition functions for Abelian $S$-type operations 
derived in \Secref{AbS}.
\begin{itemize}
\item Consider the first sequence of operations on the theory $X$. 
Let us define the variables $\{ \vec u^\beta \}$, which correspond to the $U(1)_1 \times U(1)_{n-1}$ mass parameters of the quiver $X$ as:
\be \label{defuAbEx2}
u^1=m_1, u^2=m_5,
\ee 
and implement the $S$-type operation $\CO_1$ on the relevant flavor nodes of $X$ (the permutation matrices $\{\CP_\beta\}$ 
are again trivial in this case). The partition function of the dual quiver $\wt{\CO}_1(Y)=Y^{(1)}$ is then given by the expressions 
\eref{PFGFILinNf1}-\eref{HyperGFILinNf1} as follows:
\begin{align}
& Z^{\wt{\CO}_1(Y)}(\vec m'^{(1)}, \vec \eta'^{(1)})\nn \\
&= \int \,\prod^{4}_{\gamma'=1} \Big[d\s^{\gamma'}\Big]\, Z^{\rm hyper}_{\wt{\CO}_1(Y)}(\{\s^{\gamma'}\}, \wt{\eta}^{(1)}, \vec t)\,Z^{(Y)}_{\rm int}(\{\s^{\gamma'}\}, -\vec t,\{{u}^\beta =\mu^\beta + m^{(1)}_{F}\}, m_2,m_4), \nn \\
&=\int \,\prod^{4}_{\gamma'=1} \Big[d\s^{\gamma'}\Big]\, Z^{\rm hyper}_{\wt{\CO}_1(Y)}(\{\s^{\gamma'}\}, \wt{\eta}^{(1)}, \vec t)\,Z^{(Y)}_{\rm int}(\{\s^{\gamma'}\}, -\vec t,
\{m^{(1)}_{F}, m_2,0, m_4, m^{(1)}_{F} + \mu\}),\label{PFAbEx2gen1}
\end{align}
where for the second equality we have redefined the mass parameter $m^{(1)}_{F} \to m^{(1)}_{F} -\mu^1$ with $\mu=\mu^2-\mu^1$. Following \eref{HyperGFILinNf1},
the hypermultiplet term gives
\begin{align}
 Z^{\rm hyper}_{\wt{\CO_1}(Y)}(\{\s^{\gamma'}\}, \wt{\eta}^{(1)}, \vec t)
= Z_{\rm 1-loop}^{\rm bif}(\s^1, \s^4, -\wt{\eta}^{(1)} - \sum_\beta b^{l}_\beta t_l). \label{HyperAbEx2gen1}
\end{align}
The dual operation therefore amounts to adding a bifundamental hypermultiplet connecting the gauge nodes labelled $1$ and $4$ in quiver $Y$,
with a mass parameter $m'^{(14)}_{{\rm bif}\,1}= -\wt{\eta}^{(1)} - \sum_\beta b^{l}_\beta t_l$. Next, we implement the flavoring-gauging operation $\CO_2$ on 
the quiver $X^{(1)}=\CO_1(X)$ at the new flavor node $\alpha$ generated by $\CO_1$, i.e. we set 
\be
u^\alpha = m^{(1)}_F,
\ee
and implement the flavoring-gauging operation at $\alpha$ as before. 
The expressions \eref{HyperGFgenNf1}-\eref{PFGFgenNf1} show that the
dual quiver $\wt{\CO}_2(Y^{(1)})=Y^{(2)}$ is given by adding to the quiver gauge theory $Y^{(1)}$ one extra bifundamental hypermultiplet 
connecting the the gauge nodes labelled $1$ and $4$. Explicitly, \eref{PFGFgenNf1} gives 
\begin{align}
Z^{\wt{\CO}_2(Y^{(1)})}(\vec m'^{(2)}, \vec \eta'^{(2)})
= &\int \,\prod^{4}_{\gamma'=1}  \Big[d\s^{\gamma'}\Big]\, Z^{\rm hyper}_{\wt{\CO}_2(Y^{(1)})}(\{\s^{\gamma'}\}, \wt{\eta}^{(2)}, \vec \eta^{(1)}) \,Z^{(Y^{(1)})}_{\rm int}(\{\s^{\gamma'}\}, \vec m'^{(1)}, \vec \eta'^{(1)}), \label{PFAbEx2gen2}
\end{align}
where $\vec \eta^{(1)}=\vec \eta^{(1)}(\vec t, \wt{\eta}^{(1)})$, $\vec \eta'^{(1)}=\vec \eta'^{(1)}(u^\alpha=m^{(2)}_{F}, \mu, m_2,m_4)$.
The function $Z^{(Y^{(1)})}_{\rm int}$ is given by the integrand of the matrix integral on the RHS of \eref{PFAbEx2gen1}. The function $g_\alpha$ 
in the formula \eref{HyperGFgenNf1} for $Z^{\rm hyper}_{\wt{\CO}_2(Y^{(1)})}$ can be read off from the FI term appearing in the RHS of \eref{PFAbEx2gen1},
i.e. $g_\alpha (\s_1,\s_4)=\s_1-\s_4$. The hypermultiplet term then gives
\begin{align}
Z^{\rm hyper}_{\wt{\CO}_2(Y^{(1)})}(\{\s^{\gamma'}\}, \wt{\eta}^{(2)}, \vec \eta^{(1)})
=  Z_{\rm 1-loop}^{\rm bif}(\s^1, \s^4, -\wt{\eta}^{(1)} - \wt{\eta}^{(2)} - \sum_\beta b^{l}_\beta t_l).\label{HyperAbEx2gen2}
\end{align}
Proceeding in the same fashion, one can implement the flavoring-gauging operation successively on the new flavor node generated 
in the previous step. At the $i$-th step, 
the dual quiver $Y^{(i)}$ is given by adding a bifundamental hypermultiplet connecting the gauge nodes labelled $1$ and $4$ 
in the quiver $Y^{(i-1)}$. The partition function of the theory $Y^{(p_1-1)}$ is given as 
\begin{align}
& Z^{Y^{(p_1-1)}}(\vec m'^{(p_1-1)}, \vec \eta'^{(p_1-1)})\nn \\
&= \int \,\prod^{4}_{\gamma'=1} \Big[d\s^{\gamma'}\Big]\, \prod^{p_1-1}_{i=1}Z_{\rm 1-loop}^{\rm bif}(\s^1, \s^4, m'^{(14)}_{{\rm bif}\,i})\,Z^{(Y)}_{\rm int}(\{\s^{\gamma'}\}, -\vec t,
\{m^{(p_1-1)}_{F}, m_2,0, m_4, m^{(p_1-1)}_{F} + \mu\}),\label{PFAbEx2gen3}
\end{align}
which manifestly is the partition function for the quiver $Y^{(p_1-1)}$, as shown in the second line of the \figref{AbEx2GFI}. 
The bifundamental masses $\vec{m}'_{\rm bif}$ in the above expression are given as
\be
m'^{(14)}_{{\rm bif}\,i} = - \sum^{i}_{k=1}\wt{\eta}^{(k)}  -\sum_\beta b^{l}_\beta t_l.
\ee

\item Next, consider the second sequence of operations on the theory $X^{(p_1-1)}$, starting with the $U(1)_{l_1}$ flavor node 
shown in red in \figref{AbEx2GFI}. Let us define the mass parameter $u^\alpha$ corresponding to the $U(1)_{l_1}$ flavor node 
as
\be
u^\alpha= m_2,
\ee
and implement the flavoring-gauging operation $\CO'_1$. The partition function of the dual theory $\wt{\CO}'_1(Y^{(p_1-1)})=Y^{(p_1-1,1)}$ 
is then given as
\begin{align}
& Z^{\wt{\CO}'_1(Y^{(p_1-1)})}(\vec m'^{(p_1-1,1)}, \vec \eta'^{(p_1-1,1)}) \nn \\
&= \int \,\prod^{4}_{\gamma'=1}  \Big[d\s^{\gamma'}\Big]\, Z^{\rm hyper}_{\wt{\CO}'_1(Y^{(p_1-1)})}(\{\s^{\gamma'}\}, \xi^{(1)}, \vec \eta^{(p_1-1)}) \,Z^{(Y^{(p_1-1)})}_{\rm int}(\{\s^{\gamma'}\}, \vec m'^{(p_1-1)}, \vec \eta'^{(p_1-1)}), \label{PFAbEx2gen4}
\end{align}
where $\vec \eta^{(p_1-1)}=\vec \eta^{(p_1-1)}(\vec t, \wt{\eta}^{(i)})$, $\vec \eta'^{(p_1-1)}=\vec \eta'^{(p_1-1)}(u^\alpha=x^{(1)}_{F}, \mu, m^{(p_1-1)}_{F},m_4)$, 
and the function $Z^{(Y^{(p_1-1)})}_{\rm int}$ is given by the integrand on the RHS of \eref{PFAbEx2gen3}. The hypermultiplet term, computed as before from \eref{HyperGFgenNf1}, has the form:
\begin{align}
Z^{\rm hyper}_{\wt{\CO}'_1(Y^{(p_1-1)})}(\{\s^{\gamma'}\}, \xi^{(1)}, \vec \eta^{(p_1-1)})
= Z_{\rm 1-loop}^{\rm bif}(\s^1, \s^2, \xi^{(1)} +  b^{l}_2 \eta^{(p_1-1)}_l),\label{HyperAbEx2gen4}
\end{align}
which implies that the dual operation $\wt{\CO}'_1$ on theory $Y^{(p_1-1)}$ simply amounts to adding a bifundamental hypermultiplet 
connecting the gauge nodes labelled 1 and 2 in $Y^{(p_1-1)}$. Proceeding as before, one can now perform the flavoring-gauging operation 
on the new $U(1)$ flavor node, with mass parameter $x^{(1)}_{F}$, and repeat this operation $p_2-2$ times. The resultant dual theory 
$Y^{(p_1-1,p_2-2)}$ can be read off from the partition function:
\begin{align}
 Z^{Y^{(p_1-1,p_2-2)}}
= \int \,& \prod^{4}_{\gamma'=1} \Big[d\s^{\gamma'}\Big]\, \prod^{p_2-2}_{j=1}Z_{\rm 1-loop}^{\rm bif}(\s^1, \s^2, m'^{(12)}_{{\rm bif}\,j}) \, \prod^{p_1-1}_{i=1}Z_{\rm 1-loop}^{\rm bif}(\s^1, \s^4, m'^{(14)}_{{\rm bif}\,i})\nn \\
& \times Z^{(Y)}_{\rm int}(\{\s^{\gamma'}\}, -\vec t,\{m^{(p_1-1)}_{F}, x^{(p_2-2)}_{F},0, m_4, m^{(p_1-1)}_{F} + \mu\}),\label{PFAbEx2gen5}
\end{align}
which manifestly reproduces the quiver gauge theory $Y^{(p_1-1,p_2-2)}$ in the third line of \figref{AbEx2GFI}. The bifundamental masses $\vec{m}'^{(12)}_{\rm bif}$ in the above expression are given as
\be
m'^{(12)}_{{\rm bif}\,j} = \sum^{j}_{k=1}\xi^{(k)}  +  b^{l}_2 \eta^{(p_1-1)}_l.
\ee

\item Finally, consider the identification-gauging operation on the theory $X^{(p_1-1,p_2-2)}$, involving the two $U(1)$ flavor nodes 
shown in red in \figref{AbEx2GFI}. Let us define the variables $\{u^{\beta'}\}$ as 
\be
u^1=x^{(p_2-2)}_{F}, \, u^2 = m_4,
\ee
and implement the identification-gauging operation $\CO''$. The dual, given by the general expression \eref{dual-AbGIgen}, is an 
ungauging operation on $Y^{(p_1-1,p_2-2)}$. From \eref{PFAbEx2gen5} and \eref{dual-AbGIgen}, the partition function of the theory 
$\wt{\CO}''(Y^{(p_1-1,p_2-2)})$ is given as
\begin{align}
 Z^{\wt{\CO}''(Y^{(p_1-1,p_2-2)})}
= \int \,& \prod^{4}_{\gamma'=1} \Big[d\s^{\gamma'}\Big]\, \prod^{p_2-2}_{j=1}Z_{\rm 1-loop}^{\rm bif}(\s^1, \s^2, m'^{(12)}_{{\rm bif}\,j}) \, \prod^{p_1-1}_{i=1}Z_{\rm 1-loop}^{\rm bif}(\s^1, \s^4, m'^{(14)}_{{\rm bif}\,i})\nn \\
& \times Z^{(Y)}_{\rm int}(\{\s^{\gamma'}\}, -\vec t,\{m^{(p_1-1)}_{F}, \wt{\mu}^1,0, \wt{\mu}^2, m^{(p_1-1)}_{F} + \mu\}) \nn \\
& \times \delta \Big(\xi^{(p_2-1)} + \sum_{\beta'} b^l_{\beta'} \eta^{(p_1-1, p_2-2)}_l + \s_2+\s_4-\s_1-\s_3 \Big).\label{PFAbEx2gen6}
\end{align}
To read off the precise Lagrangian for $\wt{\CO}''(Y^{(p_1-1,p_2-2)})$, we eliminate $\s_2$ using the delta function, and make the 
following change of variables:
\begin{align}\label{covAbEx2}
\wt{\s}^1=\s^1, \, \wt{\s}^2=\s^4 -\s^3, \, \wt{\s}^3=\s^4.
\end{align}
The matrix integral \eref{PFAbEx2gen6} can then be recast into the following form:
\begin{align}
& Z^{\wt{\CO}''(Y^{(p_1-1,p_2-2)})} = \int \,\prod^{3}_{\gamma'=1} \Big[d\wt{\s}^{\gamma'}\Big]\, Z_{\rm FI}(\{\wt{\s}^{\gamma'}\},m^{(p_1-1)}_{F}, \vec{\wt{\mu}}, \mu)\,\prod^{l-l_1}_{i=1}Z_{\rm 1-loop}^{\rm bif}(\wt{\s}^1, \wt{\s}^2, m'^{(12)}_{{\rm bif}\,i})\nn \\
&\times \prod^{l_2-l}_{j=1}Z_{\rm 1-loop}^{\rm bif}(\wt{\s}^2, \wt{\s}^3, m'^{(23)}_{{\rm bif}\,j})\,
 \prod^{p_1}_{k=1}Z_{\rm 1-loop}^{\rm bif}(\wt{\s}^3, \wt{\s}^1, m'^{(31)}_{{\rm bif}\,k})\, 
 \prod^{l_1}_{i_1=1}Z_{\rm 1-loop}^{\rm fund}(\wt{\s}^1, m'^{(1)}_{{\rm fund}\,i_1})\nn \\
&\times \prod^{p_2}_{i_2=1}Z_{\rm 1-loop}^{\rm fund}(\wt{\s}^2, m'^{(2)}_{{\rm fund}\,i_2})\,
 \prod^{n-l_2}_{i_3=1}Z_{\rm 1-loop}^{\rm fund}(\wt{\s}^3, m'^{(3)}_{{\rm fund}\,i_3}),\label{PFAbEx2gen7}\\
& Z_{\rm FI}(\{\wt{\s}^{\gamma'}\},m^{(p_1-1)}_{F}, \vec{\wt{\mu}}, \mu)=e^{2\pi i m^{(p_1-1)}_F \wt{\s}^1}\, e^{2\pi i \wt\mu \wt{\s}^2}\, 
e^{-2\pi i (\mu+m^{(p_1-1)}_F) \wt{\s}^3}, \, \wt\mu= \wt\mu^2 - \wt\mu^1.
\end{align}
The matrix integral \eref{PFAbEx2gen7} manifestly reproduces the partition function of the quiver $Y'$ in the final line of \figref{AbEx2GFI}. 
The mirror map for the dual pair $(X', Y')$ can be summarized as follows. 
The mass parameters of theory $Y'$, associated with the bifundamental and fundamental hypermultiplets, can be written as linear 
functions of the FI parameters of $X'$:
\begin{align}
& m'^{(12)}_{{\rm bif }\, i} = -t_{l_1+i} +\delta +\delta_2 -\delta_1, \quad i=1,\ldots,l-l_1,\\
& m'^{(23)}_{{\rm bif }\, j} = -t_{l+j} +\delta_3 -\delta_2, \quad j=1,\ldots,l_2-l,\\
& m'^{(31)}_{{\rm bif }\, k} = \begin{cases}
(\sum^{k}_{k_1=1} \wt{\eta}^{(k_1)} + \sum_{\beta=1,5}b^l_\beta t_l) +\delta_3 -\delta_1, & \text{if } k=1,\ldots,p_1-1\\
\delta +\delta_3 -\delta_1, & \text{if } k=p_1,
\end{cases}\\
& m'^{(1)}_{{\rm fund}\,i_1} = -t_{i_1} + \delta_1, \quad i_1=1,\ldots, l_1, \\
& m'^{(2)}_{{\rm fund}\,i_2} = \begin{cases}
(\sum^{i_2}_{j_2=1} \xi^{(j_2)} + b^l_2 t_l) -\delta -\delta_2,& \text{if } i_2=1,\ldots,p_2-2\\
-\delta - \delta_2, & \text{if } i_2=p_2-1,\\
-\delta_2, & \text{if } i_2=p_2,
\end{cases}\\
& m'^{(3)}_{{\rm fund}\,i_3} = -t_{l_2+i_3} -\delta_3 , \quad i_3=1,\ldots, n-l_2,
\end{align}
where $\delta= \sum^{p_2-1}_{j} \xi^{(j)} + \sum_{\beta=2,4} b^l_\beta t_l$, and $\delta_{1,2,3}$ are chosen such that $\sum_{i_1}m'^{(1)}_{{\rm fund}\,i_1}=\sum_{i_2}m'^{(2)}_{{\rm fund}\,i_2}
=\sum_{i_3}m'^{(3)}_{{\rm fund}\,i_3}=0$. 
The FI parameters of the theory $Y'$ are also given in terms of the mass parameters of 
$X'$:
\be
\eta'^1= m^{(p-1)}_F,\, \eta'^2=\wt\mu ,\, \eta'^3=-(m^{(p-1)}_F + \mu) .
\ee
Written in this fashion, the fundamental mass parameters of $Y'$ manifestly live in the Cartan subalgebra of 
$SU(l_1) \times SU(p_2) \times SU(n-l_2)$, while the bifundamental masses parametrize the Cartan subalgebra of 
$U(l-l_1) \times U(l_2-l) \times U(p_1)$. Combined together, the masses parametrize the Cartan subalgebra of the 
Higgs branch global symmetry group of $Y'$, i.e. 
$G^{Y'}_{H} = SU(l_1) \times SU(p_2) \times SU(n-l_2) \times U(l-l_1) \times U(l_2-l) \times U(p_1)$, 
while the FI parameters live in the Cartan subalgebra of the Coulomb branch global symmetry $G^{Y'}_{C}=U(1)^3$.\\
\end{itemize}

\subsection{Examples of Non-Abelian Quiver Pairs}\label{NAbExNew}

\subsubsection{Family ${\rm III}_{[p_1,p_2,p_3]}$: A single closed loop with bifundamental matter attached to a linear quiver tail}\label{FamilyIII}
\begin{figure}[htbp]
\begin{center}
\scalebox{.65}{\begin{tikzpicture}[node distance=2cm,cnode/.style={circle,draw,thick,minimum size=8mm},snode/.style={rectangle,draw,thick,minimum size=8mm},pnode/.style={rectangle,red,draw,thick,minimum size=8mm}]
\node[cnode] (1) at (-3,0) {$2$};
\node[cnode] (2) at (-2,2) {$1$};
\node[cnode] (3) at (0,2.5) {$1$};
\node[cnode] (4) at (2,2) {$1$};
\node[cnode] (5) at (3,0) {$1$};
\node[cnode] (6) at (2,-2) {$1$};
\node[cnode] (7) at (0,-2.5) {$1$};
\node[cnode] (8) at (-2,-2) {$1$};
\node[snode] (10) at (-5,0) {$3$};
\node[cnode] (11) at (4,0) {$1$};
\node[cnode] (12) at (5,0) {$1$};
\node[cnode] (13) at (7.5,0) {$1$};
\node[snode] (14) at (9,0) {$1$};
\draw[thick] (1) to [bend left=40] (2);
\draw[thick] (2) to [bend left=40] (3);
\draw[thick,dashed] (3) to [bend left=40] (4);
\draw[thick] (4) to [bend left=40] (5);
\draw[thick] (5) to [bend left=40] (6);
\draw[dashed] (6) to [bend left=40] (7);
\draw[thick] (7) to [bend left=40] (8);
\draw[thick] (8) to [bend left=40] (1);
\draw[thick] (1) -- (10);
\draw[thick] (5) -- (11);
\draw[thick] (11) -- (12);
\draw[dashed] (12) -- (13);
\draw[thick] (13) -- (14);
\node[text width=1cm](11) at (-2,2.5) {$1$};
\node[text width=1cm](12) at (0, 3) {$2$};
\node[text width=1cm](13) at (2, 2.5) {$p_1$};
\node[text width=1cm](15) at (2,-2.5) {$p_2$};
\node[text width=1.5 cm](16) at (0,-3.2){$2$};
\node[text width=1cm](17) at (-2.5,-2.5){$1$};
\node[text width=0.1cm](18) at (4, 0.6){$1$};
\node[text width=0.1cm](19) at (5, 0.6){$2$};
\node[text width=1cm](20) at (7.5, 0.6){$p_3-1$};
\node[text width=0.1cm](30) at (0,-4){$(X')$};
\end{tikzpicture}}
\qquad
\scalebox{.65}{\begin{tikzpicture}[node distance=2cm,cnode/.style={circle,draw,thick,minimum size=8mm},snode/.style={rectangle,draw,thick,minimum size=8mm},pnode/.style={rectangle,red,draw,thick,minimum size=8mm}]
\node[cnode] (1) at (-2,4) {$1$};
\node[cnode] (2) at (-2,0) {$2$};
\node[cnode] (3) at (2,0) {$2$};
\node[cnode] (4) at (2,4) {$1$};
\node[snode] (5) at (-2,6) {$p_2$};
\node[snode] (6) at (-2,-2) {$1$};
\node[snode] (8) at (2,6) {$p_1$};
\node[snode] (7) at (2,-2) {$1$};
\draw[thick] (1) -- (2);
\draw[thick] (2) -- (3);
\draw[thick] (3) -- (4);
\draw[thick] (1) to [bend left=10] (4);
\draw[thick] (1) to [bend right=10] (4);
\draw[thick] (1) to [bend left=20] (4);
\draw[thick] (1) to [bend right=20] (4);
\draw[thick] (1) -- (5);
\draw[thick] (2) -- (6);
\draw[thick] (3) -- (7);
\draw[thick] (4) -- (8);
\draw[thick, dotted] (0,4.2) to (0,3.8);
\node[text width=1cm](15) at (0.2, 3){$p_3$};
\node[text width=1cm](30) at (0,-2){$(Y')$};
\end{tikzpicture}}
\caption{An infinite family of mirror duals labelled by the positive integers $(p_1,p_2,p_3)$ subject to the constraints
$p_1\geq 1$, $p_2 \geq 1$, and $p_3>1$.}
\label{AbEx3gen}
\end{center}
\end{figure}
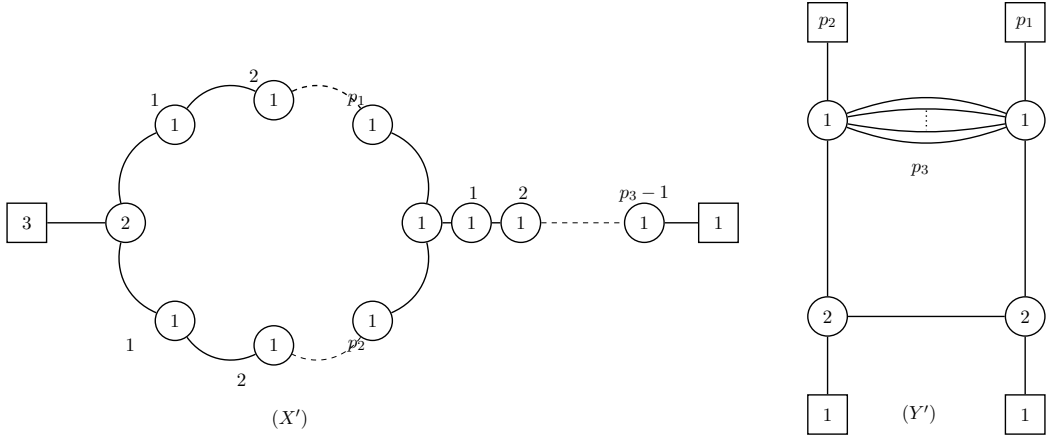

The first example in the class of non-Abelian mirrors is the infinite family of mirror duals shown in \figref{AbEx3gen}, labelled by three positive integers 
$(p_1,p_2,p_3)$ with the constraints $p_1\geq 1$, $p_2 \geq 1$, and $p_3>1$. The theory $X'$ has the shape of a single closed loop 
attached to a linear quiver tail, where the loop has a single $U(2)$ gauge node and $p_1+p_2+1$ $U(1)$ gauge nodes. 
The dual theory $Y'$ is a quiver with four gauge nodes, two of which are connected by multiple edges. 
The dimensions of the respective Higgs and Coulomb branches, and the associated global symmetries are shown in Table \ref{Tab:AbEx3gen}.\\

The quiver pair $(X', Y')$ in \figref{AbEx3gen} can be generated from the linear quiver pair $(X, Y)$ in \figref{AbEx3LIN} by a sequence of 
elementary Abelian operations on $X$ and the dual operations on $Y$, as shown in \figref{AbEx3GFI}. The Higgs branch global symmetry 
of $X$ is given by $G^{X}_{H} = (U(1)\times U(3) \times U(1))/ U(1)$. We choose to quotient with the overall $U(1)$ such that 
$G^{X}_{H} =U(1)\times SU(3) \times U(1)$.  Using the notation introduced in \Secref{LQ-IIB}, the mass parameters are 
labelled as $\vec m =(m_1, m_2, m_3,m_4,m_5)$, and the quotient by the overall $U(1)$ factor in the flavor symmetry of quiver $X$ can be 
implemented by the constraint $m_2+m_3+m_4=0$. The $U(1) \times U(1)$ subgroup of the flavor symmetry of $X$ can then be identified with the two 
terminal $U(1)$ flavor nodes, parametrized by the independent $(m_1, m_5)$.

\begin{center}
\begin{table}[htbp]
\resizebox{\textwidth}{!}{%
\begin{tabular}{|c|c|c|}
\hline
Moduli space data & Theory $X'$ & Theory $Y'$ \\
\hline \hline 
dim\,$\CM_H$ & 6 & $2+p_1+p_2 +p_3$\\
\hline
dim\,$\CM_C$ & $2+p_1+p_2 +p_3$ & 6\\
\hline
$G_H$ & $U(3) \times U(1)$ & $SU(p_1) \times SU(p_2) \times SU(p_3) \times U(1)^4$\\
\hline
$G_C$ & $SU(p_1) \times SU(p_2) \times SU(p_3) \times U(1)^4$ & $U(3) \times U(1)$ \\
\hline
\end{tabular}}
\caption{Summary table for the moduli space dimensions and global symmetries for the mirror pair in \figref{AbEx3gen}.}
\label{Tab:AbEx3gen}
\end{table}
\end{center}

\begin{figure}[htbp]
\begin{center}
\begin{tikzpicture}[
cnode/.style={circle,draw,thick,minimum size=4mm},snode/.style={rectangle,draw,thick,minimum size=8mm},pnode/.style={rectangle,red,draw,thick,minimum size=1.0cm}]
\node[cnode] (1) {$1$};
\node[cnode] (2) [right=.5cm  of 1]{$1$};
\node[cnode] (3) [right=1.5cm of 2]{1};
\node[cnode] (4) [right=.5cm of 3]{2};
\node[cnode] (5) [right=.5cm of 4]{1};
\node[cnode] (6) [right=1.5cm of 5]{1};
\node[cnode] (7) [right=0.5cm of 6]{1};
\node[snode] (10) [below=0.5cm of 1]{1};
\node[snode] (13) [below=0.5cm of 4]{3};
\node[snode] (16) [below=0.5cm of 7]{1};
\draw[-] (1) -- (2);
\draw[dashed] (2) -- (3);
\draw[-] (3) -- (4);
\draw[-] (4) --(5);
\draw[dashed] (5) --(6);
\draw[-] (6) -- (7);
\draw[-] (1) -- (10);
\draw[-] (4) -- (13);
\draw[-] (7) -- (16);
\node[text width=.5cm](9)[above=0.2cm of 3]{$1$};
\node[text width=.5cm](10)[above=0.2cm of 5]{$1$};
\node[text width=1 cm](11)[above=0.2cm of 6]{$p_1-1$};
\node[text width=1cm](12)[above=0.2cm of 2]{$p_2-1$};
\node[text width=0.1 cm](13)[above=0.2cm of 1]{$p_2$};
\node[text width=0.1 cm](14)[above=0.2cm of 7]{$p_1$};
\node[text width=1cm](20)[below=2.5cm of 4]{$(X)$};
\end{tikzpicture}
\qquad \qquad \qquad \qquad \qquad \qquad 
{\begin{tikzpicture}[
cnode/.style={circle,draw,thick, minimum size=4mm},snode/.style={rectangle,draw,thick,minimum size=8mm}]
\node[cnode] (1) {$1$};
\node[cnode] (2) [right=.75cm  of 1]{2};
\node[cnode] (3) [right=.75cm of 2]{2};
\node[cnode] (4) [right=.75cm of 3]{1};
\node[snode] (5) [below=0.5cm of 1]{$p_2$};
\node[snode] (6) [below=0.5cm of 2]{$1$};
\node[snode] (7) [below=0.5cm of 3]{$1$};
\node[snode] (8) [below=0.5cm of 4]{$p_1$};
\draw[-] (1) -- (2);
\draw[-] (2) -- (3);
\draw[-] (3) -- (4);
\draw[-] (1) --(5);
\draw[-] (2) --(6);
\draw[-] (3) -- (7);
\draw[-] (4) -- (8);
\node[text width=0.1cm](10)[above=0.2 cm of 1]{1};
\node[text width=0.1cm](11)[above=0.2 cm of 2]{2};
\node[text width=0.1cm](12)[above=0.2 cm of 3]{3};
\node[text width=0.1cm](13)[above=0.2 cm of 4]{4};
\node[text width=0.1cm](20)[below=2.5cm of 2]{$(Y)$};
\end{tikzpicture}}
\caption{Pair of linear quivers with $U(1)$ gauge groups which generate the 3d mirrors in \figref{AbEx3gen}.}
\label{AbEx3LIN}
\end{center}
\end{figure}

\begin{figure}[htbp]
\begin{center}
\begin{tabular}{ccc}
\scalebox{.6}{\begin{tikzpicture}[
cnode/.style={circle,draw,thick,minimum size=4mm},snode/.style={rectangle,draw,thick,minimum size=8mm},pnode/.style={rectangle,red,draw,thick,minimum size=1.0cm}]
\node[cnode] (1) {$1$};
\node[cnode] (2) [right=.5cm  of 1]{$1$};
\node[cnode] (3) [right=1.5cm of 2]{1};
\node[cnode] (4) [right=.5cm of 3]{2};
\node[cnode] (5) [right=.5cm of 4]{1};
\node[cnode] (6) [right=1.5cm of 5]{1};
\node[cnode] (7) [right=0.5cm of 6]{1};
\node[pnode] (10) [below=0.5cm of 1]{1};
\node[snode] (13) [below=0.5cm of 4]{3};
\node[pnode] (16) [below=0.5cm of 7]{1};
\draw[-] (1) -- (2);
\draw[dashed] (2) -- (3);
\draw[-] (3) -- (4);
\draw[-] (4) --(5);
\draw[dashed] (5) --(6);
\draw[-] (6) -- (7);
\draw[-] (1) -- (10);
\draw[-] (4) -- (13);
\draw[-] (7) -- (16);
\node[text width=.5cm](9)[above=0.2cm of 3]{$1$};
\node[text width=.5cm](10)[above=0.2cm of 5]{$1$};
\node[text width=1 cm](11)[above=0.2cm of 6]{$p_1-1$};
\node[text width=1cm](12)[above=0.2cm of 2]{$p_2-1$};
\node[text width=0.1 cm](13)[above=0.2cm of 1]{$p_2$};
\node[text width=0.1 cm](14)[above=0.2cm of 7]{$p_1$};
\node[text width=1cm](20)[below=2.5cm of 4]{$(X)$};
\end{tikzpicture}}
& \qquad 
& \scalebox{.6}{\begin{tikzpicture}[
cnode/.style={circle,draw,thick, minimum size=4mm},snode/.style={rectangle,draw,thick,minimum size=8mm}]
\node[cnode] (1) {$1$};
\node[cnode] (2) [right=.75cm  of 1]{2};
\node[cnode] (3) [right=.75cm of 2]{2};
\node[cnode] (4) [right=.75cm of 3]{1};
\node[snode] (5) [below=0.5cm of 1]{$p_2$};
\node[snode] (6) [below=0.5cm of 2]{$1$};
\node[snode] (7) [below=0.5cm of 3]{$1$};
\node[snode] (8) [below=0.5cm of 4]{$p_1$};
\draw[-] (1) -- (2);
\draw[-] (2) -- (3);
\draw[-] (3) -- (4);
\draw[-] (1) --(5);
\draw[-] (2) --(6);
\draw[-] (3) -- (7);
\draw[-] (4) -- (8);
\node[text width=0.1cm](10)[above=0.2 cm of 1]{1};
\node[text width=0.1cm](11)[above=0.2 cm of 2]{2};
\node[text width=0.1cm](12)[above=0.2 cm of 3]{3};
\node[text width=0.1cm](13)[above=0.2 cm of 4]{4};
\node[text width=0.1cm](20)[below=2.5cm of 2]{$(Y)$};
\end{tikzpicture}}\\
 \scalebox{.6}{\begin{tikzpicture}
\draw[->] (15,-3) -- (15,-5);
\node[text width=1cm](20) at (14, -4) {$\prod_i\CO_i$};
\end{tikzpicture}}
&\qquad
& \scalebox{.5}{\begin{tikzpicture}
\draw[->] (15,-3) -- (15,-5);
\node[text width=0.1cm](29) at (15.5, -4) {$\prod_i \wt{\CO}_i$};
\end{tikzpicture}}\\
\scalebox{.6}{\begin{tikzpicture}[node distance=2cm,cnode/.style={circle,draw,thick,minimum size=8mm},snode/.style={rectangle,draw,thick,minimum size=8mm},pnode/.style={rectangle,red,draw,thick,minimum size=8mm}]
\node[cnode] (1) at (-3,0) {$2$};
\node[cnode] (2) at (-2,2) {$1$};
\node[cnode] (3) at (0,2.5) {$1$};
\node[cnode] (4) at (2,2) {$1$};
\node[cnode] (5) at (3,0) {$1$};
\node[cnode] (6) at (2,-2) {$1$};
\node[cnode] (7) at (0,-2.5) {$1$};
\node[cnode] (8) at (-2,-2) {$1$};
\node[snode] (10) at (-5,0) {$3$};
\node[cnode] (11) at (4,0) {$1$};
\node[cnode] (12) at (5,0) {$1$};
\node[cnode] (13) at (7.5,0) {$1$};
\node[snode] (14) at (9,0) {$1$};
\draw[thick] (1) to [bend left=40] (2);
\draw[thick] (2) to [bend left=40] (3);
\draw[thick,dashed] (3) to [bend left=40] (4);
\draw[thick] (4) to [bend left=40] (5);
\draw[thick] (5) to [bend left=40] (6);
\draw[dashed] (6) to [bend left=40] (7);
\draw[thick] (7) to [bend left=40] (8);
\draw[thick] (8) to [bend left=40] (1);
\draw[thick] (1) -- (10);
\draw[thick] (5) -- (11);
\draw[thick] (11) -- (12);
\draw[dashed] (12) -- (13);
\draw[thick] (13) -- (14);
\node[text width=1cm](11) at (-2,2.5) {$1$};
\node[text width=1cm](12) at (0, 3) {$2$};
\node[text width=1cm](13) at (2, 2.5) {$p_1$};
\node[text width=1cm](15) at (2,-2.5) {$p_2$};
\node[text width=1.5 cm](16) at (0,-3.2){$2$};
\node[text width=1cm](17) at (-2.5,-2.5){$1$};
\node[text width=0.1cm](18) at (4, 0.6){$1$};
\node[text width=0.1cm](19) at (5, 0.6){$2$};
\node[text width=1cm](20) at (7.5, 0.6){$p_3-1$};
\node[text width=0.1cm](30) at (0,-4){$(X')$};
\end{tikzpicture}}
&\qquad 
&{\scalebox{0.6}{\begin{tikzpicture}[node distance=2cm,cnode/.style={circle,draw,thick,minimum size=8mm},snode/.style={rectangle,draw,thick,minimum size=8mm},pnode/.style={rectangle,red,draw,thick,minimum size=8mm}]
\node[cnode] (1) at (-2,4) {$1$};
\node[cnode] (2) at (-2,0) {$2$};
\node[cnode] (3) at (2,0) {$2$};
\node[cnode] (4) at (2,4) {$1$};
\node[snode] (5) at (-2,6) {$p_2$};
\node[snode] (6) at (-2,-2) {$1$};
\node[snode] (8) at (2,6) {$p_1$};
\node[snode] (7) at (2,-2) {$1$};
\draw[thick] (1) -- (2);
\draw[thick] (2) -- (3);
\draw[thick] (3) -- (4);
\draw[thick] (1) to [bend left=10] (4);
\draw[thick] (1) to [bend right=10] (4);
\draw[thick] (1) to [bend left=20] (4);
\draw[thick] (1) to [bend right=20] (4);
\draw[thick] (1) -- (5);
\draw[thick] (2) -- (6);
\draw[thick] (3) -- (7);
\draw[thick] (4) -- (8);
\node[text width=0.1cm](10)[left=0.2 cm of 1]{1};
\node[text width=0.1cm](11)[left=0.2 cm of 2]{2};
\node[text width=0.1cm](12)[right=0.2 cm of 3]{3};
\node[text width=0.1cm](13)[right=0.2 cm of 4]{4};
\draw[thick, dotted] (0,4.2) to (0,3.8);
\node[text width=1cm](15) at (0.2, 3){$p_3$};
\node[text width=1cm](30) at (0,-2){$(Y')$};
\end{tikzpicture}}}
\end{tabular}
\caption{The quiver $(X')$ and its mirror dual $(Y')$ in \figref{AbEx3gen} is generated by a sequence of 
elementary $S$-type operations on different nodes. In each step, the flavor node(s) on which the $S$-type operation 
acts is shown in red.}
\label{AbEx3GFI}
\end{center}
\end{figure}

To obtain $X'$ from $X$, we first perform an identification-flavoring-gauging operation $\CO_1$ on the flavor nodes (marked in red) at the 
two ends of the quiver $X$, with $N^\alpha_F=1$. This is followed by a sequence of flavoring-gauging operations $\{\CO_i\}$ ($i=2,\ldots,p_3$), with 
$N^\alpha_F=1$, implemented at the new flavor node obtained in the previous step. The procedure is precisely the same as the ones 
described for Family I in \Secref{FamilyI}. As before, we adopt the notation:
\begin{align}
& X^{(i)}= \CO_i \circ  \CO_{i-1} \circ \ldots \circ \CO_2 \circ \CO_1(X), \\
& Y^{(i)}= \wt\CO_i \circ  \wt\CO_{i-1} \circ \ldots \circ \wt\CO_2 \circ \wt\CO_1(Y), 
\end{align}
where $\CO_1$ is an identification-flavoring-gauging operation, and $\CO_i$ ($i > 1$) are flavoring-gauging operations. 
In this notation, we have
\be
X' = X^{(p_3)}, \, Y' = Y^{(p_3)}.
\ee
The $\CN=4$-preserving deformations of the quivers $X^{(i)}$ and $Y^{(i)}$ are denoted as $(\vec m^{(i)}, \vec \eta^{(i)})$ 
and $(\vec m'^{(i)}, \vec \eta'^{(i)})$ respectively. We will denote the deformation parameters of the dual theories 
$X'$ and $Y'$ as $(\vec m, \vec \eta)$ and $(\vec m', \vec \eta')$ respectively, which 
obviously implies
\be
(\vec m, \vec \eta)=(\vec m^{(p_3)}, \vec \eta^{(p_3)}), \quad (\vec m', \vec \eta')=(\vec m'^{(p_3)}, \vec \eta'^{(p_3)}).
\ee
Finally, the masses associated with the identification operation are parametrized as $\vec \mu$, while the mass associated with flavoring operation 
at the $i$-th step is $m^\alpha_F=m^{(i)}_F$. The FI parameter associated to the gauging operation at the $i$-th step is $\eta_\alpha=\wt{\eta}^{(i)}$.\\

Following the notation in \Secref{AbS}, we define the variables $\{ \vec u^\beta \}$, which parametrize the $U(1) \times U(1)$ 
mass parameters of the quiver $X$:
\be \label{defuAbEx1}
u^1=m_1, u^2=m_5,
\ee 
and implement the operations $\{\CO_i\}$, as shown in \figref{AbEx3GFI}. Since the computation is very similar to the one discussed 
in \Secref{FamilyI}, we simply present the answer. Following the general formulae in \eref{PFGFILinNf1}-\eref{HyperGFILinNf1} and \eref{HyperGFgenNf1}-\eref{PFGFgenNf1}, the partition function of the theory $Y^{(p_3)}$ is given as 
\begin{align}
& Z^{Y^{(p_3)}}(\vec m'^{(p_3)}, \vec \eta'^{(p_3)})\nn \\
&= \int \,\prod^{4}_{\gamma'=1} \Big[d\vec\s^{\gamma'}\Big]\, \prod^{p_3}_{i=1}Z_{\rm 1-loop}^{\rm bif}(\s^1, \s^4, m'^{(14)}_{{\rm bif}\,i})\,Z^{(Y)}_{\rm int}(\{\vec\s^{\gamma'}\}, -\vec t,
\{m^{(p_3)}_{F}, m_2, m_3, m_4, m^{(p_3)}_{F} + \mu\}),\label{PFAbEx3gen1}
\end{align}
which manifestly is the partition function for the quiver $Y'$, as shown in the second line of the \figref{AbEx3GFI}. 
The bifundamental masses $\vec{m}'_{\rm bif}$ in the above expression are given as
\be
m'^{(14)}_{{\rm bif}\,i} = - \sum^{i}_{k=1}\wt{\eta}^{(k)}  -\sum_{\beta=1,5} b^{l}_\beta t_l.
\ee

After a change of integration variables, the mirror map for the dual pair $(X', Y')$ can be summarized as follows. 
The mass parameters of theory $Y'$, associated with the bifundamental and fundamental hypermultiplets, can be written as linear 
functions of the FI parameters of $X'$:
\begin{align}
& m'^{(14)}_{{\rm bif}\,i} = - \sum^{i}_{k=1}\wt{\eta}^{(k)}  -\sum_{\beta=1,5} b^{l}_\beta t_l, \\
& m'^{(12)}_{{\rm bif}}=0,\, m'^{(23)}_{{\rm bif}}=0,\, m'^{(34)}_{{\rm bif}}=0,\\
& m'^{1}_{{\rm fund}\,i_1} = -t_{i_1} + \delta, \quad i_1=1,\ldots, p_2, \\
& m'^{2}_{{\rm fund}} = -t_{p_2+1} +\delta,\\
& m'^{3}_{{\rm fund}} = -t_{p_2+2} +\delta ,\\
& m'^{4}_{{\rm fund}\,i_2} = -t_{p_2+2+i_4} +\delta , \quad i_4=1,\ldots, p_1,
\end{align}
where we choose $\delta$ such that $\delta=\frac{\sum_{i_1}  t_{i_1}}{p_2}$. 
The FI parameters of the theory $Y'$ are also given in terms of the mass parameters of 
$X'$:
\be
\eta'^1=m^{(p_3)}_{F} -m_2,\, \eta'^2=m_2-m_3,\, \eta'^3=m_3-m_4,\, \eta'^4=m_4-(m^{(p-1)}_F + \mu).
\ee
The mass parameters of $Y'$ manifestly live in the Cartan subalgebra of the Higgs branch global symmetry 
$G^{Y'}_{H}=SU(p_2) \times U(p_3) \times U(p_1) \times U(1)^2$, while the FI parameters live in the Cartan subalgebra
of the Coulomb branch global symmetry $G^{Y'}_{C}=U(3) \times U(1)$. The Coulomb branch global symmetry can be read 
off from the observation that $Y'$ contains a balanced sub-quiver with two gauge nodes (labelled 2 and 3) which gives an $SU(3)$ 
factor, along with two unbalanced nodes which contribute a $U(1)$ factor each.

\subsubsection{Family ${\rm IV}_{[p_1,p_2,p_3]}$: A single closed loop with bifundamental and rank-2 antisymmetric matter attached to a linear quiver tail}\label{FamilyIV}

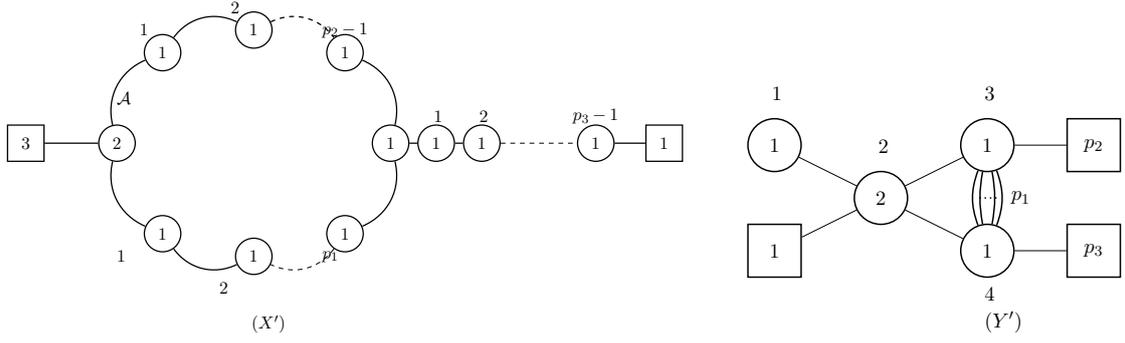
\begin{figure}[htbp]
\begin{center}
\begin{tabular}{ccc}
\scalebox{.6}{\begin{tikzpicture}[node distance=2cm,cnode/.style={circle,draw,thick,minimum size=8mm},snode/.style={rectangle,draw,thick,minimum size=8mm},pnode/.style={rectangle,red,draw,thick,minimum size=8mm}]
\node[cnode] (1) at (-3,0) {$2$};
\node[cnode] (2) at (-2,2) {$1$};
\node[cnode] (3) at (0,2.5) {$1$};
\node[cnode] (4) at (2,2) {$1$};
\node[cnode] (5) at (3,0) {$1$};
\node[cnode] (6) at (2,-2) {$1$};
\node[cnode] (7) at (0,-2.5) {$1$};
\node[cnode] (8) at (-2,-2) {$1$};
\node[snode] (10) at (-5,0) {$3$};
\node[cnode] (11) at (4,0) {$1$};
\node[cnode] (12) at (5,0) {$1$};
\node[cnode] (13) at (7.5,0) {$1$};
\node[snode] (14) at (9,0) {$1$};
\draw[thick] (1) to [bend left=40] (2);
\draw[thick] (2) to [bend left=40] (3);
\draw[thick,dashed] (3) to [bend left=40] (4);
\draw[thick] (4) to [bend left=40] (5);
\draw[thick] (5) to [bend left=40] (6);
\draw[dashed] (6) to [bend left=40] (7);
\draw[thick] (7) to [bend left=40] (8);
\draw[thick] (8) to [bend left=40] (1);
\draw[thick] (1) -- (10);
\draw[thick] (5) -- (11);
\draw[thick] (11) -- (12);
\draw[dashed] (12) -- (13);
\draw[thick] (13) -- (14);
\node[text width=1cm](11) at (-2,2.5) {$1$};
\node[text width=1cm](12) at (0, 3) {$2$};
\node[text width=1cm](13) at (2, 2.5) {$p_2-1$};
\node[text width=1cm](15) at (2,-2.5) {$p_1$};
\node[text width=1.5 cm](16) at (0,-3.2){$2$};
\node[text width=1cm](17) at (-2.5,-2.5){$1$};
\node[text width=0.1cm](18) at (4, 0.6){$1$};
\node[text width=0.1cm](19) at (5, 0.6){$2$};
\node[text width=1cm](20) at (7.5, 0.6){$p_3-1$};
\node[text width=1cm](21) at (-2.5, 1){$\CA$};
\node[text width=0.1cm](30) at (0,-4){$(X')$};
\end{tikzpicture}}
&\qquad
&\scalebox{.7}{\begin{tikzpicture}[
cnode/.style={circle,draw,thick, minimum size=1.0cm},snode/.style={rectangle,draw,thick,minimum size=1cm}]
\node[cnode] (9) at (0,1){1};
\node[snode] (10) at (0,-1){1};
\node[cnode] (11) at (2, 0){2};
\node[cnode] (12) at (4, 1){1};
\node[cnode] (13) at (4, -1){1};
\node[snode] (14) at (6, 1){$p_2$};
\node[snode] (15) at (6, -1){$p_3$};
\draw[-] (9) -- (11);
\draw[-] (10) -- (11);
\draw[-] (12) -- (11);
\draw[-] (13) -- (11);
\draw[-] (12) -- (14);
\draw[-] (13) -- (15);
\draw[thick] (12) to [bend left=10] (13);
\draw[thick] (12) to [bend right=10] (13);
\draw[thick] (12) to [bend left=20] (13);
\draw[thick] (12) to [bend right=20] (13);
\draw[thick, dotted] (3.85,0) to (4.2, 0);
\node[text width=0.1cm](20) at (4.5,0){$p_1$};
\node[text width=0.1cm](21)[above=0.2 cm of 9]{1};
\node[text width=0.1cm](22)[above=0.2 cm of 11]{2};
\node[text width=0.1cm](23)[above=0.2 cm of 12]{3};
\node[text width=0.1cm](24)[below=0.05 cm of 13]{4};
\node[text width=0.1cm](31)[below=0.5 cm of 13]{$(Y')$};
\end{tikzpicture}}
\end{tabular}
\caption{An infinite family of mirror duals labelled by the positive integers $(p_1,p_2,p_3)$ subject to the constraints
$p_1\geq 1$, $p_2 \geq 1$, and $p_3>1$. The label $\CA$ denotes a hypermultiplet transforming in a rank-2 
antisymmetric representation of $U(2)$ and having charge 1 under $U(1)_1$.
}
\label{fig: AbEx4gen}
\end{center}
\end{figure}

The second example in the class of non-Abelian mirrors is the infinite family of mirror duals shown in \figref{fig: AbEx4gen}, labelled by three positive integers 
$(p_1,p_2,p_3)$ with the constraints $p_1\geq 1$, $p_2 \geq 1$, and $p_3>1$. The theory $X'$ has the shape of a single closed loop 
attached to a linear quiver tail, where the loop has a single $U(2)$ gauge node and $p_1+p_2+1$ $U(1)$ gauge nodes. One of the hypermultiplets 
in the loop transforms in the rank-2 antisymmetric representation of $U(2)$ (i.e. it is charged $+2$ under the $U(1)$ subgroup of $U(2)$ and singlet under the 
$SU(2)$) and is charged $+1$ under an adjacent $U(1)$, while all the other hypermultiplets in the loop transform in the bifundamental representations as shown. 
 The dual theory $Y'$ is a quiver with four gauge nodes, two of which are connected by multiple edges. The dimensions of the respective Higgs and Coulomb branches, and the associated global symmetries are shown in Table \ref{Tab:AbEx4}.\\

\begin{figure}[htbp]
\begin{center}
\scalebox{0.7}{\begin{tikzpicture}[
cnode/.style={circle,draw,thick, minimum size=1.0cm},snode/.style={rectangle,draw,thick,minimum size=1cm}]
\node[cnode] (9) at (10,0){2};
\node[snode] (10) [below=1cm of 9]{4};
\draw[-] (9) -- (10);
\node[snode] (11) [right=1cm of 9]{1};
\draw[-] (9) -- (11);
\node[text width=0.1cm](12) at (10.75,0.5){$\CA$};
\node[text width=0.1cm](21)[below=0.5 cm of 10]{$(X)$};
\end{tikzpicture}}
\qquad \qquad \qquad
\scalebox{0.7}{\begin{tikzpicture}[
cnode/.style={circle,draw,thick, minimum size=1.0cm},snode/.style={rectangle,draw,thick,minimum size=1cm}]
\node[cnode] (9) at (10,1){1};
\node[snode] (10) at (10,-1){1};
\node[cnode] (11) at (12, 0){2};
\node[cnode] (12) at (14, 1){1};
\node[cnode] (13) at (14, -1){1};
\node[snode] (14) at (16, 1){1};
\draw[-] (9) -- (11);
\draw[-] (10) -- (11);
\draw[-] (12) -- (11);
\draw[-] (13) -- (11);
\draw[-] (12) -- (14);
\node[text width=0.1cm](21)[above=0.2 cm of 9]{1};
\node[text width=0.1cm](22)[above=0.2 cm of 11]{2};
\node[text width=0.1cm](23)[above=0.2 cm of 12]{3};
\node[text width=0.1cm](24)[above=0.05 cm of 13]{4};
\node[text width=0.1cm](21)[below=0.5 cm of 13]{$(Y)$};
\end{tikzpicture}}
\caption{Mirror duals involving a $D_4$ quiver on one side. This dual pair $(X,Y)$ will be our starting point for arriving 
at the duality in \figref{fig: AbEx4gen}. The dual pair can, in turn, be obtained from a pair of linear quivers as discussed 
in \Secref{AbGD4}.}
\label{fig: AbEx4D}
\end{center}
\end{figure}
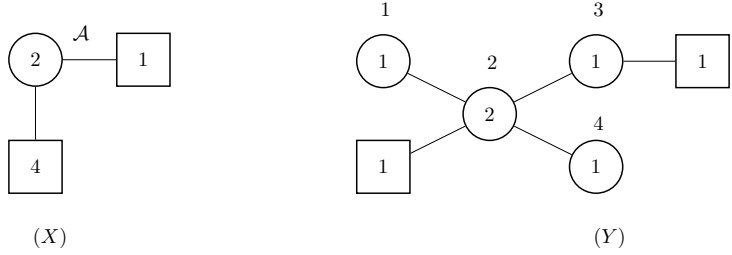

The quiver pair $(X', Y')$ in \figref{fig: AbEx4gen} can be generated from the quiver pair $(X, Y)$ in \figref{fig: AbEx4D} by a series of 
elementary Abelian $S$-type operations on $X$ and the dual operations on $Y$ (shown in \figref{AbEx4GFI}), which we shall describe 
momentarily. Note that the quiver pair $(X, Y)$ are not linear quivers, but can be easily generated from a pair of linear quivers using a 
single Abelian gauging operation, as shown in \figref{fig: AbEx4DLin} and discussed in \Secref{AbGD4} \footnote{In the notation of 
\Secref{AbGD4}, $X_{\rm here}=Y''_{\rm there}$, and $Y_{\rm here}=X'_{\rm there}$. One can also think of generating the 
dual pair in \figref{fig: AbEx4gen} from a pair of linear quivers by first implementing a gauging operation of a Coulomb branch 
global symmetry and then implementing a sequence of different Abelian $S$-type operations which gauge a subgroup of the Higgs branch global symmetry.} .\\

The Higgs branch global symmetry of $X$ is given by $G^{X}_{H} =(U(4) \times U(1)_{\CA})/U(1)$, where $U(1)_{\CA}$ is the 
flavor symmetry associated with the antisymmetric hypermultiplet, and the Coulomb branch global symmetry is $G^X_{\rm C} = U(1)$.
The fundamental mass parameters are labelled as $(m_1, m_2, m_3, m_4)$, and the mass parameter for the antisymmetric hypermultiplet 
as $m_{\CA}$. The quotient by the overall $U(1)$ factor can be implemented either by the constraint $\sum^4_{i=1} m_i=0$, or the constraint 
$m_{\CA}=0$. The Coulomb branch global symmetry of $Y$ is clearly given by $G^{Y}_{C} =SU(4) \times U(1)$, where the $SU(4)$ factor 
arises from the balanced linear subquiver of $Y$, consisting of the gauge nodes labelled 1, 2 and 3 in \figref{fig: AbEx4D}. The Cartan subalgebra 
of $G^{Y}_{C}$ is parametrized by the FI parameters $\{\eta'_l\}_{l=1,\ldots,4}$. The Higgs branch global symmetry of $X$ is given by 
$G^{Y}_{H} =(U(1)_2 \times U(1)_3)/U(1)$, where the Cartan of $U(1)_2 \times U(1)_3$ is parametrized by $(m'_2,m'_3)$.
We will implement the $U(1)$ quotient by choosing $m'_{3}=0$. The mirror map for the dual pair $(X,Y)$ relating masses and FI 
parameters across the duality is then given as follows (simply inverting the 
the relations \eref{MMD2a}-\eref{MMD2c} and interchanging the primed and unprimed parameters):
\begin{align}
& \eta'_l=-(m_l-m_{l+1}), \quad l=1,2,\label{MM4Da}\\
& \eta'_3=-m_3 -m_4 + m_{\CA}, \label{MM4Db}\\
& \eta'_4=m_4-m_3,\label{MM4Dc}\\
& m'_{2}=\eta,\, m'_{3}=0, \label{MM4Dd}
\end{align}
where it is understood that the masses of $X$ satisfy one of the constraints - $\sum^4_{i=1} m_i=0$, or $m_{\CA}=0$. 

\begin{center}
\begin{table}[htbp]
\resizebox{\textwidth}{!}{%
\begin{tabular}{|c|c|c|}
\hline
Moduli space data & Theory $X'$ & Theory $Y'$ \\
\hline \hline 
dim\,$\CM_H$ & 5 & $1+p_1+p_2 +p_3$\\
\hline
dim\,$\CM_C$ & $1+p_1+p_2 +p_3$ & 5\\
\hline
$G_H$ & $SU(3) \times U(1)^2$ & $SU(p_1) \times SU(p_2) \times SU(p_3) \times U(1)^3$\\
\hline
$G_C$ & $SU(p_1) \times SU(p_2) \times SU(p_3) \times U(1)^3$ & $SU(3) \times U(1)^2$ \\
\hline
\end{tabular}}
\caption{Summary table for the moduli space dimensions and global symmetries for the mirror pair in \figref{fig: AbEx4gen}.}
\label{Tab:AbEx4}
\end{table}
\end{center}

The theory $X'$ can be obtained from the quiver $X$ in three distinct set of steps, as shown in \figref{AbEx4GFI}, each involving a sequence of 
elementary Abelian operations which we describe below:
\begin{enumerate}
\item We impose the constraint $m_{\CA}=0$, and identify the Higgs branch global symmetry $G^{X}_{H} =U(4)$ 
with the $U(4)$ flavor node corresponding to the fundamental hypermultiplets. Then we perform an Abelian
flavoring-gauging operation $\CO_{1\,\CP}$ at the $U(4)$ flavor node, with $N^\alpha_F=1$, with the following 
choice of the permutation matrix $\CP$\footnote{One can explicitly show that, for the other choices of the matrix $\CP$, one 
either obtains the same dual Lagrangian $Y'$ in \figref{fig: AbEx4gen}, or obtains a Lagrangian which is related to $Y'$ 
by some simple field redefiniton.} : 
\be \label{P1}
\CP=\begin{pmatrix} 0 & 1 & 0 & 0\\ 0&0 & 1 & 0 \\ 0 & 0 & 0 & 1\\ 1 & 0 & 0 & 0\end{pmatrix}.
\ee
This is followed by a sequence of flavoring-gauging operations $\CO_i$ (for $i=2,\ldots,p_1$) acting at the new $U(1)$ flavor node generated 
in the previous step. At each step, the flavoring operation corresponds to $N^\alpha_F=1$. Note that only $\CO_{1\,\CP}$ 
depends on the permutation matrix $\CP$, while the rest of the operations $\CO_i$ ($i>1$) do not.
We adopt the following notation for the resultant quivers and their respective duals:
\begin{align}
& X^{(i)}_\CP= \CO_i \circ  \CO_{i-1} \circ \ldots \circ \CO_2 \circ \CO_{1\,\CP}(X), \\
& Y^{(i)}_\CP= \wt\CO_i \circ  \wt\CO_{i-1} \circ \ldots \circ \wt\CO_2 \circ \wt\CO_{1\,\CP}(Y), 
\end{align}
and denote the $\CN=4$-preserving deformations of the quivers $X^{(i)}_\CP$ and $Y^{(i)}_\CP$ as $(\vec m^{(i)}, \vec \eta^{(i)})$ 
and $(\vec m'^{(i)}, \vec \eta'^{(i)})$ respectively. The mass associated with flavoring at the $i$-th step is labelled as $m^{(i)}_F$, 
and the FI parameter associated with gauging operation at the $i$-th step is $\wt{\eta}^{(i)}$. The quivers $(X^{(p_1)}_\CP$ and 
its dual $Y^{(p_1)}_\CP)$ are shown in the second line of \figref{AbEx4GFI}.

\item The Higgs branch global symmetry of the quiver $X^{(p_1)}_\CP$ is $G^{X^{(p_1)}_\CP}_{H}=(U(3)\times U(1)_{\CA} \times U(1)) / U(1)$.
We will choose to implement the $U(1)$ quotient by constraining the masses of the hypers in the fundamental representation of $U(2)$, 
while leaving the $U(1)_{\CA}$ and $U(1)$ masses unconstrained. Then, we perform a sequence of $N^\alpha_F=1$ flavoring-gauging 
operations $\CO'_j$ ($j=1,\ldots,p_2-1$), starting at the $U(1)_{\CA}$ node, and acting on the new $U(1)$ flavor node in the subsequent 
steps. We will denote these quivers as:
\begin{align}
& X^{(p_1,j)}_\CP= \CO'_j \circ  \CO'_{j-1} \circ \ldots \circ \CO'_2 \circ \CO'_1(X^{(p_1)}_\CP), \\
& Y^{(p_1,j)}_\CP= \wt\CO'_i \circ  \wt\CO'_{i-1} \circ \ldots \circ \wt\CO'_2 \circ \wt\CO'_1(Y^{(p_1)}_\CP), 
\end{align}
and the associated $\CN=4$-preserving deformations will be denoted as $(\vec m^{(p_1,j)}, \vec \eta^{(p_1,j)})$ 
and $(\vec m'^{(p_1,j)}, \vec \eta'^{(p_1,j)})$ respectively. The mass associated with the flavoring operation at the 
$j$-th step is $x^{(j)}_F$, and the FI parameter associated with the gauging operation at the $j$-th step is $\xi^{(j)}$.
The resultant quiver $X^{(p_1,p_2-1)}_\CP$ and its dual $Y^{(p_1,p_2-1)}_\CP$ are shown in the third line of \figref{AbEx4GFI}.

\item In the final step, we implement an $N^\alpha_F=1$ identification-flavoring-gauging operation $\CO''_1$ on the two $U(1)$ flavor nodes of 
$X^{(p_1,p_2-1)}$, shown in red in the third line of \figref{AbEx4GFI}. This is again followed by a sequence of $N^\alpha_F=1$ flavoring-gauging 
operations $\CO''_k$ ($k=2,\ldots,p_3$), starting with the new $U(1)$ flavor node generated by $\CO''_1$. We will denote these quivers as:
\begin{align}
& X^{(p_1,p_2-1,k)}_\CP= \CO''_k \circ  \CO''_{k-1} \circ \ldots \circ \CO''_2 \circ \CO''_1(X^{(p_1,p_2-1)}_\CP), \\
& Y^{(p_1,p_2-1,k)}_\CP= \wt\CO''_k \circ  \wt\CO''_{k-1} \circ \ldots \circ \wt\CO''_2 \circ \wt\CO''_1(Y^{(p_1,p_2-1)}_\CP), 
\end{align}
and the associated $\CN=4$ -preserving deformations will be denoted as $(\vec m^{(p_1,p_2-1,k)}, \vec \eta^{(p_1,p_2-1,k)})$ 
and $(\vec m'^{(p_1,p_2-1,k)}, \vec \eta'^{(p_1,p_2-1,k)})$ respectively. The mass associated with the flavoring operation at the 
$k$-th step is $y^{(k)}_F$, and the FI parameter associated with the gauging operation at the $k$-th step is $\zeta^{(k)}$. The mass
associated with the identification operation is $\vec\mu$.
The resultant quiver $X^{(p_1,p_2-1,p_3)}_\CP=X'$ and its dual $Y^{(p_1,p_2-1,p_3)}_\CP=Y'$ are shown in the fourth line of \figref{AbEx4GFI}.

\end{enumerate}

\begin{figure}[htbp]
\begin{center}
\begin{tabular}{ccc}
\scalebox{.6}{\begin{tikzpicture}[
cnode/.style={circle,draw,thick, minimum size=1.0cm},snode/.style={rectangle,draw,thick,minimum size=1cm}]
\node[cnode] (1) {1};
\node[cnode] (2) [right=1cm  of 1]{2};
\node[cnode] (3) [right=1cm  of 2]{1};
\node[snode] (4) [below=1cm of 2]{2};
\node[snode] (5) [below=1cm of 3]{1};
\draw[-] (1) -- (2);
\draw[-] (2)-- (3);
\draw[-] (2)-- (4);
\draw[-] (3)-- (5);
\end{tikzpicture}}
& \qquad 
& \scalebox{.6}{\begin{tikzpicture}[
cnode/.style={circle,draw,thick, minimum size=1.0cm},snode/.style={rectangle,draw,thick,minimum size=1cm}]
\node[cnode] (1) {2};
\node[snode] (2) [below=1cm of 1]{3};
\node[cnode] (3) [right=1cm of 1]{1};
\node[snode] (4) [below=1cm of 3]{1};
\draw[-] (1) -- (2);
\draw[-] (1) -- (3);
\draw[-] (3) -- (4);
\end{tikzpicture}}\\
 \scalebox{.6}{\begin{tikzpicture}
\draw[->] (15,-3) -- (15,-5);
\node[text width=1cm](20) at (14, -4) {${\CO}^\alpha_{\vec \CP}$};
\end{tikzpicture}}
&\qquad
& \scalebox{.6}{\begin{tikzpicture}
\draw[->] (15,-3) -- (15,-5);
\node[text width=0.1cm](29) at (15.5, -4) {$\wt{\CO}_\CP$};
\end{tikzpicture}}\\
\scalebox{.6}{\begin{tikzpicture}[
cnode/.style={circle,draw,thick, minimum size=1.0cm},snode/.style={rectangle,draw,thick,minimum size=1cm}]
\node[cnode] (9) at (10,1){1};
\node[snode] (10) at (10,-1){1};
\node[cnode] (11) at (12, 0){2};
\node[cnode] (12) at (14, 1){1};
\node[cnode] (13) at (14, -1){1};
\node[snode] (14) at (16, 1){1};
\draw[-] (9) -- (11);
\draw[-] (10) -- (11);
\draw[-] (12) -- (11);
\draw[-] (13) -- (11);
\draw[-] (12) -- (14);
\end{tikzpicture}}
&\qquad 
&\scalebox{0.6}{\begin{tikzpicture}[
cnode/.style={circle,draw,thick, minimum size=1.0cm},snode/.style={rectangle,draw,thick,minimum size=1cm}]
\node[cnode] (9) at (10,0){2};
\node[snode] (10) [below=1cm of 9]{4};
\draw[-] (9) -- (10);
\node[snode] (11) [right=1cm of 9]{1};
\draw[-] (9) -- (11);
\node[text width=0.1cm](12) at (10.75,0.5){$\CA$};
\end{tikzpicture}}
\end{tabular}
\caption{The mirror dual pair $(X,Y)$ in \figref{fig: AbEx4D} can be derived from a pair of linear quivers shown in the 
top line by an Abelian gauging operation. The details of the computation can be found in \Secref{AbGD4}.}
\label{fig: AbEx4DLin}
\end{center}
\end{figure}
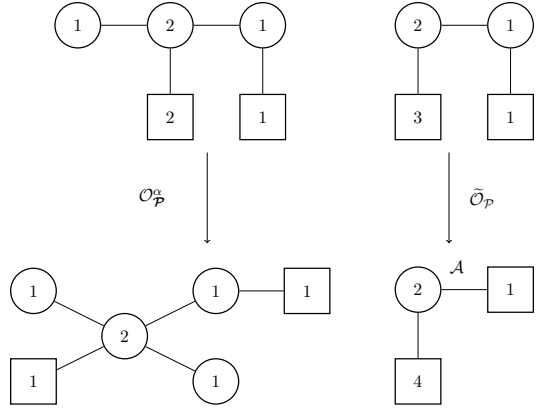

\begin{figure}[htbp]
\begin{center}
\begin{tabular}{ccc}
\scalebox{.5}{\begin{tikzpicture}[
cnode/.style={circle,draw,thick, minimum size=1.0cm},snode/.style={rectangle,draw,thick,minimum size=1cm},pnode/.style={rectangle,red,draw,thick,minimum size=8mm}]
\node[cnode] (9) at (10,0){2};
\node[pnode] (10) [below=1cm of 9]{4};
\draw[-] (9) -- (10);
\node[snode] (11) [right=1cm of 9]{1};
\draw[-] (9) -- (11);
\node[text width=0.1cm](12) at (10.75,0.5){$\CA$};
\node[text width=0.1cm](21)[below=0.5 cm of 10]{$(X)$};
\end{tikzpicture}}
& \qquad \qquad
& \scalebox{.5}{\begin{tikzpicture}[
cnode/.style={circle,draw,thick, minimum size=1.0cm},snode/.style={rectangle,draw,thick,minimum size=1cm}]
\node[cnode] (9) at (10,1){1};
\node[snode] (10) at (10,-1){1};
\node[cnode] (11) at (12, 0){2};
\node[cnode] (12) at (14, 1){1};
\node[cnode] (13) at (14, -1){1};
\node[snode] (14) at (16, 1){1};
\draw[-] (9) -- (11);
\draw[-] (10) -- (11);
\draw[-] (12) -- (11);
\draw[-] (13) -- (11);
\draw[-] (12) -- (14);
\node[text width=0.1cm](21)[above=0.2 cm of 9]{1};
\node[text width=0.1cm](22)[above=0.2 cm of 11]{2};
\node[text width=0.1cm](23)[above=0.2 cm of 12]{3};
\node[text width=0.1cm](24)[below=0.05 cm of 13]{4};
\node[text width=0.1cm](21)[below=1 cm of 11]{$(Y)$};
\end{tikzpicture}}\\
 \scalebox{.5}{\begin{tikzpicture}
\draw[->] (15,-3) -- (15,-5);
\node[text width=1cm](20) at (14, -4) {$\prod_i\CO_i$};
\end{tikzpicture}}
&\qquad \qquad
& \scalebox{.5}{\begin{tikzpicture}
\draw[->] (15,-3) -- (15,-5);
\node[text width=0.1cm](29) at (15.5, -4) {$\prod_i \wt{\CO}_i$};
\end{tikzpicture}}\\
\scalebox{.5}{\begin{tikzpicture}[
cnode/.style={circle,draw,thick, minimum size=1.0cm},snode/.style={rectangle,draw,thick,minimum size=1cm},
pnode/.style={rectangle,red,draw,thick,minimum size=8mm}]
\node[cnode] (9) at (10,0){2};
\node[pnode] (10) [below=1cm of 9]{1};
\node[snode] (17) [left=1cm of 9]{3};
\draw[-] (9) -- (10);
\draw[-] (9) -- (17);
\node[cnode] (11) [right=1cm of 9]{1};
\node[cnode] (12) [right=1cm of 11]{1};
\node[cnode] (13) [right=1cm of 12]{1};
\node[cnode] (14) [right=1.5cm of 13]{1};
\node[cnode] (15) [right=1cm of 14]{1};
\node[snode] (16) [below=1cm of 15]{1};
\draw[-] (9) -- (11);
\draw[-] (11) -- (12);
\draw[-] (12) -- (13);
\draw[thick, dashed] (13) -- (14);
\draw[-] (14) -- (15);
\draw[-] (15) -- (16);
\node[text width=0.1cm](17) at (10.25,-0.75){$\CA$};
\node[text width=0.1cm](18) [above=0.1cm of 11]{1};
\node[text width=0.1cm](19) [above=0.1cm of 12]{2};
\node[text width=0.1cm](20) [above=0.1cm of 13]{3};
\node[text width=1cm](21) [above=0.1cm of 14]{$p_1-1$};
\node[text width=1cm](22) [above=0.1cm of 15]{$p_1$};
\node[text width=0.1cm](21)[below=0.5 cm of 13]{$(X^{(p_1)})$};
\end{tikzpicture}}
&\qquad \qquad
& \scalebox{.5}{\begin{tikzpicture}[
cnode/.style={circle,draw,thick, minimum size=1.0cm},snode/.style={rectangle,draw,thick,minimum size=1cm}]
\node[cnode] (9) at (0,1){1};
\node[snode] (10) at (0,-1){1};
\node[cnode] (11) at (2, 0){2};
\node[cnode] (12) at (4, 1){1};
\node[cnode] (13) at (4, -1){1};
\node[snode] (14) at (6, 1){1};
\draw[-] (9) -- (11);
\draw[-] (10) -- (11);
\draw[-] (12) -- (11);
\draw[-] (13) -- (11);
\draw[-] (12) -- (14);
\draw[thick] (12) to [bend left=10] (13);
\draw[thick] (12) to [bend right=10] (13);
\draw[thick] (12) to [bend left=20] (13);
\draw[thick] (12) to [bend right=20] (13);
\draw[thick, dotted] (3.85,0) to (4.2, 0);
\node[text width=0.1cm](20) at (4.5,0){$p_1$};
\node[text width=0.1cm](21)[above=0.2 cm of 9]{1};
\node[text width=0.1cm](22)[above=0.2 cm of 11]{2};
\node[text width=0.1cm](23)[above=0.2 cm of 12]{3};
\node[text width=0.1cm](24)[below=0.05 cm of 13]{4};
\node[text width=0.1cm](21)[below=1 cm of 11]{$(Y^{(p_1)})$};
\end{tikzpicture}}\\
 \scalebox{.5}{\begin{tikzpicture}
\draw[->] (15,-3) -- (15,-5);
\node[text width=1cm](20) at (14, -4) {$\prod_j \CO'_j$};
\end{tikzpicture}}
&\qquad \qquad
&\scalebox{.5}{\begin{tikzpicture}
\draw[->] (15,-3) -- (15,-5);
\node[text width=0.1cm](29) at (15.5, -4) {$\prod_j \wt{\CO}'_j$};
\end{tikzpicture}}\\
\scalebox{.5}{\begin{tikzpicture}[
cnode/.style={circle,draw,thick, minimum size=1.0cm},snode/.style={rectangle,draw,thick,minimum size=1cm},pnode/.style={rectangle,red,draw,thick,minimum size=8mm}]
\node[cnode] (1) at (10,0){2};
\node[snode] (2) [left=1cm of 1]{3};
\draw[-] (1) -- (2);
\node[cnode] (3) [above=1cm of 1]{1};
\node[cnode] (4) [right=1cm of 3]{1};
\node[cnode] (5) [right=1cm of 4]{1};
\node[cnode] (6) [right=1.5cm of 5]{1};
\node[cnode] (7) [right=1cm of 6]{1};
\node[pnode] (8) [right=1cm of 7]{1};
\node[cnode] (9) [below=1cm of 1]{1};
\node[cnode] (10) [right=1cm of 9]{1};
\node[cnode] (11) [right=1cm of 10]{1};
\node[cnode] (12) [right=1.5cm of 11]{1};
\node[cnode] (13) [right=1cm of 12]{1};
\node[pnode] (14) [right=1cm of 13]{1};
\draw[-] (1) -- (3);
\draw[-] (3) -- (4);
\draw[-] (4) -- (5);
\draw[thick, dashed] (5) -- (6);
\draw[-] (6) -- (7);
\draw[-] (7) -- (8);
\draw[-] (1) -- (9);
\draw[-] (9) -- (10);
\draw[-] (10) -- (11);
\draw[thick, dashed] (11) -- (12);
\draw[-] (12) -- (13);
\draw[-] (13) -- (14);
\node[text width=0.1cm](17) at (9.5, 1){$\CA$};
\node[text width=0.1cm](18) [above=0.1cm of 3]{1};
\node[text width=0.1cm](19) [above=0.1cm of 4]{2};
\node[text width=0.1cm](20) [above=0.1cm of 5]{3};
\node[text width=1cm](21) [above=0.1cm of 6]{$p_2-2$};
\node[text width=1cm](22) [above=0.1cm of 7]{$p_2-1$};
\node[text width=0.1cm](23) [below=0.1cm of 9]{1};
\node[text width=0.1cm](24) [below=0.1cm of 10]{2};
\node[text width=0.1cm](25) [below=0.1cm of 11]{3};
\node[text width=1cm](26) [below=0.1cm of 12]{$p_1-1$};
\node[text width=0.1cm](27) [below=0.1cm of 13]{$p_1$};
\node[text width=0.1cm](28)[below=1cm of 11]{$(X^{(p_1,p_2-1)})$};
\end{tikzpicture}}
&\qquad \qquad
&\scalebox{.5}{\begin{tikzpicture}[
cnode/.style={circle,draw,thick, minimum size=1.0cm},snode/.style={rectangle,draw,thick,minimum size=1cm}]
\node[cnode] (9) at (0,1){1};
\node[snode] (10) at (0,-1){1};
\node[cnode] (11) at (2, 0){2};
\node[cnode] (12) at (4, 1){1};
\node[cnode] (13) at (4, -1){1};
\node[snode] (14) at (6, 1){$p_2$};
\draw[-] (9) -- (11);
\draw[-] (10) -- (11);
\draw[-] (12) -- (11);
\draw[-] (13) -- (11);
\draw[-] (12) -- (14);
\draw[thick] (12) to [bend left=10] (13);
\draw[thick] (12) to [bend right=10] (13);
\draw[thick] (12) to [bend left=20] (13);
\draw[thick] (12) to [bend right=20] (13);
\draw[thick, dotted] (3.85,0) to (4.2, 0);
\node[text width=0.1cm](20) at (4.5,0){$p_1$};
\node[text width=0.1cm](21)[above=0.2 cm of 9]{1};
\node[text width=0.1cm](22)[above=0.2 cm of 11]{2};
\node[text width=0.1cm](23)[above=0.2 cm of 12]{3};
\node[text width=0.1cm](24)[below=0.05 cm of 13]{4};
\node[text width=0.1cm](21)[below=1 cm of 11]{$(Y^{(p_1, p_2-1)})$};
\end{tikzpicture}}\\
\scalebox{.5}{\begin{tikzpicture}
\draw[->] (15,-3) -- (15,-5);
\node[text width=0.1cm](20) at (14.5, -4) {$\prod_k\CO''_k$};
\end{tikzpicture}}
&\qquad \qquad
& \scalebox{.5}{\begin{tikzpicture}
\draw[->] (15,-3) -- (15,-5);
\node[text width=0.1cm](29) at (15.5, -4) {$\prod_k \wt{\CO}''_k$};
\end{tikzpicture}}\\
\scalebox{0.5}{\begin{tikzpicture}[node distance=2cm,cnode/.style={circle,draw,thick,minimum size=8mm},snode/.style={rectangle,draw,thick,minimum size=8mm},pnode/.style={rectangle,red,draw,thick,minimum size=8mm}]
\node[cnode] (1) at (-3,0) {$2$};
\node[cnode] (2) at (-2,2) {$1$};
\node[cnode] (3) at (0,2.5) {$1$};
\node[cnode] (4) at (2,2) {$1$};
\node[cnode] (5) at (3,0) {$1$};
\node[cnode] (6) at (2,-2) {$1$};
\node[cnode] (7) at (0,-2.5) {$1$};
\node[cnode] (8) at (-2,-2) {$1$};
\node[snode] (10) at (-5,0) {$3$};
\node[cnode] (11) at (4,0) {$1$};
\node[cnode] (12) at (5,0) {$1$};
\node[cnode] (13) at (7.5,0) {$1$};
\node[snode] (14) at (9,0) {$1$};
\draw[thick] (1) to [bend left=40] (2);
\draw[thick] (2) to [bend left=40] (3);
\draw[thick,dashed] (3) to [bend left=40] (4);
\draw[thick] (4) to [bend left=40] (5);
\draw[thick] (5) to [bend left=40] (6);
\draw[dashed] (6) to [bend left=40] (7);
\draw[thick] (7) to [bend left=40] (8);
\draw[thick] (8) to [bend left=40] (1);
\draw[thick] (1) -- (10);
\draw[thick] (5) -- (11);
\draw[thick] (11) -- (12);
\draw[dashed] (12) -- (13);
\draw[thick] (13) -- (14);
\node[text width=1cm](11) at (-2,2.5) {$1$};
\node[text width=1cm](12) at (0, 3) {$2$};
\node[text width=1cm](13) at (2, 2.5) {$p_2-1$};
\node[text width=1cm](15) at (2,-2.5) {$p_1$};
\node[text width=1.5 cm](16) at (0,-3.2){$2$};
\node[text width=1cm](17) at (-2.5,-2.5){$1$};
\node[text width=0.1cm](18) at (4, 0.6){$1$};
\node[text width=0.1cm](19) at (5, 0.6){$2$};
\node[text width=1cm](20) at (7.5, 0.6){$p_3-1$};
\node[text width=1cm](21) at (-2.5, 1){$\CA$};
\node[text width=0.1cm](30) at (0,-4){$(X')$};
\end{tikzpicture}}
&\qquad \qquad
&\scalebox{0.5}{\begin{tikzpicture}[
cnode/.style={circle,draw,thick, minimum size=1.0cm},snode/.style={rectangle,draw,thick,minimum size=1cm}]
\node[cnode] (9) at (0,1){1};
\node[snode] (10) at (0,-1){1};
\node[cnode] (11) at (2, 0){2};
\node[cnode] (12) at (4, 1){1};
\node[cnode] (13) at (4, -1){1};
\node[snode] (14) at (6, 1){$p_2$};
\node[snode] (15) at (6, -1){$p_3$};
\draw[-] (9) -- (11);
\draw[-] (10) -- (11);
\draw[-] (12) -- (11);
\draw[-] (13) -- (11);
\draw[-] (12) -- (14);
\draw[-] (13) -- (15);
\draw[thick] (12) to [bend left=10] (13);
\draw[thick] (12) to [bend right=10] (13);
\draw[thick] (12) to [bend left=20] (13);
\draw[thick] (12) to [bend right=20] (13);
\draw[thick, dotted] (3.85,0) to (4.2, 0);
\node[text width=0.1cm](20) at (4.5,0){$p_1$};
\node[text width=0.1cm](21)[above=0.2 cm of 9]{1};
\node[text width=0.1cm](22)[above=0.2 cm of 11]{2};
\node[text width=0.1cm](23)[above=0.2 cm of 12]{3};
\node[text width=0.1cm](24)[below=0.05 cm of 13]{4};
\node[text width=0.1cm](21)[below=1 cm of 11]{$(Y')$};
\end{tikzpicture}}
\end{tabular}
\caption{The quiver $(X')$ and its mirror dual $(Y')$ in \figref{fig: AbEx4gen} is generated by a sequence of 
elementary $S$-type operations on different nodes. In each step, the flavor node(s) on which the $S$-type operation 
acts is shown in red.}
\label{AbEx4GFI}
\end{center}
\end{figure}

Let us now work out the dual quivers at each step outlined above, using the general prescriptions for the dual partition functions for 
Abelian $S$-type operations, derived in \Secref{AbS}.
\begin{itemize}
\item Consider the first sequence of operations on the theory $X$ at the $U(4)$ flavor node $\alpha$. Recall that the partition function of the theory $Y$ is given as
\footnote{Mirror symmetry implies that the partition functions for $X$ and $Y$ are related in the following fashion: $Z^{(X)}= e^{2\pi i (\sum^4_{i=1}b_i m_i + b_{\CA} m_{\CA})\,\eta}\,Z^{(Y)}$, where $\{m_i, m_\CA\}$ and $\eta$ are the mass parameters and the FI parameter of the quiver gauge theory $Y$ respectively.}
\begin{align} \label{AbEx4DPF}
&Z^{(Y)}(\vec m', \vec \eta')= \int \,\prod^{4}_{\gamma'=1} \Big[d\vec\s^{\gamma'}\Big]\, Z^{(Y)}_{\rm int}(\{\vec\s^{\gamma'}\}, \eta, \{m_1, m_2,m_3, m_4, m_\CA\})\nn \\
&= \int \,\prod^{4}_{\gamma'=1} \Big[d\vec\s^{\gamma'}\Big]\,Z^{(Y)}_{\rm FI}(\{\vec\s^{\gamma'}\}, \{m_1, m_2,m_3, m_4, m_\CA\}) \,Z^{(Y)}_{1-{\rm loop}}(\{\vec\s^{\gamma'}\}, \eta),
\end{align}
where $Z^{(Y)}_{1-{\rm loop}}$ can be read off from the quiver $Y$ in \figref{fig: AbEx4D}. The explicit form of the FI term, using the relations 
\eref{MM4Da}-\eref{MM4Dc}, is as follows:
\begin{align} \label{AbEx4DFI}
Z^{(Y)}_{\rm FI}(\{\vec\s^{\gamma'}\}, \{m_1, m_2,m_3, m_4, m_\CA\}) & = e^{-2\pi i \s_1\, (m_1-m_{2})}\,e^{-2\pi i \tr \vec\s_2 \, (m_2-m_{3})} \nn \\
& \times e^{-2\pi i \s_3\, (m_3 + m_{4}-m_{\CA})}\, e^{-2\pi i \s_4\, (m_3 - m_{4})}.
\end{align}
As discussed earlier, we will implement the overall $U(1)$ quotient for the Higgs branch global symmetry by the constraint $m_\CA=0$.
Given the choice of the permutation matrix $\CP$ in \eref{P1}, we define the variable $u^\alpha$ following the definition \eref{uvdef0}, as:
\be 
u^\alpha=m_4,
\ee 
and implement the $S$-type operation $\CO_{1 \CP}$ on $X$ first. 
The partition function of the dual quiver $\wt{\CO}_{1\,\CP}(Y)=Y^{(1)}_\CP$ is then given by the expressions 
\eref{PFGFgenNf1}-\eref{HyperGFgenNf1} as follows:
\begin{align}
& Z^{\wt{\CO}_{1\,\CP}(Y)}(\vec m'^{(1)}, \vec \eta'^{(1)})\nn \\
&= \int \,\prod^{4}_{\gamma'=1} \Big[d\vec\s^{\gamma'}\Big]\, Z^{\rm hyper}_{\wt{\CO}_{1\,\CP}(Y)}(\{\vec\s^{\gamma'}\}, \wt{\eta}^{(1)}, \eta)
\,Z^{(Y)}_{\rm int}(\{\vec\s^{\gamma'}\}, \eta, \{m_1, m_2,m_3, m_4=m^{(1)}_{F}, m_\CA=0\}), \label{PFAbEx4gen1}
\end{align}
where the hypermultiplet term is given by \eref{HyperGFgenNf1}. The function $g_\alpha(\{\vec\s^{\gamma'}\}, \CP)$ in \eref{HyperGFgenNf1} 
can be read off from the $u^\alpha=m^4$-dependent part of the FI term $Z^{(Y)}_{\rm FI}$ in \eref{AbEx4DFI}:
\be
g_\alpha(\{\vec\s^{\gamma'}\}, \CP) = \s^4 -\s^3,
\ee
which leads to the hypermultiplet term:
\begin{align}
Z^{\rm hyper}_{\wt{\CO}_{1\,\CP}(Y)}(\{\vec\s^{\gamma'}\}, \wt{\eta}^{(1)}, \eta)
= Z_{\rm 1-loop}^{\rm bif}(\s^3, \s^4, \wt{\eta}^{(1)} +  b_4 \eta). \label{HyperAbEx4gen1}
\end{align}
The dual operation therefore amounts to adding a bifundamental hypermultiplet connecting the gauge nodes labelled 
$3$ and $4$ in quiver $Y$, with a mass parameter $m'^{(34)}_{{\rm bif}\,1}= \wt{\eta}^{(1)} +  b_4 \eta$.\\

Next, we implement the flavoring-gauging operation $\CO_2$ on 
the quiver $X^{(1)}_\CP=\CO_{1\,\CP}(X)$ at the new flavor node $\alpha$ generated by $\CO_{1\,\CP}$, i.e. we set 
\be
u^\alpha = m^{(1)}_F,
\ee
and implement the flavoring-gauging operation with $N^\alpha_F=1$. Proceeding as before, one can show that the dual operation amounts to 
adding another bifundamental hypermultiplet connecting the gauge nodes labelled $3$ and $4$ in quiver $Y^{(1)}_\CP$. 
Repeating this operation $p_1-2$ times, the partition function of the dual theory $Y^{(p_1)}_\CP$ is given as 
\begin{align}
& Z^{Y^{(p_1)}_\CP}(\vec m'^{(p_1)}, \vec \eta'^{(p_1)})\nn \\
&= \int \,\prod^{4}_{\gamma'=1} \Big[d\vec\s^{\gamma'}\Big]\, \prod^{p_1}_{i=1}Z_{\rm 1-loop}^{\rm bif}(\s^3, \s^4, m'^{(34)}_{{\rm bif}\,i})\,Z^{(Y)}_{\rm int}(\{\vec\s^{\gamma'}\}, \eta,
\{m_1, m_2,m_3, m_4= m^{(p_1)}_{F}, m_\CA=0 \}),\label{PFAbEx4gen2}
\end{align}
which manifestly is the partition function for the quiver gauge theory $Y^{(p_1)}_\CP$ in the second row of the \figref{AbEx4GFI}, with
the bifundamental masses $\vec{m}'^{(34)}_{\rm bif}$ in the above expression given as
\be
m'^{(34)}_{{\rm bif}\,i} = \sum^{i}_{k=1}\wt{\eta}^{(k)} + b_4 \eta.
\ee
The Higgs branch global symmetry of the quiver $X^{(p_1)}_\CP$ is $G^{X^{(p_1)}_\CP}_{H}=(U(3)\times U(1)_{\CA} \times U(1)) / U(1)$.
We can now choose to implement the $U(1)$ quotient such that $G^{X^{(p_1)}_\CP}_{\rm H}=SU(3)\times U(1)_{\CA} \times U(1)$, where 
$(m_1, m_2,m_3)$ parametrize the Cartan subalgebra of the $SU(3)$, while the other masses $(m_\CA, m^{(p_1)}_{F})$ parametrize the 
Cartan of $U(1)_{\CA} \times U(1)$. The dual partition function can then be rewritten as
\begin{align}
& Z^{Y^{(p_1)}_\CP}(\vec m'^{(p_1)}, \vec \eta'^{(p_1)})\nn \\
&= \int \,\prod^{4}_{\gamma'=1} \Big[d\vec\s^{\gamma'}\Big]\, \prod^{p_1}_{i=1}Z_{\rm 1-loop}^{\rm bif}(\s^3, \s^4, m'^{(34)}_{{\rm bif}\,i})\,Z^{(Y)}_{\rm int}(\{\vec\s^{\gamma'}\}, \eta,
\{m_1, m_2,m_3, m^{(p_1)}_{F}, m_\CA \}),\label{PFAbEx4gen2a}
\end{align}
where $\sum^3_{i=1} m_i=0$. The FI term in the partition function of the theory $Y^{(p_1)}_\CP$ as well as the mirror map for the dual pair 
$(X^{(p_1)}_\CP,Y^{(p_1)}_\CP)$ can be read off from \eref{PFAbEx4gen2a} and \eref{AbEx4DFI}.

\item Now consider the sequence of $N^\alpha_F=1$ flavoring-gauging operations $\CO'_j$ ($j=1,\ldots,p_2-1$). The operation 
$\CO'_1$ acts on the $U(1)_{\CA}$ node of the quiver $X^{(p_1)}_\CP$, while the subsequent ones act on the new $U(1)$ flavor node 
created in the previous step. For implementing $\CO'_1$, we define:
\be
u^\alpha=m_\CA.
\ee
Following \eref{HyperGFgenNf1}-\eref{PFGFgenNf1}, the partition function of the dual theory $\wt{\CO}'_1(Y^{(p_1)}_\CP)=Y^{(p_1,1)}_\CP$ is then given as
\begin{align}
& Z^{\wt{\CO}'_1(Y^{(p_1)}_\CP)}(\vec m'^{(p_1,1)}, \vec \eta'^{(p_1,1)}) \nn \\
&= \int \,\prod^{4}_{\gamma'=1}  \Big[d\vec\s^{\gamma'}\Big]\, Z^{\rm hyper}_{\wt{\CO}'_1(Y^{(p_1)}_\CP)}(\{\vec\s^{\gamma'}\}, \xi^{(1)}, \vec \eta^{(p_1)}) \,Z^{Y^{(p_1)}_\CP}_{\rm int}(\{\vec\s^{\gamma'}\}, \vec m'^{(p_1)}, \vec \eta'^{(p_1)}), \label{PFAbEx4gen3}
\end{align}
where $\vec \eta^{(p_1)}=\vec \eta^{(p_1)}(\eta, \wt{\eta}^{(i)})$, $\vec \eta'^{(p_1)}=\vec \eta'^{(p_1)}(m_1,m_2,m_3, m^{(p_1)}_{F}, x^{(1)}_{F})$, 
and the function $Z^{(Y^{(p_1)}_\CP)}_{\rm int}$ is given by the integrand on the RHS of the \eref{PFAbEx4gen2a}. 
The hypermultiplet term can be computed as before from \eref{HyperGFgenNf1}, where the function $g_\alpha(\{\s^{\gamma'}\})$ can be read off 
from the $u^\alpha=m_\CA$-dependent part of the FI term in \eref{PFAbEx4gen2}:
\be
g_\alpha(\{\vec\s^{\gamma'}\})= \s^3.
\ee
This leads to the hypermultiplet contribution
\begin{align}
Z^{\rm hyper}_{\wt{\CO}'_1(Y^{(p_1)}_\CP)}(\{\vec\s^{\gamma'}\}, \xi^{(1)}, \vec \eta^{(p_1)})
= Z_{\rm 1-loop}^{\rm fund}(\s^3, -\xi^{(1)} -  b_\CA \eta), \label{HyperAbEx4gen3}
\end{align}
which implies that the dual operation $\wt{\CO}'_1$ amounts to adding a single fundamental hyper at the gauge node labelled 3 in the quiver 
$Y^{(p_1)}_\CP$, with a mass $m'^{(3)}_{{\rm fund}\,1}=-\xi^{(1)} -  b_\CA \eta$. 
Proceeding with the subsequent operations $\CO'_j$ ($j=2,\ldots,p_2-1$), the theory $Y^{(p_1,p_2-1)}$ can be read off from the partition function:
\begin{align}
Z^{Y^{(p_1,p_2-1)}_\CP}
= \int \,& \prod^{4}_{\gamma'=1} \Big[d\vec\s^{\gamma'}\Big]\, \prod^{p_2-1}_{j=1}Z_{\rm 1-loop}^{\rm fund}(\s^3, m'^{(3)}_{{\rm fund}\,j}) \, \prod^{p_1}_{i=1}Z_{\rm 1-loop}^{\rm bif}(\s^3, \s^4, m'^{(34)}_{{\rm bif}\,i})\nn \\
& \times Z^{(Y)}_{\rm int}(\{\vec\s^{\gamma'}\}, \eta,\{m_1,m_2,m_3, m^{(p_1)}_{F}, x^{(p_2-1)}_{F}\}),\label{PFAbEx4gen4}
\end{align}
which manifestly reproduces the quiver gauge theory $Y^{(p_1,p_2-1)}$ in the third line of \figref{AbEx4GFI}. The fundamental masses $\vec{m}'^{(3)}_{\rm fund}$ in the above expression are given as
\be
{m}'^{(3)}_{{\rm fund}\,j} = -\sum^{j}_{k=1}\xi^{(k)}  - b_\CA \eta .
\ee
The Higgs branch global symmetry of the quiver $X^{(p_1,p_2-1)}_\CP$ is $G^{X^{(p_1,p_2-1)}_\CP}_{H}=SU(3)\times U(1)_{p_1} \times U(1)_{p_2-1}$,
where $(m_1, m_2,m_3)$ parametrize the Cartan subalgebra of the $SU(3)$, while the other masses $(m^{(p_1)}_{F}, x^{(p_2-1)}_{F})$ parametrize the 
Cartan of $U(1)_{p_1} \times U(1)_{p_2-1}$ respectively.

\item Finally, let us implement the sequence of Abelian $S$-type operations $\CO''_k$ ($k=1,2,\ldots,p_3$), where 
$\CO''_1$ is an identification-flavoring-gauging operation with $N^\alpha_F=1$, acting on the $U(1)_{p_1} \times U(1)_{p_2-1}$
flavor nodes of the quiver $X^{(p_1,p_2-1)}_\CP$. The nodes are shown in red in the third line of \figref{AbEx4GFI}. 
The subsequent operations $\CO''_k$ ($k=2,\ldots,p_3$) are $N^\alpha_F=1$ flavoring-gauging operations acting on the 
new $U(1)$ flavor node generated in the previous step. Proceeding in the same way as the relevant parts 
of the computation for the dual quiver Family I, II and III, one can confirm that the dual operation $\wt{\CO}''_k$ amounts to adding 
a fundamental hypermultiplet to the gauge node labelled 4 in the quiver $Y^{(p_1,p_2-1, k-1)}_\CP$. We therefore simply state 
the final answer for the theory $Y^{(p_1,p_2-1, p_3)}_\CP$:
\begin{align}
Z^{Y^{(p_1,p_2-1,p_3)}_\CP}
= \int \,& \prod^{4}_{\gamma'=1} \Big[d\vec\s^{\gamma'}\Big]\, \prod^{p_3}_{k=1}Z_{\rm 1-loop}^{\rm fund}(\s^4, m'^{(4)}_{{\rm fund}\,k})\,
\prod^{p_2-1}_{j=1}Z_{\rm 1-loop}^{\rm fund}(\s^3, m'^{(3)}_{{\rm fund}\,j}) \nn \\ 
& \times \prod^{p_1}_{i=1}Z_{\rm 1-loop}^{\rm bif}(\s^3, \s^4, m'^{(34)}_{{\rm bif}\,i}) \, Z^{(Y)}_{\rm int}(\{\vec\s^{\gamma'}\}, \eta,\{\vec m, y^{(p_3)}_F, y^{(p_3)}_F + \mu \}),\label{PFAbEx4gen4}
\end{align}
which manifestly reproduces the quiver gauge theory $Y'=Y^{(p_1,p_2-1,p_3)}$ in the last row of \figref{AbEx4GFI}, with $\vec m=(m_1, m_2,m_3)$ 
and the fundamental masses $\vec{m}'^{(4)}_{{\rm fund}}$ are given as
\be
m'^{(4)}_{{\rm fund}\,k} = - \sum^k_{l=1} \zeta^{(l)}  -\sum^{p_2-1}_{j=1}\xi^{(j)} - \sum^{p_1}_{i=1}\wt{\eta}^{(i)} - (b_\CA + b_4) \eta.
\ee
After a shift in the integration variables, the mirror map can be read off from \eref{PFAbEx4gen4} above.
\begin{align}
& m'^{(34)}_{{\rm bif}\,i} = \sum^{i}_{k=1}\wt{\eta}^{(k)} + b_4 \eta,\quad (i=1,\ldots,p_1)\\
& {m}'^{(3)}_{{\rm fund}\,j} = \begin{cases}
-\sum^{j}_{k=1}\xi^{(k)}  - (b_\CA +1) \eta, & \text{if } j=1,\ldots,p_2-1\\
-\eta, & \text{if } j=p_2,
\end{cases}\\
& m'^{(4)}_{{\rm fund}\,k} = - \sum^k_{l=1} \zeta^{(l)}  -\sum^{p_2-1}_{j=1}\xi^{(j)} - \sum^{p_1}_{i=1}\wt{\eta}^{(i)} - (b_\CA + b_4+1) \eta, \quad (k=1,\ldots,p_3) \\
& \eta'_l= -(m_l - m_{l+1}),\quad (l=1,2),\\
& \eta'_3=-m_3  +\mu,\\
& \eta'_4=y^{(p_3)}_F-m_3, 
\end{align}
with the additional constraint $m_1+m_2+m_3=0$.
The mass parameters of $Y'$ manifestly live in the Cartan subalgebra of the Higgs branch global symmetry 
$G^{Y'}_{H}=U(p_1) \times U(p_2) \times U(p_3)$, while the FI parameters live in the Cartan subalgebra
of the Coulomb branch global symmetry $G^{Y'}_{C}=SU(3) \times U(1)^2$. The Coulomb branch global symmetry can be read 
off from the observation that $Y'$ contains a balanced sub-quiver with two gauge nodes (labelled 1 and 2) which gives an $SU(3)$ 
factor, along with two unbalanced nodes which contribute a $U(1)$ factor each.
\end{itemize}

\section{3d mirror pairs for class $\CS$ SCFTs on a circle}\label{ADMirrors}
In this section, we comment on the fact that special cases of the 3d mirror pairs, constructed in \Secref{AbMirr}, 
are related to certain 4d $\CN=2$ SCFTs of class $\CS$, which arise from the twisted compactification of a 6d (2,0) $A_{N-1}$
theory on a Riemann surface with punctures \cite{Gaiotto:2009we, Gaiotto:2009hg, Chacaltana:2010ks}. 
Putting a 4d class $\CS$ SCFT on a circle and flowing to the deep IR
yields an interacting 3d SCFT which generically may not have a Lagrangian description. For theories with non-trivial Higgs branches, 
the class $\CS$ construction can be used to argue that the 3d mirror of the aforementioned SCFT has a Lagrangian description \cite{Benini:2010uu, Xie:2012hs, boalch2008irregular}. 
For the purpose of this paper, we will restrict ourselves to 4d SCFTs that arise from a 6d (2,0) $A_{N-1}$ theory compactified on a Riemann 
sphere with either a single irregular puncture, or a single irregular puncture and a single regular puncture. The 4d SCFTs that arise from such 
compactification were classified in \cite{Gaiotto:2009hg, Xie:2012hs, Wang:2015mra, Wang:2018gvb}. A subset of these 4d theories were found to have a Lagrangian 3d mirror. 
This includes, for example, Argyres-Douglas (AD) theories of the types $(A_s, A_{(s+1)p-1})$ and $(A_s, D_{(s+1)p+2})$, for $s$ and $p$ being 
positive integers\footnote{The reduction of AD theories on a circle has a large literature. For some of the recent work on this subject, see 
\cite{Buican:2015ina, Buican:2015hsa, Buican:2017uka, Fredrickson:2017yka, Dedushenko:2019mzv, Dedushenko:2019mnd, Buican:2019kba} and the references therein. }. \\

Our strategy, will be to show that many of these 3d mirrors, obtained via the class $\CS$ construction of \cite{Xie:2012hs}, have 
Lagrangian mirrors themselves, using our findings from \Secref{AbMirr}. This will allow us to propose explicit 3d $\CN=4$ Lagrangian descriptions for the 3d SCFTs that are 
constructed by taking the related 4d SCFTs on a circle and flowing to the deep IR. This shows that the 3d SCFTs  have a pair of Lagrangian 
descriptions, which, in particular, implies that the Coulomb branch and the Higgs branch of the SCFT can both be described as \hk quotients. 
Our construction, using $S$-type operations, can be extended to the 4d SCFTs (with 3d Lagrangian mirrors) arising from more general punctures. 
Most of these cases, however, necessitate the use of non-Abelian $S$-type operations and will be discussed in a future paper \cite{Dey:2020xyz}. \\

In \Secref{AD-Gen}, we make some general comments about the 4d SCFTs of interest, which include the AD theories, and their 3d mirrors.
In \Secref{AD-A}, we discuss 3d mirror pairs associated with 4d SCFTs which arise from the compactification of the $A_{N-1}$ 6d theory on a Riemann 
sphere with a single  irregular puncture. This includes AD theories of the type $(A_s, A_{(s+1)p-1})$ for positive integers $s,p$,
of which the 3d mirror pair for the $s=1$ case was already known. We will explicitly write down the mirror pairs for $s \leq 3$, and generic $p$, 
but the generalization to higher values of $s$ can be be obtained in an analogous fashion. 
In \Secref{AD-D}, we discuss 3d mirror pairs associated with 4d SCFTs which arise from
the 6d $A_{N-1}$ theory compactified on a Riemann sphere with a single irregular puncture and a single minimal regular puncture. This includes 
AD theories of the type $(A_s, D_{(s+1)p+2})$  positive integers $s,p$, of which the 3d mirror pairs for the $s=1$ case was previously known. 
Again, we explicitly write down the mirror pairs for $s \leq 2$ and generic $p$, but the generalization to higher values $s$ can be obtained 
in a similar fashion. We also discuss an SCFT which arises from a single irregular puncture and a single maximal regular puncture, 
and present the associated 3d mirror pair. 

\subsection{AD theories, $(G,G')$ theories and Lagrangian 3d mirrors}\label{AD-Gen}
AD SCFTs were originally discovered \cite{Argyres:1995jj, Argyres:1995xn} as the IR theories that arise at special points on the Coulomb branch of 
4d $\CN=2$ $SU(N)$ SYM, where mutually non-local dyons became massless. These strongly coupled SCFTs are characterized by 
fractional scaling dimensions of BPS operators on the Coulomb branch and the presence of relevant operators.\\
 
A large class of AD-type SCFTs were discovered in \cite{Cecotti:2010fi} using geometric engineering, where the authors studied the Type IIB 
superstring theory on a singular hypersurface of $\BC^4$ given by the following locus:
\be
W_G(x_1, x_2) + W_{G'}(y_1, y_2)=0,
\ee
with $G$ and $G'$ being any pair of simply-laced groups, and $W_G,W_{G'}$ being quasi-homogeneous polynomials of the respective types.
These SCFTs are therefore labelled by a pair of ADE singularities $(G, G')$. \\

It was soon understood that a much larger class of AD-type SCFTs, which includes the $(G, G')$ SCFTs, can be realized in the 
class $\CS$ setting \cite{Gaiotto:2009we}. The construction, as discussed in \cite{Gaiotto:2009hg, Xie:2012hs}, involves compactifying a 6d $A_{N-1}$
(2,0) theory on a Riemann sphere with an irregular puncture, i.e. singularity with a higher order pole in the Higgs field of the 
associated Hitchin system. The $A$-type irregular punctures and the associated SCFTs were classified in \cite{Xie:2012hs}, and the 
construction was extended to other (2,0) theories in \cite{Wang:2015mra, Wang:2018gvb}.
Compactifying on a circle and flowing to the IR, an AD theory gives a 3d $\CN=4$ interacting SCFT, which is generically not expected 
to have a Lagrangian description. However, similar to the case for regular punctures \cite{Benini:2010uu}, the class $\CS$ construction 
predicts a Lagrangian mirror dual for certain families of AD theories \cite{Xie:2012hs}. Two sub-families of the $(G, G')$ SCFTs, namely 
the $(A_s, A_{(s+1)p-1})$ and $(A_s, D_{(s+1)p+2})$ SCFTs (for positive integers $s$ and $p$), have 3d Lagrangian mirror duals and 
will feature prominently in our discussion.

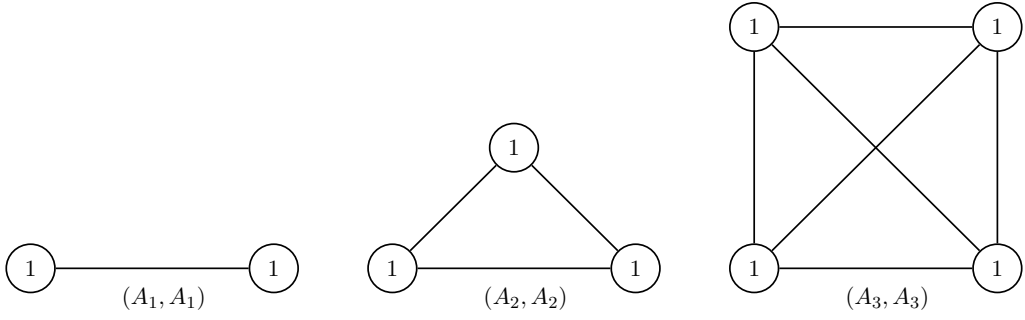
\begin{figure}[htbp]
\begin{center}
\scalebox{0.8}{\begin{tikzpicture}[node distance=2cm,cnode/.style={circle,draw,thick,minimum size=8mm},snode/.style={rectangle,draw,thick,minimum size=8mm},pnode/.style={rectangle,red,draw,thick,minimum size=8mm}]
\node[cnode] (1) at (-2,0) {$1$};
\node[cnode] (2) at (2,0) {$1$};
\draw[thick] (1) -- (2);
\node[text width=1cm](3) at (0,-0.5){$(A_{1}, A_{1})$};
\end{tikzpicture}}
\qquad
\scalebox{0.8}{\begin{tikzpicture}[node distance=2cm,cnode/.style={circle,draw,thick,minimum size=8mm},snode/.style={rectangle,draw,thick,minimum size=8mm},pnode/.style={rectangle,red,draw,thick,minimum size=8mm}]
\node[cnode] (1) at (-2,0) {$1$};
\node[cnode] (2) at (2,0) {$1$};
\node[cnode] (3) at (0,2) {$1$};
\draw[thick] (1) -- (2);
\draw[thick] (2) -- (3);
\draw[thick] (1) -- (3);
\node[text width=1cm](4) at (0,-0.5){$(A_{2}, A_{2})$};
\end{tikzpicture}}
\qquad
\scalebox{0.8}{\begin{tikzpicture}[node distance=2cm,cnode/.style={circle,draw,thick,minimum size=8mm},snode/.style={rectangle,draw,thick,minimum size=8mm},pnode/.style={rectangle,red,draw,thick,minimum size=8mm}]
\node[cnode] (1) at (-2,0) {$1$};
\node[cnode] (2) at (2,0) {$1$};
\node[cnode] (3) at (-2,4) {$1$};
\node[cnode] (4) at (2,4) {$1$};
\draw[thick] (1) -- (2);
\draw[thick] (1) -- (3);
\draw[thick] (1) -- (4);
\draw[thick] (2) -- (4);
\draw[thick] (2) -- (3);
\draw[thick] (3) -- (4);
\node[text width=1cm](4) at (0,-0.5){$(A_{3}, A_3)$};
\end{tikzpicture}}
\caption{3d mirrors from class $\CS$ construction for $(A_s, A_{(s+1)p-1})$ theories, with $s=1, 2, 3$, and $p=1$. For a given $s$, 
the 3d mirror for generic $p$ is given by the same quiver diagram with every pair of gauge nodes being connected by $p$ lines. 
The gauge group is obtained by factoring out an overall $U(1)$, which can be implemented by ungauging any one of the $U(1)$ gauge nodes.}
\label{fig: AD-AA}
\end{center}
\end{figure}

First, consider the case of $(A_s, A_{(s+1)p-1})$ SCFTs. Generally speaking, the SCFTs of the type $(A_{N-1}, A_{k-1})$ are realized by 
compactifying on a Riemann sphere with a single irregular puncture of a specific type \cite{Xie:2012hs}. In the associated Hitchin system, 
the order of the pole of the Higgs field at the singular point is a linear function of the integer $k$. The 3d SCFT and its mirror generically do not have a known Lagrangian 
description. For the sub-family $(A_s, A_{(s+1)p-1})$, for which the Coulomb branch has a non-trivial global symmetry \cite{DelZotto:2014kka}, one can 
expect a Lagrangian 3d mirror, and the precise form of the quivers can be guessed explicitly, as shown in \figref{fig: AD-AA}. Similar to the case of 
the regular punctures, one can therefore associate a quiver tail to the associated irregular puncture.

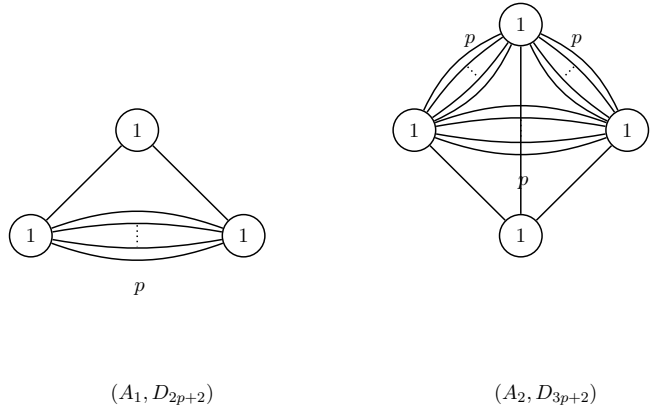
\begin{figure}[htbp]
\begin{center}
\scalebox{.7}{\begin{tikzpicture}[node distance=2cm,cnode/.style={circle,draw,thick,minimum size=8mm},snode/.style={rectangle,draw,thick,minimum size=8mm},pnode/.style={rectangle,red,draw,thick,minimum size=8mm}]
\node[cnode] (1) at (-2,0) {$1$};
\node[cnode] (2) at (2,0) {$1$};
\node[cnode] (3) at (0,2) {$1$};
\draw[thick] (1) to [bend left=10] (2);
\draw[thick] (1) to [bend right=10] (2);
\draw[thick] (1) to [bend left=20] (2);
\draw[thick] (1) to [bend right=20] (2);
\draw[thick, dotted] (0,0.2) to (0,-0.2);
\node[text width=0.1cm](15) at (0,-1){$p$};
\draw[thick] (1) -- (3);
\draw[thick] (2) -- (3);
\node[text width=1cm](16) at (0,-3){$(A_1, D_{2p+2})$};
\end{tikzpicture}}
\qquad \qquad 
\scalebox{.7}{\begin{tikzpicture}[node distance=2cm,cnode/.style={circle,draw,thick,minimum size=8mm},snode/.style={rectangle,draw,thick,minimum size=8mm},pnode/.style={rectangle,red,draw,thick,minimum size=8mm}]
\node[cnode] (1) at (-2,0) {$1$};
\node[cnode] (2) at (2,0) {$1$};
\node[cnode] (3) at (0,2) {$1$};
\node[cnode] (4) at (0,-2) {$1$};
\draw[thick] (1) -- (4);
\draw[thick] (2) -- (4);
\draw[thick] (3) -- (4);
\draw[thick] (1) to [bend left=10] (2);
\draw[thick] (1) to [bend right=10] (2);
\draw[thick] (1) to [bend left=20] (2);
\draw[thick] (1) to [bend right=20] (2);
\draw[thick, dotted] (0,0.2) to (0,-0.2);
\node[text width=0.1cm](15) at (0,-1){$p$};
\draw[thick] (1) to [bend left=10] (3);
\draw[thick] (1) to [bend right=10] (3);
\draw[thick] (1) to [bend left=20] (3);
\draw[thick] (1) to [bend right=20] (3);
\draw[thick, dotted] (-1,1.2) to (-0.8, 1.0);
\node[text width=0.1cm](15) at (-1, 1.7){$p$};
\draw[thick] (2) to [bend left=10] (3);
\draw[thick] (2) to [bend right=10] (3);
\draw[thick] (2) to [bend left=20] (3);
\draw[thick] (2) to [bend right=20] (3);
\draw[thick, dotted] (1,1.2) to (0.8, 1.0);
\node[text width=0.1cm](15) at (1, 1.7){$p$};
\node[text width=1cm](16) at (0,-5){$(A_2, D_{3p+2})$};
\end{tikzpicture}}
\caption{3d mirrors from class $\CS$ construction for $(A_s, D_{(s+1)p+2})$ theories, with $s=1, 2$, and generic $p$. The gauge group is obtained 
by factoring out an overall $U(1)$, which can be implemented by ungauging any one of the $U(1)$ gauge nodes.}
\label{fig: AD-AD}
\end{center}
\end{figure}

Let us now consider the case of $(A_s, D_{(s+1)p+2})$ SCFTs. These are realized by compactifying the 6d $A_s$ theory on a Riemann sphere with 
a single irregular puncture (the one associated with an $(A_s, A_{(s+1)p-1})$ theory) and a single minimal regular puncture. The 3d mirrors for this 
sub-family of SCFTs are also expected to have Lagrangian descriptions. The 3d mirrors are obtained by gluing the 
flavor nodes of the $A_s$ minimal puncture quiver tail to the gauge nodes of the irregular puncture quiver tail \cite{Xie:2012hs, Song:2015wta}, 
as shown in \figref{fig: AD-AD}.

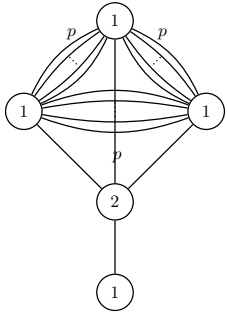
\begin{figure}[htbp]
\begin{center}
\scalebox{.6}{\begin{tikzpicture}[node distance=2cm,cnode/.style={circle,draw,thick,minimum size=8mm},snode/.style={rectangle,draw,thick,minimum size=8mm},pnode/.style={rectangle,red,draw,thick,minimum size=8mm}]
\node[cnode] (1) at (-2,0) {$1$};
\node[cnode] (2) at (2,0) {$1$};
\node[cnode] (3) at (0,2) {$1$};
\node[cnode] (4) at (0,-2) {$2$};
\node[cnode] (5) at (0,-4) {$1$};
\draw[thick] (1) -- (4);
\draw[thick] (2) -- (4);
\draw[thick] (3) -- (4);
\draw[thick] (5) -- (4);
\draw[thick] (1) to [bend left=10] (2);
\draw[thick] (1) to [bend right=10] (2);
\draw[thick] (1) to [bend left=20] (2);
\draw[thick] (1) to [bend right=20] (2);
\draw[thick, dotted] (0,0.2) to (0,-0.2);
\node[text width=0.1cm](15) at (0,-1){$p$};
\draw[thick] (1) to [bend left=10] (3);
\draw[thick] (1) to [bend right=10] (3);
\draw[thick] (1) to [bend left=20] (3);
\draw[thick] (1) to [bend right=20] (3);
\draw[thick, dotted] (-1,1.2) to (-0.8, 1.0);
\node[text width=0.1cm](15) at (-1, 1.7){$p$};
\draw[thick] (2) to [bend left=10] (3);
\draw[thick] (2) to [bend right=10] (3);
\draw[thick] (2) to [bend left=20] (3);
\draw[thick] (2) to [bend right=20] (3);
\draw[thick, dotted] (1,1.2) to (0.8, 1.0);
\node[text width=0.1cm](15) at (1, 1.7){$p$};
\end{tikzpicture}}
\caption{3d mirror for a class $\CS$ SCFT which is obtained by compactifying the 6d $A_2$ theory on a Riemann sphere with an irregular puncture 
and a regular maximal puncture. The gauge group is obtained by factoring out an overall $U(1)$, which can be implemented by ungauging any one 
of the $U(1)$ gauge nodes.}
\label{fig: I-M}
\end{center}
\end{figure}

In addition to the $(G,G')$ theories, the class $\CS$ construction gives a rich class of AD theories, some of which can also have Lagrangian 3d mirrors. 
A particular class of such SCFTs can be realized by a compactification involving an irregular puncture of the $(A_s, A_{(s+1)p-1})$ type and a maximal 
$A_s$ regular puncture. We will call this 4d SCFT $A^{{\rm{maximal}}}_{s,p}$.
The 3d mirror is obtained by gluing the flavor nodes of the $A_s$ maximal puncture quiver tail to the gauge nodes of the 
irregular puncture quiver tail \cite{Xie:2012hs}. The 3d mirror for the case $s=2$ and generic $p$ is shown in \figref{fig: I-M}.

\subsection{Mirror pairs: 4d SCFTs from a single irregular puncture}\label{AD-A}

\subsubsection{Trivial case: $(A_1, A_{2p-1})$ theories for generic $p$}
\begin{figure}[htbp]
\begin{center}
\scalebox{0.8}{\begin{tikzpicture}[node distance=2cm,cnode/.style={circle,draw,thick,minimum size=8mm},snode/.style={rectangle,draw,thick,minimum size=8mm},pnode/.style={rectangle,red,draw,thick,minimum size=8mm}]
\node[cnode] (1) at (-2,0) {$1$};
\node[snode] (2) at (-2,-2) {$p$};
\draw[thick] (1) -- (2);
\node[text width=0.1cm](16) at (-2,-3){$(X)$};
\end{tikzpicture}}
\qquad \qquad
\scalebox{0.8}{\begin{tikzpicture}[node distance=2cm,cnode/.style={circle,draw,thick,minimum size=8mm},snode/.style={rectangle,draw,thick,minimum size=8mm},pnode/.style={rectangle,red,draw,thick,minimum size=8mm}]
\node[cnode] (1) at (-2,0) {$1$};
\node[cnode] (2) at (0,0) {$1$};
\node[cnode] (3) at (2,0) {$1$};
\node[snode] (4) at (-2,-2) {$1$};
\node[cnode] (5) at (4,0) {$1$};
\node[cnode] (6) at (8,0) {$1$};
\node[snode] (7) at (8,-2) {$1$};
\draw[thick] (2) -- (3);
\draw[thick] (1) -- (2);
\draw[thick] (1) -- (4);
\draw[thick] (6) -- (7);
\draw[thick] (3) -- (5);
\draw[thick] (5) to (5,0);
\draw[thick,dashed] (5,0) to (6.5,0);
\draw[thick] (6.6,0) to (6);
\draw[thick] (6) -- (7);
\node[text width=0.1cm](16) at (3,-3){$(Y)$};
\node[text width=0.1cm](17) at (-2, 0.6){$1$};
\node[text width=0.1cm](18) at (0, 0.6){$2$};
\node[text width=1cm](19) at (2, 0.6){$3$};
\node[text width=1cm](20) at (4, 0.6){$4$};
\node[text width=1cm](21) at (8, 0.6){$p-1$};
\end{tikzpicture}}
\caption{The pair of Lagrangian theories associated with the 3d $\CN=4$ SCFT that arise from the dimensional reduction of an $(A_1, A_{2p -1})$ AD theory 
on a circle.}
\label{ADEx0}
\end{center}
\end{figure}
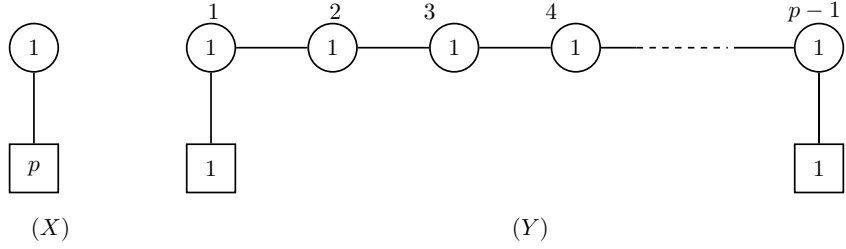

The mirror pair corresponding to the $(A_1, A_{2p-1})$ AD theories is well known. 
The class $\CS$ mirror, labelled as quiver $X$ in \figref{ADEx0}, is a linear quiver 
which has a linear mirror dual given by quiver $Y$. The dimensions of the moduli 
spaces and global symmetries are summarized in Table \ref{Tab:ADEx0}.

\begin{table}[htbp]
\begin{center}
\begin{tabular}{|c|c|c|}
\hline
Moduli space data & Theory $X$ & Theory $Y$ \\
\hline \hline 
dim\,$\CM_H$ & $p-1$ & $1$\\
\hline
dim\,$\CM_C$ & $1$ & $p-1$\\
\hline
$G_H$ & $SU(p)$ & $U(1)$ \\
\hline
$G_C$ & $U(1)$ & $SU(p)$ \\
\hline
\end{tabular}
\caption{Summary table for the moduli space dimensions and global symmetries for the mirror pair in \figref{ADEx0}.}
\label{Tab:ADEx0}
\end{center}
\end{table}

\subsubsection{$(A_2, A_{3p-1})$ theories for generic $p$}
\begin{figure}[htbp]
\begin{center}
\scalebox{0.8}{\begin{tikzpicture}[node distance=2cm,cnode/.style={circle,draw,thick,minimum size=8mm},snode/.style={rectangle,draw,thick,minimum size=8mm},pnode/.style={rectangle,red,draw,thick,minimum size=8mm}]
\node[cnode] (1) at (-2,0) {$1$};
\node[cnode] (2) at (2,0) {$1$};
\node[snode] (3) at (-2,-2) {$p$};
\node[snode] (4) at (2,-2) {$p$};
\draw[thick] (1) -- (3);
\draw[thick] (2) -- (4);
\draw[thick,dotted] (0,0.4) to (0, -0.4);
\draw[thick] (1) to [bend left=20] (2);
\draw[thick] (1) to [bend right=20] (2);
\draw[thick] (1) to [bend left=40] (2);
\draw[thick] (1) to [bend right=40] (2);
\node[text width=0.1cm](15) at (0,1){$p$};
\node[text width=0.1cm](16) at (0,-3){$(X)$};
\end{tikzpicture}}
\qquad \qquad
\scalebox{0.8}{\begin{tikzpicture}[node distance=2cm,cnode/.style={circle,draw,thick,minimum size=8mm},snode/.style={rectangle,draw,thick,minimum size=8mm},pnode/.style={rectangle,red,draw,thick,minimum size=8mm}]
\node[cnode] (1) at (-2,0) {$1$};
\node[cnode] (2) at (-1,1) {$1$};
\node[cnode] (3) at (1,1) {$1$};
\node[cnode] (4) at (2,0) {$1$};
\node[cnode] (5) at (1,-1) {$1$};
\node[cnode] (6) at (-1,-1) {$1$};
\node[cnode] (7) at (3,0) {$1$};
\node[cnode] (8) at (4,0) {$1$};
\node[cnode] (9) at (5,0) {$1$};
\node[cnode] (10) at (7,0) {$1$};
\node[snode] (11) at (8,0) {$1$};
\node[snode] (12) at (-2,-2) {$1$};
\draw[thick] (1) to [bend left=40] (2);
\draw[dashed,thick] (2) to [bend left=40] (3);
\draw[thick] (3) to [bend left=40] (4);
\draw[thick] (4) to [bend left=40] (5);
\draw[dashed,thick] (5) to [bend left=40] (6);
\draw[thick] (6) to [bend left=40] (1);
\draw[thick] (1) -- (12);
\draw[thick] (4) -- (7);
\draw[thick] (7) -- (8);
\draw[thick] (8) -- (9);
\draw[thick, dashed] (9) -- (10);
\draw[thick] (10) -- (11);
\node[text width=1cm](20) at (-2,0.5) {$p$};
\node[text width=1cm](21) at (-1,1.5) {$p+1$};
\node[text width=1cm](22) at (1,1.5) {$2p-1$};
\node[text width=1cm](23) at (2,0.5) {$2p$};
\node[text width=0.1cm](24) at (1,-1.5) {$1$};
\node[text width=1cm](25) at (-1,-1.5) {$p-1$};
\node[text width=0.1cm](26) at (3,0.7) {$1$};
\node[text width=0.1cm](27) at (4,0.7) {$2$};
\node[text width=0.1cm](28) at (5,0.7) {$3$};
\node[text width=1 cm](29) at (7,0.7) {$p-2$};
\node[text width=0.1cm](30) at (3,-3){$(Y)$};
\end{tikzpicture}}
\caption{The pair of Lagrangian theories associated with the 3d $\CN=4$ SCFT that arise from the dimensional reduction of an $(A_s, A_{(s+1)p -1})$ AD theory 
on a circle, for $s=2$ and generic $p$ . Quiver $(X)$ was obtained using class $\CS$ construction of AD theories, while quiver $(Y)$ is the proposed 3d mirror 
dual of quiver $(X)$.}
\label{ADEx1}
\end{center}
\end{figure}
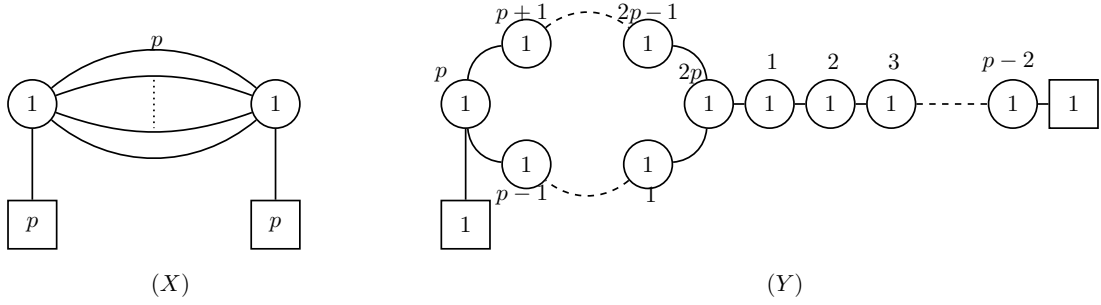

The mirror pair in this case is given by \figref{ADEx1}. The class $\CS$ mirror is given by the quiver $X$, while 
the dual theory is given by the quiver $Y$. The duality can be read off from the dual pairs of Family ${\rm{I}}_{[n,l,p]}$ in \figref{AbEx1gen},
labelled by the triplet of integers $(n,l,p)$, for the special case:
\be
n=2p, \quad l=p+1, \quad p\geq 2.
\ee 
The moduli space dimensions of the dual theories 
and the global symmetries associated with the Higgs and Coulomb branches are given in Table \ref{Tab:ADEx1}. 

\begin{center}
\begin{table}[htbp]
\resizebox{\textwidth}{!}{%
\begin{tabular}{|c|c|c|}
\hline
Moduli space data & Theory $X$ & Theory $Y$ \\
\hline \hline 
dim\,$\CM_H$ & $3p-2$ & $2$\\
\hline
dim\,$\CM_C$ & $2$ & $3p-2$\\
\hline
$G_H$ & $SU(p) \times  SU(p)  \times SU(p) \times U(1) $ & $U(1) \times U(1)$ \\
\hline
$G_C$ & $U(1) \times U(1)$ & $SU(p) \times SU(p) \times SU(p) \times U(1) $ \\
\hline
\end{tabular}}
\caption{Summary table for the moduli space dimensions and global symmetries for the mirror pair in \figref{ADEx1}.}
\label{Tab:ADEx1}
\end{table}
\end{center}

\subsubsection{$(A_3, A_{4p-1})$ theories with generic $p$}

\begin{figure}[htbp]
\begin{center}
\scalebox{.6}{\begin{tikzpicture}[node distance=2cm,cnode/.style={circle,draw,thick,minimum size=8mm},snode/.style={rectangle,draw,thick,minimum size=8mm},pnode/.style={rectangle,red,draw,thick,minimum size=8mm}]
\node[cnode] (1) at (-2,0) {$1$};
\node[cnode] (2) at (2,0) {$1$};
\node[cnode] (3) at (0,2) {$1$};
\node[snode] (4) at (-2,-2) {$p$};
\node[snode] (5) at (2,-2) {$p$};
\node[snode] (6) at (0,4) {$p$};
\draw[thick] (1) -- (4);
\draw[thick] (2) -- (5);
\draw[thick] (3) -- (6);
\draw[thick] (1) to [bend left=10] (2);
\draw[thick] (1) to [bend right=10] (2);
\draw[thick] (1) to [bend left=20] (2);
\draw[thick] (1) to [bend right=20] (2);
\draw[thick, dotted] (0,0.2) to (0,-0.2);
\node[text width=0.1cm](15) at (0,-1){$p$};
\draw[thick] (1) to [bend left=10] (3);
\draw[thick] (1) to [bend right=10] (3);
\draw[thick] (1) to [bend left=20] (3);
\draw[thick] (1) to [bend right=20] (3);
\draw[thick, dotted] (-1,1.2) to (-0.8, 1.0);
\node[text width=0.1cm](15) at (-1, 1.7){$p$};
\draw[thick] (2) to [bend left=10] (3);
\draw[thick] (2) to [bend right=10] (3);
\draw[thick] (2) to [bend left=20] (3);
\draw[thick] (2) to [bend right=20] (3);
\draw[thick, dotted] (1,1.2) to (0.8, 1.0);
\node[text width=0.1cm](15) at (1, 1.7){$p$};
\node[text width=0.1cm](16) at (0,-2){(X)};
\end{tikzpicture}}
\qquad \qquad \qquad 
\scalebox{0.6}{\begin{tikzpicture}[node distance=2cm,cnode/.style={circle,draw,thick,minimum size=8mm},snode/.style={rectangle,draw,thick,minimum size=8mm},pnode/.style={rectangle,red,draw,thick,minimum size=8mm}]
\node[cnode] (1) at (-3,0) {$1$};
\node[cnode] (2) at (-2,2) {$1$};
\node[cnode] (3) at (0,2.5) {$1$};
\node[cnode] (4) at (2,2) {$1$};
\node[cnode] (5) at (3,0) {$1$};
\node[cnode] (6) at (2,-2) {$1$};
\node[cnode] (7) at (0,-2.5) {$1$};
\node[cnode] (8) at (-2,-2) {$1$};
\node[cnode] (9) at (0,1.5) {$1$};
\node[snode] (10) at (-5,0) {$1$};
\node[cnode] (11) at (4,0) {$1$};
\node[cnode] (12) at (5,0) {$1$};
\node[cnode] (13) at (8,0) {$1$};
\node[snode] (14) at (9,0) {$1$};
\node[cnode] (15) at (0,0.5) {$1$};
\node[cnode] (16) at (0,-1.5) {$1$};
\draw[thick] (1) to [bend left=40] (2);
\draw[thick, dashed] (2) to [bend left=40] (3);
\draw[thick] (3) to [bend left=40] (4);
\draw[thick, dashed] (4) to [bend left=40] (5);
\draw[thick] (5) to [bend left=40] (6);
\draw[thick,dashed] (6) to [bend left=40] (7);
\draw[thick,dashed] (7) to [bend left=40] (8);
\draw[thick] (8) to [bend left=40] (1);
\draw[thick] (1) -- (10);
\draw[thick] (5) -- (11);
\draw[thick] (3) -- (9);
\draw[thick] (9) -- (15);
\draw[thick] (11) -- (12);
\draw[thick,dashed] (12) -- (13);
\draw[thick] (13) -- (14);
\draw[thick, dashed] (15) -- (16);
\draw[thick] (16) -- (7);
\node[text width=1cm](10) at (-3,0.5) {$2p$};
\node[text width=1cm](11) at (-2,2.5) {$2p+1$};
\node[text width=1cm](12) at (0, 3) {$3p$};
\node[text width=1cm](13) at (2, 2.5) {$3p+1$};
\node[text width=1cm](14) at (3, 0.5) {$4p$};
\node[text width=1cm](15) at (2,-2.5) {$1$};
\node[text width=0.1 cm](16) at (0,-3.2){$p$};
\node[text width=1cm](17) at (-2.5,-2.6){$2p-1$};
\node[text width=0.1cm](18) at (4, 0.6){1};
\node[text width=0.1cm](19) at (5, 0.6){2};
\node[text width=1cm](20) at (8, 0.6){$p-2$};
\node[text width=1cm](21) at (1,1.5){$p-1$};
\node[text width=1cm](22) at (1,0.5){$p-2$};
\node[text width=1cm](23) at (1,-1.5){$1$};
\node[text width=0.1cm](30) at (0,-4){(Y)};
\end{tikzpicture}}
\caption{The pair of Lagrangian theories associated with the 3d $\CN=4$ SCFT that arise from the dimensional reduction of an $(A_s, A_{(s+1)p -1})$ AD theory 
on a circle, for $s=3$ and generic $p$. Quiver $(X)$ was obtained using class $\CS$ construction of AD theories, while quiver $(Y)$ is the proposed 3d mirror 
dual of quiver $(X)$.}
\label{ADEx2}
\end{center}
\end{figure}
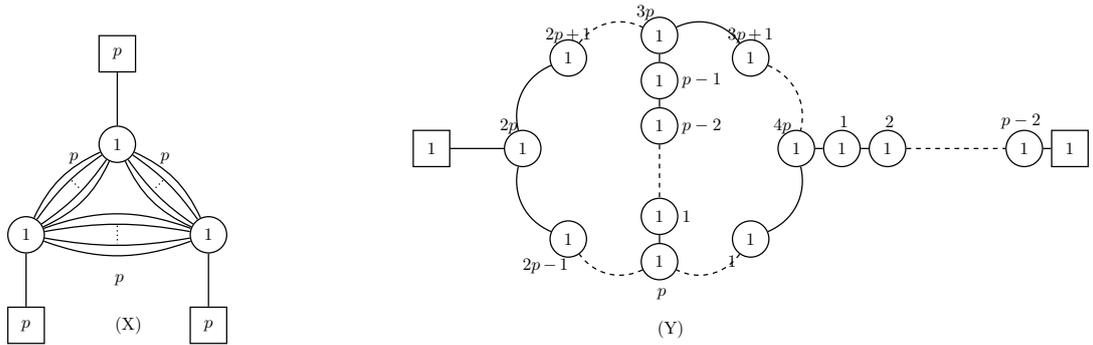

The dual pair in this case is given by \figref{ADEx2}. The class $\CS$ mirror is given by the quiver $X$, while 
the dual theory is given by the quiver $Y$. The duality can be read off from the dual pairs of Family ${\rm{II}}_{[n,l,l_1,l_2,p_1,p_2]}$ in \figref{AbEx2gen},
labelled by the set of integers $(n,l,l_1, l_2,p_1,p_2)$, for the special case:
\begin{align}
& n=4p, \quad l=2p, \quad l_1=p, \quad l_2=3p, \\
&p_1=p_2=p.
\end{align}
The moduli space dimensions of the dual theories 
and the global symmetries associated with the Higgs and Coulomb branches are given in Table \ref{Tab:ADEx2}. 

\begin{center}
\begin{table}[htbp]
\resizebox{\textwidth}{!}{%
\begin{tabular}{|c|c|c|}
\hline
Moduli space data & Theory $X$ & Theory $Y$ \\
\hline \hline 
dim\,$\CM_H$ & $6p-3$ & $3$\\
\hline
dim\,$\CM_C$ & $3$ & $6p-3$\\
\hline
$G_H$ & $SU(p)^3 \times  SU(p)^3  \times U(1)^3 $ & $U(1)^3$\\
\hline
$G_C$ & $U(1)^3$ & $SU(p)^3 \times  SU(p)^3  \times U(1)^3$\\
\hline
\end{tabular}}
\caption{Summary table for the moduli space dimensions and global symmetries for the mirror pair in \figref{ADEx2}.}
\label{Tab:ADEx2}
\end{table}
\end{center}

\subsection{Mirror pairs: 4d SCFTs from a single irregular puncture and a regular puncture}\label{AD-D}

\subsubsection{Trivial Case: $(A_1, D_{2p+2})$ theories for generic $p$}

\begin{figure}[htbp]
\begin{center}
\scalebox{0.8}{\begin{tikzpicture}[node distance=2cm,cnode/.style={circle,draw,thick,minimum size=8mm},snode/.style={rectangle,draw,thick,minimum size=8mm},pnode/.style={rectangle,red,draw,thick,minimum size=8mm}]
\node[cnode] (1) at (-2,0) {$1$};
\node[cnode] (2) at (2,0) {$1$};
\node[snode] (3) at (-2,-2) {$1$};
\node[snode] (4) at (2,-2) {$p$};
\draw[thick] (1) -- (3);
\draw[thick] (2) -- (4);
\draw[thick] (1) -- (2);
\node[text width=0.1cm](16) at (0,-3){$(X)$};
\end{tikzpicture}}
\qquad \qquad
\scalebox{0.8}{\begin{tikzpicture}[node distance=2cm,cnode/.style={circle,draw,thick,minimum size=8mm},snode/.style={rectangle,draw,thick,minimum size=8mm},pnode/.style={rectangle,red,draw,thick,minimum size=8mm}]
\node[cnode] (3) at (2,0) {$1$};
\node[snode] (4) at (2,-2) {$1$};
\node[cnode] (5) at (4,0) {$1$};
\node[cnode] (6) at (8,0) {$1$};
\node[snode] (7) at (8,-2) {$2$};
\draw[thick] (4) -- (3);
\draw[thick] (6) -- (7);
\draw[thick] (3) -- (5);
\draw[thick, dashed] (5) -- (6);
\draw[thick] (6) -- (7);
\node[text width=0.1cm](16) at (3,-3){$(Y)$};
\node[text width=0.1cm](17) at (2, 0.6){$1$};
\node[text width=0.1cm](18) at (4, 0.6){$2$};
\node[text width=0.1cm](19) at (8, 0.6){$p$};
\end{tikzpicture}}
\caption{An example of a Lagrangian realization of the dimensional reduction of an $(A_1, D_{2p +2})$ theory is given by the quiver $(Y)$ for generic $p$.}
\label{ADEx3}
\end{center}
\end{figure}
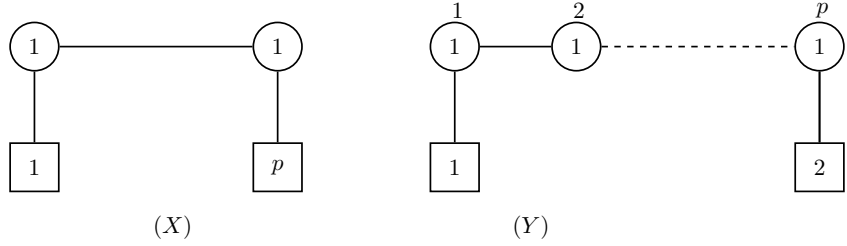

The mirror pair corresponding to the $(A_1, D_{2p+2})$ AD theories, like the $(A_1, A_{2p-1})$ case, 
is well known. The class $\CS$ mirror, labelled as quiver $X$ in \figref{ADEx3}, is a linear quiver 
which has a linear mirror dual given by quiver $Y$. The dimensions of the moduli 
spaces and global symmetries are summarized in Table \ref{Tab:ADEx3}.
\begin{table}[htbp]
\begin{center}
\begin{tabular}{|c|c|c|}
\hline
Moduli space data & Theory $X$ & Theory $Y$ \\
\hline \hline 
dim\,$\CM_H$ & $p$ & $2$\\
\hline
dim\,$\CM_C$ & $2$ & $p$ \\
\hline
$G_H$ & $SU(p) \times U(1) $ & $SU(2) \times U(1)$ \\
\hline
$G_C$ & $SU(2) \times U(1)$ & $SU(p) \times U(1) $ \\
\hline
\end{tabular}
\caption{Summary table for the moduli space dimensions and global symmetries for the mirror pair in \figref{ADEx3}.}
\label{Tab:ADEx3}
\end{center}
\end{table}

\subsubsection{$(A_2, D_{3p+2})$ theories for generic $p$}

The dual pair in this case is given by \figref{ADEx4}. The class $\CS$ mirror is given by the quiver $X$, while 
the dual theory is given by the quiver $Y$. The duality can be read off from the dual pairs of Family ${\rm{II}}_{[n,l,l_1,l_2,p_1,p_2]}$ in \figref{AbEx2gen},
labelled by the set of integers $(n,l,l_1, l_2,p_1,p_2)$, for the special case:
\begin{align}
& n=2p+2, \quad l=p+1, \quad l_1=p, \quad l_2=p+2, \\
&p_1=p,\, p_2=1.
\end{align}
The moduli space dimensions of the dual theories 
and the global symmetries associated with the Higgs and Coulomb branches are given in Table \ref{Tab:ADEx4}. 

\begin{center}
\begin{table}[htbp]
\resizebox{\textwidth}{!}{%
\begin{tabular}{|c|c|c|}
\hline
Moduli space data & Theory $X$ & Theory $Y$ \\
\hline \hline 
dim\,$\CM_H$ & $3p$ & $3$\\
\hline
dim\,$\CM_C$ & $3$ & $3p$\\
\hline
$G_H$ & $SU(p) \times  SU(p)  \times SU(p) \times U(1)^3 $ & $U(1)^3$ \\
\hline
$G_C$ & $U(1)^3$ & $SU(p) \times  SU(p)  \times SU(p) \times U(1)^3 $\\
\hline
\end{tabular}}
\caption{Summary table for the moduli space dimensions and global symmetries for the mirror pair in \figref{ADEx4}.}
\label{Tab:ADEx4}
\end{table}
\end{center}

\begin{figure}[htbp]
\begin{center}
\scalebox{.7}{\begin{tikzpicture}[node distance=2cm,cnode/.style={circle,draw,thick,minimum size=8mm},snode/.style={rectangle,draw,thick,minimum size=8mm},pnode/.style={rectangle,red,draw,thick,minimum size=8mm}]
\node[cnode] (1) at (-2,0) {$1$};
\node[cnode] (2) at (2,0) {$1$};
\node[cnode] (3) at (0,2) {$1$};
\node[snode] (4) at (-2,-2) {$p$};
\node[snode] (5) at (2,-2) {$p$};
\node[snode] (6) at (0,4) {$1$};
\draw[thick] (1) -- (4);
\draw[thick] (2) -- (5);
\draw[thick] (3) -- (6);
\draw[thick] (1) to [bend left=10] (2);
\draw[thick] (1) to [bend right=10] (2);
\draw[thick] (1) to [bend left=20] (2);
\draw[thick] (1) to [bend right=20] (2);
\draw[thick, dotted] (0,0.2) to (0,-0.2);
\node[text width=0.1cm](15) at (0,-1){$p$};
\draw[thick] (1) -- (3);
\draw[thick] (2) -- (3);
\node[text width=0.1cm](16) at (0,-2){$(X)$};
\end{tikzpicture}}
\qquad \qquad \qquad 
\scalebox{0.7}{\begin{tikzpicture}[node distance=2cm,cnode/.style={circle,draw,thick,minimum size=8mm},snode/.style={rectangle,draw,thick,minimum size=8mm},pnode/.style={rectangle,red,draw,thick,minimum size=8mm}]
\node[cnode] (1) at (-2,0) {$1$};
\node[cnode] (2) at (-1,1) {$1$};
\node[cnode] (3) at (1,1) {$1$};
\node[cnode] (4) at (2,0) {$1$};
\node[cnode] (5) at (1,-1) {$1$};
\node[cnode] (6) at (-1,-1) {$1$};
\node[cnode] (7) at (3,0) {$1$};
\node[cnode] (8) at (4,0) {$1$};
\node[cnode] (9) at (5,0) {$1$};
\node[cnode] (10) at (7,0) {$1$};
\node[snode] (11) at (8,0) {$1$};
\node[snode] (12) at (-2,-2) {$1$};
\draw[thick] (1) to [bend left=40] (2);
\draw[dashed,thick] (2) to [bend left=40] (3);
\draw[thick] (3) to [bend left=40] (4);
\draw[thick] (4) to [bend left=40] (5);
\draw[dashed,thick] (5) to [bend left=40] (6);
\draw[thick] (6) to [bend left=40] (1);
\draw[thick] (2) to [bend left=40] (6);
\draw[thick] (1) -- (12);
\draw[thick] (4) -- (7);
\draw[thick] (7) -- (8);
\draw[thick] (8) -- (9);
\draw[thick, dashed] (9) -- (10);
\draw[thick] (10) -- (11);
\node[text width=1cm](20) at (-2,0.5) {$p+1$};
\node[text width=1cm](21) at (-1,1.5) {$p+2$};
\node[text width=1cm](22) at (1,1.5) {$2p+1$};
\node[text width=1cm](23) at (2,0.5) {$2p+2$};
\node[text width=0.1cm](24) at (1,-1.5) {$1$};
\node[text width=1cm](25) at (-1,-1.5) {$p$};
\node[text width=0.1cm](26) at (3,0.7) {$1$};
\node[text width=0.1cm](27) at (4,0.7) {$2$};
\node[text width=0.1cm](28) at (5,0.7) {$3$};
\node[text width=1 cm](29) at (7,0.7) {$p-2$};
\node[text width=0.1cm](30) at (3,-3){$(Y)$};
\end{tikzpicture}}
\caption{The pair of Lagrangian theories associated with the 3d $\CN=4$ SCFT that arise from the dimensional reduction of an $(A_s, D_{(s+1)p +2})$ AD theory 
on a circle, for $s=2$ and generic $p$. Quiver $(X)$ was obtained using class $\CS$ construction of AD theories, while quiver $(Y)$ is the proposed 3d mirror 
dual of quiver $(X)$.}
\label{ADEx4}
\end{center}
\end{figure}

\subsubsection{4d SCFTs from an irregular puncture and a maximal puncture}
We consider the specific case of the 4d SCFTs $A^{{\rm{maximal}}}_{s,p}$, for $s=2$ and generic $p$. As mentioned earlier, 
these SCFTs are realized from the twisted compactification of a $A_{s=2}$ (2,0) 6d theory on a Riemann sphere with an 
irregular puncture (corresponding to the SCFT $(A_2, A_{3p-1})$) and a maximal regular puncture. 
The mirror pair in this case is given by \figref{ADEx5}. The class $\CS$ mirror is given by the quiver $X$, while 
the dual theory is given by the quiver $Y$. The duality can be read off from the dual pairs of Family ${\rm{IV}}_{[p_1,p_2, p_3]}$ in \figref{fig: AbEx4gen},
labelled by the triplet of integers $(p_1,p_2,p_3)$, for the special case:
\be
p_1=p_2=p_3=p.
\ee 
The moduli space dimensions of the dual theories 
and the global symmetries associated with the Higgs and Coulomb branches are given in Table \ref{Tab:ADEx5}. 

\begin{center}
\begin{table}[htbp]
\resizebox{\textwidth}{!}{%
\begin{tabular}{|c|c|c|}
\hline
Moduli space data & Theory $X$ & Theory $Y$ \\
\hline \hline 
dim\,$\CM_H$ & $3p+1$ & $5$\\
\hline
dim\,$\CM_C$ & $5$ & $3p+1$\\
\hline
$G_H$ & $SU(p) \times  SU(p)  \times SU(p) \times U(1)^3 $ & $SU(3) \times U(1)^2$ \\
\hline
$G_C$ & $SU(3) \times U(1)^2$ & $SU(p) \times  SU(p)  \times SU(p) \times U(1)^3 $\\
\hline
\end{tabular}}
\caption{Summary table for the moduli space dimensions and global symmetries for the mirror pair in \figref{ADEx5}.}
\label{Tab:ADEx5}
\end{table}
\end{center}

\begin{figure}[htbp]
\begin{center}
\scalebox{0.6}{\begin{tikzpicture}[
cnode/.style={circle,draw,thick, minimum size=1.0cm},snode/.style={rectangle,draw,thick,minimum size=1cm}]
\node[cnode] (9) at (0,1){1};
\node[snode] (10) at (0,-1){1};
\node[cnode] (11) at (2, 0){2};
\node[cnode] (12) at (4, 1){1};
\node[cnode] (13) at (4, -1){1};
\node[snode] (14) at (6, 1){$p$};
\node[snode] (15) at (6, -1){$p$};
\draw[-] (9) -- (11);
\draw[-] (10) -- (11);
\draw[-] (12) -- (11);
\draw[-] (13) -- (11);
\draw[-] (12) -- (14);
\draw[-] (13) -- (15);
\draw[thick] (12) to [bend left=10] (13);
\draw[thick] (12) to [bend right=10] (13);
\draw[thick] (12) to [bend left=20] (13);
\draw[thick] (12) to [bend right=20] (13);
\draw[thick, dotted] (3.85,0) to (4.2, 0);
\node[text width=0.1cm](20) at (4.5,0){$p$};
\node[text width=0.1cm](21)[above=0.2 cm of 9]{1};
\node[text width=0.1cm](22)[above=0.2 cm of 11]{2};
\node[text width=0.1cm](23)[above=0.2 cm of 12]{3};
\node[text width=0.1cm](24)[below=0.05 cm of 13]{4};
\node[text width=0.1cm](21)[below=0.5 cm of 13]{$(X)$};
\end{tikzpicture}}
\qquad \qquad 
\scalebox{0.6}{\begin{tikzpicture}[node distance=2cm,cnode/.style={circle,draw,thick,minimum size=8mm},snode/.style={rectangle,draw,thick,minimum size=8mm},pnode/.style={rectangle,red,draw,thick,minimum size=8mm}]
\node[cnode] (1) at (-3,0) {$2$};
\node[cnode] (2) at (-2,2) {$1$};
\node[cnode] (3) at (0,2.5) {$1$};
\node[cnode] (4) at (2,2) {$1$};
\node[cnode] (5) at (3,0) {$1$};
\node[cnode] (6) at (2,-2) {$1$};
\node[cnode] (7) at (0,-2.5) {$1$};
\node[cnode] (8) at (-2,-2) {$1$};
\node[snode] (10) at (-5,0) {$3$};
\node[cnode] (11) at (4,0) {$1$};
\node[cnode] (12) at (5,0) {$1$};
\node[cnode] (13) at (7.5,0) {$1$};
\node[snode] (14) at (9,0) {$1$};
\draw[thick] (1) to [bend left=40] (2);
\draw[thick] (2) to [bend left=40] (3);
\draw[thick,dashed] (3) to [bend left=40] (4);
\draw[thick] (4) to [bend left=40] (5);
\draw[thick] (5) to [bend left=40] (6);
\draw[dashed] (6) to [bend left=40] (7);
\draw[thick] (7) to [bend left=40] (8);
\draw[thick] (8) to [bend left=40] (1);
\draw[thick] (1) -- (10);
\draw[thick] (5) -- (11);
\draw[thick] (11) -- (12);
\draw[dashed] (12) -- (13);
\draw[thick] (13) -- (14);
\node[text width=1cm](11) at (-2,2.5) {$1$};
\node[text width=1cm](12) at (0, 3) {$2$};
\node[text width=1cm](13) at (2, 2.5) {$p-1$};
\node[text width=1cm](15) at (2,-2.5) {$p$};
\node[text width=1.5 cm](16) at (0,-3.2){$2$};
\node[text width=1cm](17) at (-2.5,-2.5){$1$};
\node[text width=0.1cm](18) at (4, 0.6){$1$};
\node[text width=0.1cm](19) at (5, 0.6){$2$};
\node[text width=1cm](20) at (7.5, 0.6){$p-1$};
\node[text width=1cm](21) at (-2.5, 1){$\CA$};
\node[text width=0.1cm](30) at (0,-4){$(Y)$};
\end{tikzpicture}}
\caption{An example of a Lagrangian realization of the dimensional reduction of a 4d SCFT which is constructed in the class $\CS$ picture 
from an irregular puncture and a regular maximal puncture.}
\label{ADEx5}
\end{center}
\end{figure}

\section*{Acknowledgement}
The author would like to thank Jacques Distler and Andrew Neitzke for various comments 
and discussion during the preparation of this work. The author would like to thank Vivek 
Saxena for comments on the draft and discussion on related issues.
The author would also like to thank 
Ibrahima Bah, Aswin Balasubramanian, Amihay Hanany, Gregory Moore, Wolfger Peelaers, Martin Ro\v{c}ek, and Jaewon 
Song for discussion at various stages of the project. The author is partially supported at the Johns 
Hopkins University by NSF grant PHY-1820784. A significant fraction of the work was done at the
NHETC, Rutgers, where the author's postdoctoral fellowship was supported by the DOE grant DOE-SC0010008.

\appendix

\section{Flavored $D_4$ quivers: Dependence of the dual Lagrangian on $\CP$}\label{GFI-ex}
In this section, we work out two sets of simple examples of mirror dual theories, involving D-type quivers (and their affine cousins), 
which can be obtained from linear quivers using Abelian $S$-type operations. The first set involves gauging 
operations, while the second involves flavoring-gauging operations. The resultant dual pairs have appeared in the literature of 3d mirror symmetry 
\cite{Dey:2013nf, Dey:2014tka, Fan:2019jii}, and were studied from a Type IIB point of view in \cite{Hanany:1999sj, Gaiotto:2008ak}.  We present them here to illustrate how the Abelian $S$-type operations work in simple examples involving non-Abelian gauge groups. In addition, we show that the mirror dual of a given theory can have multiple Lagrangian 
descriptions labelled by the data $\CP$. As mentioned earlier, we turn off all defects and CS interactions for these examples.

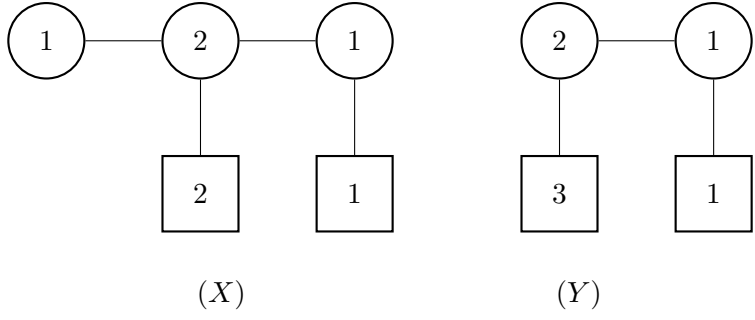
\begin{figure}[htbp]
\begin{center}
\begin{tikzpicture}[
cnode/.style={circle,draw,thick, minimum size=1.0cm},snode/.style={rectangle,draw,thick,minimum size=1cm}]
\node[cnode] (1) {1};
\node[cnode] (2) [right=1cm  of 1]{2};
\node[cnode] (3) [right=1cm  of 2]{1};
\node[snode] (4) [below=1cm of 2]{2};
\node[snode] (5) [below=1cm of 3]{1};
\node[text width=0.1cm](20)[below=0.5 cm of 4]{$(X)$};
\draw[-] (1) -- (2);
\draw[-] (2)-- (3);
\draw[-] (2)-- (4);
\draw[-] (3)-- (5);
\end{tikzpicture}
\qquad \qquad 
\begin{tikzpicture}[
cnode/.style={circle,draw,thick, minimum size=1.0cm},snode/.style={rectangle,draw,thick,minimum size=1cm}]
\node[cnode] (1) {2};
\node[snode] (2) [below=1cm of 1]{3};
\node[cnode] (3) [right=1cm of 1]{1};
\node[snode] (4) [below=1cm of 3]{1};
\draw[-] (1) -- (2);
\draw[-] (1) -- (3);
\draw[-] (3) -- (4);
\node[text width=0.1cm](20)[below=0.5 cm of 2]{$(Y)$};
\end{tikzpicture}
\caption{A pair of linear quiver theories mirror dual to each other.}
\label{fig: LQEx2}
\end{center}
\end{figure}

Consider the pair of linear quiver gauge theories, $X$ and $Y$, in \figref{fig: LQEx2}, which are mirror dual to each other.
The theory $X$ has a Higgs branch global symmetry $G^{X}_H= (U(2) \times U(1))/U(1)$, where the Cartan subalgebras of $U(2)$ and $U(1)$ are 
parametrized by the mass parameters $(m_1, m_2)$ and $m_3$ respectively. We choose to implement the $U(1)$ quotient 
by the constraint $m_3=0$, and therefore identify $G^{X}_H= U(2)$ and with the $U(2)$ flavor node of quiver $X$.
We will implement the Abelian $S$-type operations at this $U(2)$ flavor node. The $S^3$ partition function of $X$ is given as:
\begin{align}
& Z^{(X)}(\vec{m}, \vec{t})= \int \prod^3_{\gamma=1}\Big[d\vec{s}^\gamma \Big] Z^{(X)}_{\rm int} (\{\vec{s}^{\gamma'}\}, \vec{m}, \vec{t})\nn \\
= & \int Z^{(X)}_{\rm FI} (\{\vec{s}^\gamma\}, \vec t)\,Z^{\rm vector}_{\rm{1-loop}}(\vec s^2)\prod^2_{i=1} Z^{\rm fund}_{\rm{1-loop}}(\vec s^2, \vec m_i)\,Z^{\rm fund}_{\rm{1-loop}}(\vec s^3,0)\prod^{2}_{\gamma=1} Z^{\rm bif}_{\rm{1-loop}}(\vec s^\gamma, \vec s^{\gamma +1},0),
\end{align}
where $Z^{(X)}_{\rm FI}$ and the one-loop factors are given in \eref{PF-Agen1}. Similarly, the partition function of the quiver $Y$ is given as
\begin{align}
& Z^{(Y)}(\vec{t}, \vec{m})=  \int \prod^2_{\gamma'=1}\Big[d\vec{\s}^{\gamma'} \Big] Z^{(Y)}_{\rm int} (\{\vec{\s}^\gamma\}, \vec{t}, \vec{m})\nn \\
=& \int Z^{(Y)}_{\rm FI}(\{\vec \s^{\gamma'} \}, \vec m)  Z^{\rm vector}_{\rm{1-loop}}(\vec \s^1)\, \prod^3_{l=1}Z^{\rm fund}_{\rm{1-loop}}(\vec \s^{1}, t_l)\, Z^{\rm fund}_{\rm{1-loop}}(\s^{2}, t_4)\,Z^{\rm bif}_{\rm{1-loop}}(\vec \s^1, \s^2,0), 
\end{align}
where $Z^{(Y)}_{\rm FI}$ and the one-loop factors are given in \eref{PF-Bgen1}. In particular, $Z^{(Y)}_{\rm FI}$ can be explicitly written as
\be\label{FIEx1Y}
Z^{(Y)}_{\rm FI}(\{\vec \s^{\gamma'} \}, \vec m)= e^{2\pi i ({m}_1- {m}_2)\tr \vec\s^1}\, e^{2\pi i  {m}_2 \s^2}.
\ee
Mirror symmetry implies that the two partition functions are related in the following fashion:
\be \label{MSEx1}
Z^{(X)}(\vec{m}, \vec{t})=e^{2\pi i (m_1(t_1+t_2) -m_2 t_3)}\, Z^{(Y)}(-\vec{t}, \vec{m}).
\ee

 \subsection{Abelian Gauging Operations}\label{AbGD4}
\begin{figure}[htbp]
\begin{center}
\scalebox{0.7}{\begin{tikzpicture}[
cnode/.style={circle,draw,thick, minimum size=1.0cm},snode/.style={rectangle,draw,thick,minimum size=1cm}, pnode/.style={rectangle,red, draw,thick, minimum size=1.0cm}]
\node[cnode] (1) {1};
\node[cnode] (2) [right=1cm  of 1]{2};
\node[cnode] (3) [right=1cm  of 2]{1};
\node[pnode] (4) [below=1cm of 2]{2};
\node[snode] (5) [below=1cm of 3]{1};
\node[text width=0.1cm](20)[below=0.5 cm of 4]{$(X)$};
\draw[-] (1) -- (2);
\draw[-] (2)-- (3);
\draw[-] (2)-- (4);
\draw[-] (3)-- (5);
\draw[->] (5,0) -- (8,0);
\node[text width=3cm](7) at (7, 0.5){Abelian};
\node[text width=3cm](8) at (7, -0.5){Gauging}; 
\node[cnode] (9) at (10,1){1};
\node[cnode] (10) at (10,-1){1};
\node[cnode] (11) at (12, 0){2};
\node[cnode] (12) at (14, 1){1};
\node[snode] (13) at (14, -1){1};
\node[snode] (14) at (16, 1){1};
\draw[-] (9) -- (11);
\draw[-] (10) -- (11);
\draw[-] (12) -- (11);
\draw[-] (13) -- (11);
\draw[-] (12) -- (14);
\node[text width=0.1cm](21)[below=0.5 cm of 13]{$(X')$};
\end{tikzpicture}}
\caption{A $U(1)$ gauging operation on the quiver $X$ leading to the theory $X'$.}
\label{fig: GEx1a}
\end{center}
\end{figure}
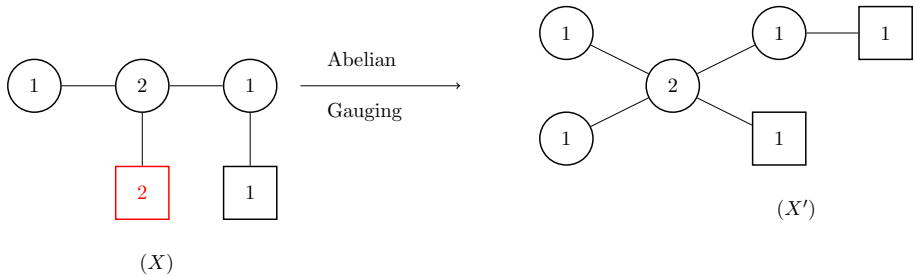

Let us implement an Abelian Gauging operation at the $U(2)$ flavor node of theory $(X, \CP)$, 
which we label as $\alpha$. 
The theory $X' = G^\alpha_{\CP}(X)$ is given by \figref{fig: GEx1a}, and is independent of the permutation 
matrix $\CP$. The Lagrangian of the dual theory, however, depends on the choice of the matrix $\CP$, 
and we will derive the dual Lagrangians explicitly for the two possible choices of $\CP$. 
Consider first the choice $\CP=\CP_1 = \mathbb{I}_{2\times 2}$, for which, following \eref{uvdef0},we define
\be 
u^\alpha=m_1, \quad v^\alpha=m_2.
\ee
The mirror symmetry relation \eref{MSEx1} then implies that
\be
Z^{(X,\CP_1)}(u^\alpha,v^\alpha, \vec{t})= e^{2\pi i (u^\alpha (t_1+t_2) -v^\alpha t_3)}\, Z^{(Y,\CP_1)}(-\vec{t}, u^\alpha,v^\alpha),
\ee
and the FI contribution to the partition function of the theory $(Y,\CP_1)$ in \eref{FIEx1Y} can be written as
\be
Z^{(Y,\CP_1)}_{\rm FI}(\{\vec \s^{\gamma'} \}, u^\alpha, v^\alpha)= e^{2\pi i u^\alpha \tr \vec\s^1}\, e^{2\pi i  v^\alpha (\s^2 - \tr \vec{\s}^1)}.
\ee

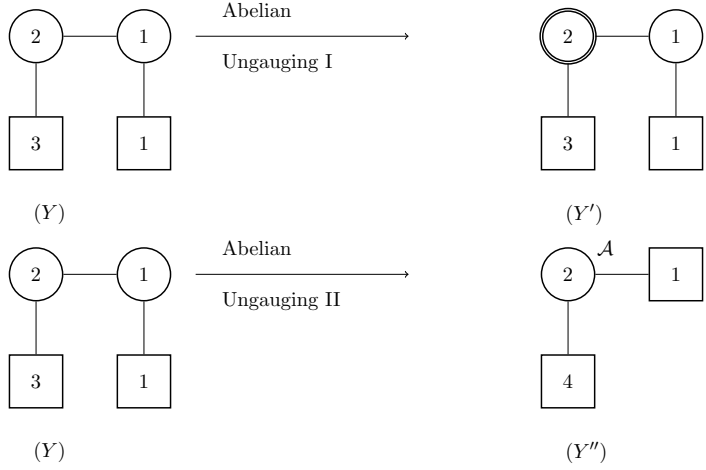
\begin{figure}[htbp]
\begin{center}
\scalebox{0.7}{\begin{tikzpicture}[
cnode/.style={circle,draw,thick, minimum size=1.0cm},snode/.style={rectangle,draw,thick,minimum size=1cm},pnode/.style={circle,double,draw,thick, minimum size=1.0cm}]
\node[cnode] (1) {2};
\node[snode] (2) [below=1cm of 1]{3};
\node[cnode] (3) [right=1cm of 1]{1};
\node[snode] (4) [below=1cm of 3]{1};
\draw[-] (1) -- (2);
\draw[-] (1) -- (3);
\draw[-] (3) -- (4);
\node[text width=0.1cm](20)[below=0.5 cm of 2]{$(Y)$};
\draw[->] (3,0) -- (7,0);
\node[text width=3cm](7) at (5, 0.5){Abelian};
\node[text width=3cm](8) at (5, -0.5){Ungauging I}; 
\node[pnode] (9) at (10,0){2};
\node[snode] (10) [below=1cm of 9]{3};
\node[cnode] (11) [right=1cm of 9]{1};
\node[snode] (12) [below=1cm of 11]{1};
\draw[-] (9) -- (10);
\draw[-] (9) -- (11);
\draw[-] (11) -- (12);
\node[text width=0.1cm](21)[below=0.5 cm of 10]{$(Y')$};
\end{tikzpicture}}
\qquad \qquad \qquad \qquad 
\scalebox{0.7}{\begin{tikzpicture}[
cnode/.style={circle,draw,thick, minimum size=1.0cm},snode/.style={rectangle,draw,thick,minimum size=1cm}]
\node[cnode] (1) {2};
\node[snode] (2) [below=1cm of 1]{3};
\node[cnode] (3) [right=1cm of 1]{1};
\node[snode] (4) [below=1cm of 3]{1};
\draw[-] (1) -- (2);
\draw[-] (1) -- (3);
\draw[-] (3) -- (4);
\node[text width=0.1cm](20)[below=0.5 cm of 2]{$(Y)$};
\draw[->] (3,0) -- (7,0);
\node[text width=3cm](7) at (5, 0.5){Abelian};
\node[text width=3cm](8) at (5, -0.5){Ungauging II}; 
\node[cnode] (9) at (10,0){2};
\node[snode] (10) [below=1cm of 9]{4};
\draw[-] (9) -- (10);
\node[snode] (11) [right=1cm of 9]{1};
\draw[-] (9) -- (11);
\node[text width=0.1cm](12) at (10.6,0.5){$\CA$};
\node[text width=0.1cm](21)[below=0.5 cm of 10]{$(Y'')$};
\end{tikzpicture}}
\caption{The figure shows the two quiver gauge theories -- $Y'$ and $Y''$-- that arise from the Abelian ungauging operation on the theory $(Y)$
for two different choices of $\CP$. The line labelled as $\CA$ denotes a hypermultiplet in the rank-2 antisymmetric 
representation of the gauge group $U(2)$ and has charge $1$ under the flavor symmetry group $U(1)$. It is charge 2 under the $U(1)$ subgroup of $U(2)$, 
and uncharged under the $SU(2)$.}
\label{fig: GEx1b}
\end{center}
\end{figure}

From the general formula \eref{G0Ab} for the dual of Abelian gauging, we then have
\begin{align}
Z^{\wt{G}_{\CP_1}(Y)}(\vec{m}'(\vec{t},\eta_\alpha); {\eta}'({v^\alpha})) =\int \prod^{2}_{\gamma'=1} \Big[d\s^{\gamma'}\Big]\,  \delta \Big(\eta_\alpha + b^{l}t_l + \tr \vec{\s}^1 \Big)\cdot Z^{(Y,\CP_1)}_{\rm int}(\{\vec \s^{\gamma'} \}, -\vec t, {u}^\alpha=0, {v}^\alpha),
\end{align}
where $b^l=1, \, l=1,2$. Shifting the integration variables, one can recast the above integral (up to some phase factors) as:
\begin{align}
Z^{\wt{G}_{\CP_1}(Y)}(\vec{m}'(\vec{t},\eta_\alpha); {\eta}'({v}^\alpha)) = &\int \prod^{2}_{\gamma'=1} \Big[d\s^{\gamma'}\Big]\,  \delta \Big(\tr \vec{\s}^1 \Big)\,Z_{\rm FI}(\s^2, v^\alpha)  Z^{\rm vector}_{\rm{1-loop}}(\vec \s^1)\, \prod^3_{l=1}Z^{\rm fund}_{\rm{1-loop}}(\vec \s^{1}, -t_l + \delta)\nn \\
&\qquad \qquad \qquad \times Z^{\rm fund}_{\rm{1-loop}}(\s^{2}, -t_4 + \delta)\,Z^{\rm bif}_{\rm{1-loop}}(\vec \s^1, \s^2,0) \nn \\
=&\int \prod^{2}_{\gamma'=1} \Big[d\s^{\gamma'}\Big]\,Z^{(Y')}_{\rm int}(\{\vec \s^{\gamma'} \}, \vec{m}'= -\vec t + \delta, \eta'=v^\alpha)  \nn \\
=& Z^{(Y')}(\vec{m}'= -\vec t + \delta, \eta'=v^\alpha),
\end{align}
where $\delta=\frac{\eta_\alpha + b^{l}t_l}{2}$, and the theory $(Y')$ is the quiver in \figref{fig: GEx1b}. The map of mass and FI parameters between 
the pair of dual theories $(X',Y')$ can be read off from the final equality.\\

Now, consider the other possible choice of $\CP=\CP_2=\begin{pmatrix} 0 & 1 \\ 1 & 0 \end{pmatrix}$, such that 
\eref{uvdef0}
\be 
v^\alpha=m_1, \quad u^\alpha=m_2.
\ee
The mirror symmetry relation \eref{MSEx1} then implies that
\be
Z^{(X,\CP_2)}(u^\alpha,v^\alpha, \vec{t})= e^{2\pi i (v^\alpha (t_1+t_2) - u^\alpha t_3)}\, Z^{(Y,\CP_2)}(-\vec{t}, u^\alpha,v^\alpha),
\ee
and the FI contribution to the partition function of the theory $(Y,\CP_2)$ in \eref{FIEx1Y} can be written as
\be
Z^{(Y,\CP_2)}_{\rm FI}(\{\vec \s^{\gamma'} \}, u^\alpha, v^\alpha)= e^{2\pi i v^\alpha \tr \vec\s^1}\, e^{2\pi i  u^\alpha (\s^2 - \tr \vec{\s}^1)}.
\ee
Using \eref{G0Ab} and following the same steps as above, we can show that (up to some phase factors)
\begin{align}
Z^{\wt{G}_{\CP_2}(Y)}(\vec{m}'(\vec{t},\eta_\alpha); {\eta}'({v^\alpha})) = &\int \frac{d\s^{1}}{2!}\,Z_{\rm FI}(\vec{\s}^1, v^\alpha)  Z^{\rm vector}_{\rm{1-loop}}(\vec \s^1)\, \prod^4_{l=1}Z^{\rm fund}_{\rm{1-loop}}(\vec \s^{1}, {m}'_l(\vec t, \eta_\alpha))\, Z^{\CA}_{\rm{1-loop}}(\vec \s^1, -t_4) \nn \\
=&\int \frac{d^2\s^{1}}{2!}\,Z^{(Y'')}_{\rm int}(\vec \s^{1}, \vec{m}'(\vec t, \eta_\alpha), \eta'=v^\alpha)  \nn \\
=& Z^{(Y'')}(\vec{m}'(\vec t, \eta_\alpha), \eta'=v^\alpha),
\end{align}
where $Z^{\CA}_{\rm{1-loop}}$ denotes the 1-loop contribution of a hypermultiplet in a rank-1 antisymmetric representation of $U(2)$,
and $Y''$ is the quiver shown in the second row of \figref{fig: GEx1b}. Note that $Z^{\CA}_{\rm{1-loop}}(\vec \s^1, -t_4)= 
Z^{\rm fund}_{\rm{1-loop}}(\tr \vec \s^1, -t_4)$. The mass parameters of $Y''$ are given as
\begin{align}
& m'_l=-t_l - \delta_0, \, \text{for }\, l=1,2,3, \label{MMD2a}\\
&m'_4=\eta_\alpha-t_3 -\delta_0, \label{MMD2b}\\
& m'_{\CA}=\eta_\alpha - t_3 - t_4 -2\delta_0, \label{MMD2c}
\end{align}
where $\delta_0$ should be chosen depending on the constraint one imposes on the masses $Y''$ - $\sum^4_{l=1} m'_l=0$,
or $m'_{\CA}=0$. The FI parameter of $Y''$ is 
\be
\eta'=v^\alpha=m_1.
\ee
We conjecture that the two theories $Y'$ and $Y''$ are different Lagrangians of the theory mirror dual to $X'$.
The global symmetries  (Higgs and Coulomb) of $Y'$ and $Y''$ match (up to a discrete group):
\begin{align}
& G^{Y'}_{H}=SO(6) \times U(1), \qquad &G^{Y'}_{C}= U(1), \nn \\
& G^{(Y'')}_{H}= SU(4) \times U(1), \qquad &G^{(Y'')}_{\rm C}=U(1). 
\end{align}
Note that the Higgs branch symmetry associated with $n$ fundamental hypers (i.e. $2n$ fundamental half-hypers) in $Y'$ is $SO(2n)$. Also, note that the 
Coulomb branch symmetry of $(Y'')$ is not enhanced because of the additional antisymmetric hyper.\\

\begin{figure}[htbp]
\begin{center}
\scalebox{0.7}{\begin{tikzpicture}[
cnode/.style={circle,draw,thick, minimum size=1.0cm},snode/.style={rectangle,draw,thick,minimum size=1cm},pnode/.style={rectangle,red, draw,thick, minimum size=1.0cm}]
\node[cnode] (1) {1};
\node[cnode] (2) [right=1cm  of 1]{2};
\node[cnode] (3) [right=1cm  of 2]{1};
\node[pnode] (4) [below=1cm of 2]{2};
\node[snode] (5) [below=1cm of 3]{1};
\node[text width=0.1cm](20)[below=0.5 cm of 4]{$(X)$};
\draw[-] (1) -- (2);
\draw[-] (2)-- (3);
\draw[-] (2)-- (4);
\draw[-] (3)-- (5);
\draw[->] (5,0) -- (8,0);
\node[text width=3cm](7) at (7, 0.5){Abelian};
\node[text width=3cm](8) at (7, -0.5){$S$-Operation}; 
\node[cnode] (9) at (10,1){1};
\node[cnode] (10) at (10,-1){1};
\node[cnode] (11) at (12, 0){2};
\node[cnode] (12) at (14, 1){1};
\node[cnode] (13) at (14, -1){1};
\node[snode] (14) at (16, 1){1};
\draw[-] (9) -- (11);
\draw[-] (10) -- (11);
\draw[-] (12) -- (11);
\draw[-] (13) -- (11);
\draw[-] (12) -- (14);
\node[text width=0.1cm](21)[below=0.5 cm of 13]{$(X')$};
\end{tikzpicture}}
\qquad \qquad \qquad \qquad 
\scalebox{0.7}{\begin{tikzpicture}[
cnode/.style={circle,draw,thick, minimum size=1.0cm},snode/.style={rectangle,draw,thick,minimum size=1cm},pnode/.style={circle,draw,double,thick, minimum size=1.0cm}]
\node[cnode] (1) {2};
\node[snode] (2) [below=1cm of 1]{3};
\node[cnode] (3) [right=1cm of 1]{1};
\node[snode] (4) [below=1cm of 3]{1};
\draw[-] (1) -- (2);
\draw[-] (1) -- (3);
\draw[-] (3) -- (4);
\node[text width=0.1cm](20)[below=0.5 cm of 2]{$(Y)$};
\draw[->] (3,0) -- (7,0);
\node[text width=3cm](7) at (5, 0.5){Dual};
\node[text width=3cm](8) at (5, -0.4){Operation}; 
\node[pnode] (9) at (10,0){2};
\node[snode] (10) [below=1cm of 9]{4};
\draw[-] (9) -- (10);
\node[snode] (11) at (12,0){1};
\node[text width=0.1cm](21)[below=0.5 cm of 10]{$(Y')$};
\end{tikzpicture}}
\caption{This figure shows the action of an Abelian $S$-type operation (involving two elementary gauging operations) on the quiver $X$ 
and the dual operation the quiver $Y$ of \figref{fig: LQEx2} respectively. The quiver $X'$ is mirror dual to the quiver $Y'$.}
\label{fig: GEx1}
\end{center}
\end{figure}
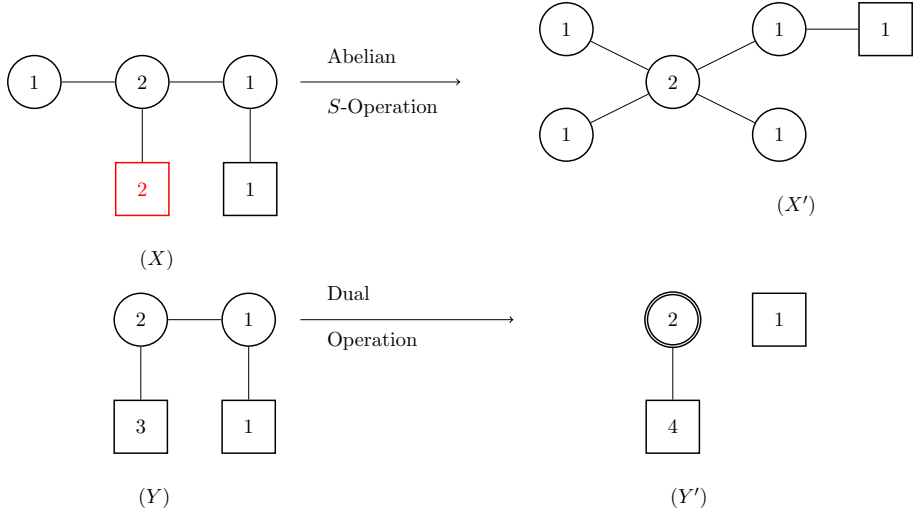

Finally, consider a sequence of two Abelian gauging operations on the quiver $X$, such that the $U(1) \times U(1)$ subgroup of the $U(2)$ 
flavor symmetry of quiver $X$ is gauged. We will denote these $S$-type operations as $G_1$ and $G_2$ respectively. 
The resultant quiver $X'=G_2\circ G_1(X)$ is an affine $D_4$ quiver with a single flavor, shown in \figref{fig: GEx1}.
In this case, the dual theory is independent of the permutation matrix $\CP$ and therefore one can drop the explicit $\CP$-dependence 
from the partition functions. Let us define:
\be
u^1=m_1, \quad u^2= m_2.
\ee

The mirror symmetry relation \eref{MSEx1} then implies that
\be
Z^{(X)}(u^1, u^2, \vec{t})= e^{2\pi i (u^1 (t_1+t_2) - u^2 t_3)}\, Z^{(Y)}(-\vec{t}, u^1, u^2),
\ee
and the FI contribution to the partition function of the theory $(Y,\CP_1)$ in \eref{FIEx1Y} can be written as
\be
Z^{(Y)}_{\rm FI}(\{\vec \s^{\gamma'} \}, u^1, u^2)= e^{2\pi i u^1 \tr \vec\s^1}\, e^{2\pi i  u^2 (\s^2 - \tr \vec{\s}^1)}.
\ee

Using the general formula \eref{G0Ab} for the dual of Abelian gauging twice, we have
\begin{align}
Z^{\wt{G}_2 \circ \wt{G}_1(Y)}(\vec{m}'(\vec{t},\vec \eta); {\eta}'({v})) =\int \prod^{2}_{\gamma'=1} \Big[d\s^{\gamma'}\Big]\,  & \delta \Big(\eta_1 + b^{1l}t_l + \tr \vec{\s}^1 \Big)\cdot 
 \delta \Big(\eta_2 + b^{2l}t_l - \tr \vec{\s}^1 + \s^2 \Big)
\nn \\ &\times Z^{(Y)}_{\rm int}(\{\vec \s^{\gamma'} \}, -\vec t, {u}_i=0),
\end{align}
where $b^{1l}=1, \, l=1,2$ and $b^{23}=-1$ are the non-vanishing components of the matrix $b^{il}$. 
Shifting the integration variables, one can recast the above integral (up to some phase factors) as:
\begin{align}
Z^{\wt{G}_2 \circ \wt{G}_1(Y)}(\vec{m}'(\vec{t}, \vec\eta); {\eta}'({v})) = &\int \frac{d^2\s^{1}}{2!}\,  \delta \Big(\tr \vec{\s}^1 \Big)\, Z^{\rm vector}_{\rm{1-loop}}(\vec \s^1)\, \prod^4_{l=1}Z^{\rm fund}_{\rm{1-loop}}(\vec \s^{1}, m_l(\vec t, \vec \eta))\,Z^{\rm free}(t_4 -2(\delta + \delta')) \nn \\
=&\int \, \frac{d^2\s^{1}}{2!}\,Z^{(Y')}_{\rm int}(\vec \s^{1}, \vec{m}'(\vec t, \vec \eta))  \nn \\
=& Z^{(Y')}(\vec{m}'(\vec t, \vec \eta)),
\end{align}
where  the theory $(Y')$ is the quiver in \figref{fig: GEx1}, and the mass parameters $m_l=-t_l + \delta$ for $l=1,2,3$, $m_4=2\delta'$, with $\delta=\frac{\eta_1 + b^{1l}t_l}{2}$ and $\delta'=\frac{\eta_2 + b^{2l}t_l}{2}$. The partition function of a free hypermultiplet of mass $m$ is $Z^{\rm free}(m)=\frac{1}{\ch{m}}$.
The map of mass and FI parameters between the pair of dual theories $(X',Y')$ can be read off from the final equality.\\

\subsection{Abelian Flavoring-Gauging Operation}\label{AbGFD4}

Consider an Abelian $S$-type operation on the quiver $X$ involving an Abelian flavoring-gauging at the $U(2)$ flavor node, followed 
by an Abelian gauging operation at the remaining $U(1)$ flavor node. Splitting the $U(2)$ flavor node as $U(2) \to U(1)_1 \times U(1)_2$, 
the $S$-type operation can be written as $\CO_{\CP}=G^2 \circ (G^1_{\CP} \circ F^1_{\CP})$. 
The quiver $X'=G^2 \circ (G^1_{\CP}\circ F^1_{\CP})(X)$ is  shown in \figref{fig: GFEx1A}. The resultant quiver $X'$  is an affine $D_4$ quiver with a single flavor. Choosing the permutation matrix $\CP$ as $\CP=\CP_1 = \mathbb{I}_{2\times 2}$, we define
\be
u^1=m_1, \quad u^2=m_2, 
\ee
where the operation $(G^1_{\CP} \circ F^1_{\CP})$ gauges the $U(1)_1$ flavor node, and $G^2$ gauges $U(1)_2$.

\begin{figure}[htbp]
\begin{center}
\scalebox{0.7}{\begin{tikzpicture}[
cnode/.style={circle,draw,thick, minimum size=1.0cm},snode/.style={rectangle,draw,thick,minimum size=1cm},pnode/.style={rectangle,red, draw,thick, minimum size=1.0cm}]
\node[cnode] (1) {1};
\node[cnode] (2) [right=1cm  of 1]{2};
\node[cnode] (3) [right=1cm  of 2]{1};
\node[pnode] (4) [below=1cm of 2]{2};
\node[snode] (5) [below=1cm of 3]{1};
\node[text width=0.1cm](20)[below=0.5 cm of 4]{$(X)$};
\draw[-] (1) -- (2);
\draw[-] (2)-- (3);
\draw[-] (2)-- (4);
\draw[-] (3)-- (5);
\draw[->] (5,0) -- (8,0);
\node[text width=3cm](7) at (7, 0.5){Abelian};
\node[text width=3cm](8) at (7, -0.5){$S$-operation}; 
\node[cnode] (9) at (10,1){1};
\node[cnode] (10) at (10,-1){1};
\node[cnode] (11) at (12, 0){2};
\node[cnode] (12) at (14, 1){1};
\node[cnode] (13) at (14, -1){1};
\node[snode] (14) at (16, 1){1};
\node[snode] (15) at (16, -1){1};
\draw[-] (9) -- (11);
\draw[-] (10) -- (11);
\draw[-] (12) -- (11);
\draw[-] (13) -- (11);
\draw[-] (12) -- (14);
\draw[-] (13) -- (15);
\node[text width=0.1cm](21)[below=0.5 cm of 13]{$(X')$};
\end{tikzpicture}}
\caption{This figure shows the action of an Abelian $S$-type operation (involving an elementary flavoring-gauging and an elementary gauging operation) 
on the quiver $X$ on the LHS. }
\label{fig: GFEx1A}
\end{center}
\end{figure}
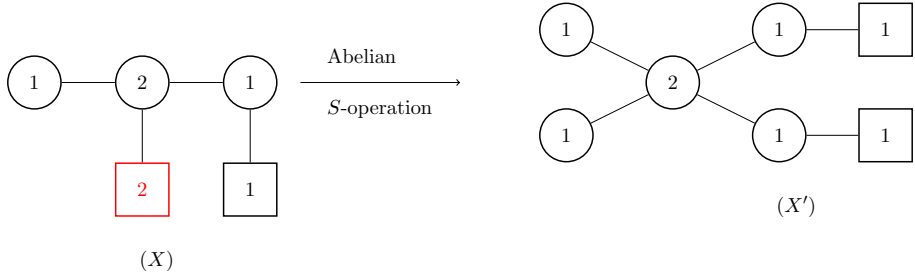

\begin{figure}[htbp]
\begin{center}
\scalebox{0.7}{\begin{tikzpicture}[
cnode/.style={circle,draw,thick, minimum size=1.0cm},snode/.style={rectangle,draw,thick,minimum size=1cm}]
\node[cnode] (1) {2};
\node[snode] (2) [below=1cm of 1]{3};
\node[cnode] (3) [right=1cm of 1]{1};
\node[snode] (4) [below=1cm of 3]{1};
\draw[-] (1) -- (2);
\draw[-] (1) -- (3);
\draw[-] (3) -- (4);
\node[text width=0.1cm](20)[below=0.5 cm of 2]{$(Y)$};
\draw[->] (3,0) -- (7,0);
\node[text width=3cm](7) at (5, 0.5){Dual};
\node[text width=3cm](8) at (5, -0.5){Operation I}; 
\node[cnode] (9) at (10,0){2};
\node[snode] (10) [below=1cm of 9]{4};
\node[snode] (11) [right=1cm of 9]{2};
\draw[-] (9) -- (10);
\draw[-] (9) -- (11);
\node[text width=0.1cm](12) at (10.75,0.5){$\CA$};
\node[text width=0.1cm](21)[below=0.5 cm of 10]{$(Y')$};
\end{tikzpicture}}
\qquad \qquad \qquad \qquad 
\scalebox{0.7}{\begin{tikzpicture}[
cnode/.style={circle,draw,thick, minimum size=1.0cm},snode/.style={rectangle,draw,thick,minimum size=1cm},pnode/.style={circle,double,draw,thick, minimum size=1.0cm}]
\node[cnode] (1) {2};
\node[snode] (2) [below=1cm of 1]{3};
\node[cnode] (3) [right=1cm of 1]{1};
\node[snode] (4) [below=1cm of 3]{1};
\draw[-] (1) -- (2);
\draw[-] (1) -- (3);
\draw[-] (3) -- (4);
\node[text width=0.1cm](20)[below=0.5 cm of 2]{$(Y)$};
\draw[->] (3,0) -- (7,0);
\node[text width=3cm](7) at (5, 0.5){Dual};
\node[text width=3cm](8) at (5, -0.5){Operation II}; 
\node[pnode] (9) at (10,0){2};
\node[snode] (10) [below=1cm of 9]{3};
\node[cnode] (11) [right=1cm of 9]{1};
\node[snode] (12) [below=1cm of 11]{2};
\draw[-] (9) -- (10);
\draw[-] (9) -- (11);
\draw[-] (11) -- (12);
\node[text width=0.1cm](21)[below=0.5 cm of 10]{$(Y'')$};
\end{tikzpicture}}
\caption{This figure shows dual operations for different choices of the permutation matrix $\CP$. The quiver $(X')$ in \figref{fig: GFEx1A} is mirror dual to the quivers $(Y')$ 
and $(Y'')$. The quiver $(Y')$ has a pair of hypermultiplets in the rank-2 antisymmetric representation of the gauge group $U(2)$. Note that the 
theories $(Y')$ and $(Y'')$ have the same Coulomb and Higgs branch symmetries.}
\label{fig: GFEx1B}
\end{center}
\end{figure}
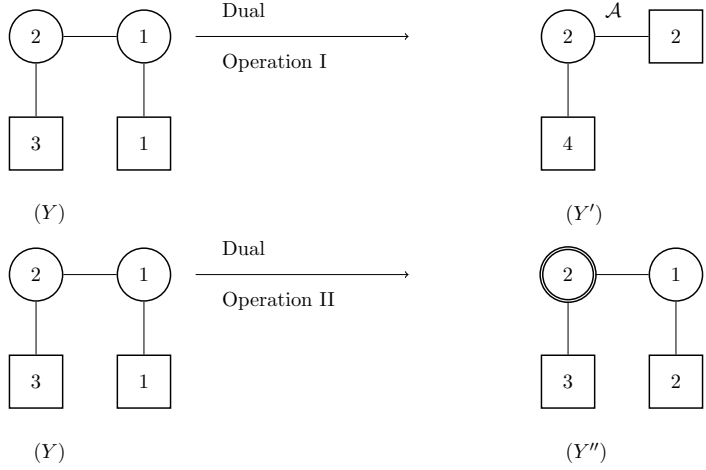

The mirror symmetry relation \eref{MSEx1} can be written as
\be
Z^{(X)}(u^1, u^2, \vec{t})= e^{2\pi i (u^1 (t_1+t_2) - u^2 t_3)}\, Z^{(Y)}(-\vec{t}, u^1, u^2),
\ee
and the FI term in the partition function of the theory $(Y,\CP_1)$ in \eref{FIEx1Y} can be written as
\be
Z^{(Y)}_{\rm FI}(\{\vec \s^{\gamma'} \}, u^1, u^2)= e^{2\pi i u^1 \tr \vec\s^1}\, e^{2\pi i  u^2 (\s^2 - \tr \vec{\s}^1)}.
\ee
The operation $\CO_{\CP_1}(X)$ gives the quiver on the RHS in \figref{fig: GFEx1A}. The partition function of the 
dual theory $\wt{\CO}_{\CP_1}(Y)$ can be obtained by first using the general formula \eref{PFGFLinNf1}-\eref{HyperGFLinNf1} and then the formula \eref{G0Ab}, 
which gives:
\begin{align}
Z^{\wt{\CO}_{\CP_1}(Y)}(\vec{m}'(\vec{t},\vec \eta), {\eta}'(m^{\alpha}_F))= &\int \prod^{2}_{\gamma'=1} \Big[d\s^{\gamma'}\Big]\,
\frac{ \delta \Big(\eta_2 + b^{2l}t_l - \tr \vec{\s}^1 + \s^2 \Big)}{\ch{(\tr\vec\s^1 + \eta_1 + b^{1l}t_l)}}
\nn \\ &\times Z^{(Y,\CP_1)}_{\rm int}(\{\vec \s^{\gamma'} \}, -\vec t, {u}^1=m^\alpha_F, u^2=0),
\end{align}
where $b^{1l}=1, \, l=1,2$ and $b^{23}=-1$ are the non-vanishing components of the matrix $b^{il}$. 
Imposing the delta function and shifting the integration variables by constants, one can recast the above integral (up to some phase factors) as:
\begin{align}
Z^{\wt{\CO}_{\CP_1}(Y)} 
= &\int \frac{d^2\s^{1}}{2!}\,Z_{\rm FI}(\vec{\s}^1,m^{\alpha}_F)  Z^{\rm vector}_{\rm{1-loop}}(\vec \s^1)\, \prod^4_{l=1}Z^{\rm fund}_{\rm{1-loop}}(\vec \s^{1}, {m}'_{l} (\vec t, \vec\eta)) \prod^2_{k=1} Z^{\CA}_{\rm{1-loop}}(\vec \s^1,\vec m'^{k}_{\CA}(\vec t, \vec\eta))\\
=&Z^{(Y')}(\vec m'(\vec{t},\vec \eta), \vec m'_{\CA}(\vec{t},\vec \eta), \eta'= m^\alpha_F),
\end{align}
where the theory $Y'$ is the quiver in \figref{fig: GFEx1B}, and $\vec m', \vec m'_{\CA}$ denote the masses for the four fundamental hypers and 
the two rank-2 antisymmetric hypers of the $U(2)$ gauge group respectively. Explicitly, the mirror map is given as
\begin{align}
&{m}'_l(\vec t, \eta)= -t_l + t_4/4 - (\delta' -\delta)/2,\, \text{for } l=1,2,3, \\
&{m}'_4(\vec t, \eta)=t_4/4 - (\delta' -\delta)/2 + 2\delta',\\
&m'^1_{\CA}=-m'^2_{\CA}=-t_4/2 + (\delta' +\delta), \\
& \eta'= m^\alpha_F.
\end{align}
where $\delta=\frac{\eta_1 + b^{1l}t_l}{2}$ and $\delta'=\frac{\eta_2 + b^{2l}t_l}{2}$. The masses, written above, parametrize the Cartan subalegbra 
of the Higgs branch global symmetry $G^{Y'}_{H}=U(4) \times SU(2)$, while the FI parameter parametrizes the Cartan of the Coulomb branch 
global symmetry $G^{Y'}_{C}= U(1)$.\\

One can proceed with the choice of the matrix $\CP=\CP_2=\begin{pmatrix} 0 & 1 \\ 1 & 0 \end{pmatrix}$ in an analogous fashion, 
and show that the dual theory is given by the quiver $(Y'')$ in \figref{fig: GFEx1B}, i.e.
\be
Z^{\wt{\CO}_{\CP_2}(Y)}(\vec{m}'(\vec{t},\vec \eta), {\eta}'(m^{\alpha}_F))= Z^{(Y'')}(\vec m'(\vec{t},\vec \eta), \eta'= m^{\alpha}_F).
\ee

 $Y'$ and $Y''$ are conjectured to be mirror dual to $X'$. The global symmetries of $Y'$ and $Y''$ match, up to a discrete group:
\begin{align}
& G^{Y'}_{H}=SU(4) \times U(2), \qquad &G^{Y'}_{C}= U(1), \nn \\
& G^{(Y'')}_{H}= SO(6) \times U(2), \qquad &G^{(Y'')}_{\rm C}=U(1). 
\end{align}

\section{Superconformal index on $S^2 \times S^1$}\label{SCI}
\subsection{Definition and localization formula}
The 3d $\CN=4$ index on the manifold $S^2 \times S^1$ \cite{Imamura:2011su, Kapustin:2011jm, Benini:2011nc, Razamat:2014pta} can be defined as:
\begin{equation}
\begin{split}
\mathcal{I}_{S^2\times S^1}= \tr_{S^2} \left[(-1)^F (\tq)^{j_2 + \frac{R_H + R_C}{2}} (\tti)^{R_H -R_C}  e^{-\beta(E-2j_2 -2R_H -R_C)} \prod_i \mu_i^{f_i}\right],
\end{split}
\end{equation}
where $j_2$ is an angular momentum operator on $S^2$, $R_H$ and $R_C$ are Cartan generators of the $su(2)_H$ and $su(2)_C$ Lie algebras respectively, 
and $f_i$ collectively denote generators associated with other global symmetries.
The index receives non-zero contributions from those states which satisfy $E-2j_2-2R_H - R_C=0$, Therefore, the 3d conformal dimension $\tE=\frac{E+R_C}{2}$ for the states 
contributing to the index can be written as
\begin{equation}
\tE=j_2 +R_H+R_C.
\end{equation}
Let us define the following parameters $(x,\tx)$ in terms of $(\tq, \tti)$ as follows:
\be
x=\tq^{1/2} \tti, \quad \tx=\tq^{1/2} \tti^{-1}.
\ee
Then the 3d index can be written as 
\begin{equation}
\begin{split}
\mathcal{I}_{S^2\times S^1} (x,\tx;\mu_i)= & \tr_{S^2} \left[(-1)^F (x)^{\tE - R_C} (\tx)^{\tE -R_H} e^{-\beta(\tE-j_2 -R_H -R_C)} \prod_i \mu_i^{f_i} \right],\\
=& \tr_{S^2} \left[(-1)^F (x)^{j_2 + R_H} (\tx)^{j_2 +R_C} e^{-\beta(\tE-j_2 -R_H -R_C)} \prod_i \mu_i^{f_i} \right].
\end{split}
\end{equation}
Two important limits of the 3d index -- the Coulomb branch index $\mathcal{I}_C$ and the Higgs branch index $\mathcal{I}_H$ -- are defined as follows:
\begin{equation}
\begin{split}
&\mathcal{I}_C(\tx;\mu_i)=\tr_{\CH_C} \left[(-1)^F(\tx)^{\tE - R_H} e^{-\beta(\tE-j_2 -R_H -R_C)} \prod_i \mu_i^{f_i} \right]=\lim_{x \to 0} \mathcal{I}_{S^2\times S^1}(x,\tx; \mu_i)\\
&\mathcal{I}_H(x;\mu_i)=\tr_{\CH_H} \left[(-1)^F(x)^{\tE - R_C} e^{-\beta(\tE-j_2 -R_H -R_C)} \prod_i \mu_i^{f_i} \right]=\lim_{\tx \to 0} \mathcal{I}_{S^2\times S^1}(x,\tx; \mu_i).
\end{split}
\end{equation}
Note that $\CH_C$ and $\CH_H$ are subspaces of the Hilbert space, where the states satisfy the constraints $\tE - R_C=0$ and $\tE -R_H=0$ respectively.
It is sensible to take these limits of the original index since unitarity dictates that
\be
\tE \geq R_{C,H}.
\ee

Let us now write down the general expression of the index on $S^2 \times S^1$ for an $\CN=4$ gauge theory with gauge group $G$ and
global symmetry group $G_{H}$, where the hypermultiplets transforms in some representation $\CR$ of $G \times G_H$.
This can be computed using localization and we refer the reader to the references \cite{Imamura:2011su, Kim:2009wb, Krattenthaler:2011da, Kapustin:2011jm} 
for details of the computation, while simply providing the answer here. 
Let $\vec z \in T_G$, where $T_G$ is the maximal torus of the group $G$, and let $\vec a$ 
be the flux associated with $G$. Similarly, $(\vec \tz, \vec \ta)$ denote the analogous pair for the global symmetry group $G_{H}$. 
We choose $\vec a$ and $\vec \ta$ to take integer values.
Also, let $\vec s \in g$, and $\vec{m} \in {G_H}$, where $g, g_H$ denote the Cartan subalgebra of $g$ and $g_H$ 
respectively, such that $\vec z = e^{2\pi i \vec s}$ and $\vec \tz = e^{2\pi i \vec{m}}$. 
In addition, we will turn on fugacities and discrete fluxes associated with the $U(1)_J$ symmetry of the theory. 
Let $\beta$ label the unitary factors in the gauge group $G$, with $(\vec{z}_{(\beta)}, \vec{a}^{(\beta)})$ representing the fugacity and flux of 
the $\beta$-th unitary factor, and $(w_\beta, n_\beta)$ denoting the fugacity and flux of a twisted $U(1)$ vector multiplet that couples to the 
$U(1)_J$ current of the $\beta$-th unitary factor.\\

Given the above data, the expression for the index can be written as a contour integral over $\vec z$ summed over the fluxes $\vec a$, and the 
answer is a function of the global symmetry fugacities and fluxes, i.e. $(\vec \tz, \vec \ta, \vec w, \vec n)$, in addition to the parameters
$\tq, \tti$.
\begin{align} \label{SCI-gen}
 \CI(\vec \tz,  \vec \ta,\vec w, \vec n; \tq,\tti)= \sum_{\{ \vec a \}} \frac{1}{W(\vec a)} \, \oint_{|z_i|=1} \prod^{{\rm rank} G}_{i=1} \frac{dz_i}{2\pi i z_i}\, &\CI_{\rm FI}(\vec w, \vec n,\vec z, \vec a)  \,\CI_{\rm vector} (\vec z, \vec a;\tq,\tti)\nn \\ & \times \CI^{\CR}_{\rm half-hyper} (\vec z, \vec a, \vec \tz, \vec \ta; \tq,\tti),
\end{align}
where $W(\vec a)$ is the order of the Weyl symmetry group left unbroken by the fluxes $\vec a$, and the contour of integration is 
$\mathcal{C}= \prod^{{\rm rank} G}_{i=1} \mathcal{C}_i$ with $\mathcal{C}_i$ being a unit circle around the origin on the $i$-th complex plane 
\footnote{A 3d $\CN=4$ hypermultiplet consists of a half-hyper in the representation $\CR$ and another half-hyper in the complex conjugate representation $\CR^*$.
For pseudo-real representations, one presents the matter content in terms of half-hypers.}.
The integrands $\CI_{\rm FI}$, $\CI_{\rm vector}$ and $\CI^{\CR}_{\rm half-hyper}$ are given as follows.

\begin{align}
& \CI_{\rm FI}(\vec w, \vec n,\vec z, \vec a)= \Big(\prod_{\beta} \prod_{i(\beta)} w^{a_{i(\beta)}}_\beta \cdot z^{n_\beta}_{i(\beta)} \Big), \label{SCIgenFI}\\
& \CI_{\rm vector} (\vec z, \vec a;\tq,\tti)= \Big( \frac{(\tti \tq^{1/2}; \tq)}{(\tti^{-1} \tq^{1/2}; \tq)} \Big)^{{\rm rank} G}\, \prod_{\alpha \in {\rm ad}(G)} (1- \tq^{|\alpha(\vec a)|/2} e^{2\pi i \alpha(\vec s)}) \nn \\
&\times (\frac{\tq^{1/2}}{\tti})^{-|\alpha(\vec a)|/2} \, \frac{(\tti \tq^{1/2+|\alpha(\vec a)|/2}\, e^{2\pi i \alpha(\vec s)}; \tq)}{(\tti^{-1} \tq^{1/2+|\alpha(\vec a)|/2} e^{2\pi i \alpha(\vec s)};\tq)}, \label{SCIgenVec}\\
& \CI^{\CR}_{\rm half-hyper} (\vec z, \vec a, \vec \tz, \vec \ta; \tq,\tti)= \prod_{\rho \in \CR(G \times G_H)} (\frac{\tq^{1/2}}{\tti})^{-|\rho(\vec a, \vec \ta)|/4}\,\frac{(\tti^{-1/2} \tq^{3/4+|\rho(\vec a, \vec \ta)|/2}\, e^{ 2\pi i \rho(\vec s, \vec{\wt{s}})}; \tq)}{(\tti^{1/2} \tq^{1/4+|\rho(\vec a, \vec \ta)|/2} e^{ 2\pi i \rho(\vec s, \vec{\wt{s}})};\tq)},\label{SCIgenMatt}
\end{align}
where $\alpha$ is a root of the Lie algebra of $G$, $\rho$ is a weight of the representation $\CR$ 
of the Lie algebra of $G \times G_H$, $(z;\tq)$ is the Pochhammer symbol defined as
\be
(z;\tq):= \prod^\infty_{l=0} (1-z\tq^l), \quad |\tq| < 1,
\ee
and, for future use, the symbol $(z^{\pm};q)$ is defined as $(z^{\pm};q)= (z;q)\cdot(z^{-1};q)$.\\

For a linear quiver gauge theory $(X)$ with unitary gauge groups and (bi)fundamental matter, the expression for the index can be written as follows. 
Let the gauge group $G=\prod^L_{\gamma=1} U(N_\gamma)$ and the flavor symmetry group $G_H=\Big(\prod^L_{\gamma=1} U(M_\gamma) \Big)/U(1)$, and let us 
denote the fugacities and fluxes associated with the gauge nodes and flavor nodes as $\{ \vec z^{(\gamma)}, \vec a^{(\gamma)} \}$ and $\{ \vec{\tz}^{(\gamma)}, \vec{\ta}^{(\gamma)} \}$ respectively. We prefer to reorganize the gauge/flavor fugacities and fluxes as follows:
\begin{align} 
&z_{\beta'} = \{z^{(1)}_{j_1}, z^{(2)}_{j_2}, \ldots, z^{(\alpha)}_{j_\alpha},\ldots, z^{(L-1)}_{j_{L-1}}, z^{(L)}_{j_L} \}, \qquad \beta'=1,\ldots, \sum_\gamma N_\gamma,\label{fugconvz}\\
&a_{\beta'} = \{a^{(1)}_{j_1}, a^{(2)}_{j_2}, \ldots, a^{(\alpha)}_{j_\alpha},\ldots, a^{(L-1)}_{j_{L-1}}, a^{(L)}_{j_L} \}, \qquad \beta'=1,\ldots, \sum_\gamma N_\gamma,\label{fugconva}\\
&\tz_\beta = \{\tz^{(1)}_{i_1}, \tz^{(2)}_{i_2}, \ldots, \tz^{(\alpha)}_{i_\alpha},\ldots, \tz^{(L-1)}_{i_{L-1}}, \tz^{(L)}_{i_L} \}, \qquad \beta=1,\ldots, L^\vee +1,\label{fugconvtz}\\
&\ta_\beta = \{\ta^{(1)}_{i_1}, \ta^{(2)}_{i_2}, \ldots, \ta^{(\alpha)}_{i_\alpha},\ldots, \ta^{(L-1)}_{i_{L-1}}, \ta^{(L)}_{i_L} \}, \qquad \beta=1,\ldots, L^\vee +1,\label{fugconvta}
\end{align}
with $i_1=1,\ldots,M_1$, $i_2=1,\ldots,M_2$,$\ldots$, $i_\alpha=1,\ldots, M_\alpha$, $i_L=1,\ldots, M_L$, and $L^\vee=\sum^L_{\gamma=1} M_\gamma -1$.
In addition, we parametrize the $U(1)_J$ fugacities and fluxes $(w_\gamma, n_\gamma)$ as follows:
\begin{align}\label{fugconyb}
w_\gamma = \frac{y_\gamma}{y_{\gamma +1}}, \qquad n_\gamma= b_\gamma - b_{\gamma +1}, \qquad \gamma=1,\ldots, L.
\end{align}
The index of $(X)$ can then be written as the following contour integral:
\be \label{SCILinA}
\begin{split}
& \CI^{(X)}(\vec{\tz},\vec{\ta}, \vec y, \vec b; \tq,\tti)= \sum_{\{\vec a^{(\gamma)}\}} \, \oint_{|z^{(\gamma)}_{i_\gamma}|=1} 
\prod^L_{\gamma=1} \Big[ \frac{d{\vec z}^{(\gamma)}}{{\vec z}^{(\gamma)}}\Big]\, \CI^{(X)}_{\rm int}(\vec{z}, \vec{a}, \vec{\tz}, \vec{\ta}, \vec y, \vec b; \tq,\tti)\\
&= \sum_{\{\vec a^{(\gamma)}\}} \, \oint_{|z^{(\gamma)}_{i_\gamma}|=1} 
\prod^L_{\gamma=1} \Big[ \frac{d{\vec z}^{(\gamma)}}{{\vec z}^{(\gamma)}}\Big]\, \CI^{(X)}_{\rm FI}(\vec y, \vec b, \{\vec{z}^{(\gamma)} \},\{ \vec{a}^{(\gamma)} \}) \, \CI^{(X)}_{\rm 1-loop}(\vec{z}, \vec{a}, \vec{\tz}, \vec{\ta}; \tq,\tti)\\
&= \sum_{\{\vec a^{(\gamma)}\}} \oint_{|z^{(\gamma)}_{i_\gamma}|=1} \Big[ \frac{d{\vec z}^{(\gamma)}}{{\vec z}^{(\gamma)}}\Big]\, \CI^{(X)}_{\rm FI}(\vec y, \vec b, \{\vec{z}^{(\gamma)} \},\{ \vec{a}^{(\gamma)} \}) \,\prod^L_{\gamma=1} \CI^{(X)}_{\rm vector} (\vec{z}^{(\gamma)}, \vec{a}^{(\gamma)};\tq,\tti)\\
& \times \prod^L_{\gamma=1} \CI^{(X)\,\rm{fund.}}_{\rm hyper} (\vec z^{(\gamma)}, \vec a^{(\gamma)}, \vec{\tz}^{(\gamma)}, \vec{\ta}^{(\gamma)}; \tq,\tti)\, 
\prod^{L-1}_{\gamma=1} \CI^{(X)\, \rm bifund.}_{\rm hyper}(\vec z^{(\gamma)}, \vec a^{(\gamma)}, \vec{z}^{(\gamma+1)}, \vec{a}^{(\gamma+1)}; \tq,\tti),
\end{split}
\ee
where $\Big[ \frac{d{\vec z}^{(\gamma)}}{{\vec z}^{(\gamma)}}\Big]=\frac{1}{W(\vec a^{(\gamma)})}\,\prod^{N_\gamma}_{i_{\gamma}=1} \frac{dz^{(\gamma)}_{i_\gamma}}{2\pi i z^{(\gamma)}_{i_\gamma}}$. The constituent functions inside the integrand are given as
\begin{align}
& \CI^{(X)}_{\rm FI}(\vec y, \vec b,  \{\vec{z}^{(\gamma)} \},\{ \vec{a}^{(\gamma)} \})= \prod^L_{\gamma=1} \Big(\prod^{N_\gamma}_{j_\gamma=1} (\frac{y_\gamma}{y_{\gamma+1}})^{a^{(\gamma)}_{j_\gamma}} \cdot (z^{(\gamma)}_{j_\gamma})^{(b_\gamma - b_{\gamma+1})} \Big),\label{SCILinAFI}\\
& \CI^{(X)}_{\rm vector} (\vec{z}^{(\gamma)}, \vec{a}^{(\gamma)};\tq,\tti) = \Big( \frac{(\tti \tq^{1/2}; \tq)}{(\tti^{-1} \tq^{1/2}; \tq)} \Big)^{N_\gamma}\,\prod_{k_{\gamma} \neq j_{\gamma}} (1- \tq^{|a^{(\gamma)}_{k_{\gamma}} - a^{(\gamma)}_{j_{\gamma}}|/2} {z}^{(\gamma)}_{k_{\gamma}}/{z}^{(\gamma)}_{j_\gamma}) \nn \\
&\times (\frac{\tq^{1/2}}{\tti})^{-|a^{(\gamma)}_{k_{\gamma}} - a^{(\gamma)}_{j_{\gamma}}|/2} \, \frac{(\tti \tq^{1/2+|a^{(\gamma)}_{k_{\gamma}} - a^{(\gamma)}_{j_{\gamma}}|/2}\, {z}^{(\gamma)}_{k_{\gamma}}/{z}^{(\gamma)}_{j_{\gamma}}; \tq)}{(\tti^{-1} \tq^{1/2+|a^{(\alpha)}_{k_{\gamma}}- a^{(\alpha)}_{j_{\gamma}}|/2} {z}^{(\gamma)}_{k_{\gamma}}/{z}^{(\gamma)}_{j_{\gamma}};\tq)}, \label{SCILinAvec}\\
& \CI^{(X)\,\rm{fund.}}_{\rm hyper} (\vec z^{(\gamma)}, \vec a^{(\gamma)}, \vec{\tz}^{(\gamma)}, \vec{\ta}^{(\gamma)}; \tq,\tti) =\prod^{M_\gamma}_{i_\gamma=1} \prod^{N_\gamma}_{j_{\gamma}=1} (\frac{\tq^{1/2}}{\tti})^{|a^{(\gamma)}_{j_{\gamma}} - \ta^{(\gamma)}_{i_{\gamma}}|/2}\,\frac{(\tti^{-1/2} \tq^{3/4+|a^{(\gamma)}_{j_{\gamma}} - \ta^{(\alpha)}_{i_{\gamma}}|/2}\,({z}^{(\gamma)}_{j_{\gamma}}/{\tz}^{(\gamma)}_{i_{\gamma}} )^{\pm}; \tq)}{(\tti^{1/2} \tq^{1/4+|a^{(\gamma)}_{j_{\gamma}} - \ta^{(\gamma)}_{i_{\gamma}}|/2}\, ({z}^{(\gamma)}_{j_{\gamma}}/{\tz}^{(\gamma)}_{i_{\gamma}})^{\pm} ;\tq)}, \label{SCILinAfund}\\
& \CI^{(X)\,\rm{bifund.}}_{\rm hyper}(\vec z^{(\gamma)}, \vec a^{(\gamma)}, \vec{z}^{(\gamma+1)}, \vec{a}^{(\gamma+1)}; \tq,\tti) \nn\\
=& \prod^{N_\gamma}_{k_{\gamma}=1} \prod^{N_{\gamma+1}}_{j_{\gamma}=1}  (\frac{\tq^{1/2}}{\tti})^{|a^{(\alpha)}_{k_{\gamma}}- a^{(\gamma+1)}_{j_\gamma}|/2}\,\frac{(\tti^{-1/2} \tq^{3/4+|a^{(\gamma)}_{k_{\gamma}} - a^{(\gamma+1)}_{j_{\gamma}}|/2}\,({z}^{(\gamma)}_{k_{\gamma}}/{z}^{(\gamma+1)}_{j_{\gamma}})^{\pm} ; \tq)}{(\tti^{1/2} \tq^{1/4+|a^{(\gamma)}_{k_{\gamma}} - a^{(\gamma+1)}_{j_\gamma}|/2}\, ({z}^{(\gamma)}_{k_{\gamma}}/{z}^{(\gamma +1)}_{j_\gamma})^{\pm};\tq)}.\label{SCILinAbifund}
\end{align}

The Coulomb and the Higgs limits of the superconformal index admit expansions in terms of characters of representations of the 
respective global symmetry groups -- $G_{H}$ and $G_{C}$. In fact, these limits coincide with the Coulomb branch Hilbert Series 
and the Higgs branch Hilbert Series respectively. Schematically, one can write
\begin{align}
& \mathcal{I}_C(\tx;\mu_i)=\sum^{\infty}_{\vec k=0}\, \chi[{\CR}_{\vec k}(G_{C})](\mu_i)\,\tx^{|\vec k|},\\
& \mathcal{I}_H(x;\mu_i)=\sum^{\infty}_{\vec k=0}\, \chi[{\CR}_{\vec k}(G_{H})](\mu_i)\, x^{|\vec k|},
\end{align}
where ${\CR}_{\vec k}$ are irreducible representations labelled by the Dynkin labels $\vec k$, 
and $\chi[{\CR}_{\vec k}(G)](\mu_i)$ denotes the associated character as a function of the fugacities $\{\mu_i\}$. 
Therefore, a series expansion of the Coulomb and the Higgs indices in terms of $\tx$ and $x$ respectively can be 
used to read off the global symmetries of the respective branches.

\subsection{3d $\CN=4$ mirror symmetry}
The basic example of 3d $\CN=4$ mirror symmetry involves an SQED with a single hypermultiplet on one side  and a twisted hypermultiplet 
on the other. 3d mirror symmetry therefore implies the following identity:
\begin{align}\label{SCIDual1a}
& \CI_{SQED_1}(w,n;\tq, \tti) = \CI_{\rm twisted\,hyper} (w,n,1,0;\tq, \tti)=\CI_{\rm hyper} (w,n,1,0;\tq, \tti^{-1}),  \nn \\
\implies & \sum_{m \in \BZ} w^m\,\oint_{|z|=1} \frac{dz\, z^n}{2\pi i z}\, \CI_{\rm vector}(\tq,\tti)\,\CI_{\rm hyper}(z,m,1,0;\tq,\tti) = \CI_{\rm hyper} (w,n,1,0;\tq, \tti^{-1}).
\end{align}
The index of a ``twisted hypermultiplet" is given by substituting $\tti \to \tti^{-1}$ in the index of a hypermultiplet.
Replacing $\tti \to \tti^{-1}$, the above identity can be written in the following form :
\begin{align}
\CI_{\rm hyper} (w,n,1,0;\tq, \tti)= \sum_{m \in \BZ} w^m\,\oint_{|z|=1}  \frac{dz\, z^n}{2\pi i z}\, \CI_{\rm vector}(\tq,\tti^{-1})\,\CI_{\rm hyper}(z,m,1,0;\tq,\tti^{-1}) .
\end{align}
Parametrizing $w=y_1/y_2$ and $n=b_1 -b_2$, this identity can be written in a form that will be useful in our analysis of Abelian $S$-type operations:
\begin{align} \label{SCIDual1b}
\CI_{\rm hyper} (y_1, b_1,y_2,b_2;\tq, \tti)= \sum_{m \in \BZ} (\frac{y_1}{y_2})^m\,\oint_{|z|=1}  \frac{dz\, z^{b_1-b_2}}{2\pi i z}\, \CI_{\rm vector}(\tq,\tti^{-1})\,\CI_{\rm hyper}(z,m,1,0;\tq,\tti^{-1}).
\end{align}

Now consider a pair of linear quivers $(X,Y)$, where the index of $X$ is given by \eref{SCILinA}. In the standard 
notation, introduced in \Secref{LQ-IIB}, the statement of 3d mirror symmetry is 
\be
\CI^{(X)}(\vec{\tz},\vec{\ta}, \vec y, \vec b; \tq,\tti) = \CI^{(Y)}(\vec{y}^{-1}, -\vec b,\vec{\tz},\vec{\ta}; \tq,\tti^{-1}), 
\ee
where the index $\CI^{(Y)}(\vec y, \vec b,\vec{\tz},\vec{\ta}; \tq,\tti)$ is given as
\begin{align}\label{SCILinB}
&\CI^{(Y)}(\vec y, \vec b,\vec{\tz},\vec{\ta}; \tq,\tti)=  \sum_{\{\vec c^{(\gamma')}\}}\, \prod^{L^\vee}_{\gamma'=1} \oint_{|\xi^{(\gamma')}_{i_{\gamma'}}|=1} 
\Big[\frac{d\xi^{(\gamma')}}{\xi^{(\gamma')}} \Big]\,\CI^{(Y)}_{\rm int} (\vec{\xi}, \vec{c}, \vec y, \vec b,\vec{\tz},\vec{\ta}; \tq,\tti)  \nn \\
&= \sum_{\{\vec c^{(\gamma')}\}}\, \prod^{L^\vee}_{\gamma'=1} \oint_{|\xi^{(\gamma')}_{i_{\gamma'}}|=1} 
\Big[\frac{d\xi^{(\gamma')}}{\xi^{(\gamma')}} \Big]\,\CI^{(Y)}_{\rm FI}(\vec \tz, \vec \ta, \{\vec{\xi}^{(\gamma')} \},\{ \vec{c}^{(\gamma')} \}) \, \CI^{(Y)}_{\rm 1-loop}(\vec{\xi}, \vec{c}, \vec{y}, \vec{b}; \tq,\tti) \nn \\
&= \sum_{\{\vec {c}^{(\gamma')}\}}\, \prod^{L^\vee}_{\gamma'=1} \oint_{|\xi^{(\gamma')}_{i_{\gamma'}}|=1} 
\Big[\frac{d\xi^{(\gamma')}}{\xi^{(\gamma')}} \Big] \, \CI^{(Y)}_{\rm FI}(\vec \tz, \vec \ta, \{\vec{\xi}^{(\gamma')} \},\{ \vec{c}^{(\gamma')} \}) \,\prod^{L^\vee}_{\gamma'=1} \CI^{(Y)}_{\rm vector} (\vec{\xi}^{(\gamma')}, \vec{c}^{(\gamma')};\tq,\tti) \nn \\
& \times \prod^L_{\gamma=1} \CI^{(Y)\,\rm{fund.}}_{\rm matter} (\vec{\xi}^{(\gamma')}, \vec{c}^{(\gamma')}, \vec{y}^{(\gamma')}, \vec{b}^{(\gamma')}; \tq,\tti)\, 
\prod^{L-1}_{\gamma=1} \CI^{(Y)\, \rm bifund.}_{\rm matter}(\vec{\xi}^{(\gamma')}, \vec{c}^{(\gamma')}, \vec{\xi}^{(\gamma' +1)}, \vec{c}^{(\gamma' +1)}; \tq,\tti).
\end{align}
The FI term in the integrand is given as
\be \label{SCILinBFI}
\CI^{(Y)}_{\rm FI}(\vec \tz, \vec \ta, \{\vec{\xi}^{(\gamma')} \},\{ \vec{c}^{(\gamma')} \}) = \prod^{L^\vee}_{\gamma'=1} \Big(\prod^{N_{\gamma'}}_{j_{\gamma'}=1} (\frac{\tz_{\gamma'}}{\tz_{\gamma'+1}})^{c^{(\gamma')}_{j_{\gamma'}}} \cdot (\xi^{(\gamma')}_{j_{\gamma'}})^{(\ta_{\gamma'} - \ta_{\gamma'+1})} \Big),
\ee
while the one-loop terms are given by expressions analogous to \eref{SCILinAvec}-\eref{SCILinAbifund}.

\section{$S$-type operations in terms of the superconformal index} \label{GFI-SCI}
In this section, we demonstrate how the elementary $S$-type operations can be implemented in terms of the superconformal index on $S^2 \times S^1$. 
For new dualities, the equality of the indices serves as another strong check. In \Appref{GFI-SCI-Gen}, we first discuss the implementation of elementary 
$S$-type operations on linear quivers, and then extend it to generic quivers in class $\CU$. We then discuss the four elementary Abelian $S$-type operations 
in \Appref{GFI-SCI-Ab}, explicitly working out the dual operation in each case, as we did in \Secref{AbS} using the $S^3$ partition function. 
All examples studied in \Secref{AbMirr} can be constructed using the expressions for the dual superconformal indices in \Appref{GFI-SCI-Ab}. 
As an illustrative example, we work out the case of a flavored $\widehat{A}_{n-1}$ quiver and the Family ${\rm I}_{[n,l,p]}$ explicitly in \Appref{GFI-exAbSCI}.

\subsection{Generic $S$-type operations and their duals}\label{GFI-SCI-Gen}
The $S$-type operations and their duals can be implemented in terms of the $S^2 \times S^1$ superconformal index, following steps 
analogous to those for the partition function on $S^3$. Consider the linear quiver gauge theory $X$ in \figref{fig: LQGen}. The index of the theory is 
given in \eref{SCILinA} as function of various global symmetry fugacities and fluxes. The fugacities and fluxes associated with a flavor node $(\gamma)$ 
is denoted as $(\vec{\tz}^{(\gamma)}, \vec{\ta}^{(\gamma)})$, where the fugacities are related to the masses as follows:
\be \label{def-tzy}
\vec{\tz}^{(\gamma)} = e^{2\pi i \vec{m}^\gamma}, \qquad \gamma=1,\ldots,L,
\ee
where $\vec{m}^\gamma$ are complex parameters. 
In addition, the $U(1)_J$ fugacities and fluxes associated with the gauge nodes are denoted as $(\{w_\gamma\}, \{n_\gamma\})$.
It is convenient to parametrize them in terms of the pair $(\vec y, \vec b)$ such that
\be
w_\gamma = \frac{y_\gamma}{y_{\gamma +1}}, \qquad n_\gamma= b_\gamma - b_{\gamma +1}, \qquad \gamma=1,\ldots, L.
\ee
The fugacities $\vec y$ are related to the deformation parameter $\vec t$ as :
\be
\vec{y} = e^{2\pi i \vec{t}},
\ee
where $\vec t$ are complex parameters. For further details on the convention of the fugacities and fluxes, the reader is referred to \Appref{SCI}. \\

Now, consider picking a flavor node $(\alpha)$ corresponding to a global symmetry group $U(M_\alpha)$ in $X$ and split 
it into two, as $U(r_\alpha) \times U(M_\alpha - r_\alpha)$\footnote{As noted earlier, we will assume that the masses associated 
with the flavor node $\alpha$ are completely unconstrained. The $U(1)$ quotient for the Higgs branch global symmetry in a linear quiver 
is implemented by imposing a constraint on the masses associated to the other flavor node(s).}. 
Let us introduce a set of fugacities $\{h^{(\alpha)}_i | i=1,\ldots, r_\alpha \}$, and 
$\{h'^{(\alpha)}_j | j=1,\ldots, M_\alpha - r_{\alpha} \}$, such that $h^{(\alpha)}_i$ are valued in the maximal torus of $U(r_\alpha)$, 
and $h'^{(\alpha)}_j$ are valued in the maximal torus of $U(M_\alpha - r_\alpha)$. We will take these fugacities to be related to the $U(r_\alpha) \times U(M_\alpha - r_\alpha)$ masses $(\vec u^\alpha, \vec v^\alpha)$ defined in \eref{uvdef0} in the following fashion (with complex $(\vec u^\alpha, \vec v^\alpha)$):
\be
\vec{h}^{(\alpha)} = e^{2\pi i \vec{u}^{\alpha}}, \qquad \vec{h}'^{(\alpha)}= e^{2\pi i \vec{v}^\alpha}.
\ee
In terms of the fugacities $\vec{\tz}^{(\alpha)}$, $\vec{h}^{(\alpha)} $ and $\vec{h}'^{(\alpha)}$ are therefore given as
\be \label{def-h}
{\tz}^{(\alpha)}_{i_\alpha} = \prod_i (h^{(\alpha)}_i)^{\CP_{i_\alpha i}}\, \prod_j (h'^{(\alpha)}_j)^{\CP_{i_\alpha r_\alpha+j}}, \quad i_\alpha=1,\ldots, M_\alpha,\quad i=1,\ldots, r_\alpha, \quad j=1,\ldots, M_\alpha - r_\alpha,
\ee
where $\CP$ is an $M_\alpha \times M_\alpha$ permutation matrix.
In addition, we introduce fluxes $(\kappa^{(\alpha)},{\kappa'}^{(\alpha)})$ for $U(r_\alpha) \times U(M_\alpha - r_\alpha)$, as follows:
\be \label{def-kappa}
\ta^{(\alpha)}_{i_\alpha} = \CP_{i_\alpha i} \, \kappa^{(\alpha)}_i + \CP_{i_\alpha \, r_\alpha + j} \, \kappa'^{(\alpha)}_j, \quad i_\alpha=1,\ldots, M_\alpha,\quad i=1,\ldots, r_\alpha, \quad j=1,\ldots, M_\alpha - r_\alpha.
\ee

The index of the quiver $(X, \CP)$ can then be written as a function of  $\vec{s}^X_\alpha = (\vec{h}^{(\alpha)}, \vec{\kappa}^{(\alpha)} )$, 
$\vec{s}'^X_\alpha =(\vec{h}'^{(\alpha)}, \vec{\kappa'}^{(\alpha)})$, $\vec{\Sigma}^X = (\{\vec{\tz}^{(\gamma)} \}, \{\vec{\ta}^{(\gamma)} \})_{\gamma \neq \alpha}$,
and $\vec{s}^X_J=(\vec y, \vec b)$, in addition to the R-symmetry fugacities, as follows:
\be
\begin{split}
& \CI^{(X,\CP)}(\vec{s}^X_\alpha, \vec{s}'^X_\alpha,\vec{\Sigma}^X, \vec{s}^X_J; \tq,\tti)=: \CI^{(X)}(\vec{\tz}^{(\alpha)}(\CP, \vec{h}^{(\alpha)},\vec{h}'^{(\alpha)}), \vec{\ta}^{(\alpha)}(\CP,\vec{\kappa}^{(\alpha)}, \vec{\kappa'}^{(\alpha)}), \vec{\Sigma}^X, \vec{s}^X_J; \tq,\tti).
\end{split}
\ee
Generalization of the above formula where multiple flavor nodes, labelled by $\beta$, are split, is given by:
\be
\begin{split}
& \CI^{(X,\{\CP_\beta\})}(\{\vec{s}^X_\beta, \vec{s}'^X_\beta\},\vec{\Sigma}^X, \vec{s}^X_J; \tq,\tti)=: \CI^{(X)}(\{\vec{\tz}^{(\beta)}(\CP_\beta, \vec{h}^{(\beta)},\vec{h}'^{(\beta)}), \vec{\ta}^{(\beta)}(\CP_\beta,\vec{\kappa}^{(\beta)}, \vec{\kappa'}^{(\beta)})\}, \vec{\Sigma}^X, \vec{s}^X_J; \tq,\tti),
\end{split}
\ee
where $\vec{\Sigma}^X$ is now defined as $\vec{\Sigma}^X=  (\{\vec{\tz}^{(\gamma)} \}, \{\vec{\ta}^{(\gamma)} \})_{\gamma \neq \beta}$, with 
$\vec{s}^X_\beta=(\vec{h}^{(\beta)}, \vec{\kappa}^{(\beta)})$ and $\vec{s}'^X_\beta=(\vec{h}'^{(\beta)}, \vec{\kappa'}^{(\beta)})$. The fluxes/fugacities
associated with the node $\beta$ are defined as before:
\begin{align}
&{\tz}^{(\beta)}_{i_\beta} = \prod_i (h^{(\beta)}_i)^{\CP_{\beta\,\, i_\beta i}}\, \prod_j (h'^{(\beta)}_j)^{\CP_{\beta\,\,i_\beta r_\alpha+j}}, \label{def-h1}\\
&\ta^{(\beta)}_{i_\beta} = \CP_{\beta\,\,i_\beta i} \, \kappa^{(\beta)}_i + \CP_{\beta\,\,i_\beta \, r_\alpha + j} \, \kappa'^{(\beta)}_j, \label{def-kappa1}
\end{align}
where $i_\beta=1,\ldots, M_\beta$, $i=1,\ldots, r_\alpha$, and $j=1,\ldots, M_\beta - r_\alpha$, with $\CP_\beta$ being an $M_\beta \times M_\beta$ permutation matrix.\\

An elementary $S$-type operation $\CO^\alpha_{\vec \CP}$ on the quiver gauge theory $X$ can then be implemented in terms of the superconformal index as follows:
\begin{flalign}\label{genSI}
\boxed{ \CI^{\CO^\alpha_{\vec \CP}(X)}
= \sum_{\vec{\kappa}^{(\alpha)}} \, \oint\limits^{}_{|h^{(\alpha)}_i|=1} 
 \Big[\frac{d\vec h^{(\alpha)}}{\vec h^{(\alpha)}}\Big]\, \sI_{\CO^\alpha_{\vec \CP}(X)}(\vec{s}^X_\alpha,\{\vec{s}^X_\beta\},\vec{s}^{\CO}_F,\vec{s}^{\CO}_J; \tq,\tti) \cdot \CI^{(X,\{\CP_\beta\})}(\{\vec{s}^X_\beta, \vec{s}'^{X}_\beta\},\vec{\Sigma}^X, \vec{s}^X_J; \tq,\tti),} &&
 \end{flalign}
where $\vec{s}^{\CO}_F$ collectively denotes all the flavor fugacities and fluxes introduced by the operation $\CO^\alpha_{\vec \CP}$, while 
$\vec{s}^{\CO}_J$ collectively denotes the new $U(1)_J$ fugacity and flux. The integration measure 
$\Big[\frac{d\vec h^{(\alpha)}}{\vec h^{(\alpha)}}\Big]=\frac{1}{ W({\kappa}^{(\alpha)})}\,\prod_i \frac{dh^{(\alpha)}_i}{2\pi i h^{(\alpha)}_i}$, 
where $|W(\vec{\kappa}^{(\alpha)})|$ is the order of the Weyl group of $U(r_\alpha)$ left unbroken by the fluxes $\vec \kappa^{(\alpha)}$.\\

For the elementary $S$-type operations discussed in \Secref{GFI-summary}, the associated operator  
$\sI_{{\CO}^\alpha_{\vec \CP}(X)}$ can be constructed from the index contributions of the gauging, flavoring, identification and defect 
operations (introduced in \Secref{GFI-def}), which are given as follows:
\begin{itemize}

\item For a gauging operation $G^\alpha_{\vec \CP}$ at a flavor node $\alpha$ of quiver $X$:
\begin{align}\label{G-basic0SCI}
&\sI_{G^\alpha_{\vec \CP}(X)}(\vec{s}^X_\alpha, \vec{s}^{\CO}_J; \tq,\tti)= \CI_{\rm FI} ({\wt{w}}, {\wt{n}}, \vec{h}^{(\alpha)}, \vec{\kappa}^{(\alpha)})\, \CI_{\rm vector}(\vec{h}^{(\alpha)}, \vec{\kappa}^{(\alpha)}; \tq,\tti),
\end{align}
where $\vec{s}^{\CO}_J=({\wt{w}}, {\wt{n}})$ denotes the fugacity and flux for the $U(1)_J$ global symmetry 
introduced by the gauging operation. 

\item For a flavoring operation $F^\alpha_{\vec \CP}$ at a flavor node $\alpha$ of $X$:
\begin{align}\label{F-basic0SCI}
\sI_{F^\alpha_{\vec \CP}(X)}(\vec{s}^X_\alpha, \vec{s}^{\CO}_F; \tq,\tti)= \CI_{\rm hyper}(\vec{h}^{(\alpha)}, \vec{\kappa}^{(\alpha)}, \vec{\wt{h}}^{(\alpha)}, \vec{\wt{\kappa}}^{(\alpha)};\tq, \tti),
\end{align}
where $ \vec{s}^{\CO}_F=(\vec{\wt h}^{(\alpha)}, \vec{\wt \kappa}^{(\alpha)})$ denotes the fugacities and fluxes for the flavor symmetry introduced by the flavoring operation.  

\item For an identification operation $I^\alpha_{\vec \CP}$, which involves $p$ nodes of the
linear quiver $X$ labelled by $\beta =\{ \gamma_1, \gamma_2, \ldots, \gamma_p \}$:
\begin{align}\label{I-basic0SCI}
\sI_{ I^\alpha_{\vec\CP}(X)}(\vec s^X_\alpha, \{\vec{s}^X_\beta\}_{\beta \neq \alpha}, \vec \lambda, \vec \delta; \tq,\tti)
= \prod_{\beta} \prod^{r_\alpha}_{i=1} \oint_{C_{\vec \lambda}} \frac{dh^{(\beta)}_i}{2\pi i\,(h^{(\beta)}_i - \lambda_\beta h^{(\alpha)}_i)} \cdot
\sum_{\{\vec{\kappa}^{(\beta)}\}}\delta_{\vec{\kappa}^{(\beta)}, \vec{\kappa}^{(\alpha)} + \delta^\beta},
\end{align}
where $\lambda_\beta = e^{2\pi i \mu^{\beta}}$, $\delta^\beta$ is a set of integers, and the integration is performed over an infinitesimal closed contours 
$C_{\vec \lambda}$ around the simple poles $h^{(\beta)}_i = \lambda_\beta h^{(\alpha)}_i$. By redefining $\vec{h}^{(\alpha)}$ and $\vec{\kappa}^{(\alpha)}$, 
one can set $ \lambda_{\beta'}=1$ and $\delta^{\beta'}=0$ for a chosen $\beta' \in \{ \gamma_1, \gamma_2, \ldots, \gamma_p \}$.\\

\item For a defect operation $D^\alpha_{\vec \CP}$ at a flavor node $\alpha$ of quiver $X$:
\be 
\sI_{D^\alpha_{\vec \CP}(X)}(\vec s^X_\alpha, \CD) = \CI_{\rm defect}(\vec s^X_\alpha, \CD),
\ee
where $\CD$ denotes the additional data associated with the defect introduced, and $\CI_{\rm defect}$ denotes the contribution of the defect to the index.
\end{itemize}
The explicit operator $\sI_{\CO^\alpha_{\vec \CP}(X)}$ can be constructed using the expressions for $\sI_{G^\alpha_{\vec \CP}(X)}$, $\sI_{F^\alpha_{\vec \CP}(X)}$, 
$\sI_{ I^\alpha_{\CP}(X)}$ and $\sI_{D^\alpha_{\vec \CP}(X)}$ following the composition rule \eref{Sbasic-def}:
\be \label{Sbasic-def-SCI}
\sI_{\CO^\alpha_{\vec \CP}(X)} =\sI_{G^\alpha_{\vec \CP}(X)} \cdot \Big(\sI_{F^\alpha_{\vec \CP}(X)}\Big)^{n_3} \cdot \Big(\sI_{I^\alpha_{\vec \CP}(X)}\Big)^{n_2} \cdot \Big(\sI_{D^\alpha_{\vec \CP}(X)}\Big)^{n_1}.
\ee

Now, let $Y$ be the linear quiver which is mirror dual to the quiver $X$. Three-dimensional mirror symmetry implies:
\begin{align} 
&\CI^{(X)}(\vec{\tz},\vec{\ta}, \vec y, \vec b; \tq,\tti) = \CI^{(Y)}(\vec{y}^{-1}, -\vec b,\vec{\tz},\vec{\ta}; \tq,\tti^{-1}) \\
\implies & \CI^{(X,\{\CP_\beta\})}(\{\vec{s}^X_\beta, \vec{s}'^X_\beta\},\vec{\Sigma}^X, \vec y, \vec b; \tq,\tti) = \CI^{(Y,\{\CP_\beta\})}(\vec{y}^{-1}, -\vec b, \{\vec{s}^X_\beta, \vec{s}'^X_\beta\},\vec{\Sigma}^X; \tq,\tti^{-1}).
\end{align}
where the expression for $\CI^{(Y,\{\CP_\beta\})}$ can be read off from the index of quiver $Y$ given in \eref{SCILinB},
using the relations \eref{def-tzy}, \eref{def-h}, and \eref{def-kappa}.
Proceeding in an analogous fashion as the round sphere partition function 
analysis in \Secref{GFIdual-summary}, one can write down the operation on quiver $Y$ which is dual to the operation 
$\CO^\alpha_{\vec \CP}$ on the quiver $X$. Let $\vec S$ and $\vec S_J$ denote the fugacities/fluxes associated with $G_H$
and $G_C$ respectively of the dual theory $\wt{\CO}^\alpha_{\vec \CP}(Y)$. Then, the index of the dual theory is
\begin{empheq}[box=\vwidefbox]{align}\label{genSCId1}
&\CI^{\wt{\CO}^\alpha_{\vec \CP} (Y)}(\vec S(\vec{s}^X_J, \vec{s}^{\CO}_J), \vec S_J(\{\vec{s}'^X_\beta\},\vec{s}^{\CO}_F,\vec{\Sigma}^X); \tq,\tti^{-1}) \nn \\
&= \sum_{\{\vec c^{(\gamma')}\}} \prod^{L^\vee}_{\gamma'=1} \oint\limits^{}_{|\xi^{(\gamma')}_{i_{\gamma'}}|=1} 
\Big[\frac{d\xi^{(\gamma')}}{\xi^{(\gamma')}} \Big]\,
\sI_{\wt{\CO}^\alpha_{\vec \CP}(Y)}(\vec{\xi}, \vec{c},\vec{s}^{\CO}_J,\vec{s}^{\CO}_F; \tq,\tti) \,\cdot
\CI^{(Y,\{\CP_\beta\})}_{\rm int}(\vec{\xi}, \vec{c},\vec{y}^{-1}, -\vec b, \{\vec{s}^{(0)\, X}_\beta, \vec{s}'^X_\beta\},\vec{\Sigma}^X; \tq,\tti^{-1}),
\end{empheq}
where $\vec{s}^{(0)\,X}_\beta = (\vec{1}, \vec{0})_\beta$ (i.e. setting the fugacities for $U(r_\alpha)_\beta$ to 1 and the fluxes to zero for all 
$\beta$), and the function $\CI^{(Y,\{\CP_\beta\})}_{\rm int}$ is the integrand for the index of $(Y, \{\CP_\beta\})$ as defined in \eref{SCILinB}.
The function $\sI_{\wt{\CO}^\alpha_{\vec \CP}(Y)}$ can be written in terms of the function $\sI_{\CO^\alpha_{\vec \CP}(X)}$ appearing in \eref{genSI} as follows:
\begin{align}\label{genSCId2}
 \sI_{\wt{\CO}^\alpha_{\vec \CP}(Y)}(\vec{\xi}, \vec{c},\vec{s}^{\CO}_J,\vec{s}^{\CO}_F; \tq,\tti)
=\sum_{\vec{\kappa}^{(\alpha)}} \, \oint_{|h^{(\alpha)}_i|=1} 
 \Big[\frac{d\vec h^{(\alpha)}}{\vec h^{(\alpha)}}\Big]\, & \sI_{\CO^\alpha_{\vec \CP}(X)}(\vec{s}^X_\alpha,\{\vec{s}^X_\beta\},\vec{s}^{\CO}_F,\vec{s}^{\CO}_J; \tq,\tti) \nn \\
& \times {\CI^{(0)}}^{(Y,\{\CP_\beta \})}_{\rm FI}(\vec{\xi}, \vec{c},\{\vec{s}^X_\beta\}), 
\end{align}
where ${\CI^{(0)}}^{(Y,\{\CP_\beta \})}_{\rm FI}$ is the part of the FI contribution which depends on 
$\{\vec{s}^{X}_\beta\}=\{ \vec{h}^{(\beta)}, \vec{\kappa}^{(\beta)} \}$, i.e.
\begin{align}
& {\CI^{(0)}}^{(Y,\{\CP_\beta \})}_{\rm FI}(\vec{\xi}, \vec{c},\{\vec{s}^X_\beta\})= \prod_\beta \prod^{r_\alpha}_{i=1} 
\Big[ \Big(f^{(\beta)}_{i} (\vec{\xi}, \CP_\beta) \Big)^{\kappa^{(\beta)}_{i}} \cdot \Big( h^{(\beta)}_{i} \Big)^{g_{(\beta)}^{i}(\vec{c},\CP_\beta)} \Big],\\
& g_{(\beta)}^{i} (\vec{c}, \CP_\beta)=- \sum_{i_{\alpha'-1}}c^{(\alpha'-1)}_{i_{\alpha'-1}} + \sum_{i_{\alpha'}} c^{(\alpha')}_{i_{\alpha'}},\\
& f^{(\beta)}_{i}(\vec{\xi},\CP_\beta)= \frac{\prod_{i_{\alpha'}} \xi^{(\alpha')}_{i_{\alpha'}}}{\prod_{i_{\alpha' -1}} \xi^{(\alpha'-1)}_{i_{\alpha'-1}}} , \quad \alpha'= M_1+\ldots+ M_{\beta -1} +k_\beta, \quad 1 \leq k_\beta \leq M_\beta,
\end{align}
where for a fixed $i$, $k_\beta$ is determined by the condition that $\CP_{\beta\,\, i_\beta i} =1$ for some $i_\beta=k_\beta$ and vanishes otherwise.
If the dual theory $\wt{\CO}^\alpha_{\vec \CP}(Y)$ is Lagrangian, one should be able to manipulate the RHS of \eref{genSCId1} to rewrite it in the standard form \eref{SCI-gen} 
and read off the gauge group and matter content.\\

Now let us generalize the above formulae for the case of a Lagrangian dual pair $(X,Y)$, where $X$ is in class $\CU$ and not necessarily a linear quiver. 
The $U(r_\alpha) \times U(M_\alpha - r_\alpha)$ 
fugacities/fluxes associated with a node $(\alpha)$ of $X$ (or the $U(r_\alpha) \times U(M_\beta - r_\alpha)$ fugacities/fluxes for multiple nodes labelled by $\beta$) 
are defined as before by \eref{def-h}-\eref{def-kappa} (or by \eref{def-h1}-\eref{def-kappa1}). The $U(1)_J$ fugacities/fluxes are parametrized by 
$\vec{s}^{X}_J=(\{w_\gamma \},\{n_\gamma \})$, where $\gamma$ labels the gauge nodes of quiver $X$ (note that the fugacities/fluxes can be trivial for 
$\gamma$ corresponding to a non-unitary gauge node). The $S$-type operation $\CO^\alpha_{\vec \CP}$ is then implemented on the 
quiver $X$ by \eref{genSI}, where the operator $\sI_{{\CO}^\alpha_{\vec \CP}(X)}$ is constructed from \eref{Sbasic-def-SCI}.\\

Mirror symmetry of $X$ and $Y$ again relates the indices of the theories in the following fashion:
\be
\CI^{(X,\{\CP_\beta\})}(\{\vec{s}^X_\beta, \vec{s}'^X_\beta\},\vec{\Sigma}^X, \vec{s}^{X}_J; \tq,\tti) = \CI^{(Y,\{\CP_\beta\})}(\vec s^{Y}(\vec{s}^{X}_J), \vec{s}^Y_J(\{\vec{s}^X_\beta, \vec{s}'^X_\beta\},\vec{\Sigma}^X); \tq,\tti^{-1}),
\ee
where $\vec s^{Y}$ and $\vec{s}^Y_J$ denote the fugacities/fluxes for the Higgs branch global symmetry and the Coulomb branch global symmetry of $Y$ 
respectively. The dual index is then given as:
\begin{empheq}[box=\vwidefbox]{align}\label{genSCId3}
\CI^{\wt{\CO}^\alpha_{\vec \CP} (Y)}(\vec S(\vec{s}^X_J, \vec{s}^{\CO}_J), \vec S_J(\{\vec{s}'^X_\beta\},\vec{s}^{\CO}_F,\vec{\Sigma}^X); \tq,\tti^{-1})
&= \sum_{\{\vec c^{(\gamma')}\}} \prod_{\gamma'} \oint\limits^{}_{|\xi^{(\gamma')}_{i_{\gamma'}}|=1} 
\Big[\frac{d\xi^{(\gamma')}}{\xi^{(\gamma')}} \Big]\,\sI_{\wt{\CO}^\alpha_{\vec \CP}(Y)}(\vec{\xi}, \vec{c},\vec{s}^{\CO}_J,\vec{s}^{\CO}_F; \tq,\tti) \nn \\
& \times \CI^{(Y,\{\CP_\beta\})}_{\rm int}(\vec{\xi}, \vec{c}, \vec s^{Y}(\vec{s}^{X}_J),  \vec{s}^Y_J(\{\vec{s}^{(0)\, X}_\beta, \vec{s}'^X_\beta\},\vec{\Sigma}^X); \tq,\tti^{-1}),
\end{empheq}
where $\vec S$ and $\vec S_J$ denote the fugacities/fluxes associated with $G_H$
and $G_C$ respectively of the dual theory $\wt{\CO}^\alpha_{\vec \CP}(Y)$, and $\gamma'$ labels the gauge nodes of the 
quiver $Y$ (which is not a linear quiver). The function $\sI_{\wt{\CO}^\alpha_{\vec \CP}(Y)}$ is given as before
\begin{align}\label{genSCId4}
 \sI_{\wt{\CO}^\alpha_{\vec \CP}(Y)}(\vec{\xi}, \vec{c},\vec{s}^{\CO}_J,\vec{s}^{\CO}_F; \tq,\tti)
=\sum_{\vec{\kappa}^{(\alpha)}} \, \oint_{|h^{(\alpha)}_i|=1} 
 \Big[\frac{d\vec h^{(\alpha)}}{\vec h^{(\alpha)}}\Big]\, & \sI_{\CO^\alpha_{\vec \CP}(X)}(\vec{s}^X_\alpha,\{\vec{s}^X_\beta\},\vec{s}^{\CO}_F,\vec{s}^{\CO}_J; \tq,\tti) \nn \\
& \times {\CI^{(0)}}^{(Y,\{\CP_\beta \})}_{\rm FI}(\vec{\xi}, \vec{c},\{\vec{s}^X_\beta\}).
\end{align}
The function ${\CI^{(0)}}^{(Y,\{\CP_\beta \})}_{\rm FI}$ above can be read off from the mirror map of masses and FI parameters between the dual theories $X$ and $Y$, and 
can be parametrized as follows:
\begin{align}
& {\CI^{(0)}}^{(Y,\{\CP_\beta \})}_{\rm FI}(\vec{\xi}, \vec{c},\{\vec{s}^X_\beta\})= \prod_\beta \prod^{r_\alpha}_{i=1} 
\Big[ \Big(f^{(\beta)}_{i} (\vec{\xi}, \CP_\beta) \Big)^{\kappa^{(\beta)}_{i}} \cdot \Big( h^{(\beta)}_{i} \Big)^{g_{(\beta)}^{i}(\vec{c},\CP_\beta)} \Big],\\
& g_{(\beta)}^{i} (\vec{c}, \CP_\beta)=\sum_{\gamma'} e^{\gamma'}_i (\CP_\beta) \sum_{j_{\gamma'}} c^{(\gamma')}_{j_{\gamma'}},\\
& f^{(\beta)}_{i}(\vec{\xi},\CP_\beta)= \prod_{\gamma'} \Big(\prod_{j_{\gamma'}} \xi^{(\gamma')}_{j_{\gamma'}}\Big)^{e^{\gamma'}_i (\CP_\beta)} \,,
\end{align}
where $\{e^{\gamma'}_i (\CP_\beta)\}$ are integers completely determined by the mirror map of $(X,Y)$ and the permutation matrices $\{\CP_\beta\}$.

\subsection{Abelian elementary $S$-type operations}\label{GFI-SCI-Ab}
We now discuss the Abelian versions of the four elementary $S$-type operations on a dual pair of quiver gauge theories $(X,Y)$, 
with $X$ being in class $\CU$. In particular, we explicitly derive the formulae for the dual superconformal indices, and demonstrate
that each such operation (we only consider flavoring by hypermultiplets with gauge charge 1) leads to a new pair of Lagrangian dual theories.
Similar to the analysis of the round sphere partition function, this leads us to the following conclusion. If $(X,Y)$ is a dual pair of quiver gauge 
theories, and $X'$ is a quiver gauge theory that can be obtained by a series of elementary $S$-type operations on $X$, then the theory $Y'$ 
(i.e. the dual of $X'$) is guaranteed to be a Lagrangian theory. In addition, the Lagrangian for $Y'$ can be read off, by implementing the 
formulae presented below, for each elementary $S$-type operation.\\

For our presentation of the flavoring operations below, we will restrict ourselves to the case of $N^\alpha_F=1$, which 
is sufficient for constructing all the examples in this paper. The extension to the case of $N^\alpha_F > 1$ is straightforward 
and can be dealt with in a fashion analogous to the round sphere partition function analysis in \Secref{AbS}.\\

We will need two identities for manipulating some of the expressions that will appear below:
\begin{align}
& \oint_{|z|=1}\,\frac{d z}{2\pi i z} \, z^n = \delta_{{n},0}, \label{SCI-Id1}\\
& \oint_{|z|=1}\,\frac{dz}{2\pi i z}\,\Big(\sum_{\vec{\kappa} \in \BZ}(\frac{z}{a})^{{\kappa}} \Big)\, F(z)= F(z)|_{\frac{z}{a}=1} =F(a). \label{SCI-Id2}
\end{align}

\subsubsection{Gauging} 

The Abelian gauging operation $G^\alpha_{\CP}$ at a flavor node $\alpha$ of the quiver gauge theory $X$ is implemented as:
\begin{align}
& \CI^{G^\alpha_{\CP}(X)}
= \sum_{{\kappa}^{(\alpha)}} \, \oint\limits^{}_{|h^{(\alpha)}|=1} 
 \Big[\frac{d h^{(\alpha)}}{ h^{(\alpha)}}\Big]\, \sI_{G^\alpha_{\CP}(X)}(\vec{s}^X_\alpha,\vec{s}^{\CO}_J; \tq,\tti) \cdot \CI^{(X, \CP)}(\vec{s}^X_\alpha, \vec{s}'^X_\alpha,,\vec{\Sigma}^X, \vec{s}^X_J; \tq,\tti),\\
& \sI_{G^\alpha_{\CP}(X)}(\vec{s}^X_\alpha,\vec{s}^{\CO}_J; \tq,\tti)= \CI_{\rm FI} ({\wt{w}}, {\wt{n}}, {h}^{(\alpha)}, {\kappa}^{(\alpha)})\, \CI_{\rm vector}(\tq,\tti)
= {\wt{w}}^{{\kappa}^{(\alpha)}} {{h}^{(\alpha)}}^{\wt{n}} \CI_{\rm vector}(\tq,\tti),
\end{align}
where $\CI_{\rm vector}(\tq,\tti)$ is the index of a $U(1)$ vector multiplet. From \eref{genSCId4}, the function $\sI_{\wt{G}^\alpha_{\CP}(Y)}$ is then given as
\begin{align}
& \sI_{\wt{G}^\alpha_{\CP}(Y)}(\vec{\xi}, \vec{c},\vec{s}^{\CO}_J; \tq,\tti)
=\sum_{{\kappa}^{(\alpha)}} \, \oint_{|h^{(\alpha)}|=1} 
 \Big[\frac{d h^{(\alpha)}}{h^{(\alpha)}}\Big]\, \sI_{G^\alpha_{\CP}(X)}(\vec{s}^X_\alpha, \vec{s}^{\CO}_J; \tq,\tti)
\cdot {\CI^{(0)}}^{(Y, \CP)}_{\rm FI}(\vec{\xi}, \vec{c}, \vec{s}^X_\alpha),
\end{align}
where ${\CI^{(0)}}^{(Y, \CP)}_{\rm FI}$ is the $(h^{(\alpha)}, \kappa^{(\alpha)})$ part of the FI term in the index of quiver $Y$. 
Explicitly, this function can be written as:
\begin{align}
& {\CI^{(0)}}^{(Y, \CP)}_{\rm FI}(\vec{\xi}, \vec{c}, \vec{s}^X_\alpha)=
 \Big(f^{(\alpha)} (\vec{\xi}, \CP) \Big)^{\kappa^{(\alpha)}} \cdot \Big( h^{(\alpha)} \Big)^{g_{(\alpha)}(\vec{c},\CP)}, \label{FI-hk}\\
& g_{(\alpha)} (\vec{c}, \CP)=\sum_{\gamma'} e^{\gamma'} (\CP) \sum_{j_{\gamma'}} c^{(\gamma')}_{j_{\gamma'}},  \label{FI-hk1}\\
& f^{(\alpha)} (\vec{\xi},\CP)= \prod_{\gamma'} \Big(\prod_{j_{\gamma'}} \xi^{(\gamma')}_{j_{\gamma'}}\Big)^{e^{\gamma'} (\CP)} \, , \label{FI-hk2}
\end{align}
where $\{ e^{\gamma'} (\CP) \}$ are integers completely determined by the mirror map between $X$ and $Y$, and the permutation matrix $\CP$.
Following the general equation \eref{genSCId3} and implementing the integration over $h^{(\alpha)}$ and the sum over $\kappa^{\alpha}$
(using the identities \eref{SCI-Id1} and \eref{SCI-Id2} respectively), the index of the dual theory $\wt{G}^\alpha_{\CP} (Y)$ is given as:
\begin{align}\label{GAbgen-SCI}
&\CI^{\wt{G}^\alpha_{\CP} (Y)}(\vec S(\vec{s}^X_J, \vec{s}^{\CO}_J), \vec S_J(\vec{s}'^X_\alpha, \vec{\Sigma}^X); \tq,\tti^{-1}) \nn \\
&= \sum_{\{\vec c^{(\gamma')}\}} \prod_{\gamma'} \oint\limits^{}_{|\xi^{(\gamma')}_{i_{\gamma'}}|=1} 
\Big[\frac{d\vec{\xi}^{(\gamma')}}{\vec{\xi}^{(\gamma')}} \Big]\, \frac{\Big[ \CI^{(Y, \CP)}_{\rm int}(\vec{\xi}, \vec{c}, \vec s^{Y}(\vec{s}^{X}_J),  \vec{s}^Y_J(\vec{s}^{(0)\, X}_\alpha, \vec{s}'^X_\alpha,\vec{\Sigma}^X); \tq,\tti^{-1})\Big]_{\begin{subarray}{l} \wt{w} f^{(\alpha)} =1, \\ g_{(\alpha)}=-\wt{n} \end{subarray}}}{\CI_{\rm vector}(\tq,\tti^{-1})}.
\end{align}
The factor $\CI_{\rm vector}(\tq,\tti^{-1})$ cancels with the index of a single $U(1)$ vector multiplet in $\CI^{(Y, \CP)}_{\rm int}$. Together with the conditions 
$ \wt{w} f^{(\alpha)} (\vec{\xi},\CP) =1, g_{(\alpha)} (\vec{c},\CP)=-\wt{n}$, this removes a single $U(1)$ factor from the gauge group of $Y$. The precise $U(1)$ being ungauged is determined by the functions $ f^{(\alpha)}, g_{(\alpha)}$, or equivalently the integers $\{ e^{\gamma'} (\CP) \}$. The theory $\wt{G}^\alpha_{\CP} (Y)$ is therefore a Lagrangian 
theory.

\subsubsection{Flavoring-gauging} 

The Abelian flavoring-gauging operation $\CO^\alpha_{ \CP}(X) = G^\alpha_{\CP} \circ F^\alpha_{ \CP}(X)$ with $N^\alpha_F=1$ 
can be implemented following the general expression in \eref{genSI}:
\begin{align}
\CI^{\CO^\alpha_{\CP}(X)}
= \sum_{{\kappa}^{(\alpha)}} \, \oint\limits^{}_{|h^{(\alpha)}|=1} 
 \Big[\frac{d h^{(\alpha)}}{ h^{(\alpha)}}\Big]\, \sI_{\CO^\alpha_{\CP}(X)}(\vec{s}^X_\alpha,\vec{s}^{\CO}_F, \vec{s}^{\CO}_J; \tq,\tti) \cdot \CI^{(X, \CP)}(\vec{s}^X_\alpha, \vec{s}'^X_\alpha,,\vec{\Sigma}^X, \vec{s}^X_J; \tq,\tti),
\end{align}
where the function $ \sI_{\CO^\alpha_{\CP}(X)}$ can be read off from \eref{Sbasic-def}, 
\begin{align}\label{basic-3dMS}
& \sI_{\CO^\alpha_{\CP}(X)}(\vec{s}^X_\alpha, \vec{s}^{\CO}_F, \vec{s}^{\CO}_J; \tq,\tti) 
=  \CI_{\rm FI} ({\wt{w}}, {\wt{n}}, {h}^{(\alpha)}, {\kappa}^{(\alpha)})\, \CI_{\rm vector}(\tq,\tti)\, \CI_{\rm hyper}({h}^{(\alpha)}, {\kappa}^{(\alpha)}, {\wt{h}}^{(\alpha)}, {\wt{\kappa}}^{(\alpha)};\tq, \tti) \nn \\
& = \sum_{\kappa} \oint\limits^{}_{|h|=1} \frac{dh}{2\pi i h}\, (h^{(\alpha)})^{\kappa+\wt{n}}\, (\wt{w} h)^{\kappa^{(\alpha)}} \, ({\wt{h}}^{(\alpha)})^{-\kappa} \, h^{- \wt{\kappa}^{(\alpha)}}\,
\CI_{\rm hyper}(h, {\kappa}, 1, 0;\tq, \tti^{-1}),
\end{align}
where for the second equality we have used 3d mirror symmetry between a free hyper and a $U(1)$ gauge theory with a single hypermultiplet of charge 1 
(as given in the identity \eref{SCIDual1b}). From \eref{genSCId4}, the function $\sI_{\wt{\CO}^\alpha_{\CP}(Y)}$ is given as
\begin{align}
\sI_{\wt{\CO}^\alpha_{\CP}(Y)}(\vec{\xi}, \vec{c}, \vec{s}^{\CO}_F , \vec{s}^{\CO}_J; \tq,\tti)
=\sum_{{\kappa}^{(\alpha)}} \, \oint_{|h^{(\alpha)}|=1} 
 \Big[\frac{d h^{(\alpha)}}{h^{(\alpha)}}\Big]\, \sI_{\CO^\alpha_{\CP}(X)}(\vec{s}^X_\alpha, \vec{s}^{\CO}_F, \vec{s}^{\CO}_J; \tq,\tti)
\cdot {\CI^{(0)}}^{(Y, \CP)}_{\rm FI}(\vec{\xi}, \vec{c}, \vec{s}^X_\alpha),
\end{align}
where the function ${\CI^{(0)}}^{(Y, \CP)}_{\rm FI}(\vec{\xi}, \vec{c}, \vec{s}^X_\alpha)$ is given in \eref{FI-hk}, \eref{FI-hk1} and \eref{FI-hk2}.
To simplify the above expression, we first substitute the expression for $\sI_{\CO^\alpha_{\CP}(X)}$ from \eref{basic-3dMS}, and change the order of integration and sum between 
the variables $(h^{(\alpha)}, \kappa^{(\alpha)})$ and $(h,\kappa)$.
Finally, implementing the integration and sum over $(h^{(\alpha)}, \kappa^{(\alpha)})$ first and then over $(h,\kappa)$ (using the identities \eref{SCI-Id1} and \eref{SCI-Id2} respectively), we obtain
\begin{align}
\sI_{\wt{\CO}^\alpha_{\CP}(Y)}= ({\wt{h}}^{(\alpha)})^{\wt{n}} \, (\wt{w})^{\wt{\kappa}^{(\alpha)}} \, ({\wt{h}}^{(\alpha)})^{g_{(\alpha)}}\, (f^{(\alpha)})^{\wt{\kappa}^{(\alpha)}}\,
\CI_{\rm hyper}(h, {\kappa}, 1, 0;\tq, \tti^{-1})|_{\begin{subarray}{l} h \wt{w} f^{(\alpha)} =1\\  \kappa + g_{(\alpha)} + \wt{n}=0 \end{subarray}} ,
\end{align}
where the function $\CI_{\rm hyper}(h, {\kappa}, 1, 0;\tq, \tti^{-1})|_{\begin{subarray}{l} h \wt{w} f^{(\alpha)} =1\\  \kappa + g_{(\alpha)} + \wt{n}=0 \end{subarray}}$ denotes 
the index of a single hypermultiplet charged under various $U(1)$ subgroups of the gauge group of $Y$. The precise $U(1)$ subgroups and the respective charges are 
encoded in the functions $f^{(\alpha)}, g_{(\alpha)}$, or equivalently in the integers $\{ e^{\gamma'} (\CP) \}$. Following the general equation \eref{genSCId3}, the 
index of the dual theory $\wt{\CO}^\alpha_{\CP} (Y)$ is then given as
\begin{align}\label{GFAbgen-SCI}
&\CI^{\wt{\CO}^\alpha_{\CP} (Y)}(\vec S(\vec{s}^X_J, \vec{s}^{\CO}_J), \vec S_J(\vec{s}'^X_\alpha, \vec{\Sigma}^X, \vec{s}^{\CO}_F); \tq,\tti^{-1}) \nn \\
&= \sum_{\{\vec c^{(\gamma')}\}} \prod_{\gamma'} \oint\limits^{}_{|\xi^{(\gamma')}_{i_{\gamma'}}|=1} 
\Big[\frac{d\vec{\xi}^{(\gamma')}}{\vec{\xi}^{(\gamma')}} \Big]\,  ({\wt{h}}^{(\alpha)})^{\wt{n}} \, (\wt{w})^{\wt{\kappa}^{(\alpha)}}\,\CI_{\rm hyper}(h, {\kappa}, 1, 0;\tq, \tti^{-1})|_{\begin{subarray}{l} h \wt{w} f^{(\alpha)} =1, \\  \kappa + g_\alpha + \wt{n}=0 \end{subarray}}  \nn \\
& \qquad \qquad \times  \CI^{(Y, \CP)}_{\rm int}(\vec{\xi}, \vec{c}, \vec s^{Y}(\vec{s}^{X}_J),  \vec{s}^Y_J(\{ {h}^{(\alpha)}= {\wt{h}}^{(\alpha)}, {\kappa}^{(\alpha)} ={\wt{\kappa}}^{(\alpha)} \}, \ldots); \tq,\tti^{-1}).
\end{align}
The factor $({\wt{h}}^{(\alpha)})^{\wt{n}} \, (\wt{w})^{\wt{\kappa}^{(\alpha)}}$ can be absorbed by redefining some of the fugacities and fluxes 
$(\vec{\xi}^{(\gamma')}, \vec{c}^{(\gamma')})$, such that $g_{(\alpha)} \to g_{(\alpha)} - \wt{n}$, 
and $f^{(\alpha)} \to \frac{1}{\wt{w}} f^{(\alpha)}$ . The Lagrangian for the theory $\wt{\CO}^\alpha_{\CP} (Y)$ can be read off from the index -- it involves adding a single hypermultiplet to the quiver gauge theory $Y$, where the said hypermultiplet is charged under various $U(1)$ subgroups of the gauge group of $Y$, as noted 
above.

\subsubsection{Identification-gauging} 

The Abelian identification-gauging operation $\CO^\alpha_{\vec \CP}(X) = G^\alpha_{\vec \CP} \circ I^\alpha_{\vec \CP}(X)$ 
can be implemented following the general expression in \eref{genSI}:
\begin{align}
\CI^{\CO^\alpha_{\vec \CP}(X)}
= \sum_{{\kappa}^{(\alpha)}} \, \oint\limits^{}_{|h^{(\alpha)}|=1} 
 \Big[\frac{d h^{(\alpha)}}{h^{(\alpha)}}\Big]\, & \sI_{\CO^\alpha_{\vec \CP}(X)}(\vec{s}^X_\alpha,\{\vec{s}^X_\beta\}, \vec{s}^{\CO}_F, \vec{s}^{\CO}_J; \tq,\tti) \nn \\ \times & \, \CI^{(X,\{\CP_\beta\})}(\{\vec{s}^X_\beta, \vec{s}'^{X}_\beta\},\vec{\Sigma}^X, \vec{s}^X_J; \tq,\tti),
 \end{align}
where the operator $\sI_{\CO^\alpha_{\vec \CP}(X)}$ is given as (from \eref{Sbasic-def}) :
\begin{align}
 \sI_{\CO^\alpha_{\vec \CP}(X)}
=  \CI_{\rm FI} ({\wt{w}}, {\wt{n}}, {h}^{(\alpha)}, {\kappa}^{(\alpha)})\, \CI_{\rm vector}(\tq,\tti) \,
\prod_{\beta}\oint_{C_{\vec \lambda}} \frac{dh^{(\beta)}}{2\pi i\,(h^{(\beta)} - \lambda_\beta h^{(\alpha)})} \cdot \sum_{\{\vec{\kappa}^{(\beta)}\}}\delta_{{\kappa}^{(\beta)}, {\kappa}^{(\alpha)} + \delta^\beta}.
\end{align}
From \eref{genSCId4}, the function $\sI_{\wt{\CO}^\alpha_{\vec\CP}(Y)}$ is given as:
\begin{align}
 \sI_{\wt{\CO}^\alpha_{\vec\CP}(Y)}(\vec{\xi}, \vec{c},\vec{s}^{\CO}_F, \vec{s}^{\CO}_J; \tq,\tti)
=\sum_{{\kappa}^{(\alpha)}} \, \oint_{|h^{(\alpha)}|=1} 
 \Big[\frac{d h^{(\alpha)}}{h^{(\alpha)}}\Big]\, & \sI_{\CO^\alpha_{\CP}(X)}(\vec{s}^X_\alpha,\{\vec{s}^X_\beta\}, \vec{s}^{\CO}_F, \vec{s}^{\CO}_J; \tq,\tti) \nn \\
& \times {\CI^{(0)}}^{(Y, \{\CP_\beta\})}_{\rm FI}(\vec{\xi}, \vec{c}, \{\vec{s}^X_\beta\}),
\end{align}
with the $\{ h^{(\beta)}, \kappa^{(\beta)}\}$-dependent part of the FI term of $Y$ can be explicitly written as:
\begin{align}
& {\CI^{(0)}}^{(Y,\{\CP_\beta \})}_{\rm FI}(\vec{\xi}, \vec{c},\{\vec{s}^X_\beta\})= \prod_\beta 
\Big[ \Big(f^{(\beta)} (\vec{\xi}, \CP_\beta) \Big)^{\kappa^{(\beta)}} \cdot \Big( h^{(\beta)} \Big)^{g_{(\beta)}(\vec{c},\CP_\beta)} \Big], \label{FI-hk4}\\
& g_{(\beta)} (\vec{c}, \CP_\beta)=\sum_{\gamma'} e^{\gamma'} (\CP_\beta) \sum_{j_{\gamma'}} c^{(\gamma')}_{j_{\gamma'}}, \label{FI-hk5}\\
& f^{(\beta)}(\vec{\xi},\CP_\beta)= \prod_{\gamma'} \Big(\prod_{j_{\gamma'}} \xi^{(\gamma')}_{j_{\gamma'}}\Big)^{e^{\gamma'} (\CP_\beta)} \,. \label{FI-hk6}
\end{align}
Finally, from the general expression of \eref{genSCId3}, and implementing the integration over $h^{(\alpha)}$ and the sum over $\kappa^{\alpha}$
(using the identities \eref{SCI-Id1} and \eref{SCI-Id2} respectively), we obtain the index for the dual theory $\wt{\CO}^\alpha_{\vec \CP} (Y)$:
\begin{align}\label{GIAbgen-SCI}
&\CI^{\wt{\CO}^\alpha_{\vec\CP} (Y)}(\vec S(\vec{s}^X_J, \vec{s}^{\CO}_J), \vec S_J(\{\vec{s}'^X_\beta\},\vec{s}^{\CO}_F,\vec{\Sigma}^X); \tq,\tti^{-1}) \nn \\
=& \sum_{\{\vec c^{(\gamma')}\}}  \prod_{\gamma'} \oint\limits^{}_{|\xi^{(\gamma')}_{i_{\gamma'}}|=1} 
\Big[\frac{d\vec{\xi}^{(\gamma')}}{\vec{\xi}^{(\gamma')}} \Big]\, \frac{\Big[ \CI^{(Y,\{\CP_\beta\})}_{\rm int}(\vec{\xi}, \vec{c}, \vec s^{Y}(\vec{s}^{X}_J),  
\vec{s}^Y_J(\{h^{(\beta)}=\lambda_\beta, \kappa^{(\beta)} = \delta_\beta \},\ldots); \tq,\tti^{-1}) \Big]_{\begin{subarray}{l} \wt{w} \prod_\beta f^{(\beta)} =1, \\ \sum_\beta g_\beta=-\wt{n}\end{subarray}}}{\CI_{\rm vector}(\tq,\tti^{-1})}.
\end{align}
Similar to the case of the gauging operation, the factor $\CI_{\rm vector}(\tq,\tti^{-1})$ cancels with the index of a single $U(1)$ vector multiplet in $\CI^{(Y, \CP)}_{\rm int}$. 
Together with the conditions $ \wt{w}\prod_\beta f^{(\beta)} (\vec{\xi},\CP_\beta) =1, \prod_\beta g_{(\beta)} (\vec{c},\CP_\beta)=-\wt{n}$, this removes a single $U(1)$ 
factor from the gauge group of $Y$. The precise $U(1)$ being ungauged is determined by the functions $\prod_\beta f^{(\beta)}, \sum_\beta g_{(\beta)}$, or equivalently the 
integers $\{ e^{\gamma'} (\CP_\beta) \}$. The theory $\wt{\CO}^\alpha_{\vec\CP} (Y)$ is therefore a Lagrangian theory.

\subsubsection{Identification-flavoring-gauging} 

The Abelian identification-flavoring-gauging operation $\CO^\alpha_{\vec \CP}(X) = G^\alpha_{\vec \CP} \circ F^\alpha_{\vec \CP} \circ I^\alpha_{\vec \CP}(X)$ can be implemented following the general expression in \eref{genSI}:
\begin{align}
\CI^{\CO^\alpha_{\vec \CP}(X)}
= \sum_{{\kappa}^{(\alpha)}} \, \oint\limits^{}_{|h^{(\alpha)}|=1} 
 \Big[\frac{d h^{(\alpha)}}{h^{(\alpha)}}\Big]\, & \sI_{\CO^\alpha_{\vec \CP}(X)}(\vec{s}^X_\alpha,\{\vec{s}^X_\beta\}, \vec{s}^{\CO}_F, \vec{s}^{\CO}_J; \tq,\tti) \nn\\ 
 & \times \CI^{(X,\{\CP_\beta\})}(\{\vec{s}^X_\beta, \vec{s}'^{X}_\beta\},\vec{\Sigma}^X, \vec{s}^X_J; \tq,\tti),
 \end{align}
where the operator $\sI_{\CO^\alpha_{\vec \CP}(X)}$ is given as (from \eref{Sbasic-def}) : 
\begin{align}
 \sI_{\CO^\alpha_{\vec \CP}(X)} &
=  \CI_{\rm FI} ({\wt{w}}, {\wt{n}}, {h}^{(\alpha)}, {\kappa}^{(\alpha)})\, \CI_{\rm vector}(\tq,\tti)\, \CI_{\rm hyper}({h}^{(\alpha)}, {\kappa}^{(\alpha)}, {\wt{h}}^{(\alpha)}, {\wt{\kappa}}^{(\alpha)};\tq, \tti) \nn \\
& \times \prod_{\beta}\oint_{C_{\vec \lambda}} \frac{dh^{(\beta)}}{2\pi i\,(h^{(\beta)} - \lambda_\beta h^{(\alpha)})} \cdot \sum_{\{\vec{\kappa}^{(\beta)}\}}\delta_{{\kappa}^{(\beta)}, {\kappa}^{(\alpha)} + \delta^\beta} \\
& = \sum_k \oint\limits^{}_{|h|=1} \frac{dh}{2\pi i h}\, (h^{(\alpha)})^{\kappa+\wt{n}}\, (\wt{w} h)^{\kappa^{(\alpha)}} \, ({\wt{h}}^{(\alpha)})^{-\kappa} \, h^{- \wt{\kappa}^{(\alpha)}}\,
\CI_{\rm hyper}(h, {\kappa}, 1, 0;\tq, \tti^{-1}) \nn \\
& \times \prod_{\beta}\oint_{C_{\vec \lambda}} \frac{dh^{(\beta)}}{2\pi i\,(h^{(\beta)} - \lambda_\beta h^{(\alpha)})} \cdot \sum_{\{\vec{\kappa}^{(\beta)}\}}\delta_{{\kappa}^{(\beta)}, {\kappa}^{(\alpha)} + \delta^\beta}, \label{basic-3dMS1}
\end{align}
where, for the second equality, we have used the basic 3d mirror symmetry, as given in \eref{SCIDual1b}. 
From \eref{genSCId4}, the function $\sI_{\wt{\CO}^\alpha_{\vec\CP}(Y)}$ is given as:
\begin{align}
 \sI_{\wt{\CO}^\alpha_{\vec\CP}(Y)}(\vec{\xi}, \vec{c},\vec{s}^{\CO}_F, \vec{s}^{\CO}_J; \tq,\tti)
=\sum_{{\kappa}^{(\alpha)}} \, \oint_{|h^{(\alpha)}|=1} 
 \Big[\frac{d h^{(\alpha)}}{h^{(\alpha)}}\Big]\, & \sI_{\CO^\alpha_{\vec\CP}(X)}(\vec{s}^X_\alpha,\{\vec{s}^X_\beta\}, \vec{s}^{\CO}_F, \vec{s}^{\CO}_J; \tq,\tti) \nn \\
& \times {\CI^{(0)}}^{(Y, \{\CP_\beta\})}_{\rm FI}(\vec{\xi}, \vec{c}, \{\vec{s}^X_\beta\}),
\end{align}
with ${\CI^{(0)}}^{(Y, \{\CP_\beta\})}_{\rm FI}$, i.e. the $\{ h^{(\beta)}, \kappa^{(\beta)}\}$-dependent part of the FI term of $Y$ is given in \eref{FI-hk4}, \eref{FI-hk5} and \eref{FI-hk6}. 
The above expression can be simplified by first substituting the expression for $\sI_{\CO^\alpha_{\CP}(X)}$ from \eref{basic-3dMS1}, and change the order of integration and sum between the variables $(h^{(\alpha)}, \kappa^{(\alpha)})$ and $(h,\kappa)$.
Finally, implementing the integration and sum over $(h^{(\alpha)}, \kappa^{(\alpha)})$ first and then over $(h,\kappa)$ (using the identities \eref{SCI-Id1} and \eref{SCI-Id2} respectively), 
we obtain
\begin{align}
 \sI_{\wt{\CO}^\alpha_{\vec\CP}(Y)}
 = ({\wt{h}}^{(\alpha)})^{\wt{n}} \, (\wt{w})^{\wt{\kappa}^{(\alpha)}} \, \prod_\beta ({\wt{h}}^{(\alpha)}\lambda_\beta)^{g_{(\beta)}}\, \prod_\beta (f^{(\beta)})^{\delta^\beta+ \wt{\kappa}^{(\alpha)}}\,
\CI_{\rm hyper}(h, {\kappa}, 1, 0;\tq, \tti^{-1})|_{\begin{subarray}{l} h \wt{w} \prod_\beta f^{(\beta)} =1 \\  \kappa + \sum_\beta g_{(\beta)} + \wt{n}=0 \end{subarray}}
 \end{align}
where the function $\CI_{\rm hyper}(h, {\kappa}, 1, 0;\tq, \tti^{-1})|_{\begin{subarray}{l} h \wt{w} \prod_\beta f^{(\beta)} =1 \\  \kappa + \sum_\beta g_{(\beta)} + \wt{n}=0 \end{subarray}}$ denotes the index of a single hypermultiplet charged under various $U(1)$ subgroups of the gauge group of $Y$. The precise $U(1)$ subgroups and the respective charges are 
encoded in the functions $\prod_\beta f^{(\beta)}, \sum_\beta g_{(\beta)} $, or equivalently in the integers $\{ e^{\gamma'} (\CP_\beta) \}$. Following the general equation \eref{genSCId3}, the index of the dual theory $\wt{\CO}^\alpha_{\vec \CP} (Y)$ is then given as 
\begin{align}\label{GFIAbgen-SCI}
&\CI^{\wt{\CO}^\alpha_{\vec\CP} (Y)}(\vec S(\vec{s}^X_J, \vec{s}^{\CO}_J), \vec S_J(\{\vec{s}'^X_\beta\},\vec{s}^{\CO}_F,\vec{\Sigma}^X); \tq,\tti^{-1}) \nn \\
=& \sum_{\{\vec c^{(\gamma')}\}}  \prod_{\gamma'} \oint\limits^{}_{|\xi^{(\gamma')}_{i_{\gamma'}}|=1} 
\Big[\frac{d\vec{\xi}^{(\gamma')}}{\vec{\xi}^{(\gamma')}} \Big]\, ({\wt{h}}^{(\alpha)})^{\wt{n}} \, (\wt{w})^{\wt{\kappa}^{(\alpha)}}\,\CI_{\rm hyper}(h, {\kappa}, 1, 0;\tq, \tti^{-1})|_{\begin{subarray}{l} h \wt{w} \prod_\beta f^{(\beta)} =1 \\  \kappa + \sum_\beta g_{(\beta)} + \wt{n}=0 \end{subarray}}  \nn \\
& \qquad \qquad \times  \CI^{(Y, \{\CP_\beta \})}_{\rm int}(\vec{\xi}, \vec{c}, \vec s^{Y}(\vec{s}^{X}_J),  \vec{s}^Y_J(\{ {h}^{(\beta)}= {\wt{h}}^{(\alpha)}\lambda_\beta, {\kappa}^{(\beta)} =\delta^\beta + {\wt{\kappa}}^{(\alpha)} \}, \ldots); \tq,\tti^{-1}).
\end{align}
As mentioned in our discussion on \eref{I-basic0SCI}, one has the freedom to choose $\lambda_{\beta'} =1$, $\delta_{\beta'}=0$, for some $\beta'\in \{ \gamma_1, \gamma_2, \ldots, \gamma_p \}$. Redefining some of the fugacities and fluxes $(\vec{\xi}^{(\gamma')}, \vec{c}^{(\gamma')})$, such that $g_{(\beta')} \to g_{(\beta')} - \wt{n}$, 
and $f^{(\beta')} \to \frac{1}{\wt{w}} f^{(\beta')}$, the factor $({\wt{h}}^{(\alpha)})^{\wt{n}} \, (\wt{w})^{\wt{\kappa}^{(\alpha)}}$ can be absorbed in the integrand. 
The Lagrangian for the theory $\wt{\CO}^\alpha_{\vec \CP} (Y)$ can be read off from the index -- it involves adding a single hypermultiplet 
to the quiver gauge theory $Y$, where the said hypermultiplet is charged under various $U(1)$ subgroups of the gauge group of $Y$, as noted 
above.

\subsection{Sample Computation: Flavored $\widehat{A}_{n-1}$ quiver and Family ${\rm I}_{[n,l,p]}$}\label{GFI-exAbSCI}
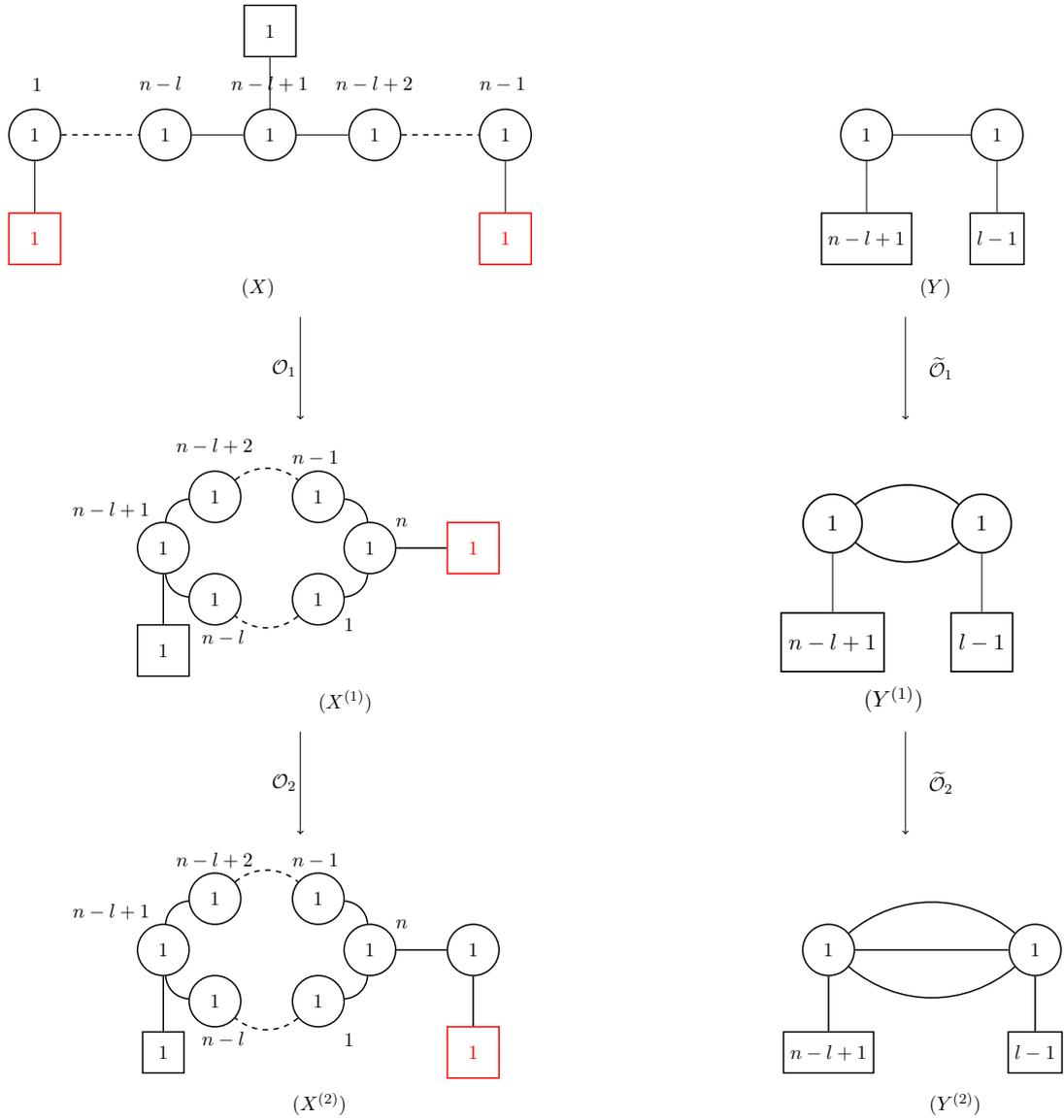
\begin{figure}[htbp]
\begin{center}
\begin{tabular}{ccc}
\scalebox{.7}{\begin{tikzpicture}[node distance=2cm,
cnode/.style={circle,draw,thick, minimum size=1.0cm},snode/.style={rectangle,draw,thick,minimum size=1cm}, pnode/.style={red,rectangle,draw,thick, minimum size=1.0cm},]
\node[cnode] (1) {1};
\node[cnode] (2) [right=1.5cm  of 1]{1};
\node[cnode] (3) [right= 1 cm  of 2]{1};
\node[snode] (4) [above=1 cm of 3]{1};
\node[cnode] (5) [right=1 cm  of 3]{1};
\node[cnode] (6) [right=1.5cm  of 5]{1};
\node[pnode] (7) [below=1 cm of 1]{1};
\node[pnode] (8) [below=1 cm of 6]{1};
\node[text width=1.5cm](9)[above=0.2cm of 3]{$n-l+1$};
\node[text width=1.5cm](10)[above=0.2cm of 5]{$n-l+2$};
\node[text width=1 cm](11)[above=0.2cm of 6]{$n-1$};
\node[text width=1cm](12)[above=0.2cm of 2]{$n-l$};
\node[text width=0.1 cm](13)[above=0.2cm of 1]{$1$};
\draw[thick, dashed] (1) -- (2);
\draw[-] (2)-- (3);
\draw[-] (3)-- (4);
\draw[-] (3)-- (5);
\draw[thick, dashed] (5)-- (6);
\draw[-] (1)-- (7);
\draw[-] (8)-- (6);
\node[text width=6cm](20) at (7, -3) {$(X)$};
\end{tikzpicture}}
& \qquad 
& \scalebox{.7}{\begin{tikzpicture}[node distance=2cm,
cnode/.style={circle,draw,thick, minimum size=1.0cm},snode/.style={rectangle,draw,thick,minimum size=1cm}, pnode/.style={red,rectangle,draw,thick, minimum size=1.0cm},]\node[cnode] (25) at (16,0){1};
\node[snode] (26) [below=1cm of 25]{$l-1$};
\node[cnode] (27) [left=1.5 cm of 25]{1};
\node[snode] (28) [below=1cm of 27]{$n-l+1$};
\draw[-] (25) -- (26);
\draw[-] (25) -- (27);
\draw[-] (27) -- (28);
\node[text width=1cm](30) at (15, -3) {$(Y)$};
\end{tikzpicture}}\\
 \scalebox{.7}{\begin{tikzpicture}
\draw[->] (15,-3) -- (15,-5);
\node[text width=0.1cm](20) at (14.5, -4) {$\CO_1$};
\end{tikzpicture}}
&\qquad
& \scalebox{.7}{\begin{tikzpicture}
\draw[->] (15,-3) -- (15,-5);
\node[text width=0.1cm](29) at (15.5, -4) {$\wt{\CO}_1$};
\end{tikzpicture}}\\
\scalebox{.7}{\begin{tikzpicture}[node distance=2cm,cnode/.style={circle,draw,thick,minimum size=1.0cm},snode/.style={rectangle,draw,thick,minimum size=1.0cm},pnode/.style={rectangle,red,draw,thick,minimum size=1.0cm}, nnode/.style={circle, red, draw,thick,minimum size=1.0cm}]
\node[cnode] (1) at (2,0) {$1$};
\node[cnode] (2) at (3,1) {$1$};
\node[cnode] (3) at (5,1) {$1$};
\node[cnode] (4) at (6,0) {$1$};
\node[cnode] (5) at (5,-1) {$1$};
\node[cnode] (6) at (3,-1) {$1$};
\node[pnode] (7) at (8,0) {$1$};
\node[snode] (9) at (2,-2) {$1$};
\draw[thick] (1) to [bend left=40] (2);
\draw[dashed,thick] (2) to [bend left=40] (3);
\draw[thick] (3) to [bend left=40] (4);
\draw[thick] (4) to [bend left=40] (5);
\draw[dashed,thick] (5) to [bend left=40] (6);
\draw[thick] (6) to [bend left=40] (1);
\draw[thick] (1) -- (9);
\draw[thick] (4) -- (7);
\node[text width=1.5cm](10) at (1,0.75) {$n-l+1$};
\node[text width=1.5cm](11) at (3,2) {$n-l+2$};
\node[text width=1cm](12) at (5,1.75) {$n-1$};
\node[text width=1cm](13) at (7,0.5) {$n$};
\node[text width=1cm](14) at (6,-1.5) {$1$};
\node[text width=1cm](15) at (3.25,-1.75) {$n-l$};
\node[text width=2cm](20) at (6, -3) {$(X^{(1)})$};
\end{tikzpicture}}
&\qquad 
& \scalebox{.8}{\begin{tikzpicture}[node distance=2cm,cnode/.style={circle,draw,thick,minimum size=1.0cm},snode/.style={rectangle,draw,thick,minimum size=1.0cm},pnode/.style={rectangle,red,draw,thick,minimum size=1.0cm}, nnode/.style={circle, red, draw,thick,minimum size=1.0cm}]
\node[] (24) at (10,0){};
\node[cnode] (25) at (16,0){1};
\node[snode] (26) [below=1cm of 25]{$l-1$};
\node[cnode] (27) [left=1.5 cm of 25]{1};
\node[snode] (28) [below=1cm of 27]{$n-l+1$};
\draw[-] (25) -- (26);
\draw[thick] (25) to [bend left=40] (27);
\draw[thick] (25) to [bend right=40] (27);
\draw[-] (27) -- (28);
\node[text width=6cm](29) at (17, -3) {$(Y^{(1)})$};
\end{tikzpicture}}\\
\scalebox{.7}{\begin{tikzpicture}
\draw[->] (15,-3) -- (15,-5);
\node[text width=0.1cm](20) at (14.5, -4) {$\CO_2$};
\end{tikzpicture}}
&\qquad
& \scalebox{.7}{\begin{tikzpicture}
\draw[->] (15,-3) -- (15,-5);
\node[text width=0.1cm](29) at (15.5, -4) {$\wt{\CO}_2$};
\end{tikzpicture}}\\
\scalebox{.7}{\begin{tikzpicture}[node distance=2cm,cnode/.style={circle,draw,thick,minimum size=1.0cm},snode/.style={rectangle,draw,thick,minimum size=8mm},pnode/.style={rectangle,red,draw,thick,minimum size=1.0cm}, nnode/.style={circle, red, draw,thick,minimum size=1.0cm}]
\node[cnode] (1) at (-4,0) {$1$};
\node[cnode] (2) at (-3,1) {$1$};
\node[cnode] (3) at (-1,1) {$1$};
\node[cnode] (4) at (0,0) {$1$};
\node[cnode] (5) at (-1,-1) {$1$};
\node[cnode] (6) at (-3,-1) {$1$};
\node[cnode] (7) at (2,0) {$1$};
\node[pnode] (8) at (2,-2) {$1$};
\node[snode] (9) at (-4,-2) {$1$};
\node[text width=1cm](16) at (-1,-3) {$(X^{(2)})$};
\draw[thick] (1) to [bend left=40] (2);
\draw[dashed,thick] (2) to [bend left=40] (3);
\draw[thick] (3) to [bend left=40] (4);
\draw[thick] (4) to [bend left=40] (5);
\draw[dashed,thick] (5) to [bend left=40] (6);
\draw[thick] (6) to [bend left=40] (1);
\draw[thick] (1) -- (9);
\draw[thick] (7) -- (8);
\draw[thick] (4) -- (7);
\node[text width=1.5cm](10) at (-5,0.75) {$n-l+1$};
\node[text width=1.5cm](11) at (-3,1.75) {$n-l+2$};
\node[text width=1cm](12) at (-1,1.75) {$n-1$};
\node[text width=1cm](13) at (1,0.5) {$n$};
\node[text width=1cm](14) at (0,-1.75) {$1$};
\node[text width=1cm](15) at (-2.75,-1.75) {$n-l$};
\end{tikzpicture}}
&\qquad
&\scalebox{.7}{\begin{tikzpicture}[node distance=2cm,cnode/.style={circle,draw,thick,minimum size=1.0cm},snode/.style={rectangle,draw,thick,minimum size=8mm},pnode/.style={rectangle,red,draw,thick,minimum size=1.0cm}, nnode/.style={circle, red, draw,thick,minimum size=1.0cm}]\node[cnode] (21) at (7,0) {$1$};
\node[cnode] (22) at (11,0) {$1$};
\node[snode] (23) at (7,-2) {$n-l+1$};
\node[snode] (24) at (11,-2) {$l-1$};
\draw[thick] (21) -- (23);
\draw[thick] (22) -- (24);
\draw[thick] (21) -- (22);
\draw[thick] (21) to [bend left=40] (22);
\draw[thick] (21) to [bend right=40] (22);
\node[text width=0.1cm](16) at (9,-3) {$(Y^{(2)})$};
\end{tikzpicture}}
\end{tabular}
\caption{The quiver $(X')$ and its mirror dual $(Y')$ in \figref{AbEx1gen} (for $p=3$) is generated by a sequence of 
elementary $S$-type operations on different nodes. In each step, the flavor node(s) on which the $S$-type operation 
acts is shown in red. The intermediate step involves a flavored $\widehat{A}_{n-1}$ quiver and its dual.}
\label{SimpAbEx1GFIApp}
\end{center}
\end{figure}

In this section, we will demonstrate concrete examples of how an elementary $S$-type operation and its dual may be 
implemented in terms of the superconformal index, to construct new pairs of dual theories. We first work out the 
example of a flavored $\widehat{A}_{n-1}$ quiver (discussed in \Secref{GFI-exAbPF} in terms of the $S^3$ partition function), 
followed by the Family ${\rm I}_{[n,l,p]}$ (discussed in \Secref{FamilyI}). The quiver operation is shown in \figref{SimpAbEx1GFIApp}. 
The indices of the dual theories can be simply read off from the general expressions \eref{GFIAbgen-SCI} (identification-flavoring-gauging) and 
\eref{GFAbgen-SCI} (flavoring-gauging) respectively, but we work out the first example in details to familiarize the reader with the computation.\\

Given the dual linear quiver pair $(X,Y)$, their superconformal indices can be read off from the general expressions \eref{SCILinA} 
and \eref{SCILinB} respectively. The Higgs branch global symmetry for $X$ is $G^{X}_{H}= U(1)^3/U(1)$, where we choose 
to impose the $U(1)$ quotient such that $G^{X}_{H}=U(1)_1\times U(1)_{n-1}$. The fugacities and GNO fluxes 
$\{h^\beta, \kappa^\beta\}$ for $G^{X}_{H}$ are then defined as:
\begin{align}
&\tz_1=h^1,\, \tz_2=1, \, \tz_3=h^2,\\
& \ta_1=\kappa^1,\,  \ta_2=0,\, \ta_3= \kappa^2,
\end{align}
where the fugacities $\vec{\tz}$ and the fluxes $\vec{\ta}$ for a linear quiver are defined in \eref{fugconvtz} and in \eref{fugconvta} respectively.
Using the above choice, the FI term in the SCI of the quiver $Y$, following \eref{SCILinBFI}, is given as
\begin{align}
\CI^{(Y)}_{\rm FI}(\vec \tz, \vec \ta, \{\vec{\xi}^{(\gamma')} \},\{ \vec{c}^{(\gamma')} \}) =& \prod^2_{i=1}(\frac{\tz_{i}}{\tz_{i+1}})^{c^{i}} \cdot (\xi^{i})^{(\ta_{i} - \ta_{i+1})}
= (h^1)^{c^1}\, (\xi^1)^{\kappa^1}\,(h^2)^{-c^2}\, (\xi^2)^{-\kappa^2} \label{SCIYFI}.
\end{align}

Now, let us implement an identification-flavoring-gauging operation $\CO_1$ at the $U(1)_1\times U(1)_{n-1}$ flavor nodes of $X$. The  
superconformal index of the resultant theory $\CO_1(X)$ is given by \eref{genSI}, i.e.
\begin{align}
\CI^{\CO_1(X)}
= \sum_{{\kappa}^{(\alpha)}} \, \oint_{|h^{(\alpha)}|=1}\,\frac{dh^{(\alpha)}}{2\pi i h^{(\alpha)}} \, \sI_{\CO_1(X)}(s^X_\alpha, \{s^X_\beta\},\vec{s}^{\CO_1}_F,\vec{s}^{\CO_1}_J; \tq,\tti) \cdot \CI^{(X)}(s^X_\beta,\vec{s}^X_J; \tq,\tti), 
\end{align}
where $\CI^{(X)}$ is given by \eref{SCILinA}. The operator $\sI_{\CO_1(X)}$, corresponding to an Abelian identification-flavoring-gauging operation 
with $N^\alpha_F=1$, can be constructed from \eref{G-basic0SCI}-\eref{I-basic0SCI} and \eref{Sbasic-def-SCI} as follows:
\begin{align}\label{IOX-An}
\sI_{\CO_1(X)}= \CI_{\rm FI}(\wt{w},\wt{n},h^{(\alpha)},\kappa^{(\alpha)})\,\CI_{\rm vector}(&{h}^{(\alpha)}, {\kappa}^{(\alpha)}; \tq,\tti)\,
\CI_{\rm hyper}({h}^{(\alpha)}, {\kappa}^{(\alpha)}, {\wt{h}}^{(\alpha)}, {\wt{\kappa}}^{(\alpha)};\tq, \tti) \nn \\
\times & \prod^2_{\beta=1} \oint_{C_{\vec \lambda}} \frac{dh^{\beta}}{2\pi i\,(h^{\beta} - \lambda_\beta h^{(\alpha)})} \cdot
\sum_{\{{\kappa}^{\beta}\}} \prod^2_{\beta=1} \delta_{{\kappa}^{\beta}, {\kappa}^{(\alpha)} + \delta^\beta}.
\end{align}
We will eventually set $\lambda_1=1,\, \delta_1=0$, which can be chosen by an appropriate reparametrization of $h^{(\alpha)}$.
The FI term and the vector multiplet contribution for a $U(1)$ gauge group can be read off from \eref{SCIgenFI}-\eref{SCIgenVec}:
\begin{align}
& \CI_{\rm FI}(\wt{w},\wt{n},h^{(\alpha)},\kappa^{(\alpha)})=\wt{w}^{\kappa^{(\alpha)}}\,(h^{(\alpha)})^{\wt{n}}, \label{AbEx1SCI-FI}\\
& \CI_{\rm vector}(\vec{h}^{(\alpha)}, \vec{\kappa}^{(\alpha)}; \tq,\tti)=\CI_{\rm vector}(\tq,\tti)=\Big( \frac{(\tti \tq^{1/2}; \tq)}{(\tti^{-1} \tq^{1/2}; \tq)} \Big).
\label{AbEx1SCI-vector}
\end{align}
The contribution of the single free hypermultiplet, given by \eref{SCIgenMatt}, obeys the identity:
\begin{align}
& \CI_{\rm hyper}({h}^{(\alpha)}, {\kappa}^{(\alpha)}, {\wt{h}}^{(\alpha)}, {\wt{\kappa}}^{(\alpha)};\tq, \tti)=(\frac{\tq^{1/2}}{\tti})^{|(\kappa^{(\alpha)} - {\wt{\kappa}}^{(\alpha)})|/2}\,\frac{(\tti^{-1/2} \tq^{3/4+|(\kappa^{(\alpha)} - {\wt{\kappa}}^{(\alpha)})|/2}\, (\frac{h^{(\alpha)}}{\wt{h}^{(\alpha)}})^{\pm 1}; \tq)}{(\tti^{1/2} \tq^{1/4+|(\kappa^{(\alpha)} - {\wt{\kappa}}^{(\alpha)})|/2} (\frac{h^{(\alpha)}}{\wt{h}^{(\alpha)}})^{\pm 1};\tq)} \label{AbEx1SCI-Hyper1}\\
& = \sum_{\kappa} \oint_{|h|=1}\,\frac{dh}{2\pi i h}\, (\frac{h^{(\alpha)}}{\wt{h}^{(\alpha)}})^{\kappa}\, h^{(\kappa^{(\alpha)} - {\wt{\kappa}}^{(\alpha)})}\,
 \CI_{\rm vector}({h},\kappa; \tq,\tti^{-1})\, \CI_{\rm hyper}({h}, {\kappa}, 1, 0;\tq, \tti^{-1})\\
&= \sum_{\kappa} \oint_{|h|=1}\,\frac{dh}{2\pi i h}\, (\frac{h^{(\alpha)}}{\wt{h}^{(\alpha)}})^{\kappa}\, h^{(\kappa^{(\alpha)} - {\wt{\kappa}}^{(\alpha)})} \,
 \CI_{\rm vector}(\tq,\tti^{-1})\, \CI_{\rm hyper}({h}, {\kappa},1,0;\tq, \tti^{-1}), \label{AbEx1SCI-Hyper2}
\end{align}
where for the second equality we have used the 3d mirror symmetry relation \eref{SCIDual1b} between a single twisted hypermultiplet and a 
$U(1)$ gauge theory with a single hypermultiplet.\\

Given the operator $\sI_{\CO_1(X)}$ in \eref{IOX-An}, the SCI of the dual theory $\wt{\CO}_1(Y)$ can be computed using \eref{genSCId1}-\eref{genSCId2}.
Let us first compute the function $\sI_{\wt{\CO}_1(Y)}(\vec{\xi}, \vec{c}, \vec{s}^{\CO_1}_F, \vec{s}^{\CO_1}_J; \tq,\tti)$ using \eref{genSCId2}, which gives:
\begin{align}
&\sI_{\wt{\CO}_1(Y)}
=\sum_{\vec{\kappa}^{(\alpha)}} \, \oint_{|h^{(\alpha)}|=1}\,
\frac{d h^{(\alpha)}}{2\pi i h^{(\alpha)}} \,\sI_{\CO_1(X)}(\vec{s}_\alpha,\{\vec{s}_\beta\},\vec{s}^{\CO_1}_F,\vec{s}^{\CO_1}_J; \tq,\tti)\,
{\CI^{(0)}}^{(Y)}_{\rm FI}(\vec{\xi}, \vec{c},\{\vec{s}_\beta\}), \\
& {\CI^{(0)}}^{(Y)}_{\rm FI}(\vec{\xi}, \vec{c},\{\vec{s}_\beta\})= (h^1)^{c^1}\, (\xi^1)^{\kappa^1}\,(h^2)^{-c^2}\, (\xi^2)^{-\kappa^2}, 
\end{align}
where we have read off $\{h^\beta, \kappa^\beta\}$-dependent ${\CI^{(0)}}^{(Y)}_{\rm FI}$ from \eref{SCIYFI}. Using \eref{AbEx1SCI-FI}, 
\eref{AbEx1SCI-vector}, and \eref{AbEx1SCI-Hyper2}, and interchanging the order of integration and sum over fluxes, we get 
\begin{align}
 \sI_{\wt{\CO}_1(Y)}= &\sum_{\kappa} \oint_{|h|=1}\,\frac{dh}{2\pi i h}\,\Big(\sum_{\vec{\kappa}^{(\alpha)}}(\frac{\wt{w} h \xi^1}{\xi^2})^{{\kappa}^{(\alpha)}} \Big)
\oint_{|h^{(\alpha)}|=1}\,\frac{d h^{(\alpha)}}{2\pi i h^{(\alpha)}} \, (h^{(\alpha)})^{c^1-c^2+\kappa+\wt{n}} \nn \\
&\times  (h^{-\wt{k}^\alpha}\,(\wt{h}^{(\alpha)})^{-\kappa}\,\lambda^{c^1}_1\,\lambda^{-c^2}_2\,(\xi^1)^{\delta_1}\, (\xi^2)^{-\delta_2})\,\CI_{\rm hyper}({h}, {\kappa},1,0;\tq, \tti^{-1}).
\end{align}
Using the identities,
\begin{align}
& \oint_{|h^{(\alpha)}|=1}\,\frac{d h^{(\alpha)}}{2\pi i h^{(\alpha)}} \, (h^{(\alpha)})^{c^1-c^2+\kappa+\wt{n}} = \delta_{c^1-c^2+\kappa+\wt{n},0}, \\
& \oint_{|h|=1}\,\frac{dh}{2\pi i h}\,\Big(\sum_{\vec{\kappa}^{(\alpha)}}(\frac{\wt{w} h \xi^1}{\xi^2})^{{\kappa}^{(\alpha)}} \Big)\, F(h,\kappa)= F(\frac{\xi^2}{\xi^1 \wt{w}},\kappa),
\end{align}
we have the following expression
\begin{align}
\sI_{\wt{\CO}_1(Y)}=& (\wt{h}^{(\alpha)})^{c^1-c^2+ \wt{n}}(\lambda^{c^1}_1\,\lambda^{-c^2}_2\,(\xi^1)^{\delta_1}\, (\xi^2)^{-\delta_2})\, (\frac{\xi^2}{\xi^1 \wt{w}})^{-\wt{k}^\alpha}\,\CI_{\rm hyper}(\frac{\xi^2}{\xi^1 \wt{w}}, -c^1 + c^2 -\wt{n},1,0;\tq, \tti^{-1})\\
=& (\wt{h}^{(\alpha)})^{c^1+\wt{n}}\,(\wt{h}^{(\alpha)}\,\lambda)^{-c^2}\, (\xi^1 \wt{w})^{\wt{k}^\alpha}\,(\xi^2)^{-\wt{k}^\alpha -\delta}\,\CI^{\rm bif}_{\rm hyper}(\xi^2, c^2, \xi^1\wt{w}, c^1+\wt{n}; \tq, \tti^{-1}),
\end{align}
where for the second equality, we have set $\lambda_1=1, \delta_1=0$, as mentioned earlier, and $\lambda_2=\lambda, \delta_1=\delta$.
We also identify the hypermultiplet term as the 1-loop contribution of 
a bifundamental hypermultiplet, with fugacities and fluxes as shown in the argument of $\CI^{\rm bif}_{\rm hyper}$. The index of the dual theory is then given 
by:
\begin{align} \label{AbEx1SCI-dual}
\CI^{\wt{\CO}_1(Y)}
= \sum_{c^1,c^2}\, \oint_{|\xi^{i}|=1}\, \prod^{2}_{i=1}\,\Big[\frac{d\xi^{i}}{2\pi i \xi^{i}} \Big]\,&
\CI^{\rm bif}_{\rm hyper}(\xi^2, c^2, \xi^1, c^1; \tq, \tti^{-1}) \,\cdot \CI^{(Y)}_{\rm FI}(\vec \xi, \vec c,\vec \tz(\wt{h}^{(\alpha)}, \lambda), \vec \ta(\wt{k}^\alpha,\delta)) \nn\\
\times & \Big[\CI^{(Y)}_{\rm 1-loop}(\vec \xi, \vec c, \vec s^X_J; \tq,\tti^{-1})\Big]_{\xi^1\to \xi^1/\wt{w}, c^1\to c^1-\wt{n}},
\end{align}
where we have performed the change of variables $\xi^1\to \xi^1/\wt{w}, c^1\to c^1-\wt{n}$. The fugacities and fluxes 
$(\vec \tz, \vec \ta)$ are explicitly given as 
\begin{align}
& \tz_1=\wt{h}^{(\alpha)}, \tz_2=1, \tz_3=\wt{h}^{(\alpha)}\,\lambda, \\
& \ta_1=\wt{k}^\alpha,\,  \ta_2=0,\, \ta_3=\wt{k}^\alpha + \delta.
\end{align}
The dual theory $\wt{\CO}_1 (Y)$ can now be read off from the RHS of the expression \eref{AbEx1SCI-dual} for the index, and manifestly reproduce 
the quiver $Y^{(1)}$ above. Note that, we could have directly arrived at the result \eref{AbEx1SCI-dual} from the general expression for an 
identification-flavoring-gauging operation in \eref{GFIAbgen-SCI}, with the input \eref{SCIYFI}.\\

As discussed in \Secref{SCI}, the Higgs branch and the Coulomb branch global symmetries for $(X^{(1)},Y^{(1)})$ can be read off from the respective limits of the index. 
In particular, we would like to point out that the Coulomb branch symmetry for $X^{(1)}$ is $G^{X^{(1)}}_{\rm C} = G^{Y^{(1)}}_{\rm H}=U(1) \times SU(n-l+1) \times SU(l-1) \times SU(2)$. The $SU(n-l+1) \times SU(l-1)$ subgroup of $G^{X^{(1)}}_{\rm C}$ is manifest from the balanced linear subquivers inside of $X^{(1)}$, but the remaining 
$U(1)_{n-l+1} \times U(1)_n$ gets enhanced to $U(1) \times SU(2)$. This can be directly seen from the Coulomb branch index/ Hilbert Series of $X^{(1)}$. Let us compute 
the refined index for the theory labelled by $n=6$, $l=4$. The global symmetry in this case is $G^{X^{(1)}}_{\rm C} = U(1) \times SU(3)_1 \times SU(3)_2 \times SU(2)$. 
With $\tx=t^2$, the character expansion of the Coulomb branch Hilbert Series is given as:
\begin{align}\label{SCI-AbEx1X'}
\CI^{(X^{(1)})}_{\rm C}= 1 + ([1,1]_1 + [1,1]_2 + [2] +1) t^2 + ([1,0]_1\, [0,1]_2\,[1] + [0,1]_1\, [1,0]_2\,[1]) t^3 + \ldots ,
\end{align}
where $[m,n]_i$ denotes the character of the representation with Dynkin labels $[m,n]$ for $SU(3)_i$, and $[l]$ denotes the spin-$l/2$ 
representation of $SU(2)$. \\

Now let us implement the flavoring-gauging operation $\CO_2$ on the quiver gauge theory $X^{(1)}$ at the flavor node shown in red in \figref{SimpAbEx1GFIApp}. 
The fugacity and flux associated with this node is $(\wt{h}^{(\alpha)}, \wt{k}^{(\alpha)})$. Following the notation of \Appref{GFI-SCI-Gen}, we will label this node as $(\alpha)$,
and set 
\be
{h}^{(\alpha)}= \wt{h}^{(\alpha)}, \quad {k}^{(\alpha)} = \wt{k}^{(\alpha)}.
\ee
From \eref{AbEx1SCI-dual}, the FI-term contribution to the index of $Y^{(1)}$ is given as
\begin{align}
\CI^{(Y^{(1)})}_{\rm FI}(\vec \xi, \vec c,\vec \tz(\wt{h}^{(\alpha)}, \lambda), \vec \ta(\wt{k}^\alpha,\delta))=& (\wt{h}^{(\alpha)})^{c^1}\,(\wt{h}^{(\alpha)}\,\lambda)^{-c^2}\, (\xi^1)^{\wt{k}^{(\alpha)}}\,(\xi^2)^{-\wt{k}^\alpha -\delta} \nn \\
= & ({h}^{(\alpha)})^{c^1-c^2}\, (\frac{\xi_1}{\xi_2})^{{k}^{(\alpha)}}\,\lambda^{-c^2}\,(\xi^2)^{-\delta},
\end{align}
which implies that the $({h}^{(\alpha)}, {\kappa}^{(\alpha)})$-dependent part of the FI term for the theory $Y^{(1)}$ is
\be
{\CI^{(0)}}^{(Y^{(1)})}_{\rm FI}= ({h}^{(\alpha)})^{c^1-c^2}\, (\frac{\xi_1}{\xi_2})^{{k}^{(\alpha)}}.
\ee
In addition, we parametrize the $U(1)$ flavor fugacity/flux and the $U(1)_J$ topological fugacity/flux associated with the $S$-type operation $\CO_2$ 
as follows:
\be
s^{\CO_2}_F=(\wt{h}'^{(\alpha)}, \wt{k}'^{(\alpha)}), \quad s^{\CO_2}_J=(\wt{w}',\wt{n}').
\ee 
The index for the dual theory $\wt{\CO}_2(Y^{(1)})$ can then be read off from the general expression \eref{GFAbgen-SCI} as follows:
\begin{align} \label{AbEx2SCI-dual}
\CI^{\wt{\CO}_2(Y^{(1)})}
= \sum_{c^1,c^2}\, & \oint_{|\xi^{i}|=1}\, \prod^{2}_{i=1}\, \Big[\frac{d\xi^{i}}{2\pi i \xi^{i}} \Big]\,\CI^{\rm bif}_{\rm hyper}(\xi^2, c^2, \xi^1, c^1 ; \tq, \tti^{-1})\,
\CI^{\rm bif}_{\rm hyper}(\xi^2, c^2, \frac{\xi^1}{\wt{w}'}, c^1-\wt{n}'; \tq, \tti^{-1}) \nn\\
\times & \, \CI^{(Y)}_{\rm FI}(\vec \xi, \vec c,\vec \tz(\wt{h}'^{(\alpha)}, \lambda), \vec \ta(\wt{k}'^\alpha,\delta))\cdot \Big[\CI^{(Y)}_{\rm 1-loop}(\vec \xi, \vec c, \vec s^X_J; \tq,\tti^{-1})\Big]_{\xi^1\to \xi^1/(\wt{w}\,\wt{w}'), c^1\to c^1-\wt{n}-\wt{n}'}. 
\end{align}
The Lagrangian for the dual theory $\wt{\CO}_2(Y^{(1)})$ can be read from the above index -- it involves adding two hypermultiplets in the bifundamental representation 
of the $U(1) \times U(1)$ gauge group to the quiver $Y$, and is therefore given by the quiver gauge theory $Y^{(2)}$ in \figref{SimpAbEx1GFIApp}. The fugacities and fluxes 
$(\vec \tz, \vec \ta)$ are explicitly given as 
\begin{align}
& \tz_1=\wt{h}'^{(\alpha)}, \tz_2=1, \tz_3=\wt{h}'^{(\alpha)}\,\lambda, \\
& \ta_1=\wt{k}'^\alpha,\,  \ta_2=0,\, \ta_3=\wt{k}'^\alpha + \delta.
\end{align}
Implementing a sequence of flavoring-gauging operations at the new flavor node generates the infinite family of dual theories in ${\rm I}_{[n,l,p]}$, as 
discussed in \Secref{FamilyI} in terms of the sphere partition function.

\bibliography{cpn1-1}
\bibliographystyle{JHEP}

\end{document}